\documentclass[aps,prd,twocolumn,showpacs,tightenlines,floatfix,nofootinbib,superscriptaddress]{revtex4-1}
\usepackage{natbib}
\usepackage{graphicx,subfigure}
\usepackage{caption}
\usepackage{tabularx,dcolumn,booktabs}
\usepackage{multirow}
\usepackage{float}
\usepackage{mathtools}
\usepackage{multirow}
\usepackage{bm}
\usepackage{hyperref}
\usepackage{setspace}
\usepackage{caption} 
\newcommand{\Chi}{{\large{\chi}}}
\begin{document}

\title{NNLO nuclear parton distribution functions with electroweak-boson production data from the LHC}

\author{Ilkka Helenius}
\email{ilkka.m.helenius@jyu.fi}
\affiliation{University of Jyvaskyla, Department of Physics, P.O. Box 35, FI-40014 University of Jyvaskyla, Finland}%
\affiliation{Helsinki Institute of Physics, P.O. Box 64, FI-00014 University of Helsinki, Finland}%

\author{Marina Walt}
\affiliation{Currently: HQS Quantum Simulations GmbH, Haid-und-Neu-Straße 7, 76131 Karlsruhe, Germany.}
\affiliation{Previously: Institute for Theoretical Physics, University of T\"ubingen, Auf der Morgenstelle 14, 72076 T\"ubingen, Germany}%

\author{Werner Vogelsang}
\email{werner.vogelsang@uni-tuebingen.de}
\affiliation{Institute for Theoretical Physics, University of T\"ubingen, Auf der Morgenstelle 14, 72076 T\"ubingen, Germany}%

\date{\today}

\begin{abstract}
We present new sets of nuclear parton distribution functions (nPDFs) at next-to-leading order and next-to-next-to-leading order in perturbative QCD. Our analyses are based on deeply inelastic scattering data with charged-lepton and neutrino beams on nuclear targets, and experimental data from measurements of $W^{\pm},\,Z$ boson production in p+Pb collisions at the LHC. In addition, a set of proton baseline PDFs is fitted within the same framework and with the same theoretical assumptions. The results of our global QCD analysis are compared to existing nPDF sets and to the previous nPDF set TUJU19 which was based on DIS data only. Our work is performed using an open-source tool, \textsc{xFitter}, and the required extensions of the code are discussed as well. We find good agreement with the data included in the fit and a lower value for $\Chi^2/N_{\mathrm{dp}}$ when performing the fit at next-to-next-to-leading order. We apply the resulting nuclear PDFs to electro-weak boson production in Pb+Pb collisions at the LHC and compare the results to the most recent data from ATLAS and CMS.
\end{abstract}

\maketitle
{
\tableofcontents
}
\section{Introduction}

In the framework of collinear factorization \cite{Collins:1989gx}, the differential cross section
for a given process in hadronic collisions can be calculated in terms of
convolutions of process-independent parton distribution functions (PDFs) and process-specific perturbatively calculable coefficient functions. By combining data for inclusive deep-inelastic scattering (DIS) from HERA, fixed-target experiments, or neutrino scattering, with various LHC data, including e.g. (di-)jet, top-pair and electroweak-boson production, proton PDFs have by now reached accuracies at the level of a few percent or better over wide kinematic regions. This is vital for precision studies of processes in the Standard-Model and beyond (see \cite{Gao:2017yyd} and references therein). 

One may also apply the same factorization to processes involving high-energy nuclei, such as e+A or p+A. This provides information on nuclear parton distribution functions (nPDFs). Being fundamental properties of atomic nuclei, nPDFs are of much importance for our general understanding of the strong interactions. At  the same time, extracting ``cold-nuclear matter'' effects from collisions with a small projectile and a target nucleus also provides a baseline for quark-gluon plasma studies. 

The backbone for analyses of nPDFs have been fixed-target DIS e+A data, often accompanied by Drell-Yan (DY) dilepton production data. In addition to DIS with charged lepton beams, also a good amount of neutrino DIS data are nowadays available that provide some additional sensitivity to the flavor dependence of nuclear effects. In recent years, data from p+Pb collisions from the LHC have fulfilled the promise of extending the kinematic reach and further constraining the flavor dependence of the nPDFs. Prime examples are the $\mathrm{D}$-meson production at forward rapidities measured by LHCb \cite{LHCb:2017yua} which is particularly sensitive to small-$x$ gluon nPDFs \cite{Eskola:2019bgf}, and $W^{\pm}$ and $Z$ boson production data from ATLAS \cite{Aad:2015gta} and CMS \cite{Khachatryan:2015hha, Khachatryan:2015pzs}, which are primarily sensitive to different quark-flavor combinations at moderate values of $x$ and provide potential constraints for flavor dependence \cite{Paukkunen:2010qg} and also for strange quarks and gluons \cite{Kusina:2020lyz}. Furthermore, dijet production \cite{CMS:2018jpl} can offer further constraints on gluon nPDFs in the intermediate $x$-region \cite{Eskola:2019dui}. Other possible observables include inclusive direct-photon and pion production in p+Pb collisions at the LHC. The former has been recently measured by ATLAS \cite{ATLAS:2019ery} and for the latter there are already several data sets available from ALICE \cite{ALICE:2018vhm, ALICE:2016dei, ALICE:2021est}.

Numerous nPDF analyses are available by now which can be classified in terms of the data used in the analysis and of their perturbative precision. Early ``global'' analyses include the EPS09~\cite{Eskola:2009uj} and DSSZ~\cite{deFlorian:2011fp} sets, which were based on charged-lepton and neutrino DIS and DY data from fixed-target experiments, and inclusive pion production data from d+Au collisions at RHIC. The EPPS16 analysis \cite{Eskola:2016oht} was the first to include also data from p+Pb collisions at the LHC by analyzing Run~I data for dijets and $W^{\pm}$ and $Z$ boson production. The original nCTEQ15 analysis \cite{Kovarik:2015cma} did not include any LHC data, but has recently been extended to include e.g. Run~II $W^{\pm}$ and $Z$ data from p+Pb collisions (nCTEQ15wz analysis \cite{Kusina:2020lyz}). Similar data sets have been considered also by the NNPDF collaboration in their most recent analysis nNNPDF2.0 \cite{AbdulKhalek:2020yuc}. The nCTEQ15 analysis has been recently updated to contain also single-inclusive hadron production \cite{Duwentaster:2021ioo} and also other active groups have recently prepared new analyses. The EPPS21 \cite{Eskola:2021nhw} and nNNPDF3.0 \cite{Khalek:2022zqe} analyses both include a significant amount of new LHC data including e.g. dijet, $W^{\pm}$ and $Z$ boson and inclusive D-meson production from the LHCb. All analyses mentioned so far have been performed at the next-to-leading-order (NLO) of perturbative QCD. There are a few recent analyses that have been performed at next-to-next-to-leading order (NNLO): nNNPDF1.0 \cite{AbdulKhalek:2019mzd}, TUJU19 \cite{Walt:2019slu} and KSASG20 \cite{Khanpour:2020zyu}. So far, these have only considered charged-lepton and neutrino DIS, and fixed-target DY data.

In this work we present new nuclear PDF analyses at NLO and NNLO, using the same framework as in our previous analysis, TUJU19, but now including in addition to neutral-current DIS with a lepton beam and charged-current neutrino DIS data also new electroweak (EW)-boson production data from the LHC. This is the first time where LHC data are employed in a full NNLO analysis for nuclear PDFs. We include the LHC data also for a proton ``baseline'' fit. In this way, we obtain a fully consistent way of computing cross sections for p+$A$ collisions at NLO or NNLO. The resulting nPDFs may also be readily used for further cross section calculations at NNLO in nuclear collisions. As an application we study EW-boson production in Pb+Pb collisions and compare our results to recent ATLAS and CMS data \cite{ATLAS:2019ibd, ATLAS:2019maq,CMS:2021kvd} which, contrary to expectations, have previously been found to be difficult to describe within a factorization-based NNLO calculation \cite{Eskola:2020lee}. 

Our paper is organized as follows. We describe the theoretical framework for our analysis in Sec. \ref{sec:theory}, then discuss the analysis procedure in Sec. \ref{sec:analysis} and the selection of experimental data in Sec. \ref{sec:expdata}. The results of the analysis are presented in Sec. \ref{sec:results}. We discuss the application to Pb+Pb collisions at the LHC in Sec. \ref{sec:applications}. Finally, we summarize our work in Sec. \ref{sec:summary}, where also an outlook toward future developments is given.

\section{Theoretical framework}
\label{sec:theory}

The theoretical framework is very similar to that adopted in our earlier analysis, TUJU19 \cite{Walt:2019slu}, which includes also a detailed description of neutral-current (NC) and charged-current (CC) DIS processes. Here we give an overview of the calculational framework that we have set up to use the new LHC data.

\subsection{Drell-Yan and $W^{\pm}$, $Z$ boson production processes}
\label{sec:crosssections}

As new constraints we include LHC data for inclusive hadroproduction of electro-weak (EW) bosons, $W^{\pm}$ and $Z/\gamma^{*}$. Often such processes are referred to as Drell-Yan (DY) processes, originally relating to reactions where two hadrons collide and form a highly virtual photon that decays to a lepton pair \cite{Drell:1970wh}. The signature of such an event is a lepton pair ($l\bar{l}$ with $l=e,\,\mu,\,\tau$) of the same flavor with a large invariant mass. In leading order (LO) such a process can take place only by annihilation of a quark-antiquark pair of the same flavor,
\begin{equation}
q\bar{q} \rightarrow \gamma^* \rightarrow l\bar{l}\,.
\end{equation}
 
At high enough energies, similar scattering may form also other EW bosons, namely $Z$ and $W^{\pm}$. As the former has the same quantum numbers as a photon, the production and decay channels are the same as in traditional DY. In the latter case, however, the LO process requires e.g. a $u\bar{d}$ initial state and the leptonic decay channel to include production of a charged lepton and a corresponding neutrino,
\begin{align}
q\bar{q} &\rightarrow Z \rightarrow l\bar{l}\,,\quad \text{and} \notag \\
q_i\bar{q_j} &\rightarrow W^{\pm} \rightarrow l\nu_{l}\,.
\end{align}
As the (anti)neutrinos from the $W^{-(+)}$ decays cannot be directly measured in the detectors and only the momentum of the charged lepton is known, some further kinematical cuts are required to increase the sensitivity to the signal process.

A factorization theorem for DY has been explicitly proved to all orders in perturbation theory \cite{Bodwin:1984hc, Collins:1985ue, Collins:1988ig}. In order to later be able to assess the impact of DY (or EW-boson production) data on our analysis, it is useful to write down the double differential DY cross section at LO \cite{ellis_qcd_1996, sterman_handbook_2001}:
\begin{equation}
\frac{\mathrm{d}^2\sigma}{\mathrm{d}M^2 \mathrm{d}y} = \frac{4\pi\alpha^2 \tau}{3 N_C M^4} \sum_q e_q^2 \left[ q(x_1) \bar{q}(x_2)+\bar{q}(x_1) q(x_2) \right]
\label{eq-dy}
\end{equation}
where $N_C$ is the number of colors, $\alpha$ the fine structure constant, $\tau=M^2/s$ (with $\sqrt{s}$ the center-of-mass energy), and we have introduced the variables
\begin{align}
M^2 &= (p_1+p_2)^2 \equiv \hat{s} = x_1 x_2 s & &(\text{invariant mass})\,, \notag \\
y &= \frac{1}{2}\log\frac{x_1}{x_2} & &(\text{rapidity}).
\label{eq-dy-variables}
\end{align}
Here $p_{1,2}$ are momenta of the incoming partons carrying momentum fraction $x_{1,2}$ of the total hadron momentum. In terms of $M^2$ and $y$ we have (again, to LO):
\begin{align}
&x_1=\frac{M}{\sqrt{s}}\exp(y) & &\mathrm{and} & &x_2=\frac{M}{\sqrt{s}}\exp(-y)\,.
\label{eq-dy-variables2}
\end{align}
These relations can be used to estimate the sensitivity to different regions of momentum fraction when studying EW boson production at specific kinematics. Often in the experimental analysis, however, pseudorapidity $\eta$ is used instead of $y$ as the former does not require information of the mass of the given system. 

In Eq.~(\ref{eq-dy}), $q(x_1)$ and $\bar{q}(x_2)$ are the quark and antiquark PDFs, whose scale dependence we have omitted. Comparing to the LO DIS cross section formula, we notice that in DY we always have a combination of quark and antiquark PDFs. Thus we expect DY processes to provide more information on the sea quark densities.

The well-known factorized all-order expression for the Drell-Yan cross section is based on convolutions of the PDFs with perturbative hard-scattering functions. Beyond LO, there are also contributions from processes other than $q\bar{q}^{(\prime)}$ annihilation. For example, starting from NLO, also contributions from gluon-(anti)quark initial states arise, and at NNLO and beyond gluon-gluon scattering participates. Denoting the partonic hard-scattering function for a given partonic channel $ab\to {\mathrm{EW\,boson}}+X$ by $\omega_{ab}$, we have the perturbative expansion
\begin{align}
\omega_{ab}\,=\,\omega^{(0)}_{ab}+\frac{\alpha_s}{\pi}\,\omega^{(1)}_{ab}
+\left(\frac{\alpha_s}{\pi}\right)^2 \omega^{(2)}_{ab}+{\cal O}(\alpha_s^3)\,,
\end{align}
where $\alpha_s$ is the strong coupling evaluated at some renormalization scale $\mu_R$.
The NLO coefficient functions $\omega^{(1)}_{ab}$ have been known for a long time~\cite{Altarelli:1979ub}. The NNLO corrections $\omega^{(2)}_{ab}$ are also fully available~\cite{Hamberg:1990np,Harlander:2002wh,Anastasiou:2003yy,Anastasiou:2003ds,Catani:2009sm,Gavin:2010az}, and their inclusion is a new feature of our TUJU21 analysis.

\subsection{Input parameterization}

Apart from the perturbative hard-scattering functions, another key ingredient in a global analysis of collinear PDFs is their (Dokshitzer–Gribov–Lipatov–Altarelli–Parisi) DGLAP evolution \cite{Gribov:1972ri, Lipatov:1974qm,  Altarelli:1977zs, Dokshitzer:1977sg}, which describes how the PDFs depend on the factorization scale. The perturbative order of the splitting kernels to be used in the evolution equations needs to be chosen in accordance with the order in the hard-scattering functions. As only the perturbative scale evolution of the PDFs is given by the evolution equations, a non-perturbative input at some initial scale is required to obtain a PDF set. In this analysis the baseline parton distributions of a proton are parametrized similarly as in \cite{Walt:2019slu}:
\begin{equation}
xf^{p}_i\left(x,Q_0^2 \right) = c_0\,x^{c_1} (1-x)^{c_2} \left(1+c_3\,x + c_4\,x^2 \right),
\label{pdf-parameterization}
\end{equation}
where the index $i$ runs over parton flavor $i=g, \,d_{\mathrm{v}}, \,u_{\mathrm{v}}, \,\bar{u}, \,\bar{d}, \,\bar{s}$, with the subscript ``$\mathrm{v}$'' refering to the up or down valence-quark contribution. We have kept the flavour dependence in the valence sector but for the sea quarks we had to assume $\bar{u} = \bar{d} = \bar{s} = s$ as it turned out that within the applied data and the adopted framework it is difficult to have a converged fit without such a condition. We define the input parameterization at the charm mass threshold, $Q_0^2=m_c^2=1.69\,\mathrm{GeV^2}$. We do not include any intrinsic charm content at this scale and apply the FONLL-A(C) general-mass variable-flavor-number-scheme \cite{Cacciari:1998it, Forte:2010ta} to calculate the heavy-quark cross sections at $Q>Q_0$ in our NLO (NNLO) analysis. 

To obtain the PDFs for protons bound in a nucleus, we add dependence on the nuclear mass number $A$ to the parameters $c_i$ by re-parameterizing them as
\begin{equation}
c_k\,\rightarrow c_k(A) = c_{k,0}+c_{k,1}\left( 1 - A^{-c_{k,2}} \right),
\label{coeff-A}
\end{equation}
where $k={0,\dots,4}$. A similar form has been used also in the nCTEQ15 analysis \cite{Kovarik:2015cma, Kusina:2020lyz}. It is worth pointing out that with $A=1$ the $A$-dependent right-hand part of Eq.~(\ref{coeff-A}) vanishes and the free proton PDFs are recovered when indentifying $c_{k,0} = c_{k}$. In order to obtain the full nuclear PDF, we need to add also the contribution from neutrons in the nucleus. The PDFs of neutrons are not fitted separately, but are determined from the proton PDFs based on isospin symmetry, giving $u^{\mathrm{n}/A} = d^{\mathrm{p}/A}$ and $d^{\mathrm{n}/A} = u^{\mathrm{p}/A}$, and likewise for the light antiquarks. Then, an average PDF for a nucleon bound in a nucleus with $Z$ protons and $A-Z$ neutrons is obtained as
\begin{equation}
f_i^{\,N/A} \left( x,Q^{\,2} \right) = \frac{Z\cdot f_i^{\,p/A}+ (A-Z)\cdot f_i^{\,n/A}}{A}\,.
\label{nucleon}
\end{equation}

\section{Analysis procedure}
\label{sec:analysis}

\subsection{Minimization procedure}

At the heart of PDF fitting is the minimization of the difference between the experimental data and the calculated cross sections, providing the optimal PDF set within the adopted framework. In this work, as well as in our previous analysis \cite{Walt:2019slu}, this is done by minimizing $\Chi^2$ defined as
\begin{align}
\Chi^2\left(\textbf{m},\,\textbf{b}\right) =& \sum_i \, \frac{\left[ m_i - \Sigma_{\alpha}\,\gamma^i_{\alpha}\,\mu_i\,b_{\alpha} - \mu_i \right]^2}{\left( \delta_{i,\mathrm{stat}} \sqrt{\mu_i m_i} \right)^2 + \left( \delta_{i,\mathrm{uncorr}} m_i \right)^2 } \notag \\
+& \sum_{\alpha} b^2_{\alpha}.
\label{eq-chi2-scaled}
\end{align}
Here, $\mu_i$ is the value of the measured data point for a given observable, whereas $m_i$ is the actual theoretical value calculated using DGLAP-evolved PDFs with given parameters $\{c_k\}$. The uncertainties are represented by the $\delta_{i,\mathrm{stat}}$ and $\delta_{i,\mathrm{uncorr}}$, which are the relative statistical and uncorrelated systematic uncertainties, respectively, as well as by the $\gamma^{i}_{\alpha} \mu_i$, which are the correlated errors. Furthermore, the $b_{\alpha}$ are the so-called nuisance parameters that are determined during the fitting. To account for the correlated systematic uncertainties for each data set, we allow for shifts of the calculated cross section within the quoted uncertainty, penalizing such shifts by the additional $b^2_{\alpha}$ contributions to $\Chi^2$.

The minimization of $\Chi^2$, Eq.~(\ref{eq-chi2-scaled}), provides a central set of PDFs with the parameter values allowing the best description of the used data. Since the experimental data always contain several uncertainties, as listed above, a separate error analysis needs to be performed to study how these uncertainties propagate into the fitted PDFs. In this QCD analysis the Hessian method \cite{Pumplin:2001ct,Pumplin:2000vx} is used for the analysis of the uncertainties. The Hessian error analysis is performed assuming a quadratic expansion of the function $\Chi^2 = \Chi_0^2 + \Delta \Chi^2$ around its global minimum. Here, $\Chi_0^2$ is the value of the function at the global minimum (with the best-fit parameters $\{k_0\}$) and $\Delta \Chi^2$ is the displacement from the minimum \cite{Pumplin:2001ct,Pumplin:2000vx}. During the performed error analysis $\Delta \Chi^2$ defines the tolerance criterion determining the allowed growth of $\Chi^2$. As argued in our previous work \cite{Walt:2019slu}, for the proton baseline with $13$ free fit parameters we select $\Delta \Chi^2 = 20$ and for the nuclear PDF error analysis we choose $\Delta \Chi^2 = 50$ for our 16 free parameters.

\subsection{The fitting framework}

Our global analyses of the baseline proton and nuclear PDFs are performed with the \textsc{xFitter} \cite{Zenaiev:2016jnq,Bertone:2017tig} tool. The main goal of the \textsc{xFitter} project is to provide an open-source tool to fit proton PDFs with varied theoretical assumptions. In order to perform a nuclear PDF analysis several modifications of the code were performed in the context of the work presented in Ref.~\cite{Walt:2019slu}. 

When going from DIS to collisions between two hadrons where cross section calculations involve convolutions of more than one PDF, the required calculations become computationally expensive, and even more so when increasing the perturbative precision to NNLO. Therefore, to implement data for DY and EW-boson production beyond LO, it is necessary to prepare fast interpolation grids allowing for efficient comparisons with the data when the PDF parameters are iterated. Standard tools to handle such convolutions with interpolation grids, such as \textsc{APPLgrid} \cite{Carli:2010rw}, can be linked with \textsc{xFitter}, facilitating fast cross section calculations using the pre-computed grids. To prepare the actual grids used in the fitting, a code capable of calculating the given production cross section needs to be linked with an interpolation code. Depending on the available data files, further processing in \textsc{xFitter} requires grids in a ROOT format \cite{Brun:1997pa}. It is also possible to use $K$-factors when going from NLO to NNLO cross sections to reduce the computing effort. We have used such $K$-factors in our proton baseline fit in cases where they were available in xFitter for a given data set. The settings and PDFs used to calculate the $K$-factors can be found from the references provided in Table~\ref{tab-expdatady-proton}.

\begin{figure*}[h!]
\centering
\includegraphics[width=0.9\textwidth]{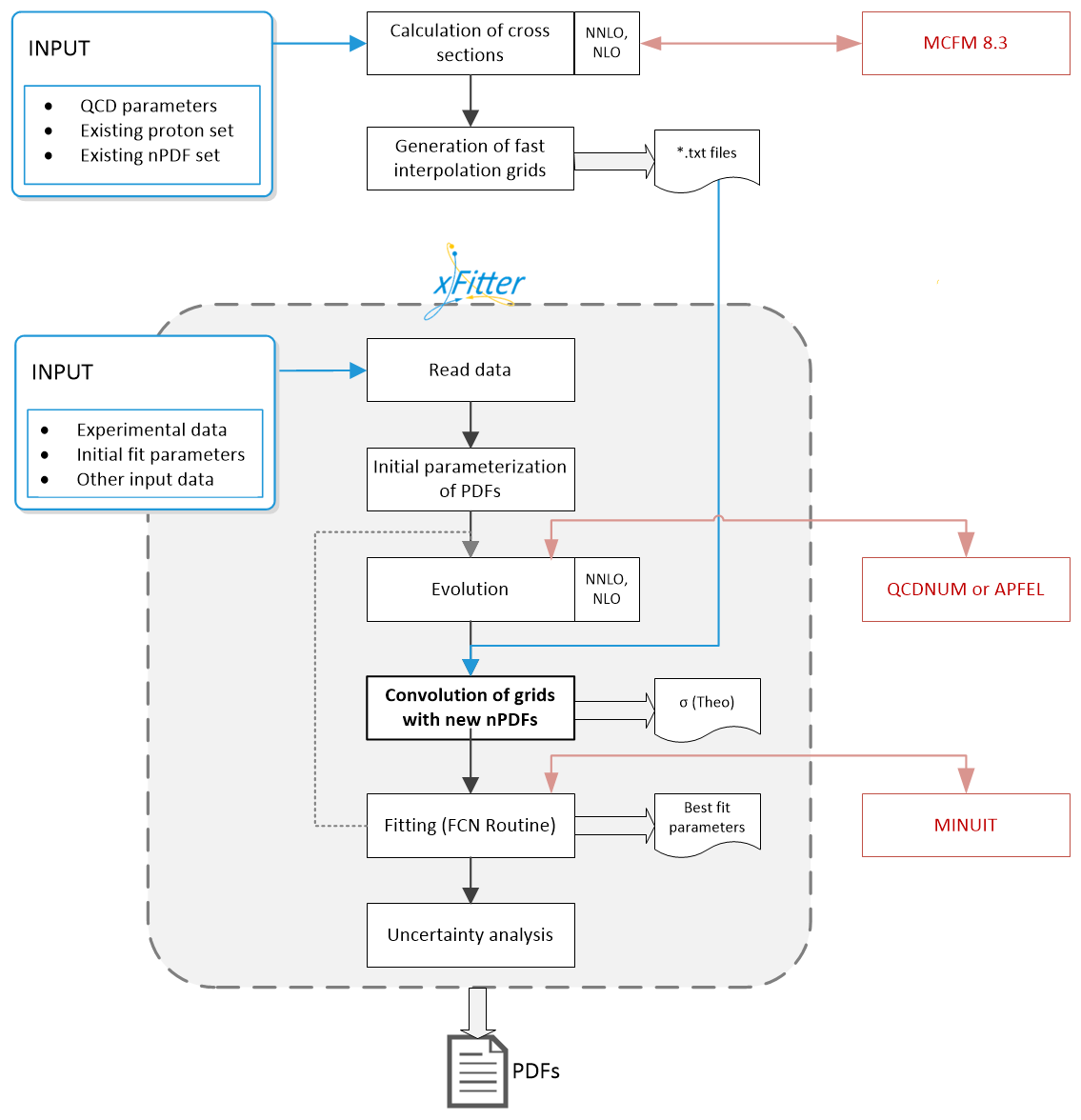}
\caption{Schematic view of the high-level \textsc{xFitter} functionalities as relevant for Drell-Yan and $W^{\pm}$, $Z$ boson production processes, with the newly implemented convolution routine in \textsc{xFitter} as used for the nPDF fit TUJU21. \textsc{xFitter} logo credited from~\cite{xfitter_link}.}
\label{fig-dy-nPDF}
\end{figure*}

The schematic overview of the fitting routine and the required tool set (\textsc{xFitter} and all other modules) as applied for the \textit{nuclear} PDF fit is shown in Fig.~\ref{fig-dy-nPDF}. For this, a new convolution routine had to be implemented in \textsc{xFitter} where one first needs to prepare fast interpolation grids by calculating the observables in the relevant kinematic region, using an existing PDF set, e.g. in the LHAPDF6 format \cite{Buckley:2014ana}. In this work, MCFM 8.3 \cite{Falkowski:2012cu, Campbell:2015qma, Boughezal:2016wmq} has been used to calculate the interpolation grids for EW boson production after suitable modification to handle asymmetric collisions as needed for p+Pb. The obtained grids were then used as the input for the optimization routine. This setup provides the possibility to use fast interpolation grids in a plain text format generated with \textsc{MCFM} up to order NNLO, without relying on approximative $K$-factors. During the fitting procedure the actual theoretical predictions were obtained by convoluting the fitted nPDFs based on updated parameters with the pre-calculated differential cross sections provided in form of grids. For the convolution step one needs to specify the proton PDF baseline that was used for the generation of the fast interpolation grids in MCFM. In our case, we are adopting our own proton PDF baseline prepared in the form of an LHAPDF6 library. 

\subsection{Experimental data}
\label{sec:expdata}

We build upon our previous analysis, TUJU19, including all the same charged-lepton and neutrino DIS data as before. On top of these we include data for DY and EW boson production for both our proton baseline and the nuclear PDFs. 

The DIS data used for the proton baseline fit are summarized in Table~\ref{tab-expdata-proton}, also showing the resulting $\Chi^2$ values at NLO and NNLO obtained in the analysis. The newly included experimental data from DY and $W^{\pm}$ and $Z$-boson production for the proton baseline are listed in Table \ref{tab-expdatady-proton}. For the experimental proton data used, the fast interpolation grids were publicly available and the details of grid generation for each proton PDF data set can be found from the references provided in Table~\ref{tab-expdata-proton}. These data include high- and low-mass DY data from ATLAS at $\sqrt{s} = 7~\text{TeV}$ \cite{Aad:2013iua, Aad:2014qja}, EW boson production data from ATLAS at $\sqrt{s} = 7~\text{TeV}$ \cite{Aad:2012sb}, and increased-luminosity data for these from ATLAS at $\sqrt{s} = 7~\text{TeV}$ \cite{Aaboud:2016btc}. In addition, also $W^{\pm}$ production cross sections from CMS at $\sqrt{s} = 8~\text{TeV}$ have been included. In total the newly included data sets consist of 134 data points.
\begin{center}
\renewcommand{\arraystretch}{1.25}
\begin{table}[tb!]
\caption{Summary of experimental DIS data used to determine our proton PDF baseline. In the last two columns the $\Chi^2$ values at NLO and NNLO obtained in our analysis are provided.}
\label{tab-expdata-proton}
\scriptsize
\begin{tabular}{l|l|c|r|c|r|r}
\toprule
  Exp. & Data set & Year & Ref. & $N_{\mathrm{dp}}$  & $\Chi^2$ NLO  & $\Chi^2$ NNLO  \\ \hline \hline
  BCDMS & F2p 100GeV & 1996 & \cite{Benvenuti:1989rh} & 83 & 96.30  & 93.69 \\
        & F2p 120GeV & & & 90   & 70.54 & 68.70 \\ 
        & F2p 200GeV & & & 79   & 91.81  & 86.32 \\ 
        & F2p 280GeV  & & & 75   & 67.52  & 69.71 \\ \hline
  HERA 1+2 & NCep 920 & 2015 & \cite{Abramowicz:2015mha} & 377 & 459.71 & 482.23 \\\ 
           & NCep 820 & & &	70    & 72.91 & 73.47\\
           & NCep 575 & & &	254   & 222.64 & 231.35 \\
           & NCep 460 & & &	204   & 218.84 & 225.68 \\
           & NCem & & &	159   & 227.80  & 232.79 \\
           & CCep & & &	39   & 46.59  & 43.17 \\
           & CCem & & &	42   & 60.49 & 63.60 \\ \hline
  NMC-97 & NCep & 1997 & \cite{Arneodo:1996qe}  & 100 & 117.72 & 111.31 \\ \hline \hline            
   \multicolumn{4}{r|}{\textbf{In total:}} & 1559 &  &  \\
\toprule  
\end{tabular} 
\end{table}
\end{center}

\begin{center}
\renewcommand{\arraystretch}{1.25}
\begin{table}[tb!]
\caption{Summary of experimental DY and $W^{\pm},\,Z$ boson production data used to determine our proton PDFs. In the last two columns the $\Chi^2$ values at NLO and NNLO obtained in our analysis are provided.}
\label{tab-expdatady-proton}
\scriptsize
\begin{tabular}{l|l|c|r|c|r|r}
\toprule
  Exp. & Data set & Year & Ref. & $N_{\mathrm{dp}}$  & $\Chi^2$ NLO  & $\Chi^2$ NNLO  \\ \hline \hline
  ATLAS DY & high mass DY & 2013 & \cite{Aad:2013iua} & 13 & 10.91  & 11.54 \\
        & low mass DY & 2014 & \cite{Aad:2014qja} &	 8  & 22.86  & 8.68 \\ \hline  
  ATLAS $W^{\pm},\,Z$ &  $W^+$ lepton $\eta$ & 2012 & \cite{Aad:2012sb} & 11 & 15.77 & 14.00 \\ 
            & $W^-$ lepton $\eta$ & & & 11  & 7.98 & 8.54 \\ 
             & $Z$ $y$ & & &	8  & 4.10 & 2.71 \\
        & $W^+$ lepton $y$ & 2016 & \cite{Aaboud:2016btc} & 11  & 19.57 & 10.71 \\  
             & $W^-$ lepton $y$ & & &	 11 & 11.11 & 11.82 \\ 
             & high mass CC $Z$ $y$ & & &	6 & 7.61 & 6.05 \\ 
            & high mass CF $Z$ $y$ & & &	6 & 3.92 & 3.95  \\ 
            &  low mass $Z$ $y$ & & &	6 & 41.32 & 23.84 \\
            & peak CC $Z$ $y$ & & & 12 & 45.76 & 14.40 \\
           & peak CF $Z$ $y$ & & & 9 & 21.85 & 7.55 \\ \hline
  CMS $W^{\pm}$ & $W^+$ $\sigma$ 8 TeV & 2016 & \cite{Khachatryan:2016pev} & 11 & 10.64 & 4.81 \\ 
           &  $W^-$ $\sigma$ 8 TeV & & &	11   & 8.14 & 9.27 \\ \hline \hline
   \multicolumn{4}{r|}{\textbf{In total:}} & 134 &  &    \\
\toprule  
\end{tabular} 
\end{table}
\end{center}

Table \ref{tab-expdata} provides a list of nuclear-DIS data as also used in the TUJU19 analysis, but with the $\Chi^2$ values obtained in this analysis. The input data files and the fast interpolation grids used for the fitting procedure laid out in Fig.~\ref{fig-dy-nPDF} for nPDFs have been collected and prepared as part of this analysis for the newly included data points summarized in Table \ref{tab-expdatady-nPDFs}. The added data include Run~I measurements of $Z$ boson production in p+Pb collisions at the LHC by ATLAS \cite{Aad:2015gta} and CMS \cite{Khachatryan:2015pzs} at $\sqrt{s_{\mathrm{NN}}} = 5.02~\text{TeV}$ and the more recent Run~II measurement of $W^{\pm}$ boson production in p+Pb collisions at $\sqrt{s_{\mathrm{NN}}} = 8.16~\text{TeV}$ by CMS \cite{CMS:2019leu}. In total these add 74 data points to the nPDF analysis. In addition to these, there are more EW-boson data available from the LHC experiments. In particular, there are data from ALICE \cite{ALICE:2016rzo} and LHCb \cite{LHCb:2014jgh} that extend the kinematic reach but have only two data points per set and suffer from large statistical uncertainties. There are also {Run-I} data for $W^{\pm}$ production from CMS \cite{CMS:2015ehw} at the same kinematics than the more recent data but with significantly larger uncertainties. There are also fixed-target DY data available e.g. from E772 \cite{Alde:1990im}, E866 \cite{NuSea:1999egr} and CLAS \cite{CLAS:2019vsb} experiments that have been used previously in similar analyses. As the grid computations at NNLO are computationally very heavy, we included only the datasets that we expect to provide the strongest constraints for the nPDFs in the selected set up. In section \ref{sec:DYpPb} we consider the recent Run-II CMS measurement for $Z/\gamma^*$ production for which it has proven difficult to obtain $\chi^2/N_{\mathrm{dp}}$ values close to unity in a NLO QCD analysis \cite{Eskola:2021nhw, Khalek:2022zqe}. As we have discussed in the context of the TUJU19 nPDF analysis, some authors have found tension between the neutrino- and charged-lepton DIS data. To check for potential tension with the new LHC data we have also performed fits without any neutrino-DIS data and found that the new data was equally well described as when the neutrino data were included. Thus we do not find any tension between these data sets.
\begin{center}
\renewcommand{\arraystretch}{1.25}
\begin{table}[tb!]
\caption{{Summary of experimental DIS data used to determine the nuclear PDFs. In the last two columns the $\Chi^2$ values at NLO and NNLO obtained in our analysis are provided.}}
\label{tab-expdata}
\scriptsize
\begin{tabular}{l|l|c|r|c|r|r}
\toprule
  Nucleus & Exp. & Year & Ref. & $N_{\mathrm{dp}}$  & $\Chi^2$ NLO  & $\Chi^2$ NNLO  \\   
\hline
\hline
  D & NMC 97 & 1996  & \cite{Arneodo:1996qe} &  120 & 151.61  & 121.52  \\ 
       & EMC 90 & 1989  & \cite{Arneodo:1989sy} & 21 & 24.31 & 22.89  \\ \hline
  He/D  & HERMES  & 2002    & \cite{Airapetian:2002fx} & 7 &  6.79 & 8.92 \\ 
      & NMC 95, re.  & 1995   & \cite{Amaudruz:1995tq} &  13 & 10.67  & 10.42  \\   
      & SLAC E139 & 1994  & \cite{Gomez:1993ri} & 11 & 6.47  & 4.42 \\ \hline
   Li/D  & NMC 95 & 1995  & \cite{Arneodo:1995cs} &  12 & 9.10  & 9.00  \\ \hline
   Be/D  &  SLAC E139 & 1994  & \cite{Gomez:1993ri} & 10 & 11.58  & 11.51  \\ \hline    
   Be/C   & NMC 96 & 1996  & \cite{Arneodo:1996rv} & 14 & 13.56  & 16.06  \\ \hline
   C    & EMC 90 & 1989  & \cite{Arneodo:1989sy} & 17 &  13.41 & 13.44  \\
   C/D  & FNAL E665 & 1995   & \cite{Adams:1995is} & 3  & 2.00  & 1.83 \\ 
        & SLAC E139 & 1994   & \cite{Gomez:1993ri} & 6 &  20.69 & 13.86  \\
        & EMC 88 & 1988  & \cite{Ashman:1988bf} & 9 & 3.70  & 4.22  \\  
        & NMC 95, re.  & 1995    & \cite{Amaudruz:1995tq} & 13 &  34.96 & 19.49  \\ \hline
   C/Li  & NMC 95, re. & 1995    & \cite{Amaudruz:1995tq} & 10 & 7.77  & 10.18 \\ \hline
   N/D  & HERMES &  2002    & \cite{Airapetian:2002fx} & 1 &  0.95 & 2.08  \\ \hline
   Al/D     & SLAC E139 & 1994  & \cite{Gomez:1993ri} &  10 & 18.49  &  9.49 \\\hline    
   Al/C    & NMC 96 &  1996   & \cite{Arneodo:1996rv} & 14   & 7.29  & 6.23   \\\hline 
   Ca &    EMC 90  & 1989  & \cite{Arneodo:1989sy} & 19 & 13.41  & 13.44  \\ 
   Ca/D  & NMC 95, re. & 1995    & \cite{Amaudruz:1995tq} & 12 & 34.75  & 21.82 \\ 
     &    FNAL E665 &  1995  & \cite{Adams:1995is} & 3  &  1.84 & 2.41  \\
     &    SLAC E139  &  1994  & \cite{Gomez:1993ri} & 6  & 16.74  & 9.01 \\                   
   Ca/Li  & NMC 95, re. & 1995  & \cite{Amaudruz:1995tq} &  10 & 1.45  &  1.33  \\ \hline
   Ca/C  & NMC 95, re. & 1995  & \cite{Amaudruz:1995tq}  & 10  & 9.35  & 8.00 \\ 
     &  NMC 96 &  1996  & \cite{Arneodo:1996rv} & 14  & 8.45  &  6.42  \\ \hline 
   Fe  &    SLAC E140  & 1993 & \cite{Dasu:1993vk} & 2  & 0.14  & 0.04 \\
   Fe/D  &    SLAC E139  & 1994  & \cite{Gomez:1993ri} & 14 & 44.53  & 32.07  \\
   Fe/C  & NMC 96 & 1996  & \cite{Arneodo:1996rv} & 14  &  11.17 & 9.93 \\ \hline     
   $\nu$ Fe    & CDHSW  & 1991 & \cite{Berge:1989hr}  &  464  & 404.26  &  358.19  \\ 
   $\bar{\nu}$ Fe    & CDHSW  & 1991  & \cite{Berge:1989hr} & 462  & 439.53  &  395.99  \\ \hline
   Cu/D  & EMC 88  &  1988 & \cite{Ashman:1988bf}  & 9  & 8.38  & 5.71  \\
        & EMC 93 & 1993   & \cite{Ashman:1992kv} & 19  & 26.38  &  12.58 \\ \hline
   Kr/D  & HERMES  & 2002   & \cite{Airapetian:2002fx}  & 1   & 2.02  &  2.02 \\ \hline   
   Ag/D  & SLAC E139  & 1994  & \cite{Gomez:1993ri} & 6  & 21.37  & 18.80 \\ \hline   
   Sn/D  & EMC 88 &  1988 & \cite{Ashman:1988bf}  & 8 & 14.37 & 13.98 \\ \hline   
   Sn/C  & NMC 96 & 1996 & \cite{Arneodo:1996rv}  & 14 & 6.48  & 8.52 \\
        & NMC 96, $Q^2$ dep. & 1996  & \cite{Arneodo:1996ru} & 134 & 76.13  & 75.03  \\ \hline     
   Xe/D   & FNAL E665 & 1992  & \cite{Adams:1992nf} &  3 & 1.64  & 1.34 \\ \hline
   Au/D  & SLAC E139  & 1994  & \cite{Gomez:1993ri} &  11 & 16.89  &  18.66  \\ \hline 
   Pb/D  & FNAL E665  & 1995  & \cite{Adams:1995is} & 2  &  8.28 & 7.72   \\ \hline 
   Pb/C   & NMC 96 &  1996  & \cite{Arneodo:1996rv} & 14   & 8.32  &  5.42  \\ \hline 
   $\nu$ Pb   & CHORUS  & 2005 & \cite{Onengut:2005kv} &  405   & 259.48  &  237.85 \\ 
   $\bar{\nu}$ Pb   & CHORUS  & 2005  & \cite{Onengut:2005kv} & 405  & 356.01  & 352.09  \\ \hline \hline 
   \multicolumn{4}{r|}{\textbf{In total:}} & 2336 &  &  \\
\toprule
\end{tabular}
\end{table}
\end{center}

\begin{center}
\renewcommand{\arraystretch}{1.25}
\begin{table}[tb!]
\caption{Summary of experimental $W^{\pm},\,Z$ boson production data from LHC p+Pb collisions in Run I and Run II used to determine our nuclear PDFs. In the last two columns the $\Chi^2$ values at NLO and NNLO obtained in our analysis are provided.}
\label{tab-expdatady-nPDFs}
\scriptsize
\begin{tabular}{l|l|l|c|r|c|r|r}
\toprule
  Nucleus & Proc. & Exp. & Year & Ref. & $N_{\mathrm{dp}}$  & $\Chi^2$ NLO  & $\Chi^2$ NNLO  \\ 
\hline
\hline
pPb & $Z$  & LHC Run I ATLAS  &  2015 & \cite{Aad:2015gta}  & 14  &  16.40 & 12.82 \\  
& $Z$  & LHC Run I CMS  & 2015  & \cite{Khachatryan:2015pzs} & 12  &  8.76 & 7.30 \\
& $W^{-}$  & LHC Run II CMS  &  2019 & \cite{CMS:2019leu}  & 24  & 39.58 & 42.83 \\   
& $W^{+}$  & LHC Run II CMS  &   &   & 24  &  41.08 & 39.07 \\ \hline \hline  
   \multicolumn{5}{r|}{\textbf{In total:}} & 74 &  &    \\ 
\end{tabular}
\end{table}
\end{center}

\section{Results}
\label{sec:results}

\subsection{Proton baseline}
\label{sec:protonpdfs}

Analyses of nuclear PDFs have often been performed by using an existing proton PDF set as a baseline for the nuclear modifications. In this work, however, we have fitted the proton PDFs using the same setup as for the nuclear PDFs. This ensures that all assumptions like sum rules, parton flavor decomposition, etc., as well as all parameters like coupling constants and quark masses, and also further settings like e.g. the heavy flavor mass scheme, are applied in a consistent way. Furthermore, this paves the way for a future combined analysis of proton and nuclear PDFs.

\begin{figure*}[tb!]
     \begin{center}
          \subfigure{
              \includegraphics[width=0.237\linewidth]{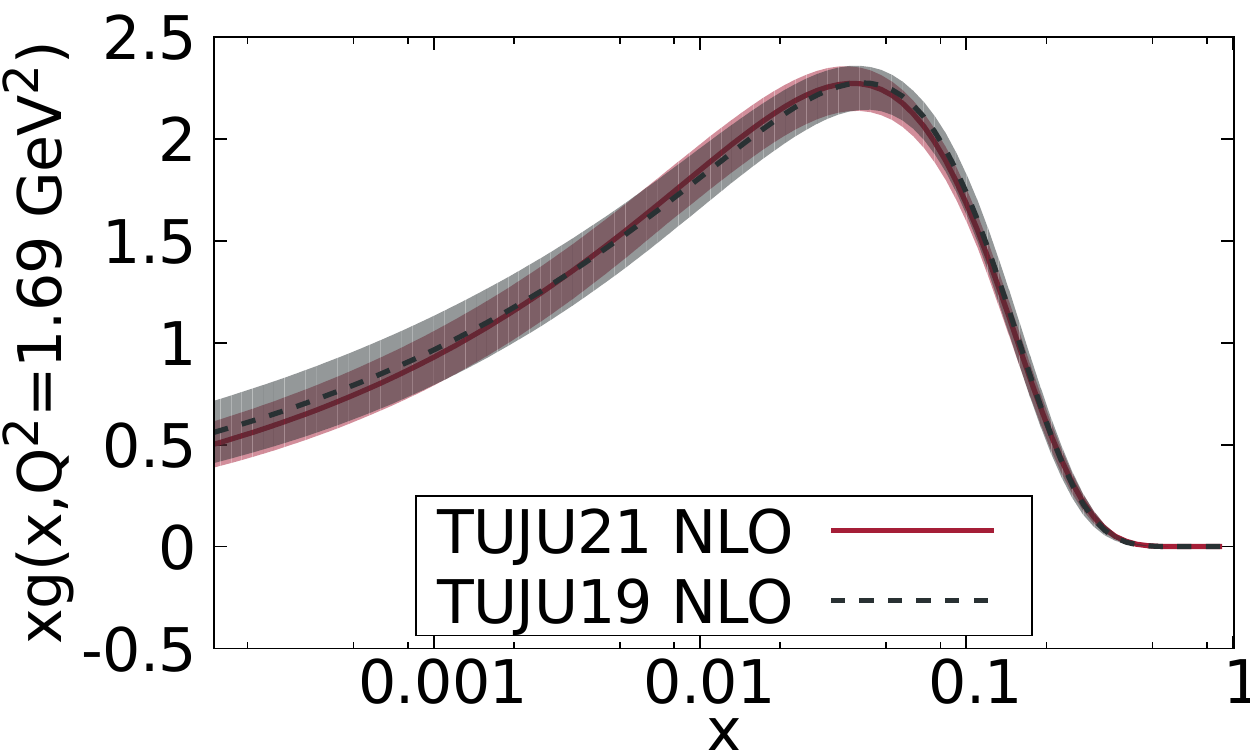}} 
          \subfigure{    
              \includegraphics[width=0.237\linewidth]{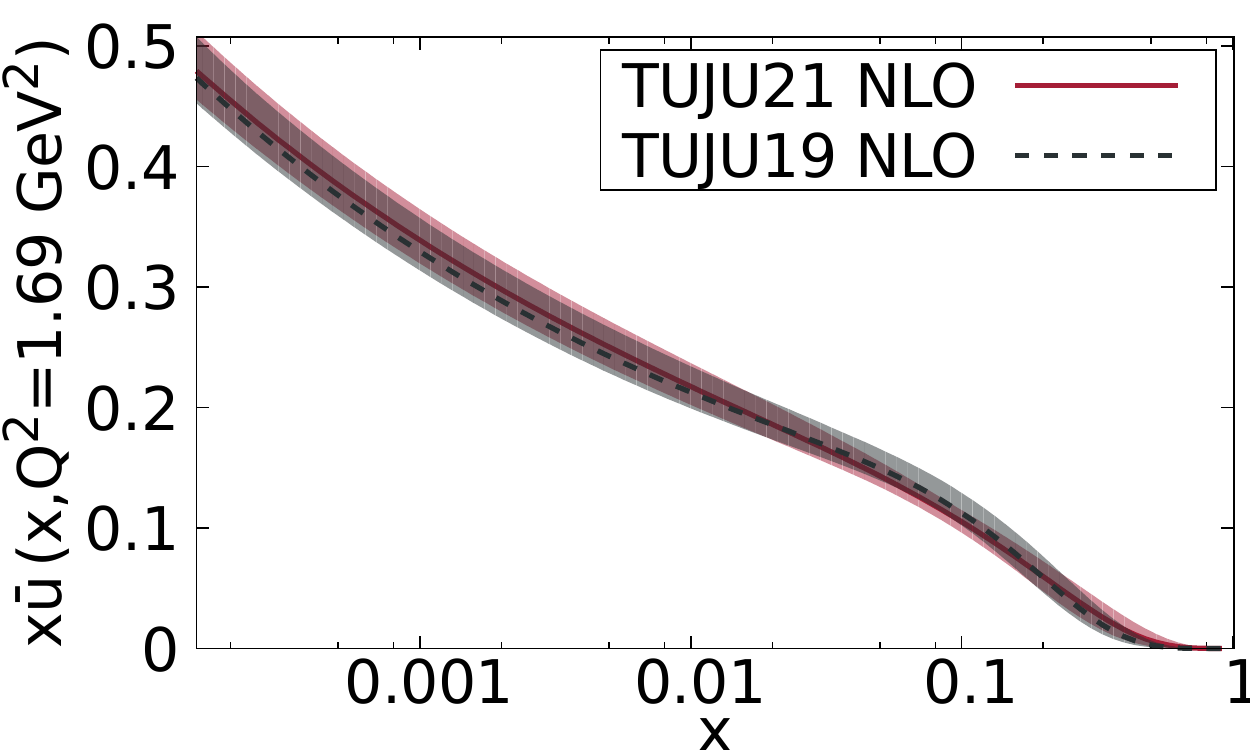}} 
          \subfigure{                           
              \includegraphics[width=0.237\linewidth]{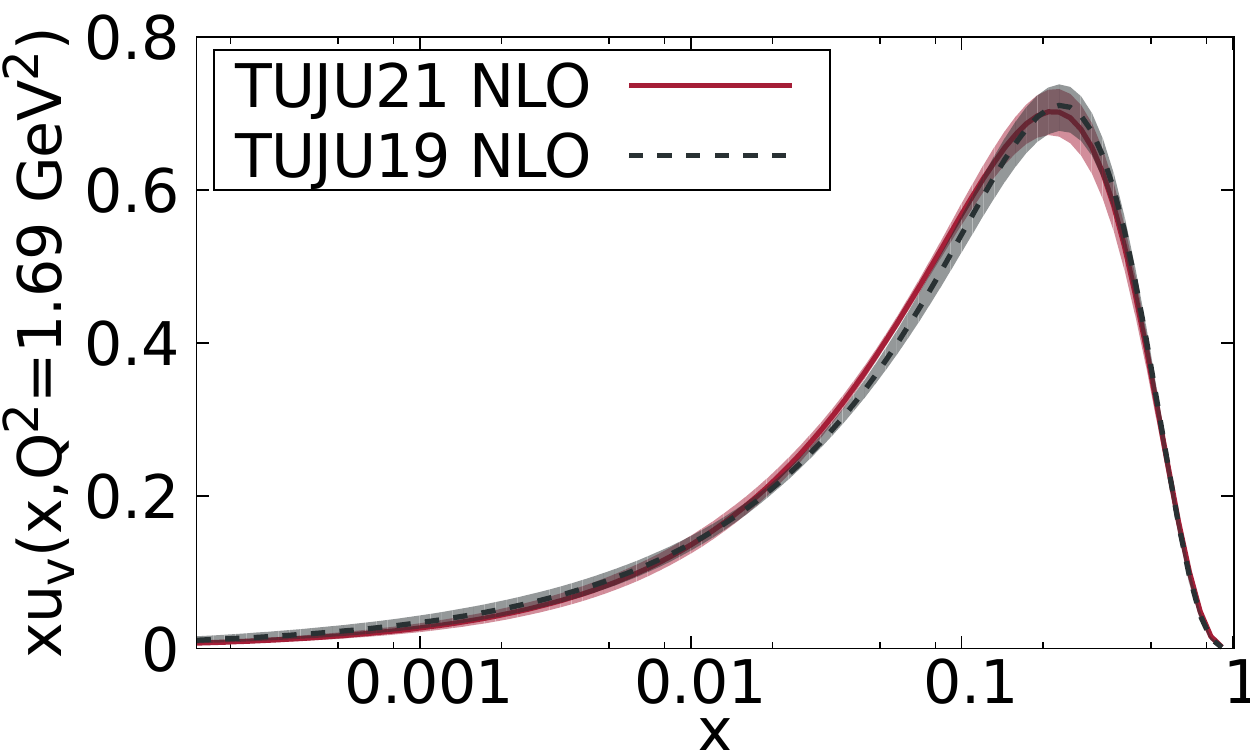}} 
          \subfigure{                                
              \includegraphics[width=0.237\linewidth]{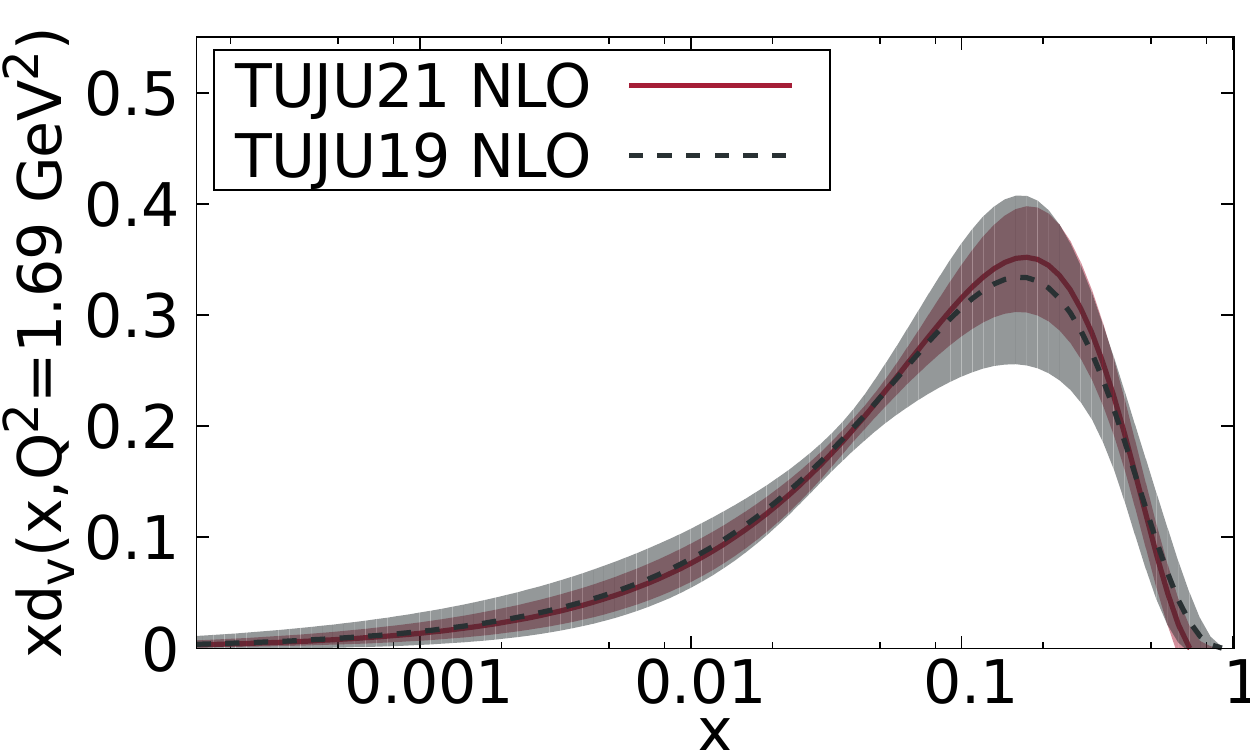}} 
          \subfigure{              
              \includegraphics[width=0.237\textwidth]{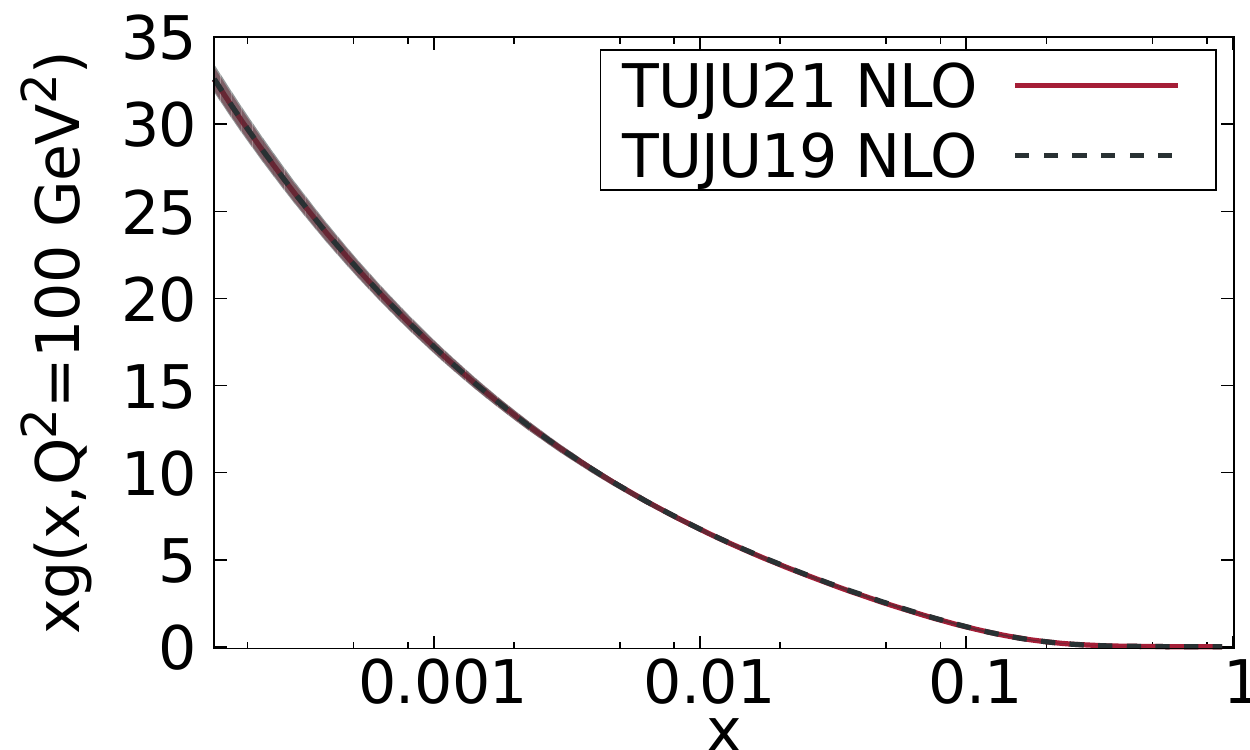}} 
          \subfigure{              
              \includegraphics[width=0.237\textwidth]{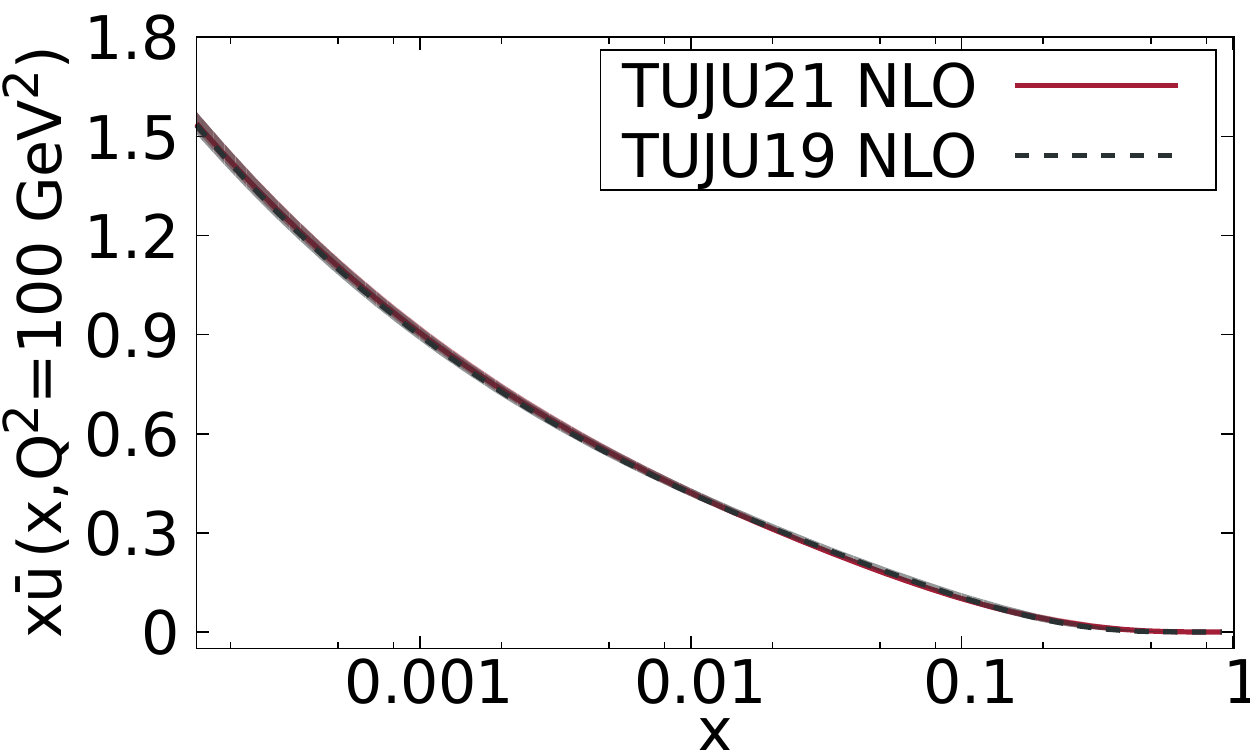}} 
          \subfigure{                        
              \includegraphics[width=0.237\textwidth]{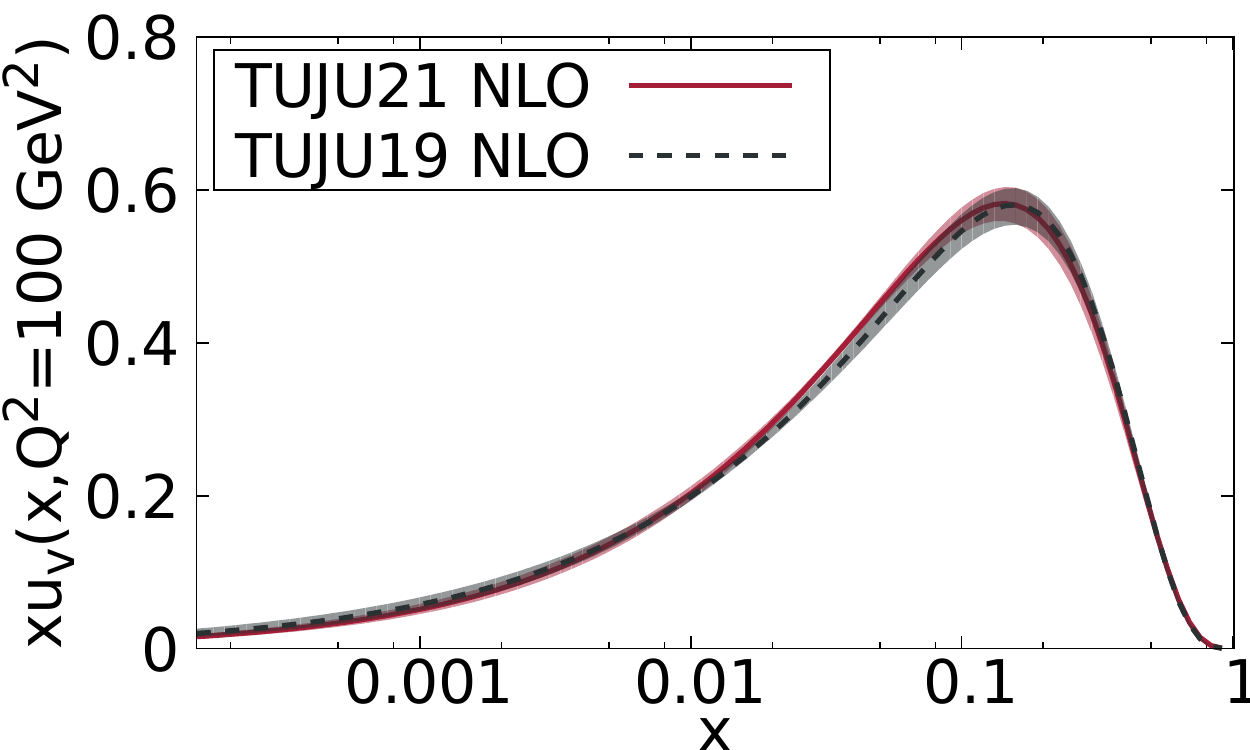}} 
          \subfigure{                                   
              \includegraphics[width=0.237\textwidth]{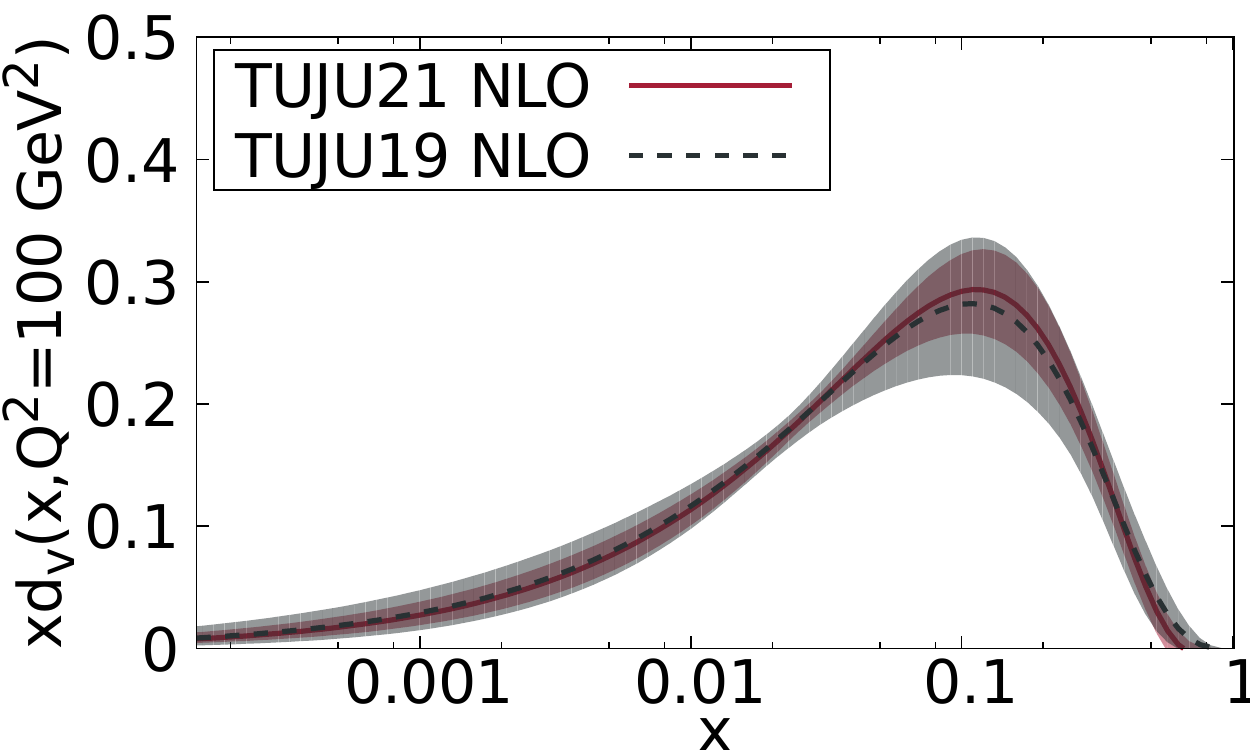}} 
          \end{center} 
    \caption{Proton baseline PDFs in TUJU21 at NLO compared to the previous TUJU19 results, shown at the initial scale $Q_0^2=1.69\,\mathrm{GeV}^2$ (upper panels) and at $Q^2=100\,\mathrm{GeV}^2$ (lower panels) after DGLAP evolution.}
\label{proton_TUJU21_NLO}    
    \end{figure*}

\begin{figure*}[tb!]
     \begin{center}
          \subfigure{
              \includegraphics[width=0.237\textwidth]{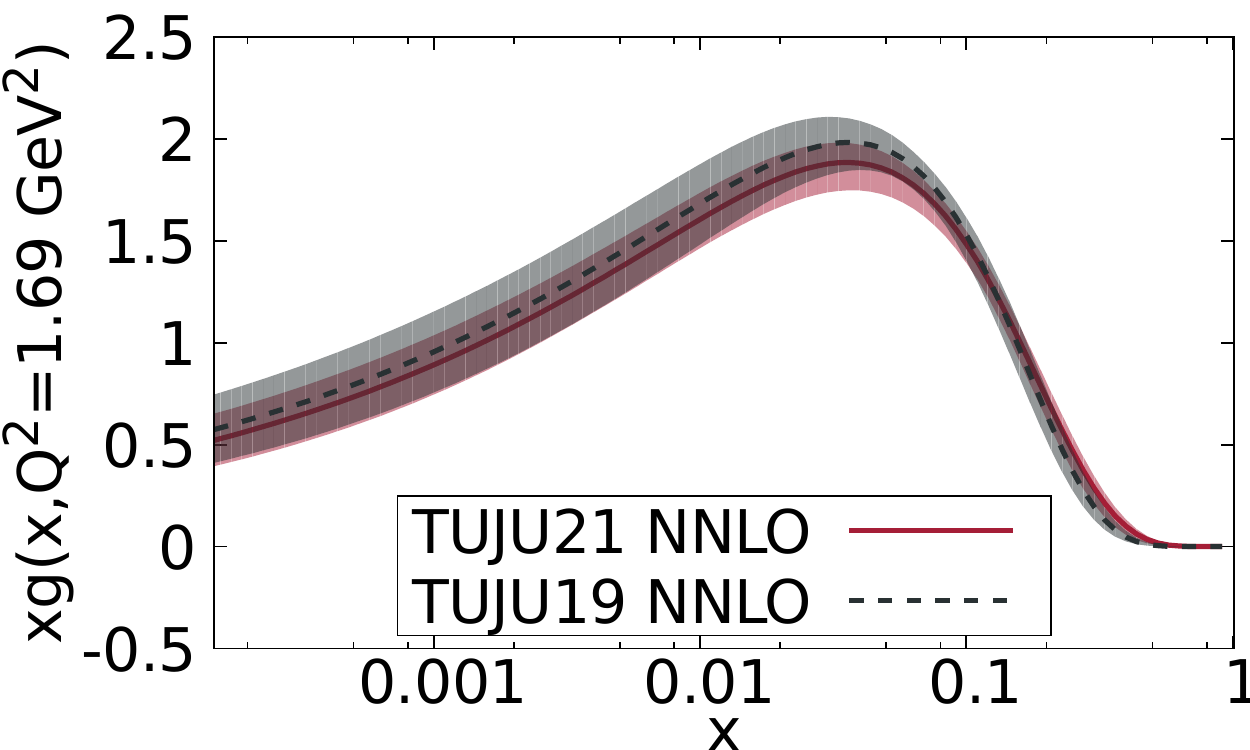}} 
          \subfigure{    
              \includegraphics[width=0.237\textwidth]{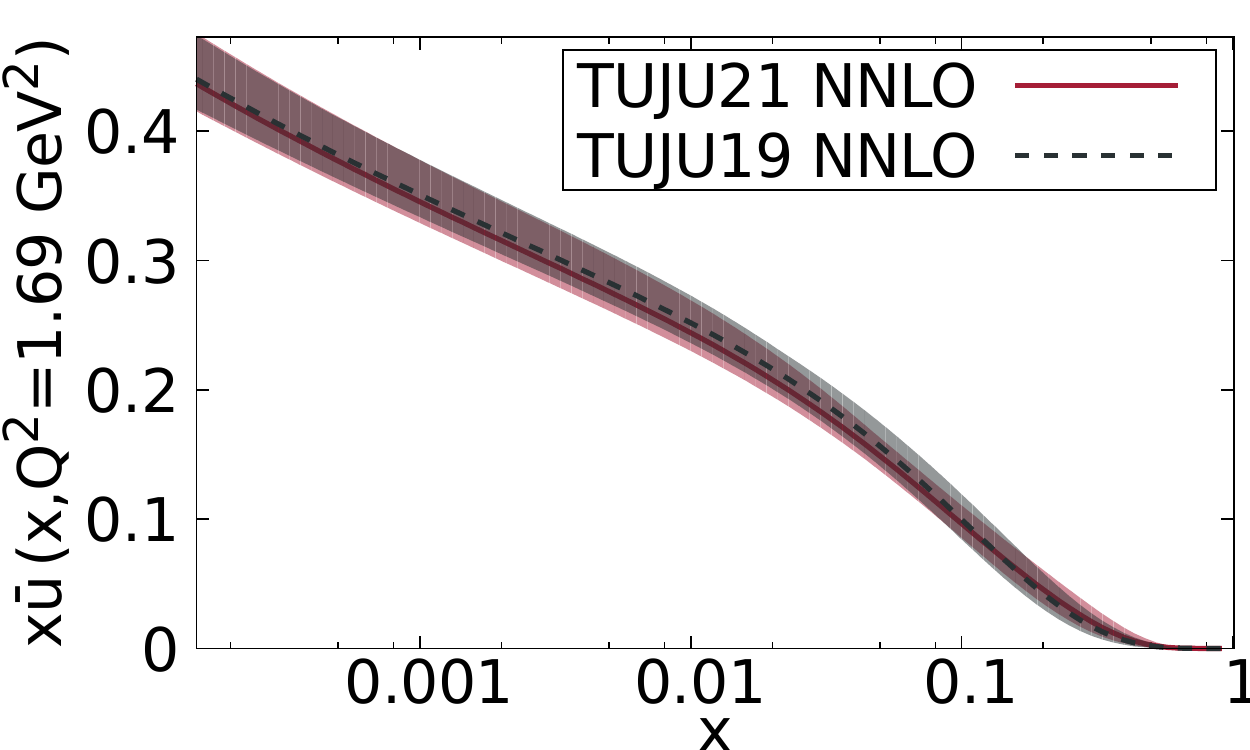}} 
          \subfigure{                           
              \includegraphics[width=0.237\textwidth]{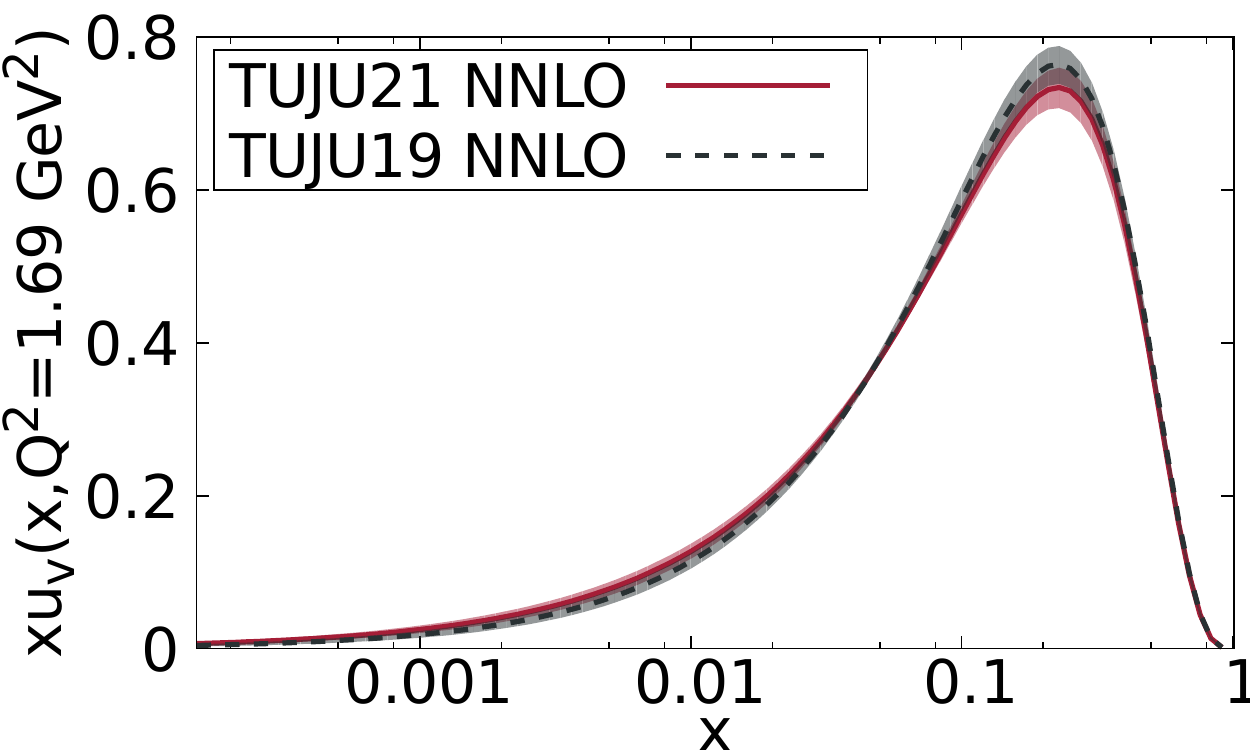}} 
          \subfigure{                                
              \includegraphics[width=0.237\textwidth]{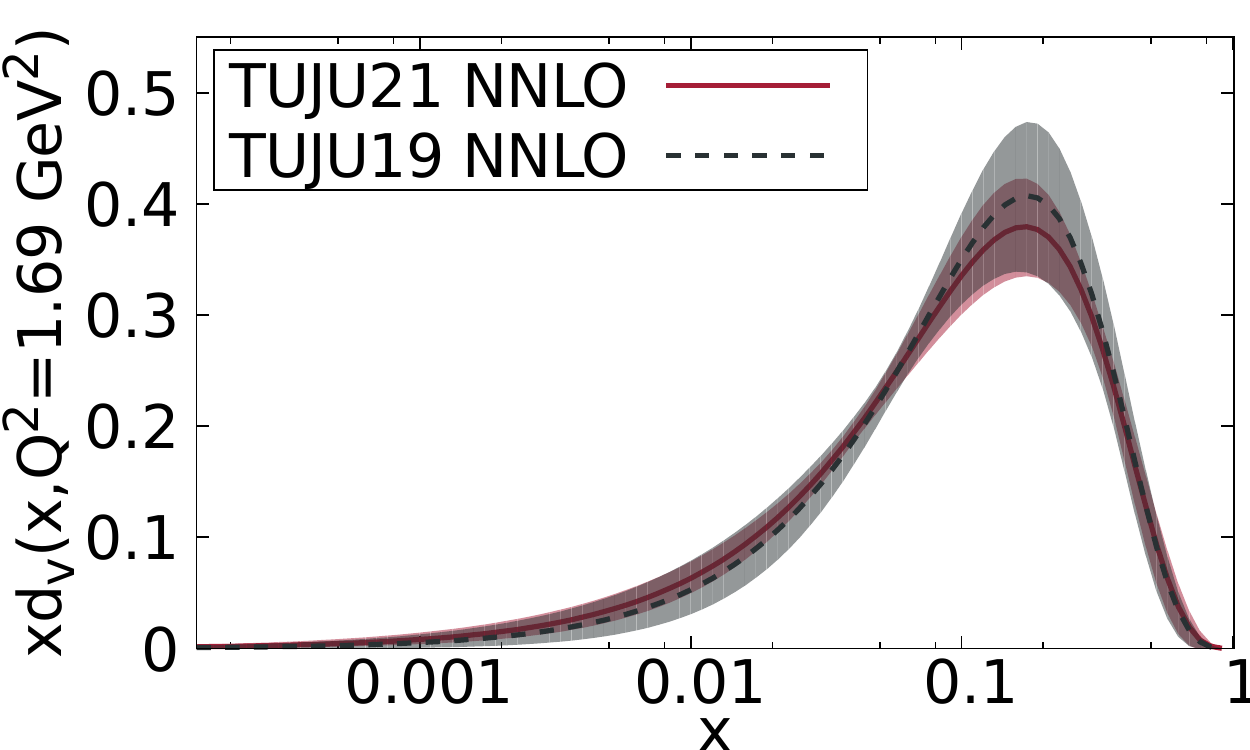}} 
          \subfigure{              
              \includegraphics[width=0.237\textwidth]{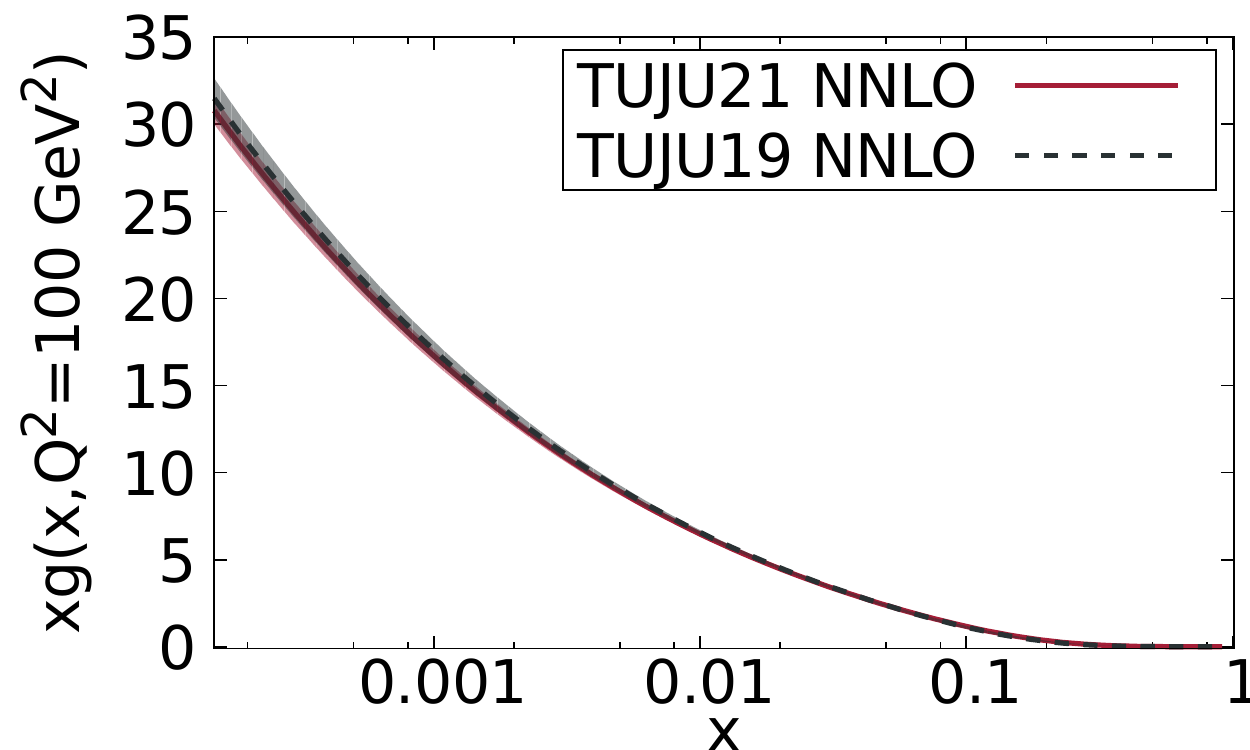}} 
          \subfigure{              
              \includegraphics[width=0.237\textwidth]{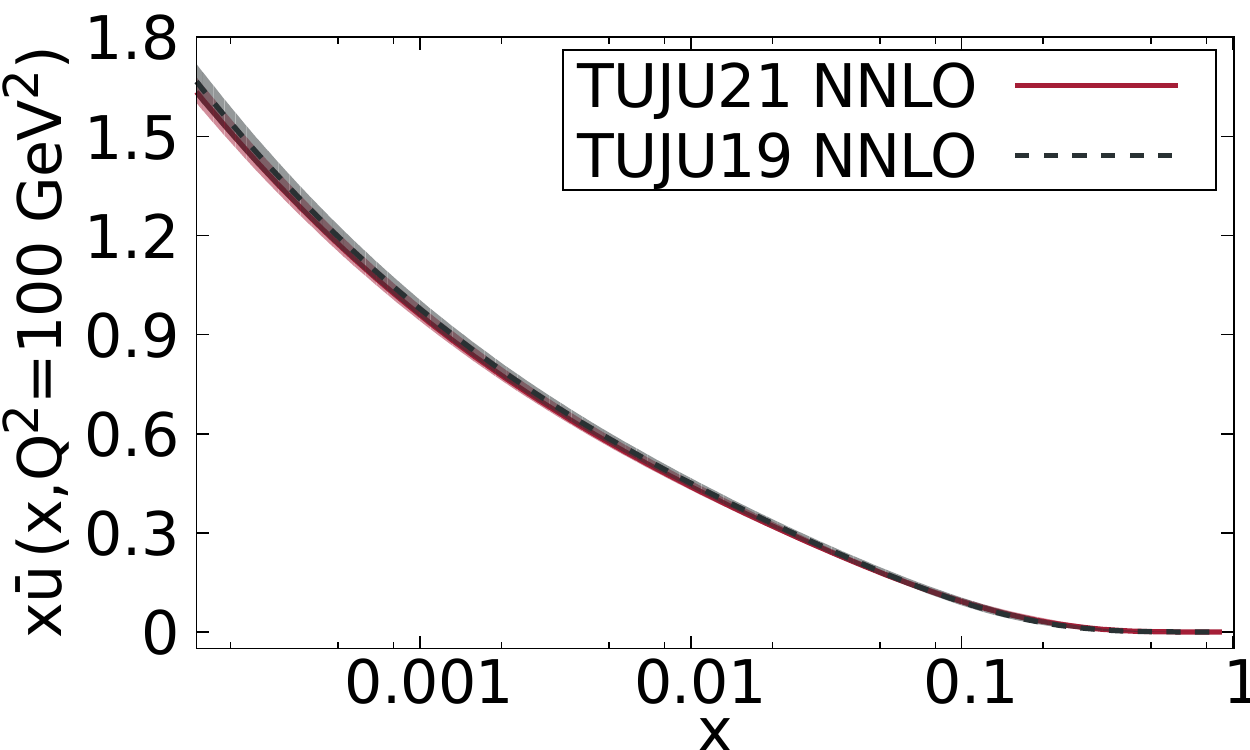}} 
          \subfigure{                        
              \includegraphics[width=0.237\textwidth]{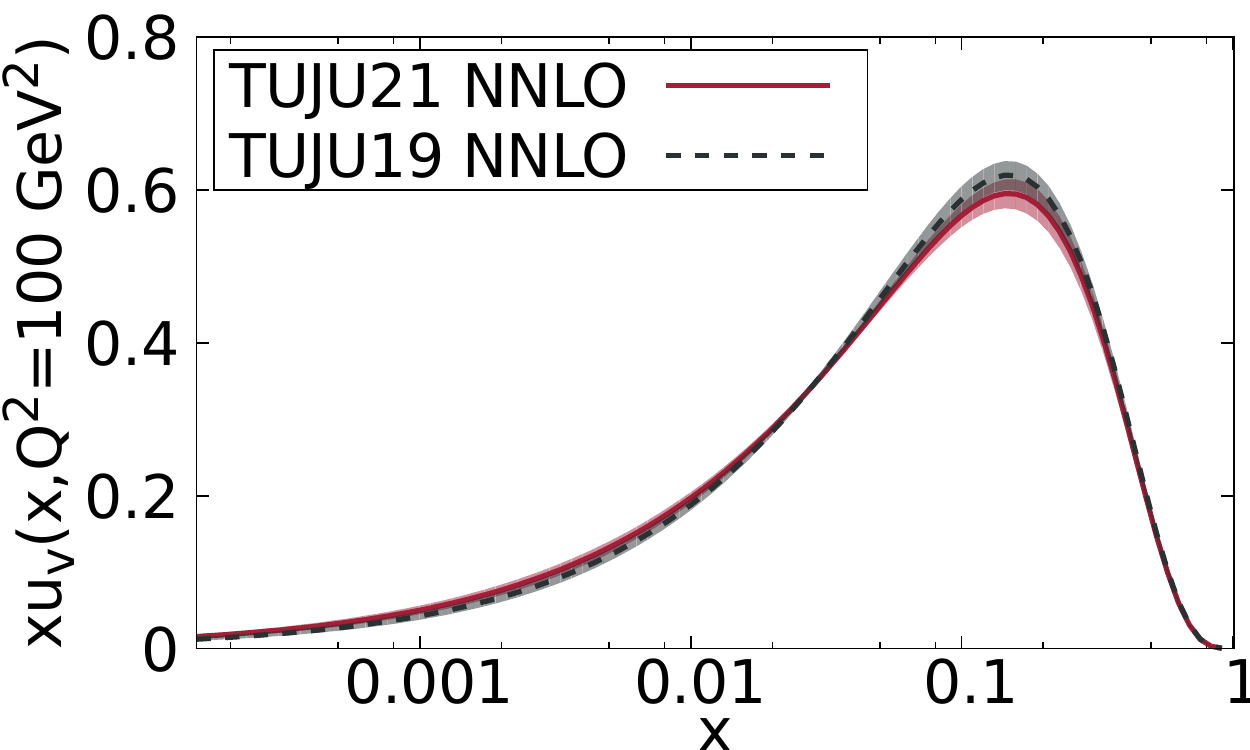}} 
          \subfigure{                                   
              \includegraphics[width=0.237\textwidth]{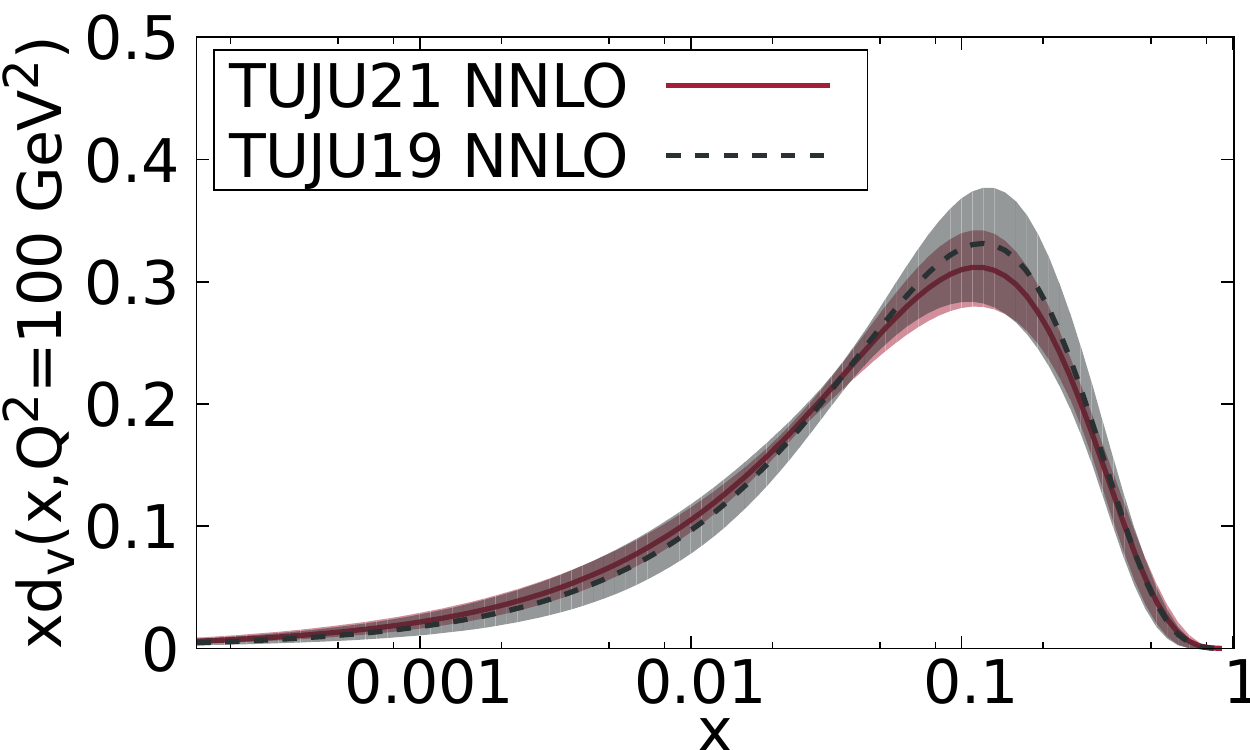}} 
          \end{center} 
    \caption{Same as for Fig. \ref{proton_TUJU21_NLO}, but at NNLO.}
\label{proton_TUJU21_NNLO}    
    \end{figure*}

In this section the updated free proton PDF sets in TUJU21 are compared to those of our earlier TUJU19 analysis, which were determined using DIS data only. The free proton PDFs used as a baseline for the nuclear part of the QCD analysis were updated by including experimental data for DY, $W^{\pm},\,Z$ boson production processes taken by the ATLAS and CMS collaborations at the LHC; see Sec. \ref{sec:expdata} for details.
The comparisons are presented in Fig. \ref{proton_TUJU21_NLO} for NLO and in Fig. \ref{proton_TUJU21_NNLO} for NNLO.

The impact of the newly added LHC data is rather mild at NLO. We observe that the uncertainties of the PDFs have become slightly smaller in some cases, especially for the valence quarks. At NNLO, the resulting distributions for the valence quarks, and especially for gluons are slightly decreased with respect to our previous analysis. The results obtained for the updated free proton PDF baseline confirm that DY, and $W^{\pm},\,Z$ boson production data can be accommodated together with the DIS data and provide further constraints in a global analysis. Using these data to pin down the proton PDFs in the same framework as the nuclear ones will ensure that the baseline is well constrained in the region where new data are included for the nPDFs.

The parameters for the input distributions for our best fit of the proton baseline are collected in Appendix~\ref{app-pdf-params}.

\subsection{Nuclear PDFs}

\begin{figure*}[b!]
     \begin{center}
          \subfigure{
              \includegraphics[width=0.237\linewidth]{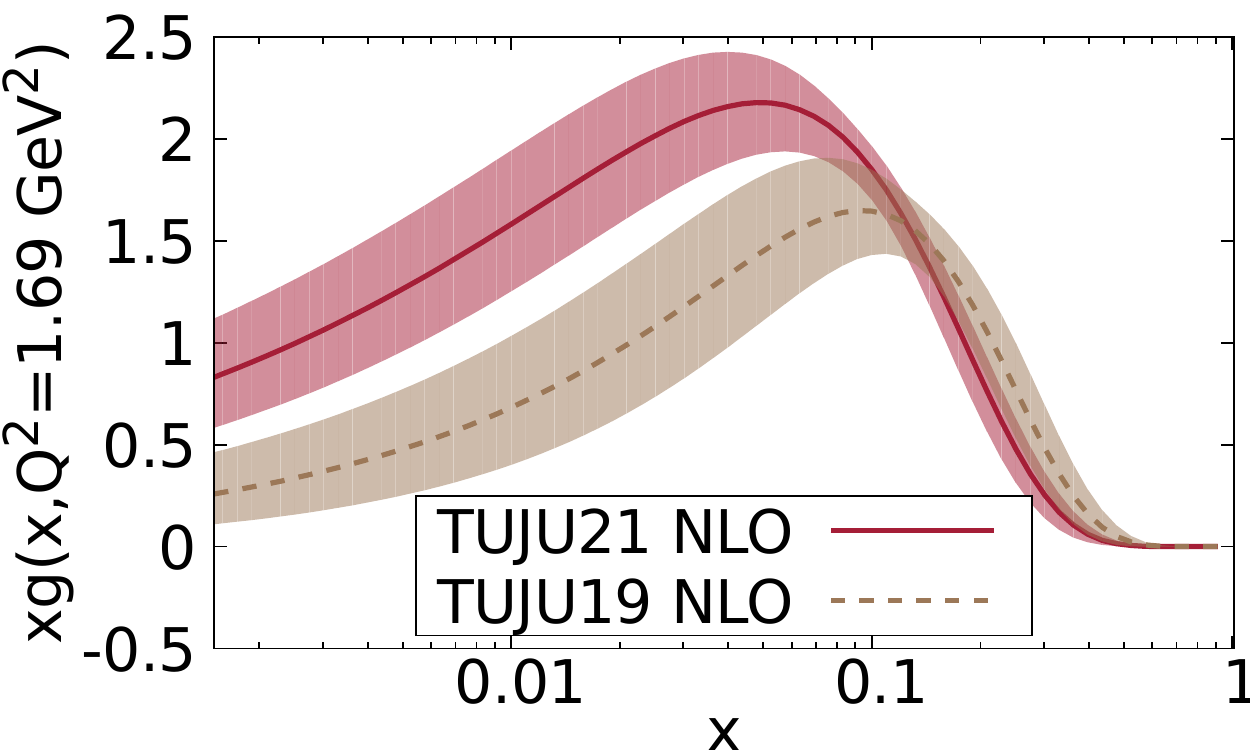}} 
          \subfigure{    
              \includegraphics[width=0.237\linewidth]{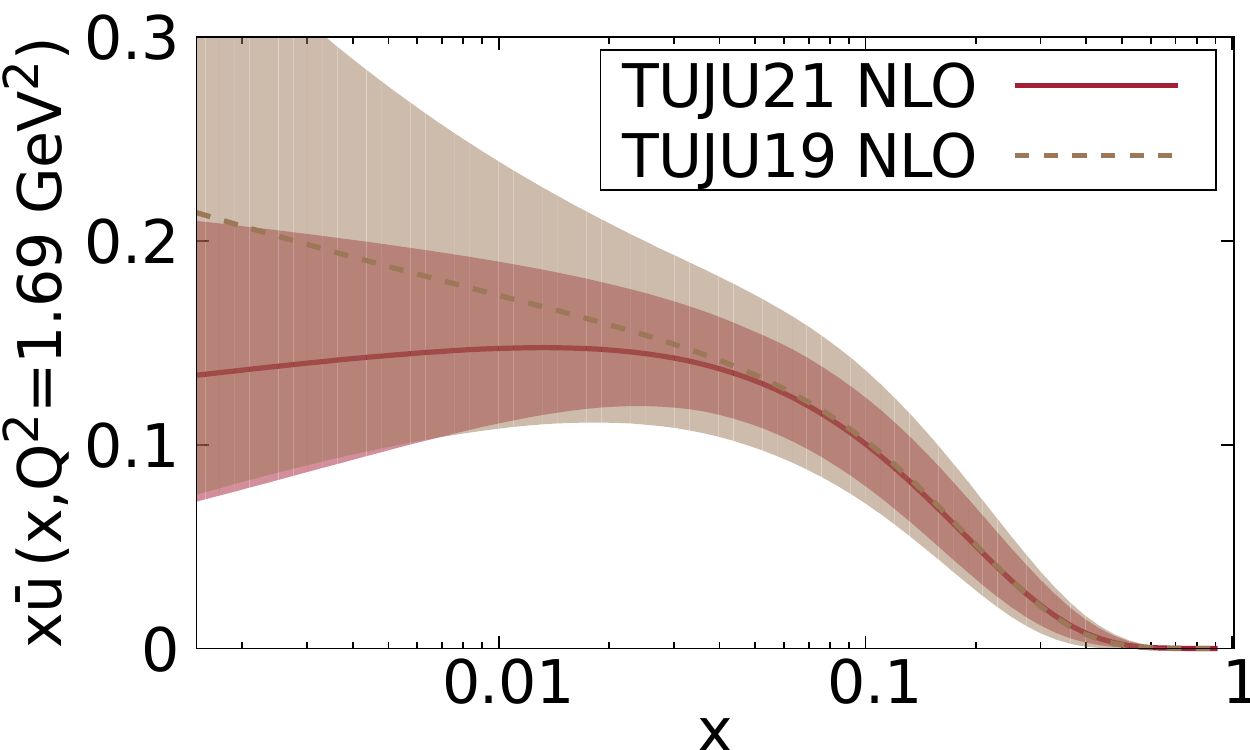}} 
          \subfigure{                           
              \includegraphics[width=0.237\linewidth]{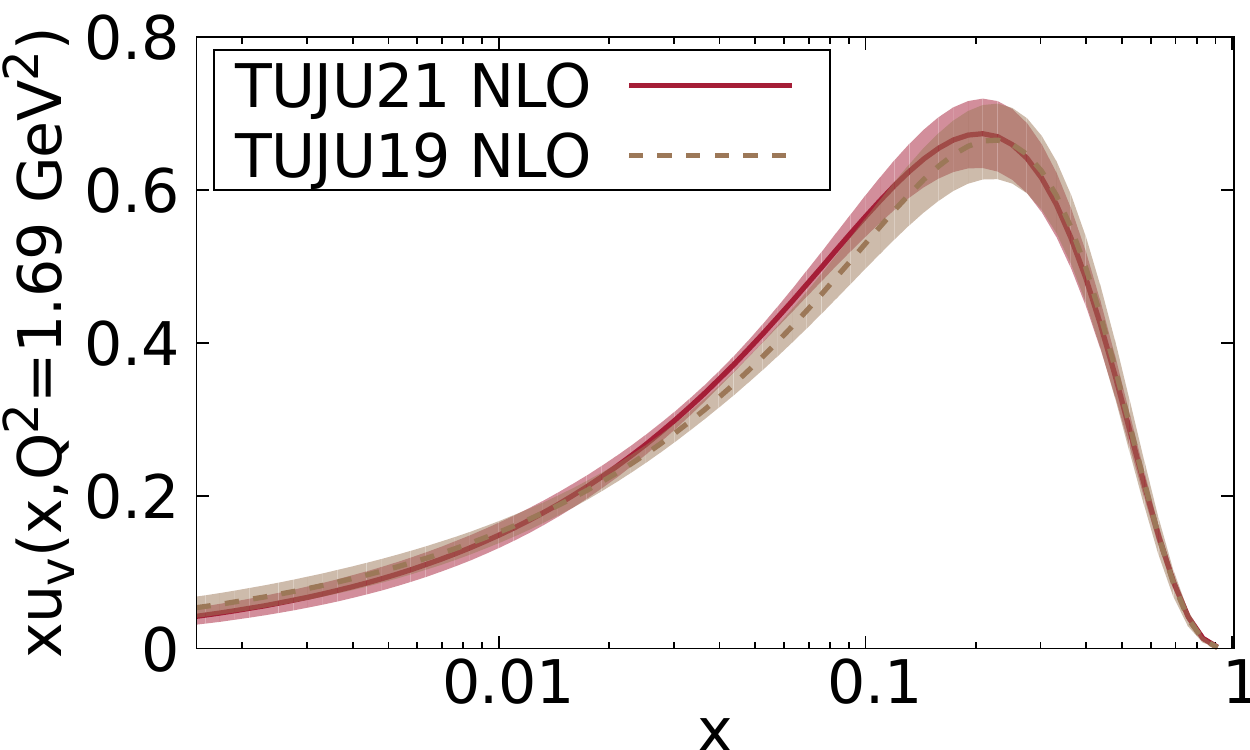}} 
          \subfigure{                                
              \includegraphics[width=0.237\linewidth]{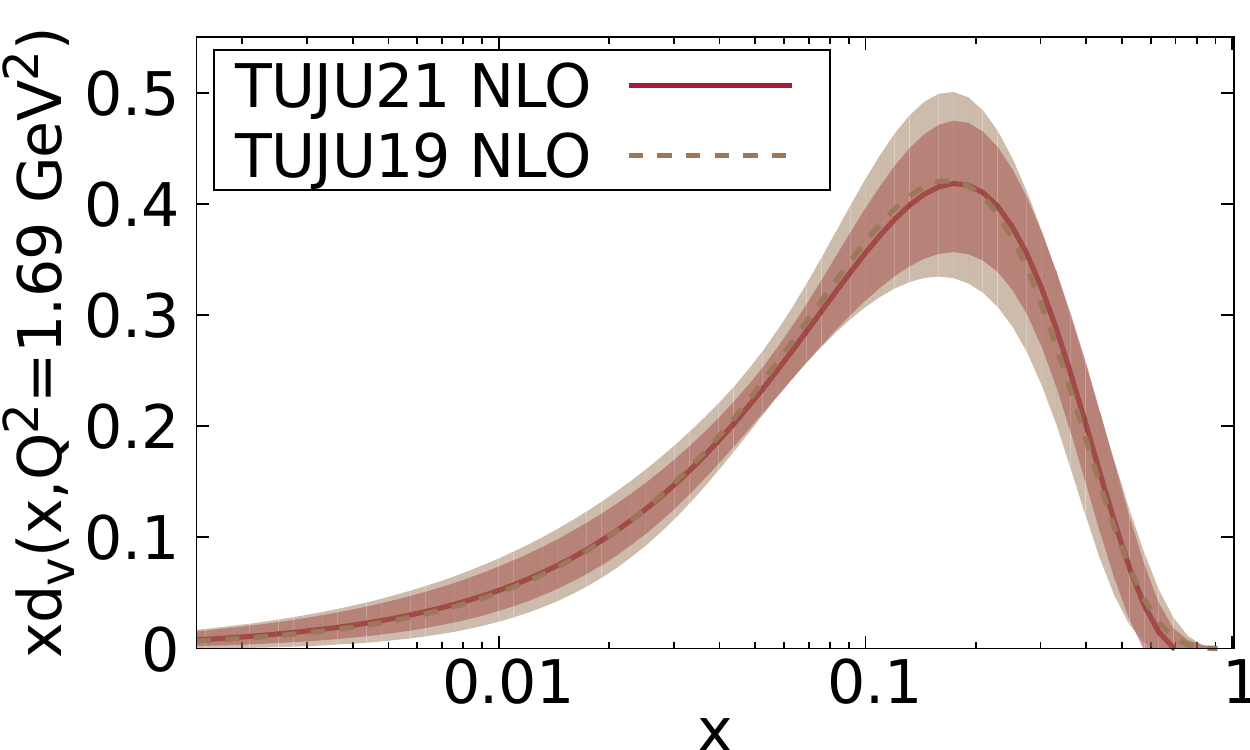}} 
          \subfigure{              
              \includegraphics[width=0.237\textwidth]{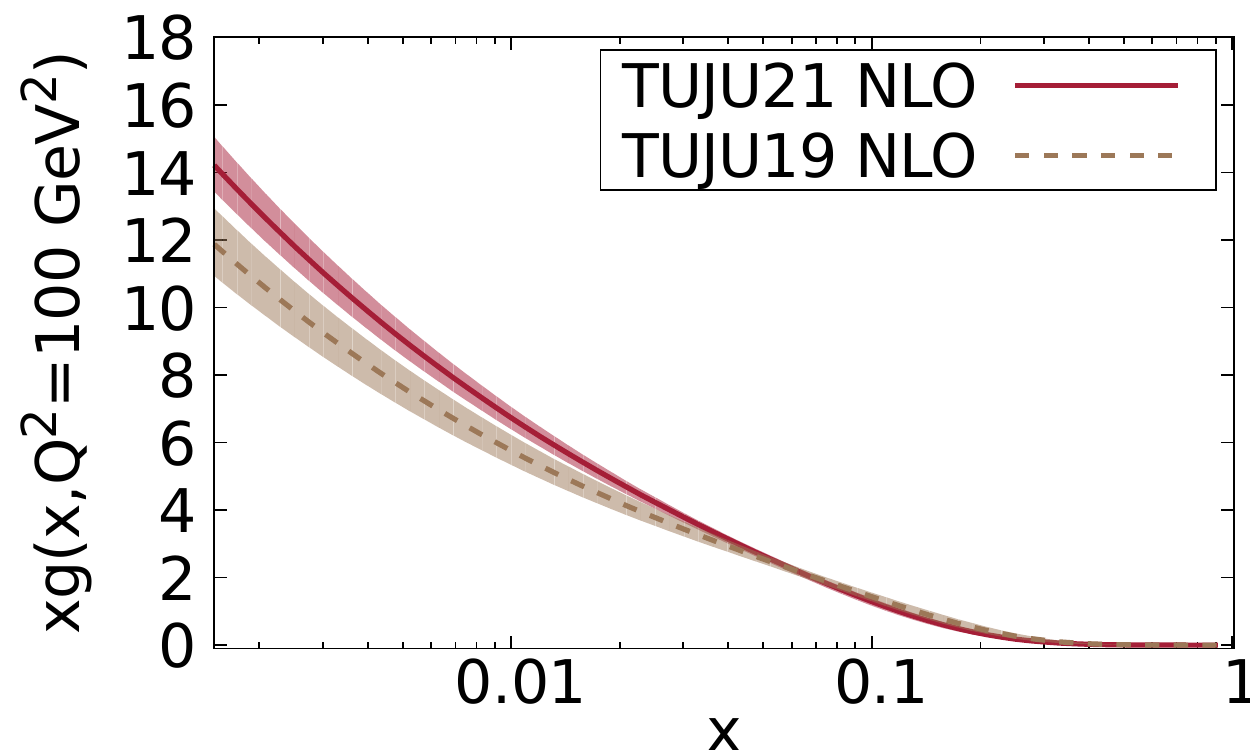}} 
          \subfigure{              
              \includegraphics[width=0.237\textwidth]{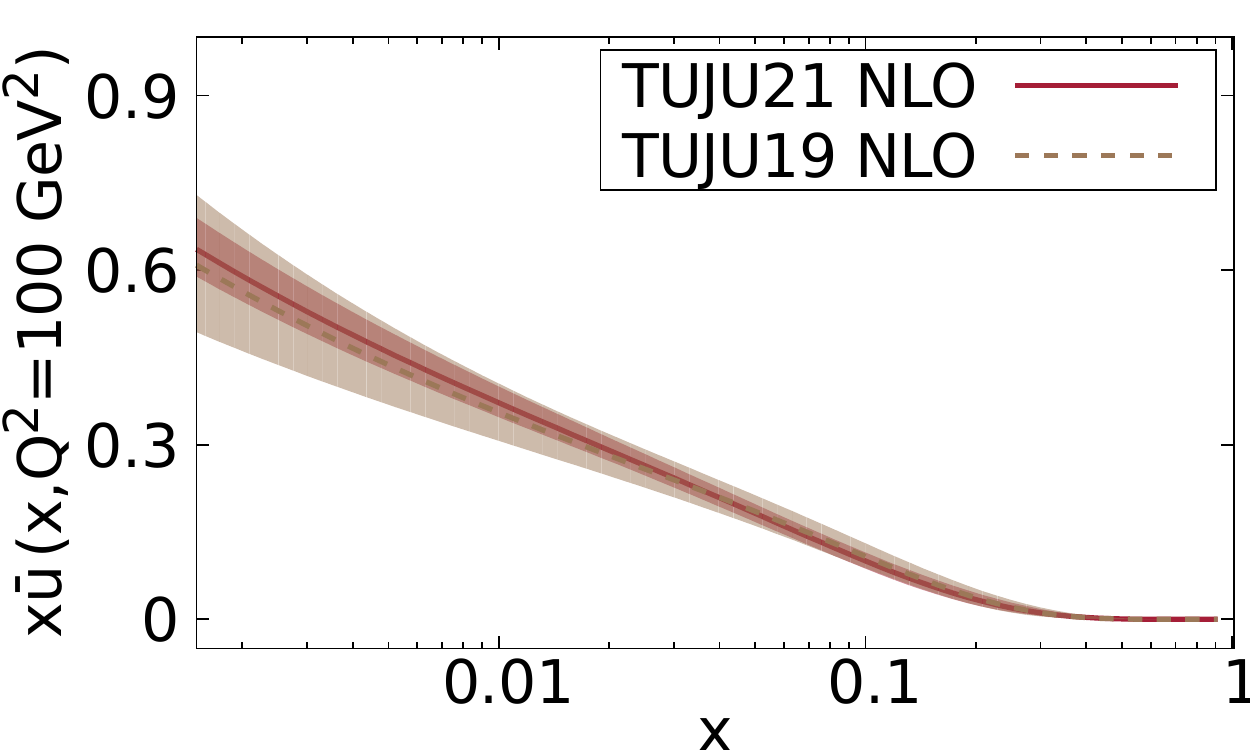}} 
          \subfigure{                        
              \includegraphics[width=0.237\textwidth]{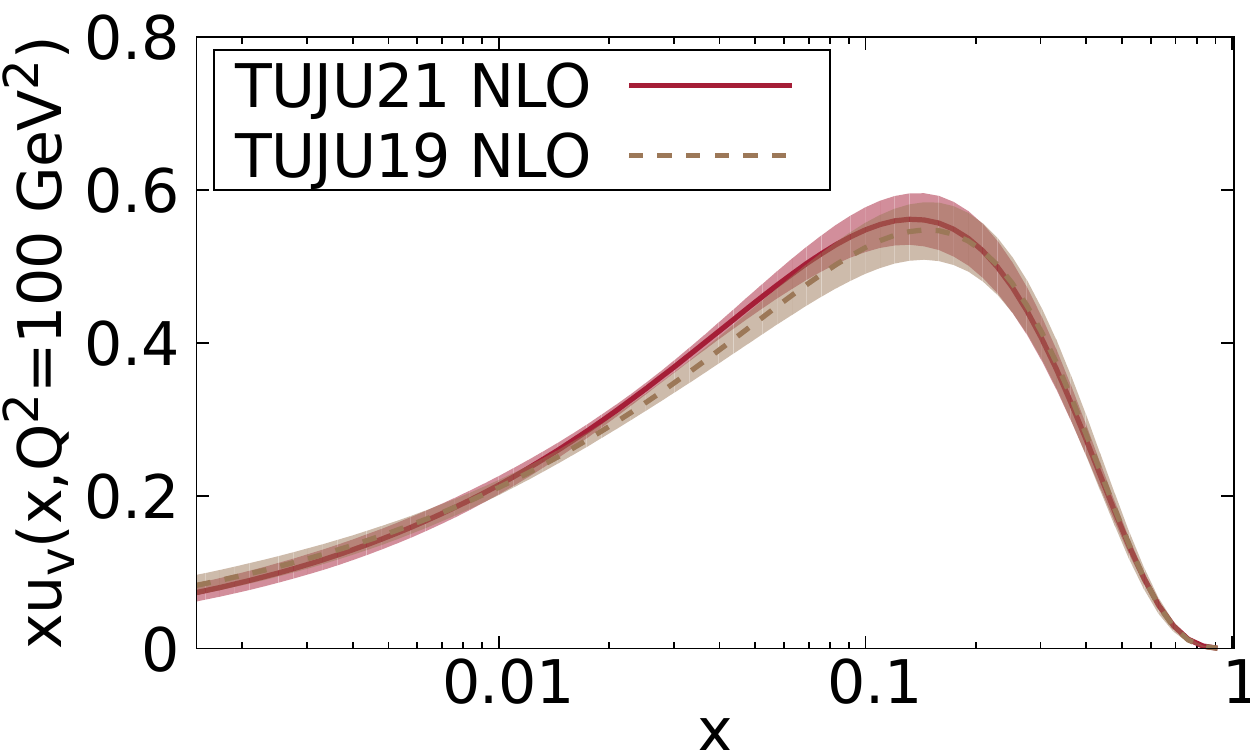}} 
          \subfigure{                                   
              \includegraphics[width=0.237\textwidth]{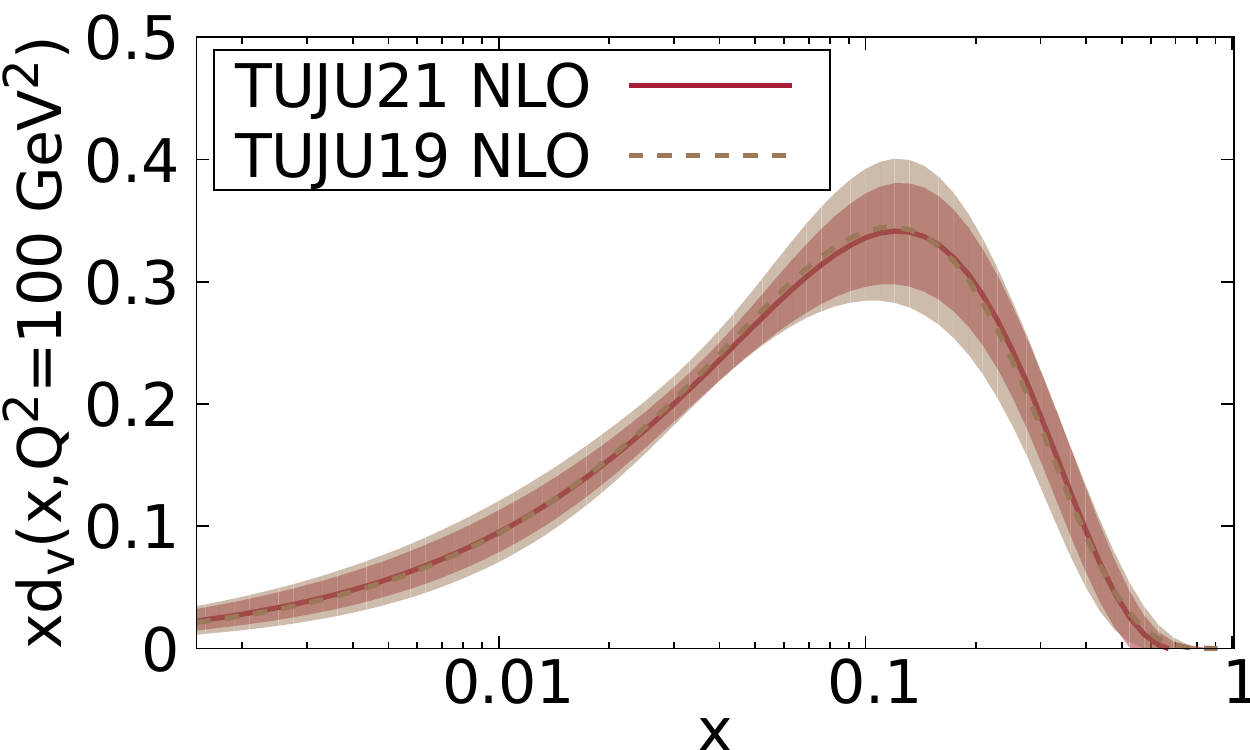}} 
          \subfigure{              
              \includegraphics[width=0.237\textwidth]{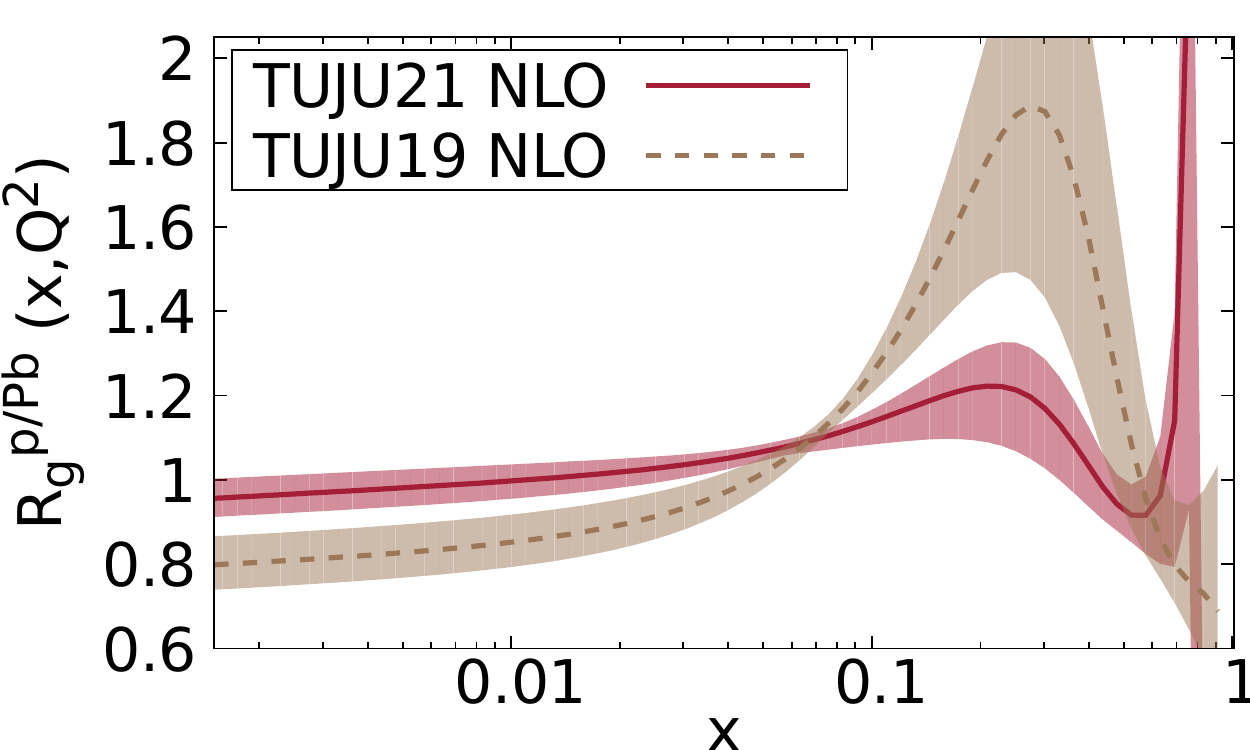}} 
          \subfigure{              
              \includegraphics[width=0.237\textwidth]{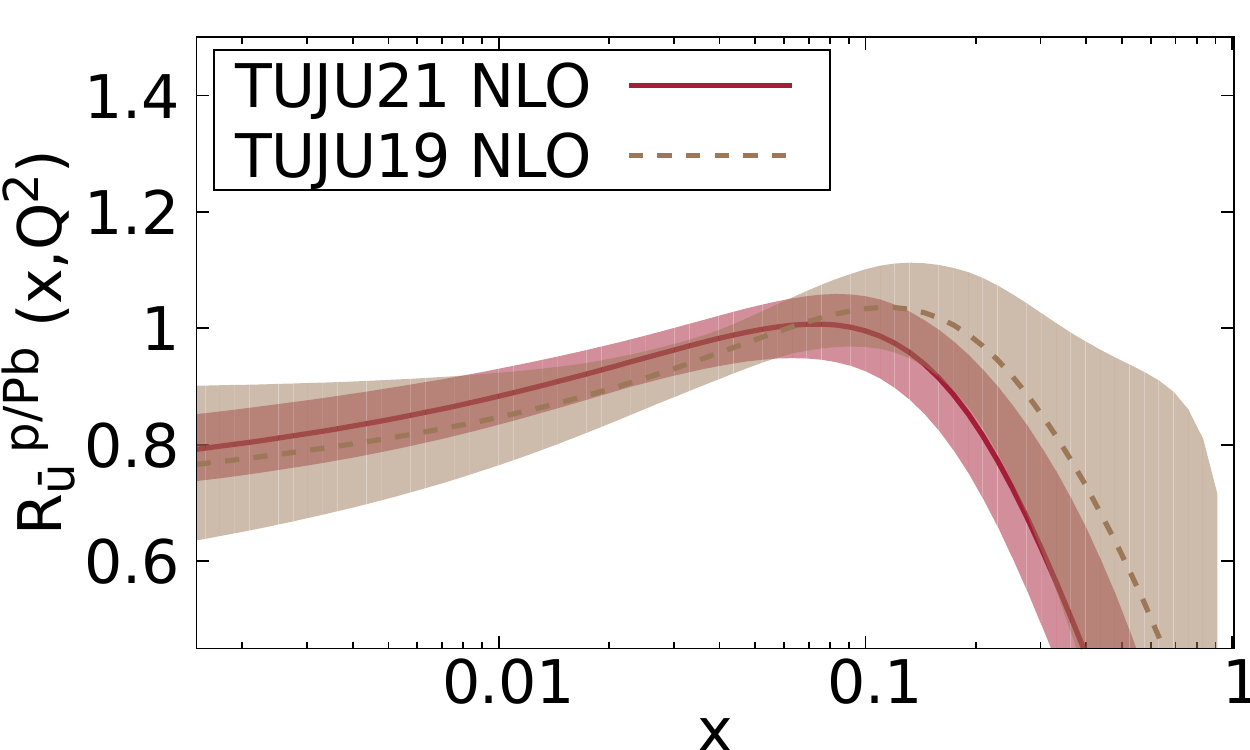}} 
          \subfigure{                        
              \includegraphics[width=0.237\textwidth]{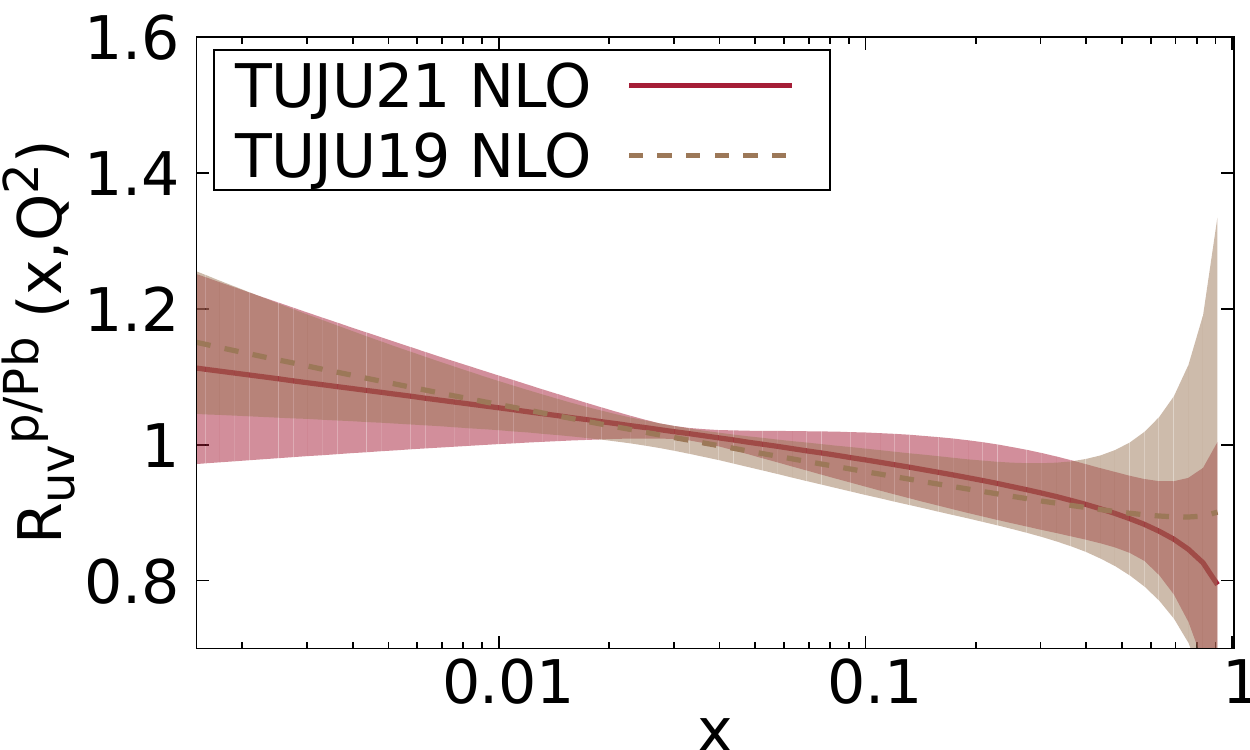}} 
          \subfigure{                                   
              \includegraphics[width=0.237\textwidth]{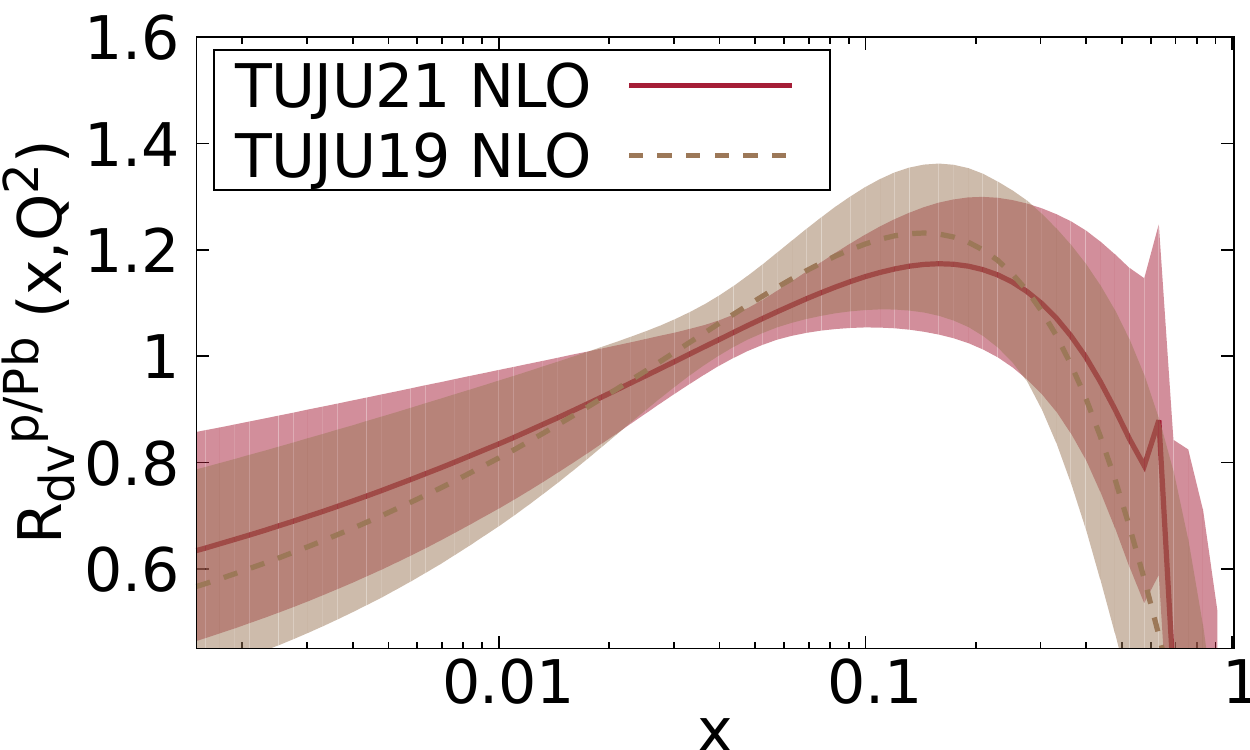}} 
          \end{center} 
    \caption{NLO nuclear parton distribution functions in TUJU21 for a lead nucleus, compared to the previous TUJU19 results, shown at the initial scale $Q_0^2=1.69\,\mathrm{GeV}^2$ (upper panels) and at $Q^2=100\,\mathrm{GeV}^2$ after DGLAP evolution (center panels). The lower panels show
    the corresponding ratios of PDFs for a proton bound in lead over the free proton PDFs.}
\label{Pb_TUJU21_NLO}    
    \end{figure*}

\begin{figure*}[b!]
     \begin{center}
          \subfigure{
              \includegraphics[width=0.237\textwidth]{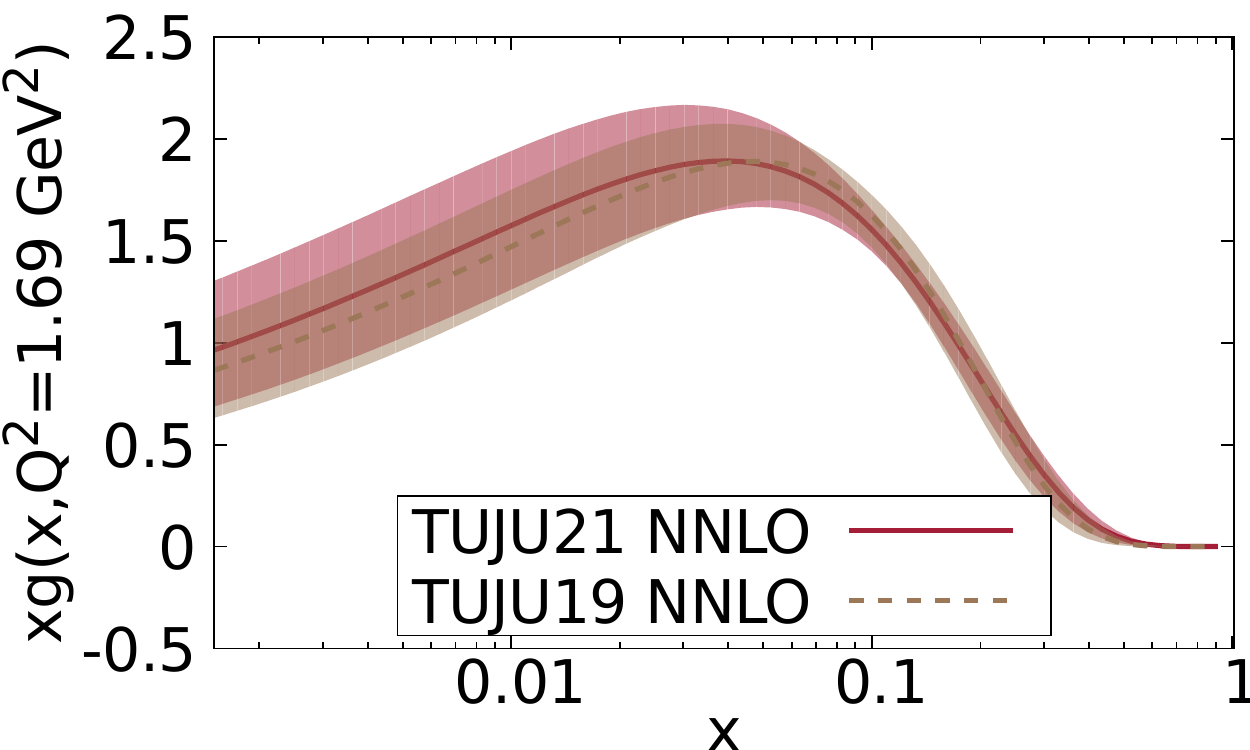}} 
          \subfigure{    
              \includegraphics[width=0.237\textwidth]{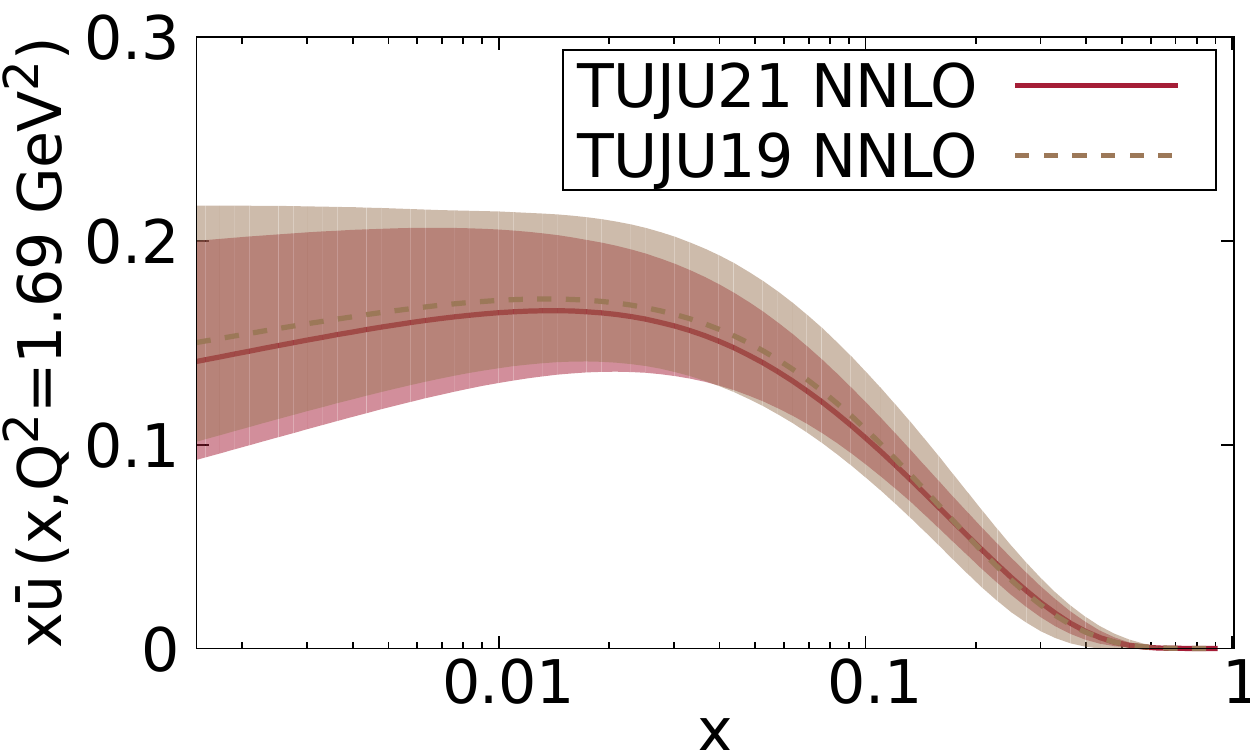}} 
          \subfigure{                           
              \includegraphics[width=0.237\textwidth]{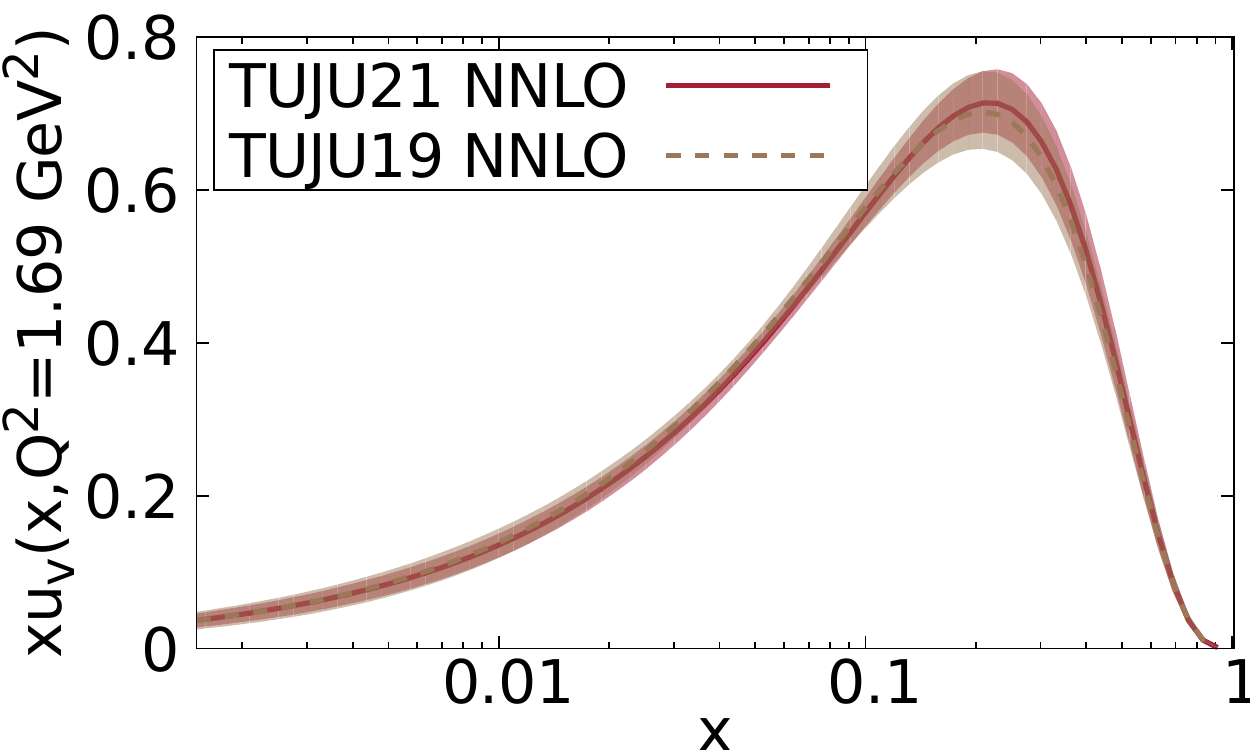}} 
          \subfigure{                                
              \includegraphics[width=0.237\textwidth]{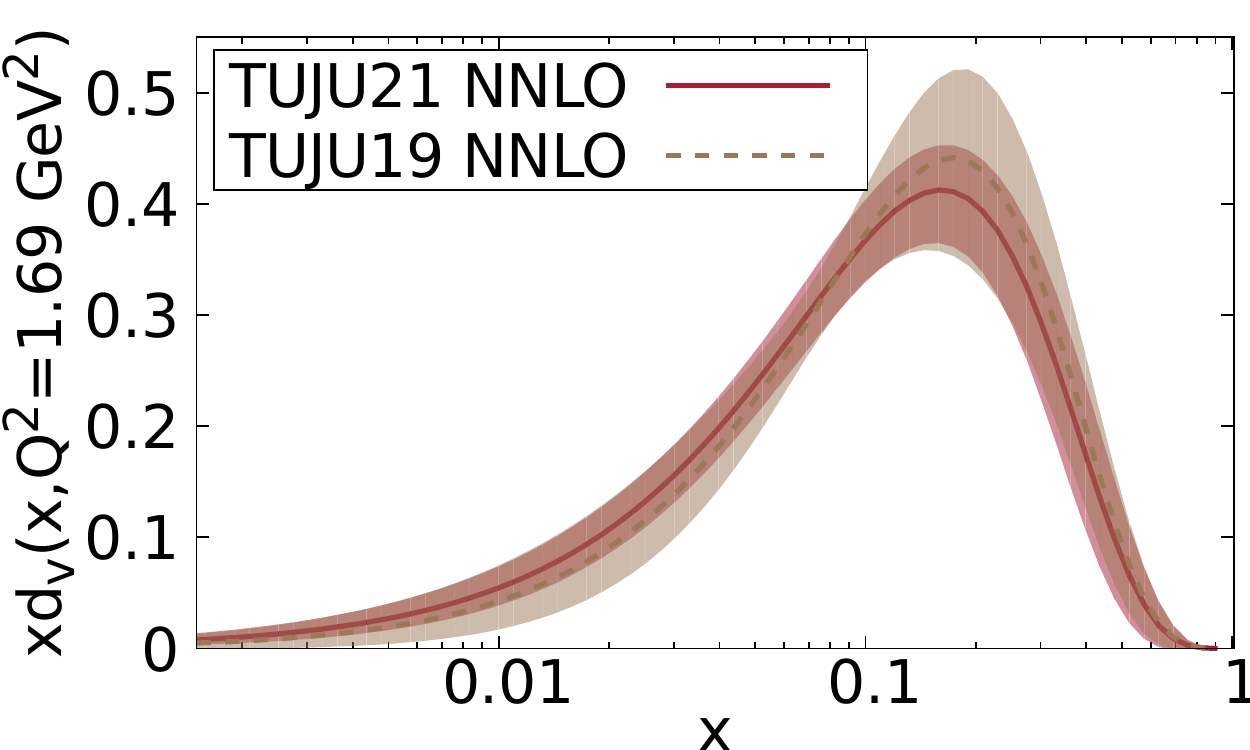}} 
          \subfigure{              
              \includegraphics[width=0.237\textwidth]{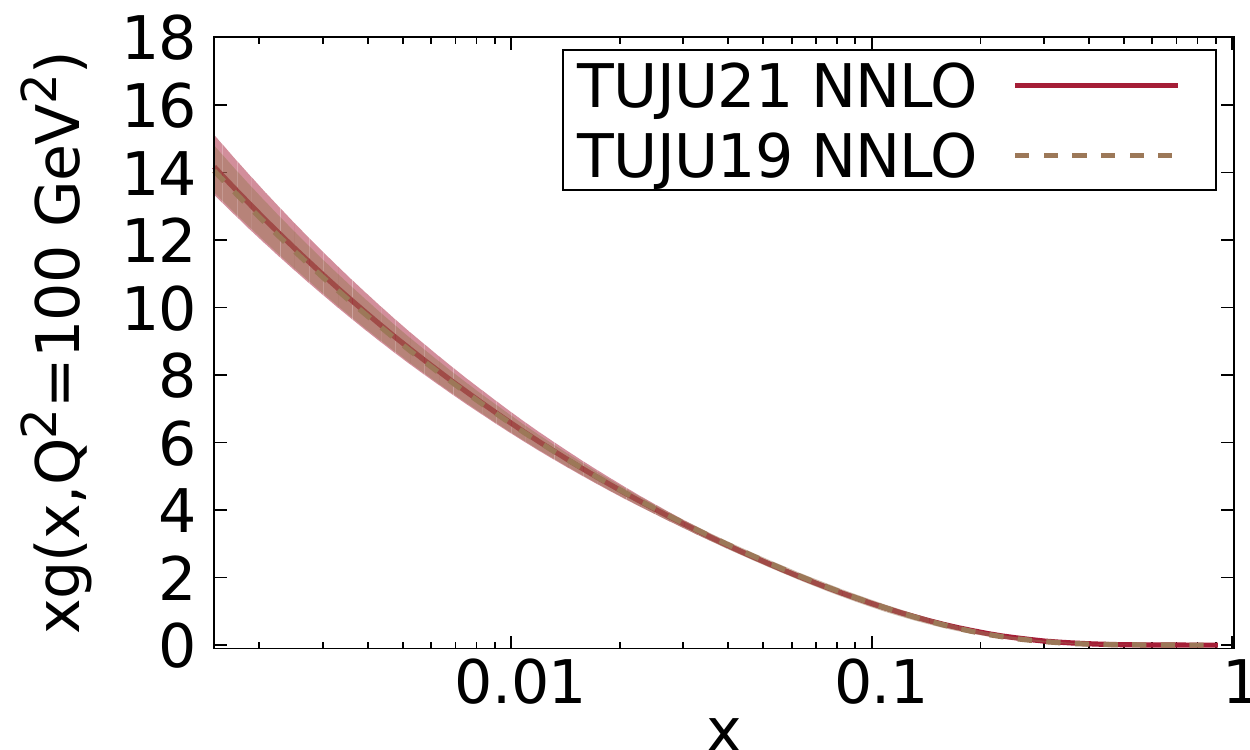}} 
          \subfigure{              
              \includegraphics[width=0.237\textwidth]{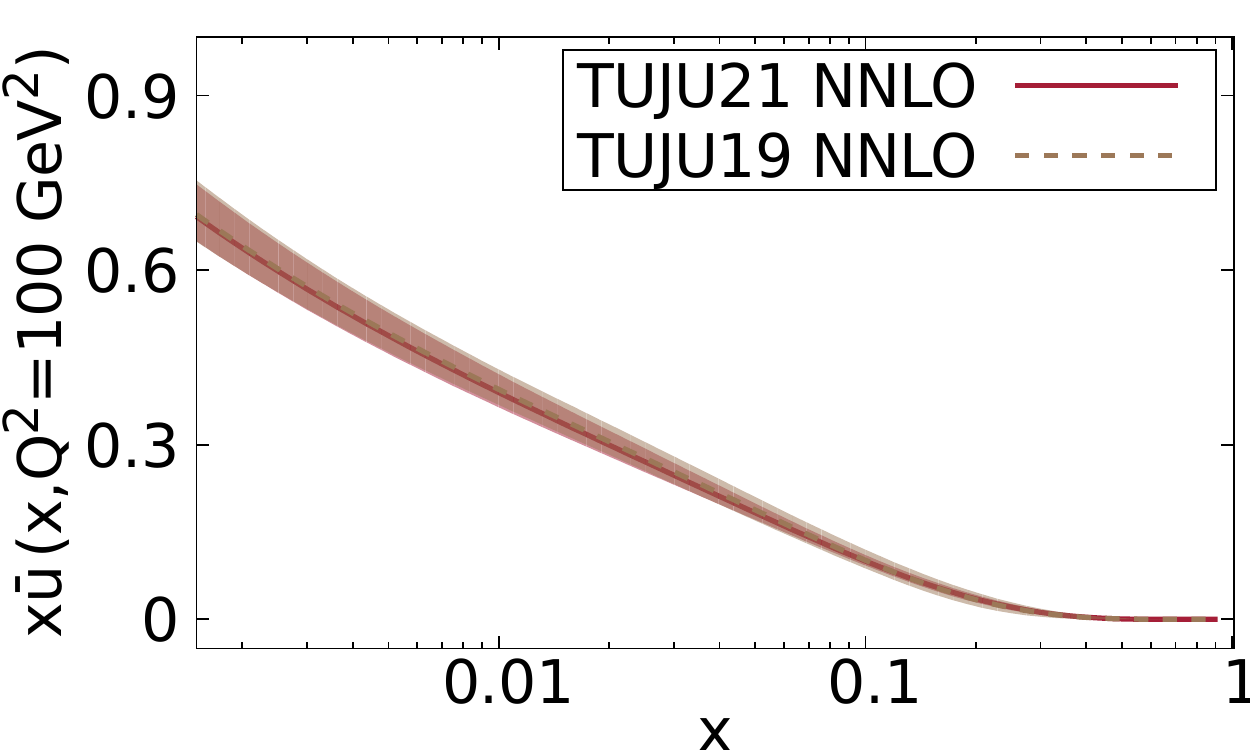}} 
          \subfigure{                        
              \includegraphics[width=0.237\textwidth]{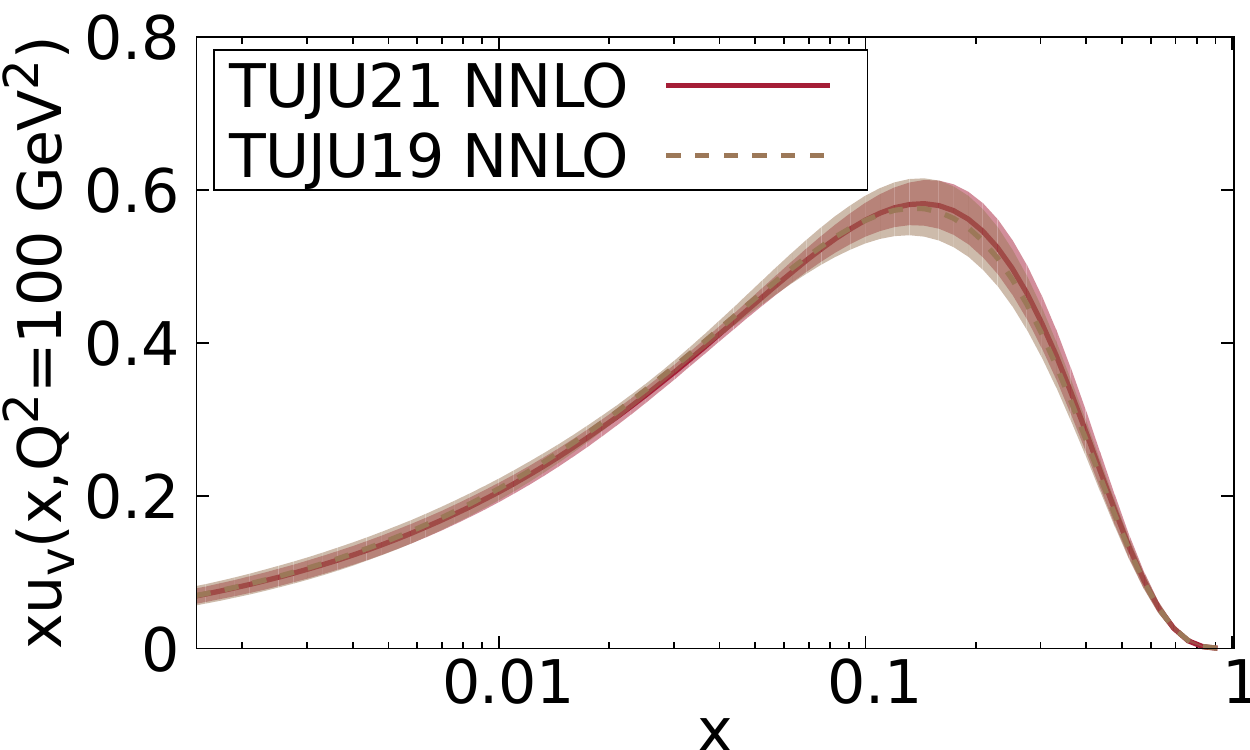}} 
          \subfigure{                                   
              \includegraphics[width=0.237\textwidth]{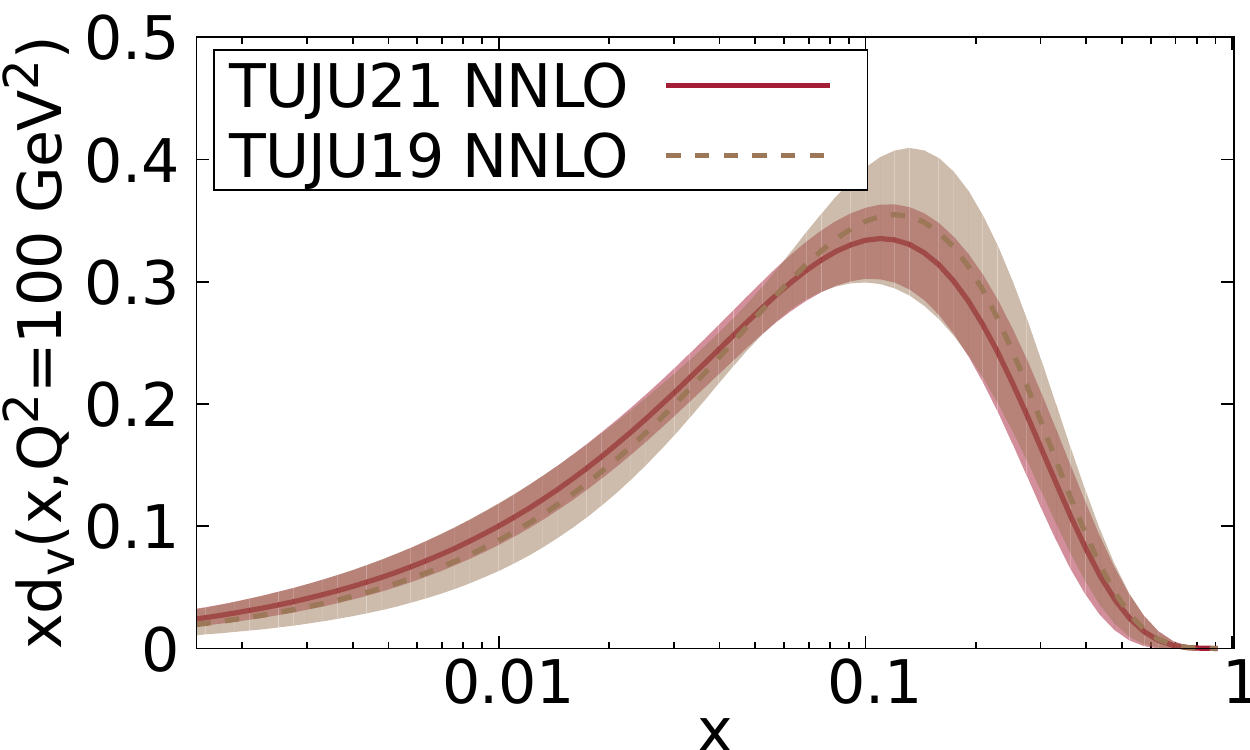}} 
          \subfigure{              
              \includegraphics[width=0.237\textwidth]{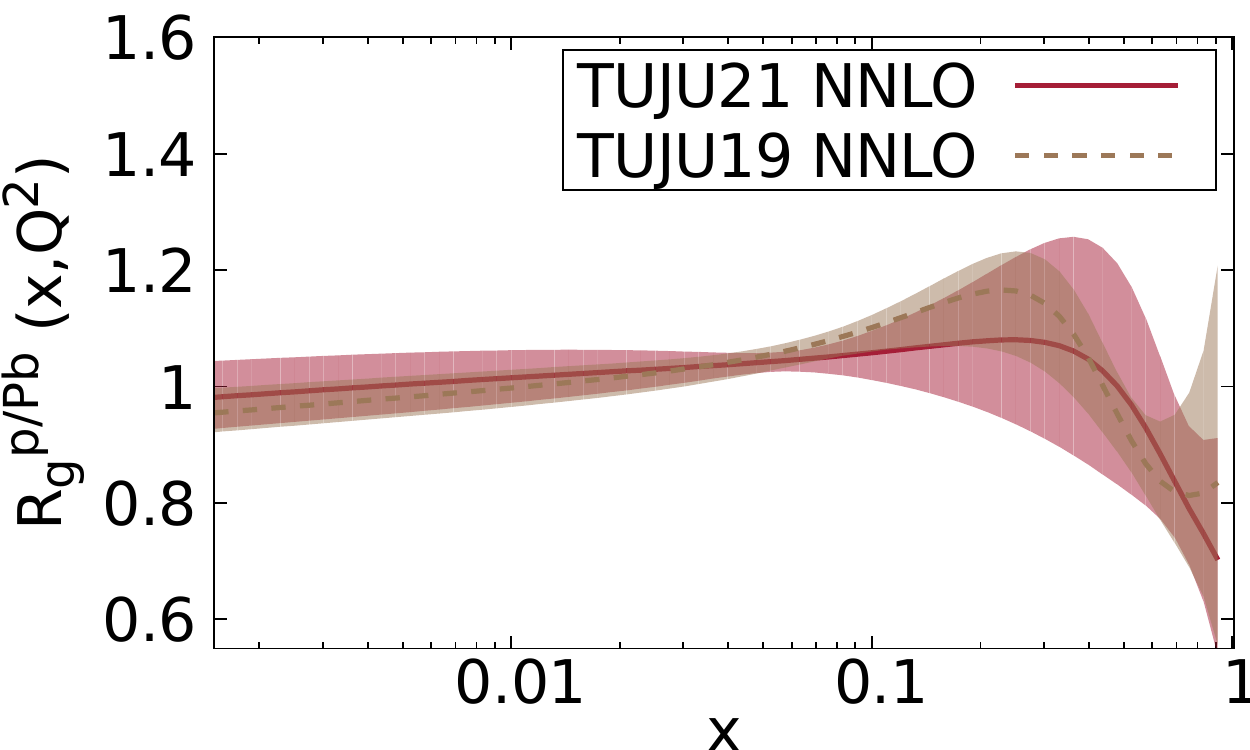}} 
          \subfigure{              
              \includegraphics[width=0.237\textwidth]{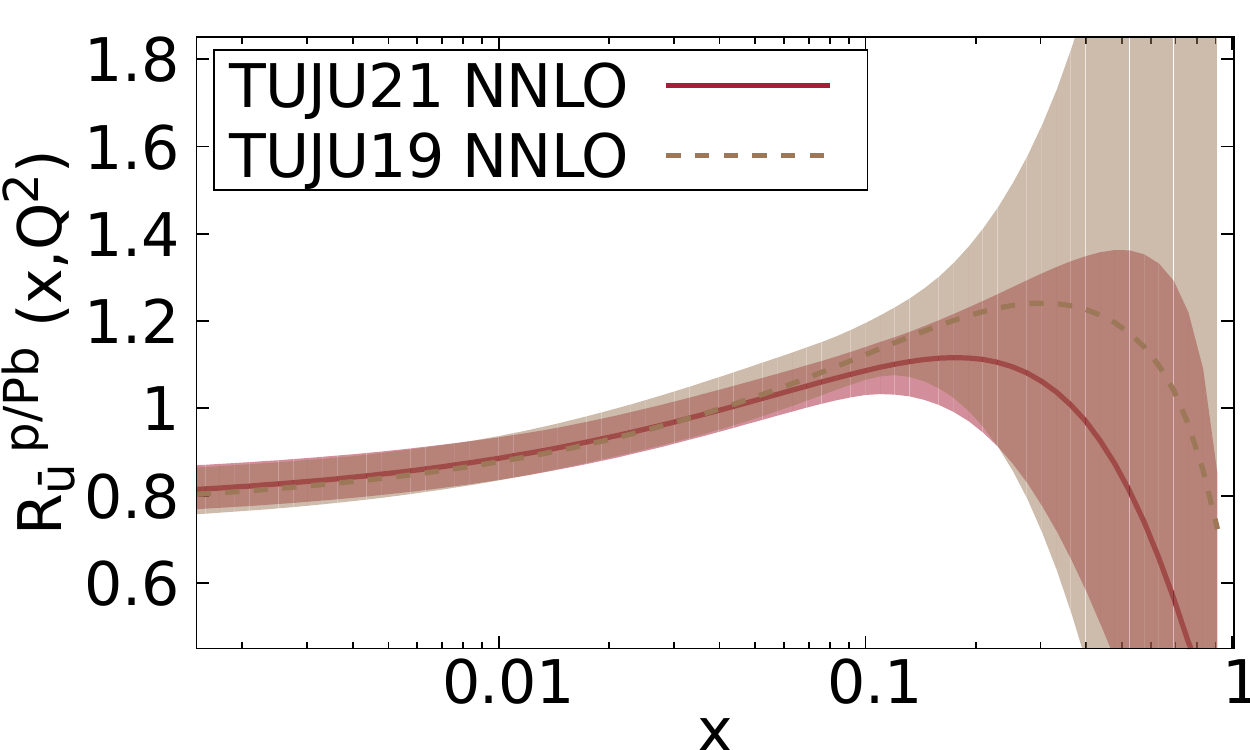}} 
          \subfigure{                        
              \includegraphics[width=0.237\textwidth]{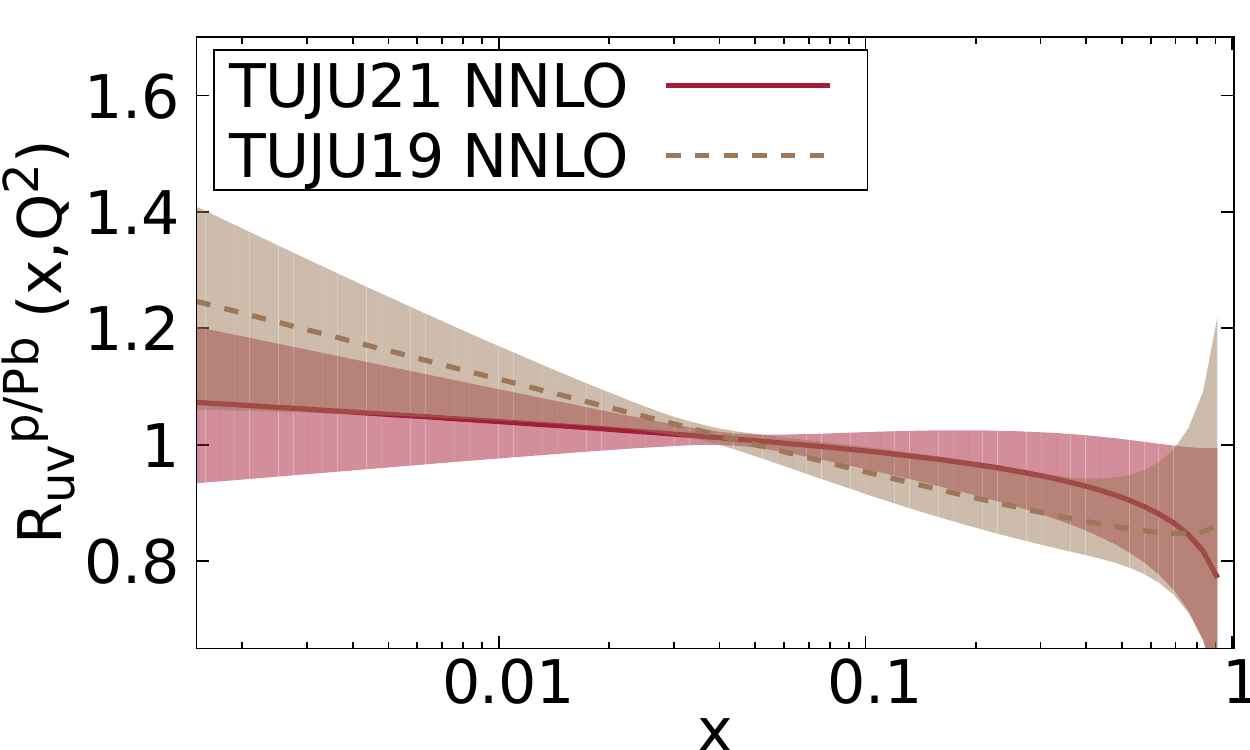}} 
          \subfigure{                                   
              \includegraphics[width=0.237\textwidth]{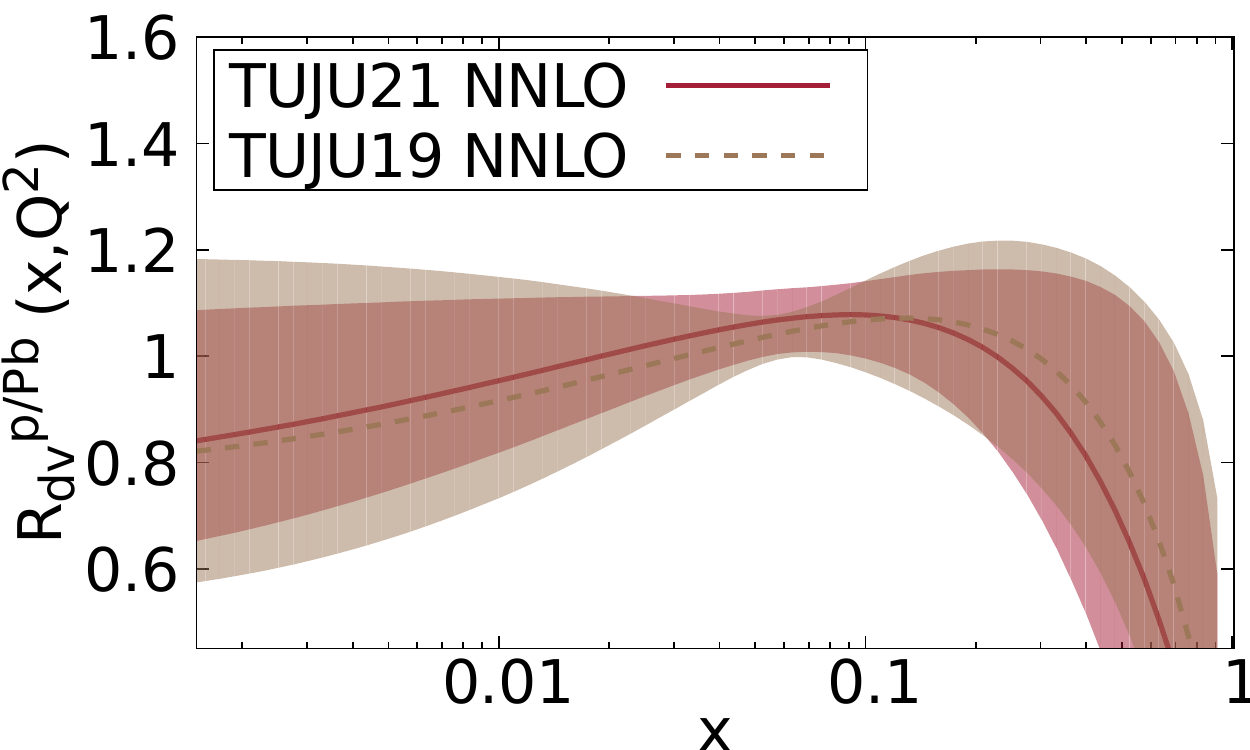}} 
          \end{center} 
    \caption{Same as for Fig. \ref{Pb_TUJU21_NLO}, but at NNLO.}
\label{Pb_TUJU21_NNLO}    
    \end{figure*}

The resulting nuclear PDFs, referred to as TUJU21, are presented in Figs. \ref{Pb_TUJU21_NLO} at NLO and \ref{Pb_TUJU21_NNLO} at NNLO for the bound-proton-in-lead-nucleus PDFs, including also ratios to our baseline free proton PDFs. We also compare to  the nPDFs of our previous DIS-only analysis TUJU19. At NLO, the largest differences between the two analyses occur for gluons and sea quarks. For gluons the small-$x$ suppression is significantly milder than in TUJU19, along with a slightly reduced uncertainty. Also, the strong antishadowing enhancement at intermediate values of $x$ we found previously is now tamed to a more moderate $\sim 10\%$ effect. At the initial scale of the fit the sea quark nPDFs are now slightly lower in the small-$x$ region, with a somewhat smaller uncertainty band, but have remained very similar at larger $x$. Because of the larger gluon nPDF in the updated fit, the sea quark distributions become larger at higher scales through scale evolution. Overall, the gluon uncertainties are likely still underestimated due to the rather rigid form of the input parameterization. For the valence quarks the resulting nPDFs are very similar as in our previous analysis, suggesting that the added EW-boson data do not provide significant constraints for the valence sector.

\begin{figure*}[tb!]
\begin{center}
\includegraphics[width=\textwidth]{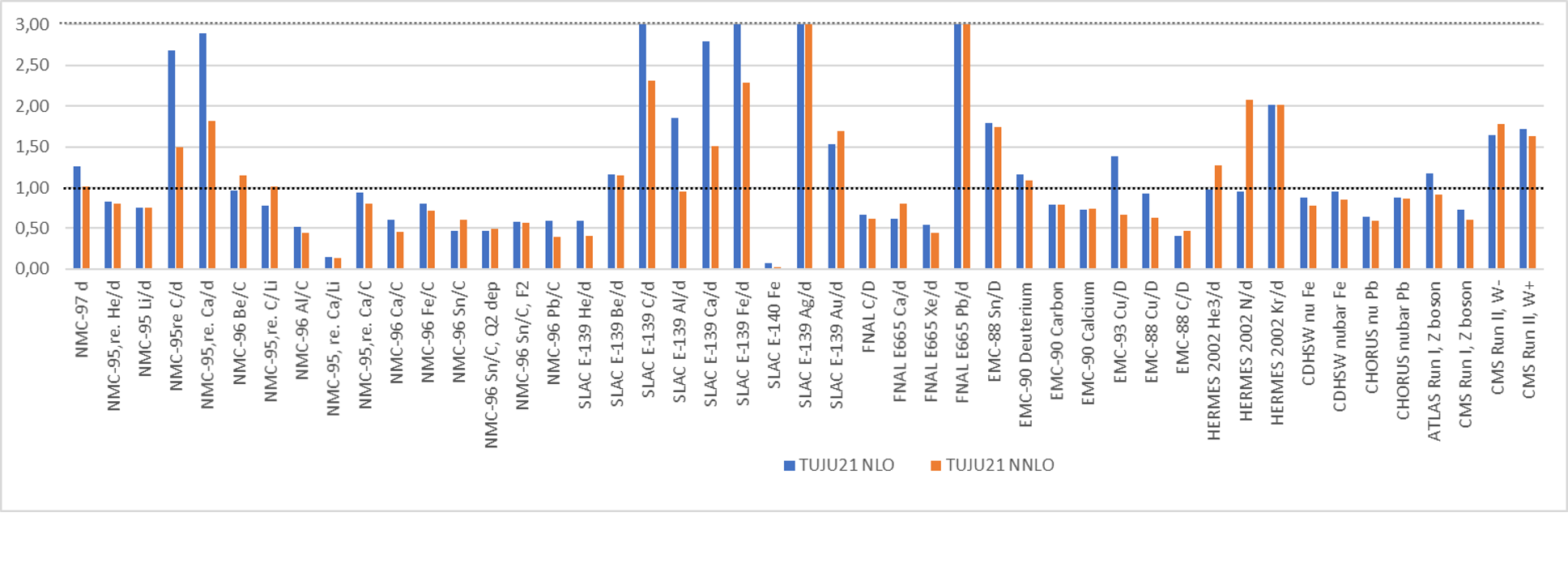}
\caption{Comparison of $\Chi^2$ values divided by the individual number of data points per data set, $N_{\mathrm{dp}}$, at NLO and NNLO. The ``ideal'' value $\Chi^2/N_{\mathrm{dp}}\,=\,1.0$ is marked by the horizontal black dotted line. The bars in the diagram corresponding to $\Chi^2/N_{\mathrm{dp}}\,> \,3.0$ have been truncated for the purpose of a clearer representation, which is symbolised by the dashed light-grey line. The newly included data for $Z$ and $W^{\pm}$ boson production from LHC Run I and Run II are shown on the far right-hand-side.}
\label{fig-Chi2}
\end{center}
\end{figure*}

At NNLO the changes with respect to our previous analysis are clearly milder. The uncertainties have now become slightly larger for gluons and smaller for sea quarks, but otherwise these are well consistent with our previous analysis. Also for the valence quarks the differences are small, with uncertainties slightly reduced. The previously observed opposite behavior of the nuclear modifications for $u_{\mathrm{v}}$ and $d_{\mathrm{v}}$ is now less pronounced. Even though there is no significant reduction in the resulting uncertainty bands, the mutual agreement between the nuclear effects found in our NLO and NNLO fits suggests that such effects are now better captured than in the DIS-only fit.

The parameters for the input distributions for our best fit of nPDFs are also collected in Appendix~\ref{app-pdf-params}. The error sets, covering the allowed modifications of each parameter within the quoted tolerance, are included as part of the resulting LHAPDF grids. In case of nuclear PDFs, the first 33 sets reflect the central result and the uncertainties in the nuclear PDF fit and the last 26 quantify the uncertainty in the underlying proton baseline analysis.

\subsection{Comparison to data}

An overview of the resulting $\Chi^2$ values, divided by the number of data points, $N_\mathrm{dp}$, is shown in Fig. \ref{fig-Chi2} for NLO and NNLO. Since the optimal values for the PDF parameters are obtained by minimizing $\Chi^2$, its value is an indicator for the quality of the fit, with $\Chi^2/N_\mathrm{dp} \approx 1$ in the optimal case. We have recalled the definition of $\Chi^2$ used in this work in Eq. (\ref{eq-chi2-scaled}); further details can be found in Ref.~\cite{Walt:2019slu}. Values above $\Chi^2/N_{\mathrm{dp}}>3.0$ have been truncated in Fig. \ref{fig-Chi2} for better representation, but the actual numbers are given in Table \ref{tab-expdata}.

\begin{figure*}[bt!]
     \begin{center}                           
          \subfigure{\includegraphics[width=0.4\textwidth]{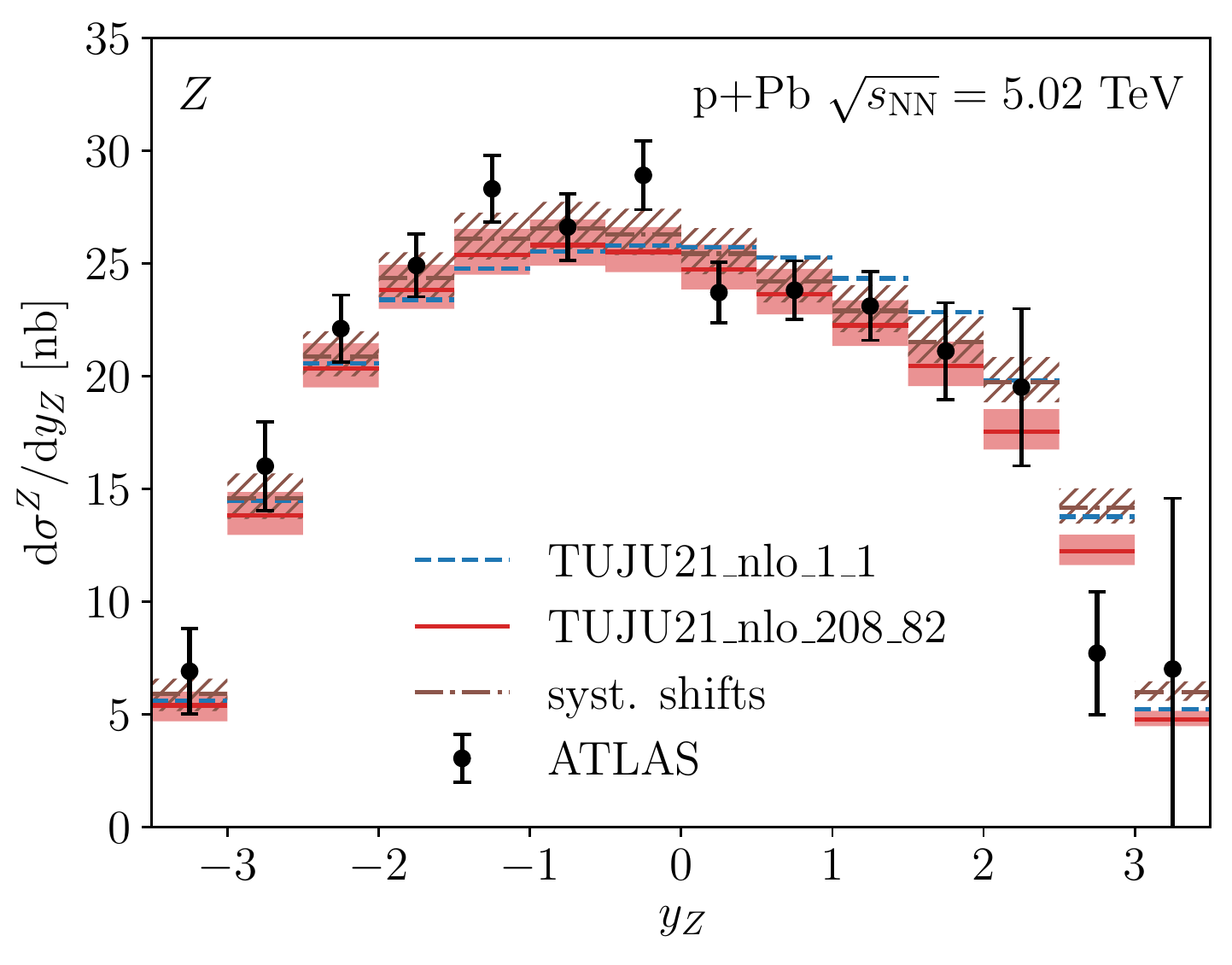}} 
          \subfigure{\includegraphics[width=0.4\textwidth]{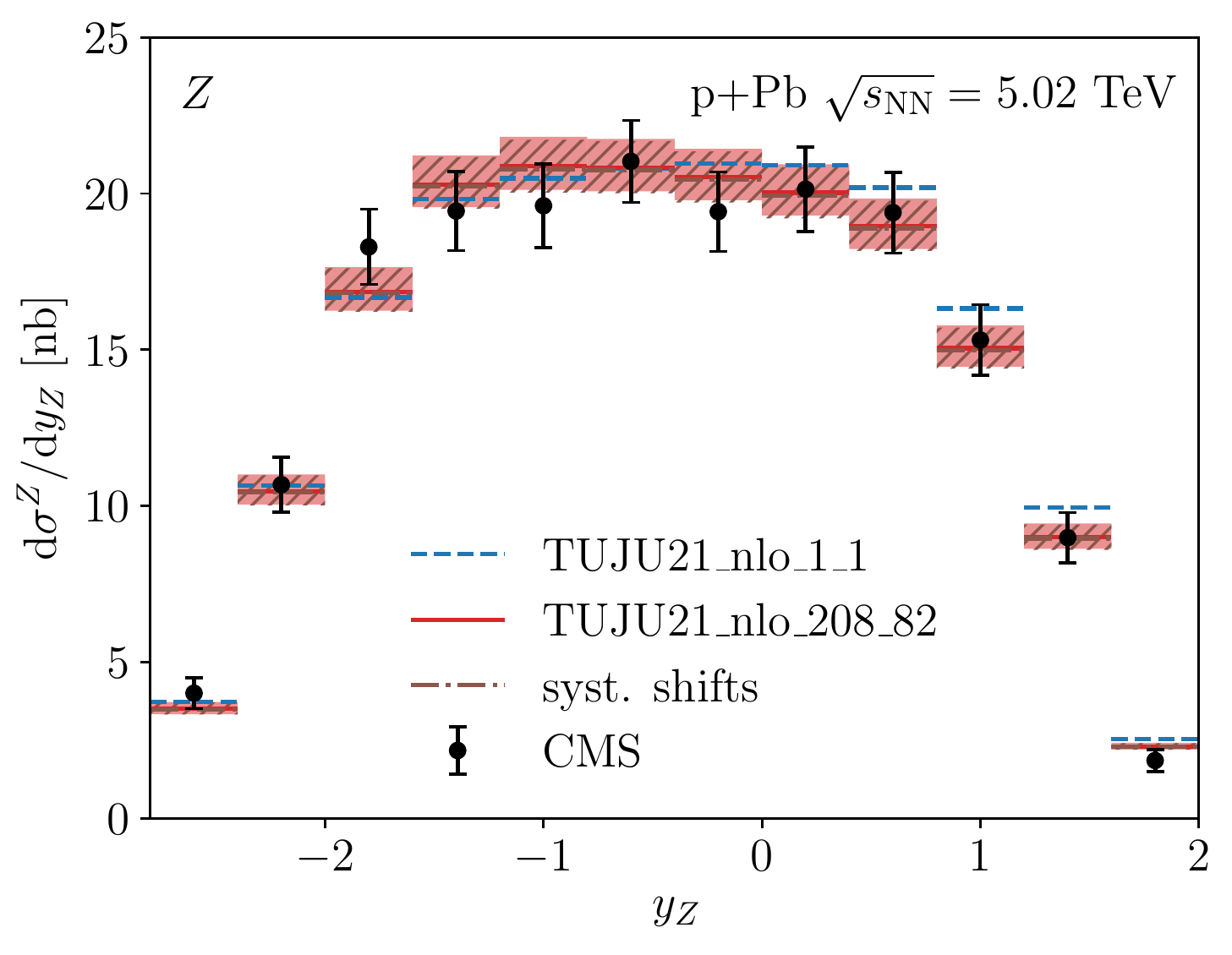}}
          \subfigure{\includegraphics[width=0.4\textwidth]{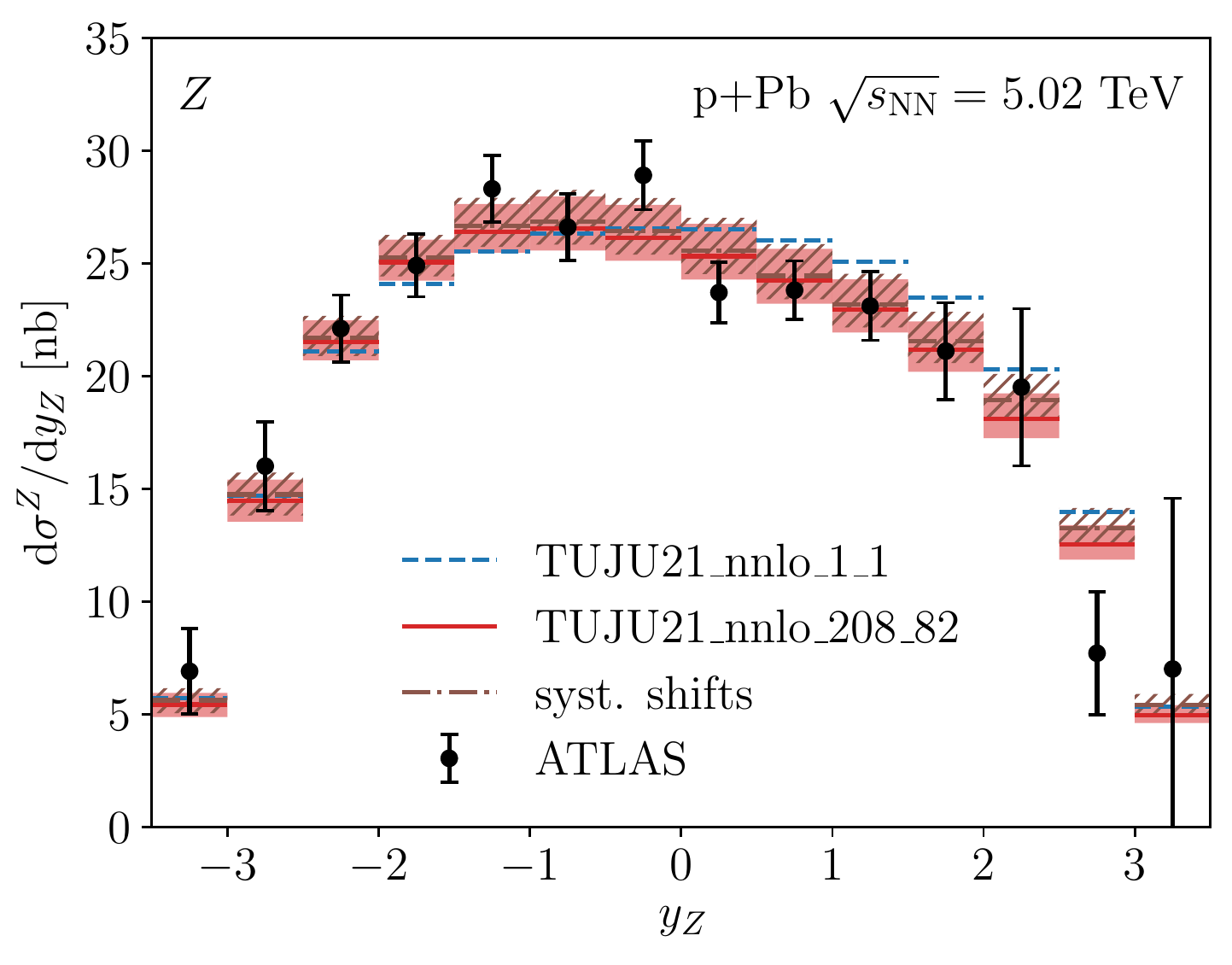}}
          \subfigure{\includegraphics[width=0.4\textwidth]{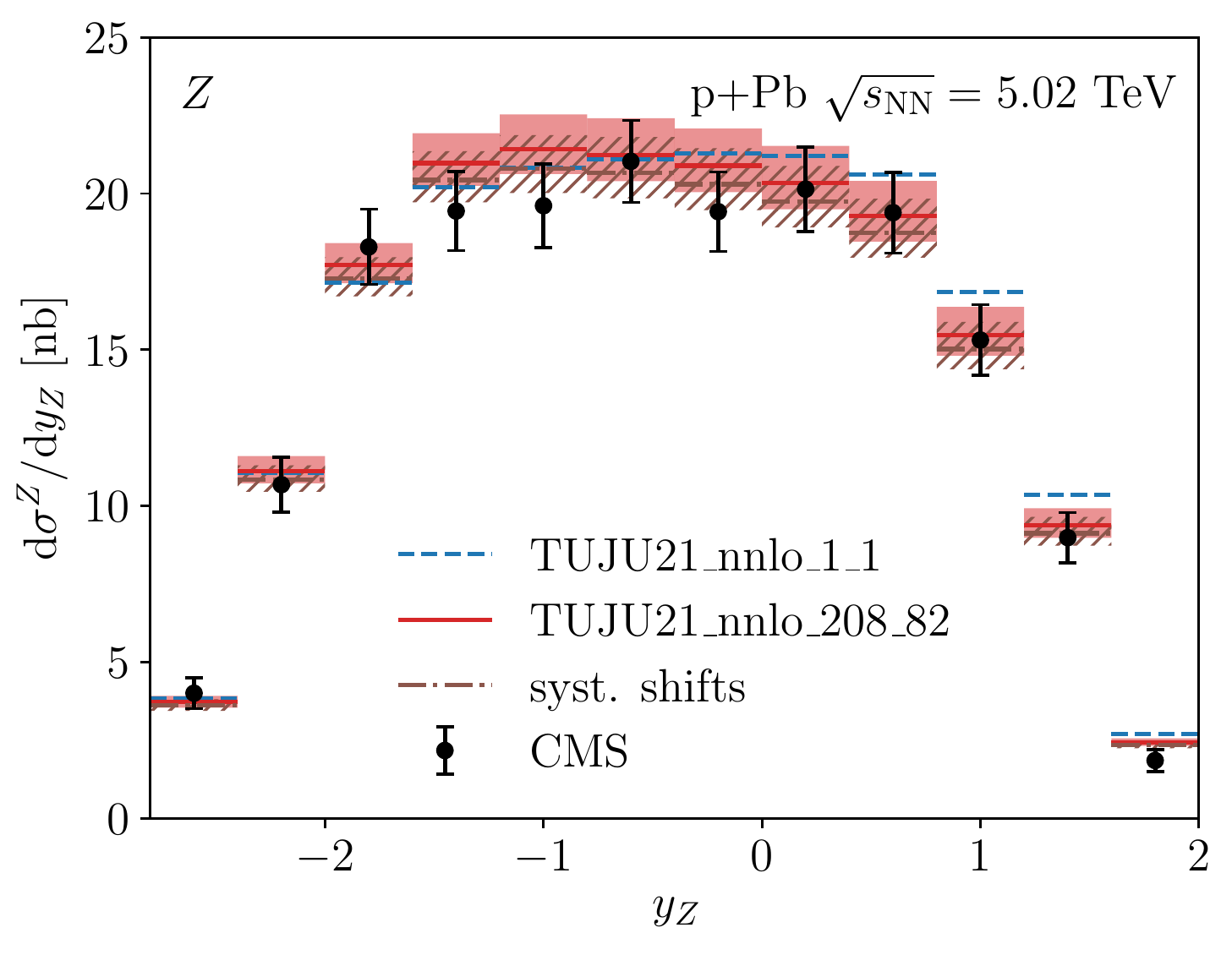}}
          \end{center} 
    \caption{Comparison to $Z$-boson production data from the ATLAS \cite{Aad:2015gta} and CMS experiments \cite{Khachatryan:2015pzs} in p+Pb collisions in LHC Run~I (solid and dotted). For the solid lines, we just use our fitted nuclear PDFs to compute the cross section, while for the brown dash-dotted lines with hatched uncertainty band we also include the normalization shift of the theoretical cross section that we obtained during the fitting procedure and that resulted in the optimal $\Chi^2$ value. For illustration, we also show the cross sections computed by using free-proton PDFs for the lead nucleus (dashed lines).}
\label{fig-Zboson}
\end{figure*}

\begin{figure*}[bt!]
     \begin{center}                           
          \subfigure{\includegraphics[width=0.4\textwidth]{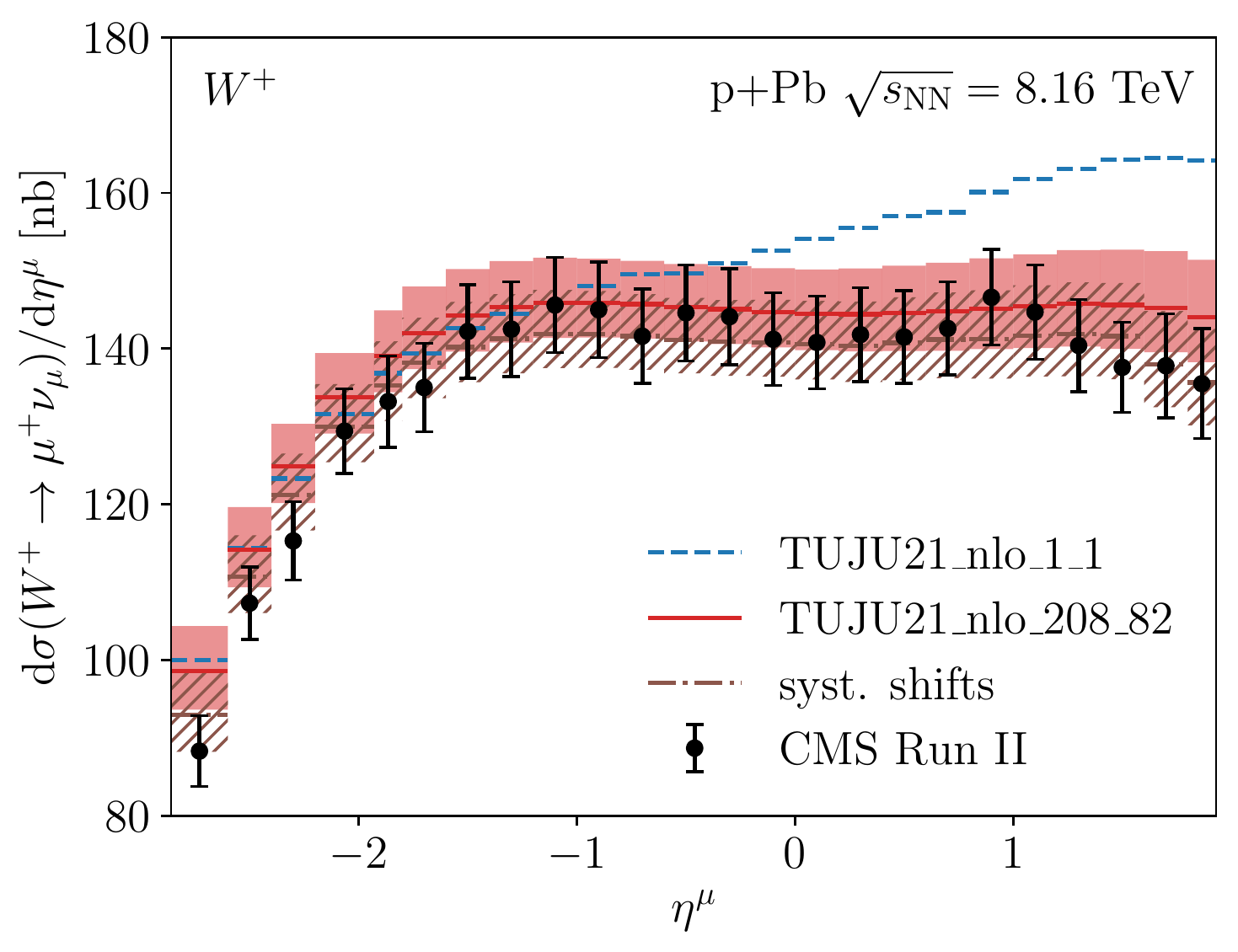}} 
          \subfigure{\includegraphics[width=0.4\textwidth]{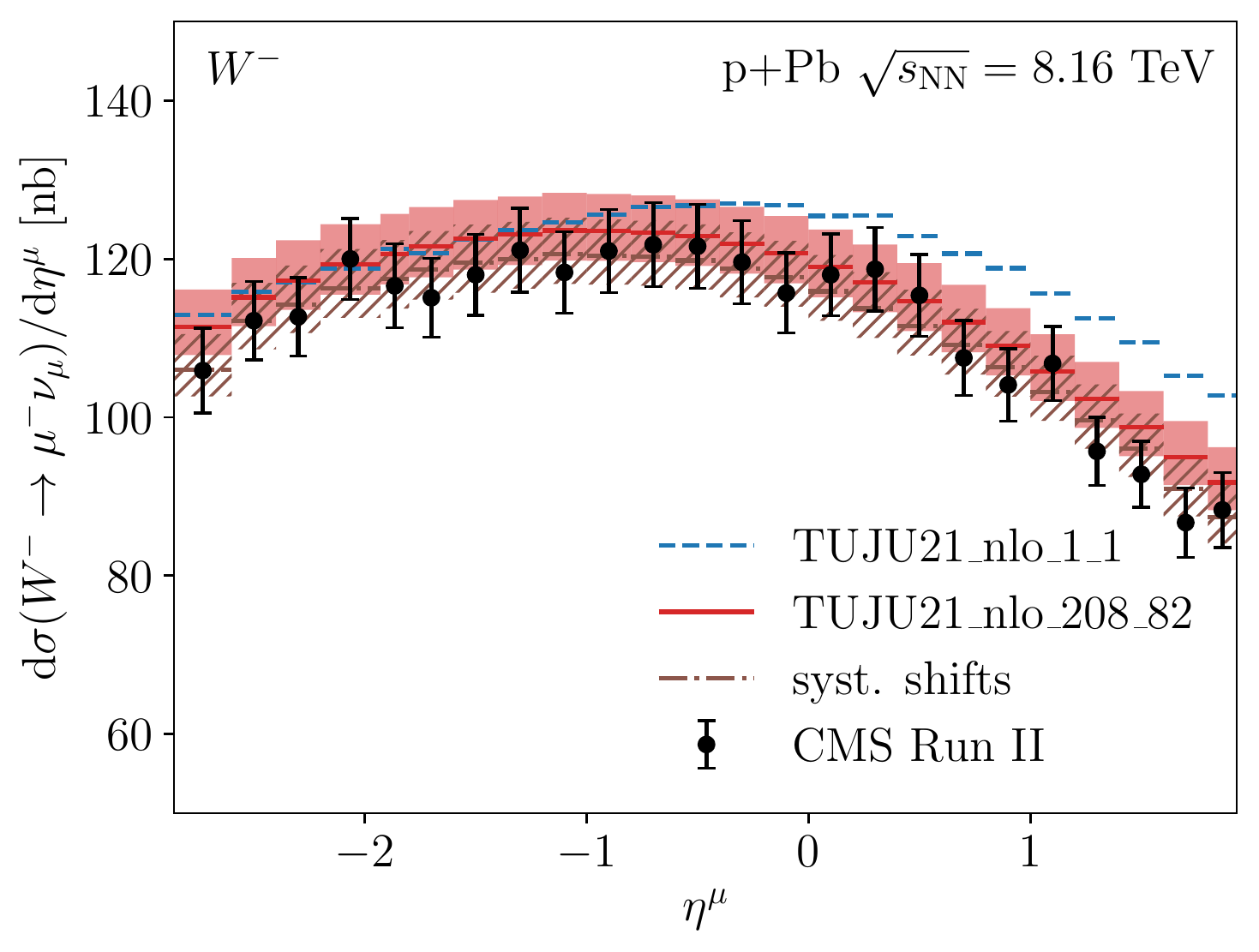}}
          \subfigure{\includegraphics[width=0.4\textwidth]{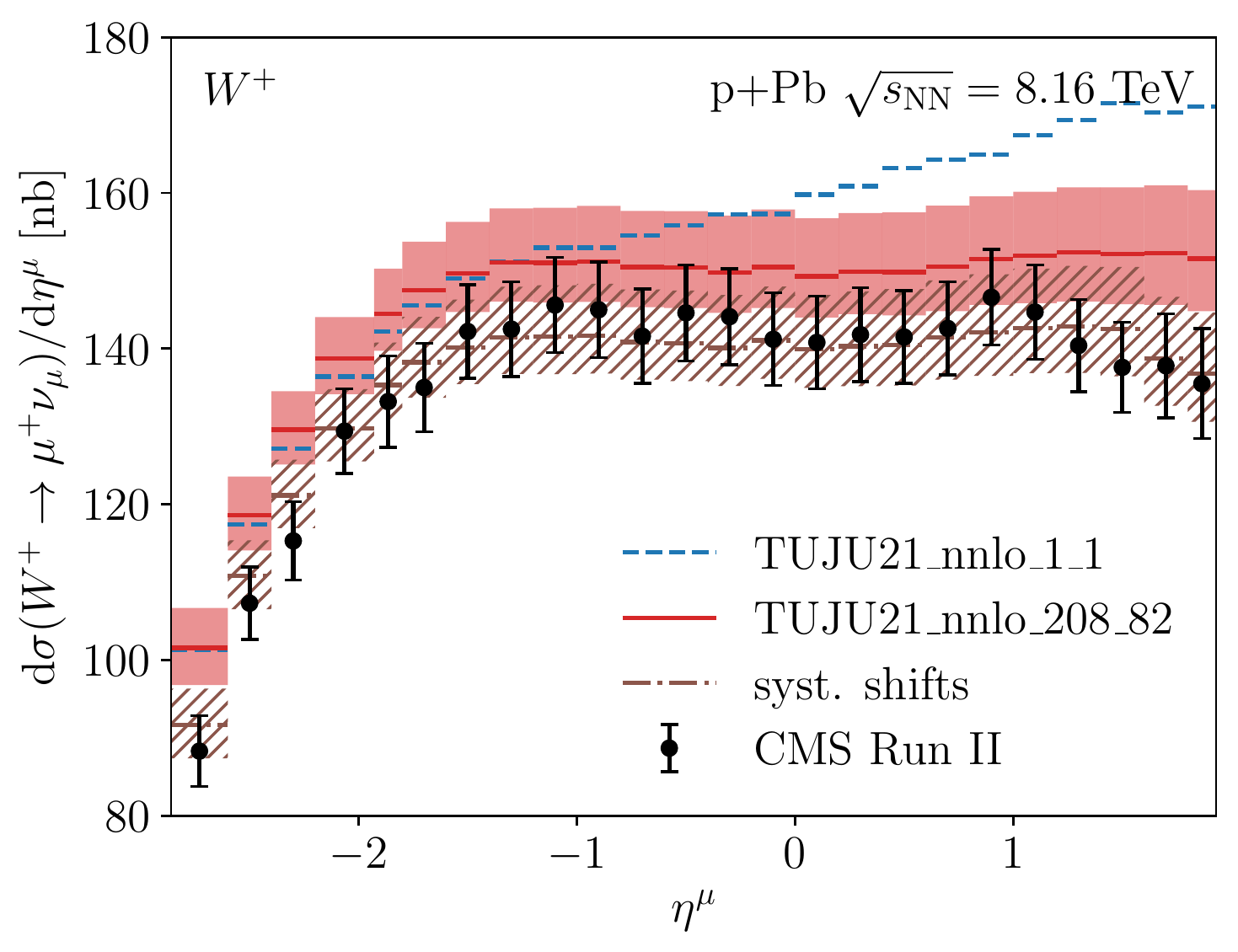}} 
          \subfigure{\includegraphics[width=0.4\textwidth]{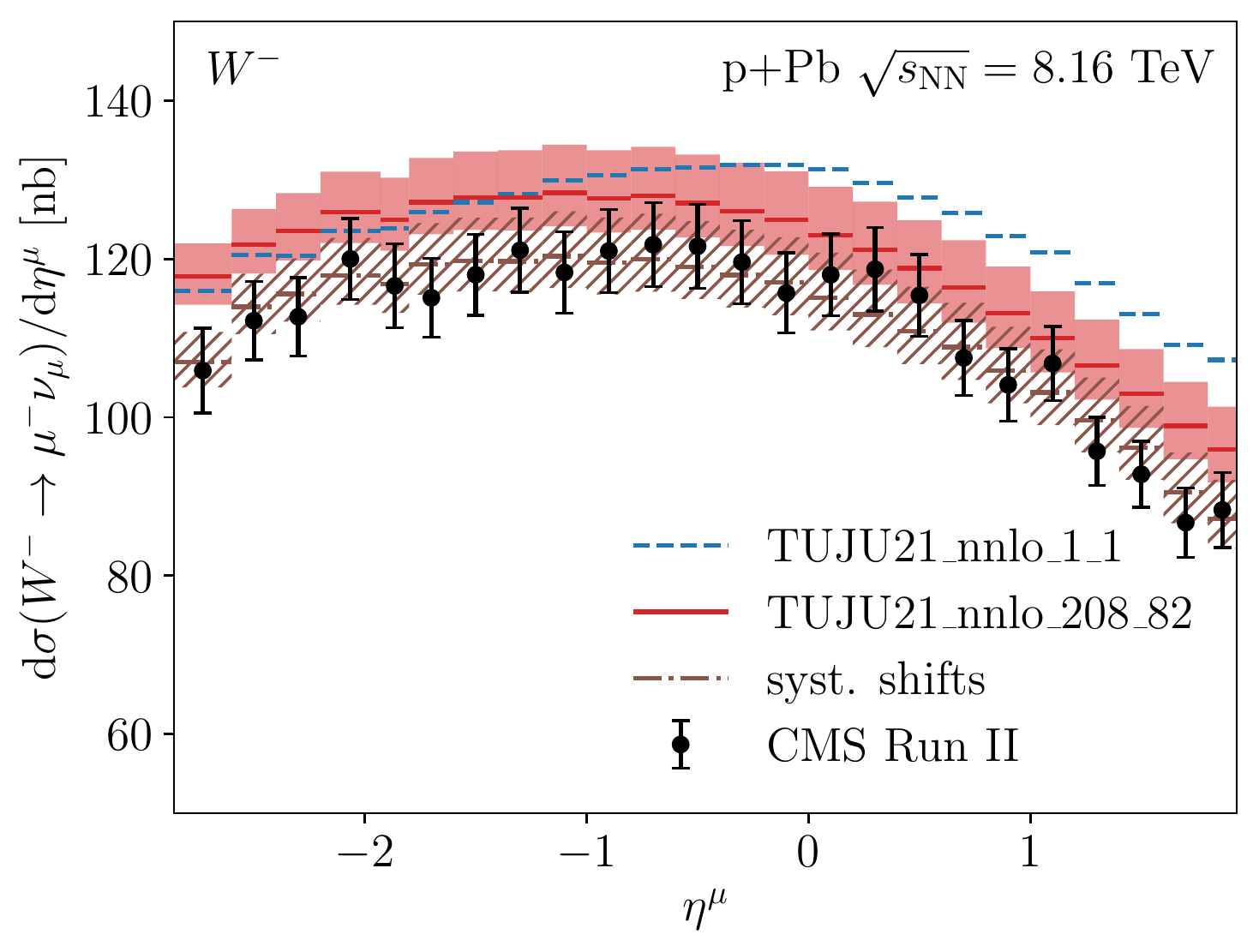}}
          \end{center}
    \caption{Comparison to $W^{\pm}$-boson production data obtained by the CMS experiment \cite{CMS:2019leu} in p+Pb collisions in LHC Run~II. The lines are as in Fig.~\ref{fig-Zboson}.}
\label{fig-Wboson}    
    \end{figure*}

\begin{figure*}[bt!]
     \begin{center}                           
          \subfigure{\includegraphics[width=0.4\textwidth]{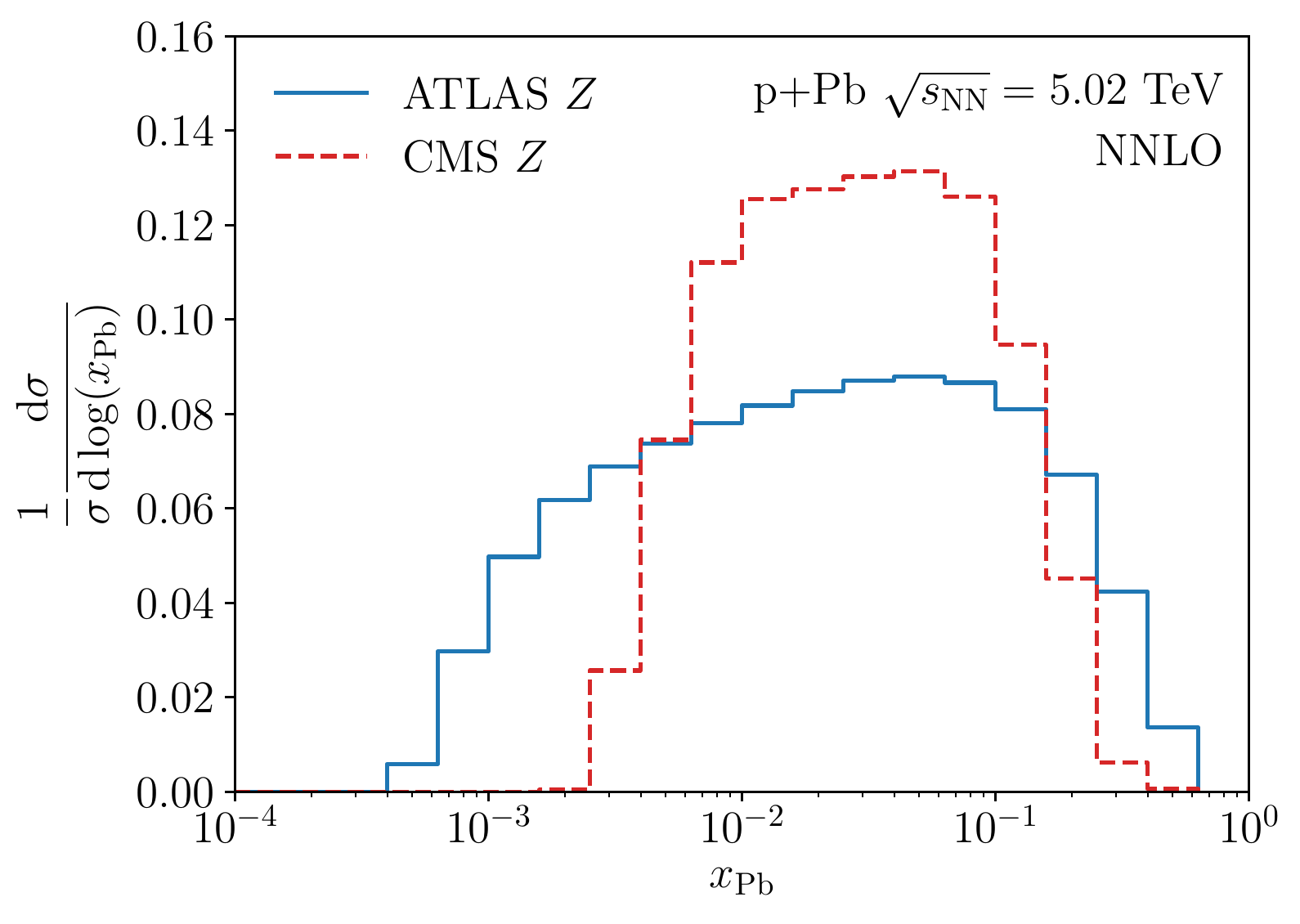}} 
          \subfigure{\includegraphics[width=0.4\textwidth]{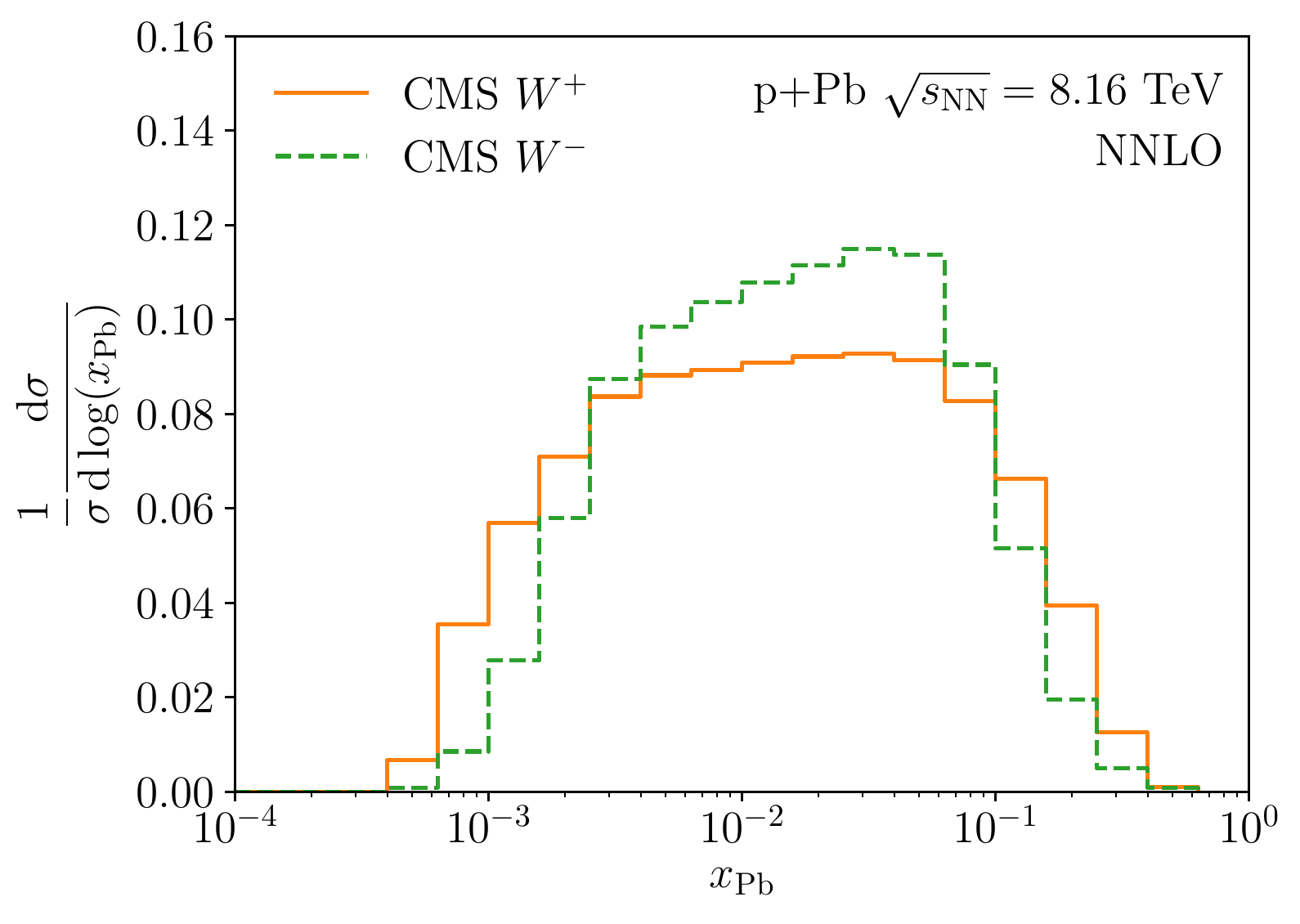}}
      \end{center} 
    \caption{Distribution of Pb momentum fraction ($x_{\mathrm{Pb}}$) values probed by the EW-boson production data in p+Pb collisions in 
    Run I for $Z$ bosons and Run II for $W$ bosons, obtained for our NNLO calculation.}
\label{fig-WZboson-data}    
\end{figure*}

As for the (neutrino) DIS data the fit results are very similar to those shown in our earlier analysis, we limit in the present paper the comparisons to experimental data to the new EW boson data. Our results are shown in Figs.~\ref{fig-Zboson} and \ref{fig-Wboson}. In all cases the experimental error bars are the quadratic combinations of statistical and systematical (correlated and uncorrelated) uncertainties. Figure~\ref{fig-Zboson} shows the comparisons to $Z$ boson production data measured by ATLAS \cite{Aad:2015gta} and CMS \cite{Khachatryan:2015pzs} in the p+Pb run at $\sqrt{s_{\mathrm{NN}}} = 5.02~\text{TeV}$ as a function of $Z$ boson rapidity, $y_Z$, in the center-of-mass frame, at NLO and NNLO. In each case we find a good fit, although at NLO the comparison to the ATLAS data results in $\Chi^2/N_{\mathrm{dp}}$ slightly above unity. The Pb momentum fraction ($x_{\mathrm{Pb}}$) regions probed by the data can be estimated using LO kinematics for the process, see Eq.~(\ref{eq-dy-variables2}). To exhibit them in the context of our NNLO calculation, we plot in Fig. \ref{fig-WZboson-data} the normalized NNLO cross section as a function of $x_{\mathrm{Pb}}$, integrated over the rapidity ranges relevant for ATLAS and CMS. We notice that the ATLAS data cover a somewhat broader range in $x_{\mathrm{Pb}}$ due to broader acceptance in $y_Z$ and larger fiducial phase-space. In case of CMS the decay leptons are accepted only if $p_{\mathrm{T}}^l > 20~\text{GeV}$ and $|\eta_{\mathrm{lab}}^l| < 2.4$ and the studied mass window for the accepted dileptons is $60 < m_Z < 120~\text{GeV}$. In the ATLAS case the data have been corrected for the limited acceptance in the lepton reconstruction, so the only remaining kinematical cut is the mass window of the dilepton pair, $66 < m_Z < 116~\text{GeV}$, explaining also the larger cross section compared to the CMS data. Using Eq.~(\ref{eq-dy-variables2}) and turning back to Fig.~\ref{fig-Zboson}, we see that at forward rapidities with respect to the proton beam ($\eta > 0$, corresponding to $x_{\mathrm{Pb}} \lesssim 0.02$), both data sets clearly favor a suppression of the nPDFs relative to the proton PDFs that is well captured by the fits. In the backward direction ($\eta < 0$, $x_{\mathrm{Pb}} \gtrsim 0.02$) both data sets may have a slight tendency toward an enhancement (anti-shadowing) of p+Pb over p+p. Overall, the NNLO results are slightly higher than the NLO ones, and the shifts by the systematic uncertainties needed to obtain optimal agreement with the data are small and partly compensate for the differences between NLO and NNLO.

The CMS data for $W^{\pm}$ boson production from the p+Pb run at $\sqrt{s_{\mathrm{NN}}} = 8.16~\text{TeV}$ offer a significantly increased precision compared to the earlier data from Run~I. As the measured cross section is now given as a function of the pseudorapidity of the charged decay muon, $\eta^{\mu}$, the LO kinematics does not provide an immediate estimate for the kinematic reach of the data. Nevertheless, in Fig. \ref{fig-WZboson-data} we plot the kinematic reach in $x_{\mathrm{Pb}}$ for $W^+$ and $W^-$ production. One can see that there is sensitivity down to $x_{\mathrm{Pb}}\sim 10^{-3}$. Figure \ref{fig-Wboson} shows the comparisons of our NLO and NNLO $W^+$ and $W^-$ cross section to the ATLAS and CMS data. Again we also compute the cross sections using free-proton PDFs instead of the ones for lead. We find very clear suppression with respect to this proton baseline at $\eta^{\mu} > 0$, which is well in line with the shadowing observed in the resulting nuclear PDFs. At $\eta^{\mu} < 0$ we find only modest effects from our nuclear PDFs, which partly follows from rather modest anti-shadowing and partly since the cross section can get contributions both from regions of $x_{\mathrm{Pb}}$ where nPDF effects provide enhancement or suppression. The shapes of the resulting cross sections are well in line with the data for $W^+$ and $W^-$. This holds for both, NLO and NNLO, but in case of NNLO somewhat larger shifts ($\sim 7\%$) are required to match the data than at NLO ($\sim 3\%$), both being below or around the quoted correlated uncertainties. In all cases a reasonably good agreement, $\Chi^2/N_{\mathrm{dp}} \sim 1.7$, is observed.
    
In our previous TUJU19 analysis which included only DIS data for the proton baseline and the nuclear PDFs, we found that the resulting $\Chi^2/N_{\mathrm{dp}}$ values were rather similar for the NLO and the NNLO analysis. Interestingly, when we add the EW boson data into the analysis, the resulting $\Chi^2/N_{\mathrm{dp}}$ values are clearly smaller at NNLO than at NLO. For the proton baseline fit we obtain $\Chi^2/N_{\mathrm{dp}} = 1.30$ at NLO and 1.24 at NNLO. In case of the nuclear PDF analysis, the resulting values are 0.94 (NLO) and 0.84 (NNLO). This demonstrates that even though there are still significant uncertainties in the nuclear effects in large parts of the relevant kinematical regions, it will be necessary to go beyond NLO in such analyses when more high-precision LHC data are included. It is, however, still possible that using a different parameterization or employing only normalized nuclear cross sections in the fit could result in more similar $\Chi^2/N_{\mathrm{dp}}$ values at NLO and NNLO.

\subsection{Comparisons to other recent nPDF sets}

\begin{figure*}[bt!]
     \begin{center}                           
          \subfigure{\includegraphics[width=0.32\textwidth]{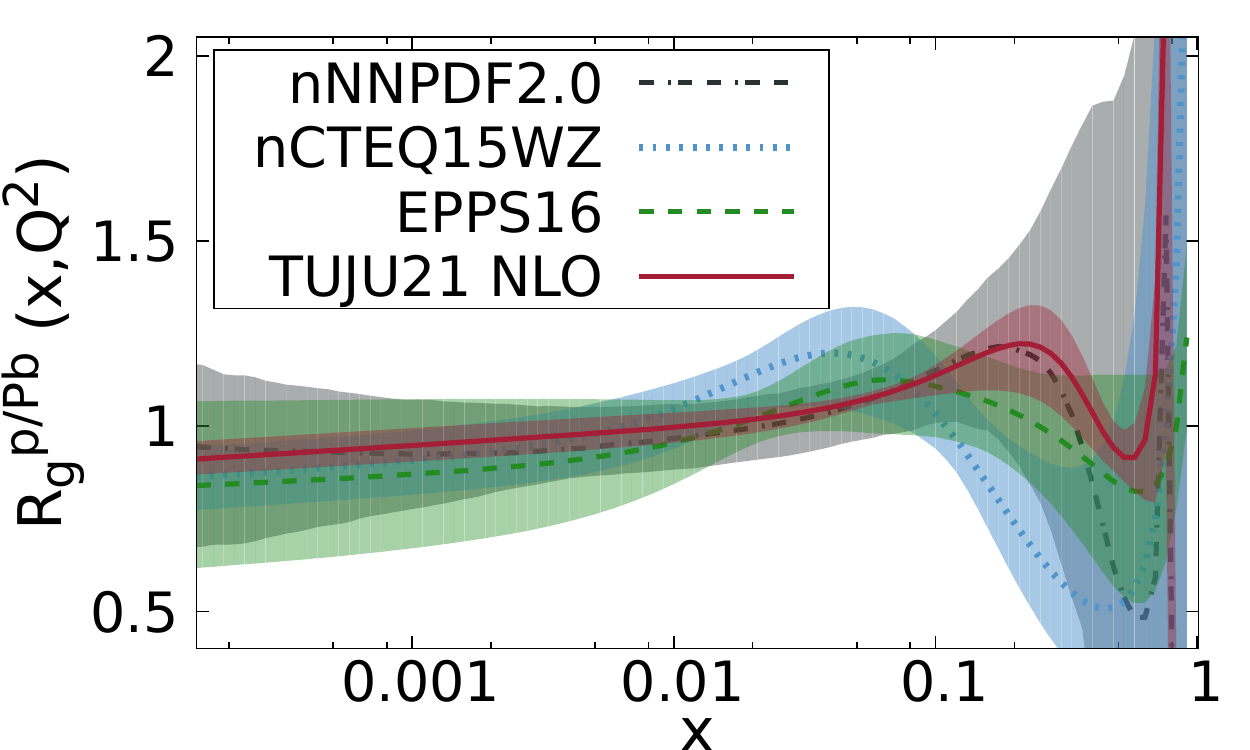}} 
          \subfigure{\includegraphics[width=0.32\textwidth]{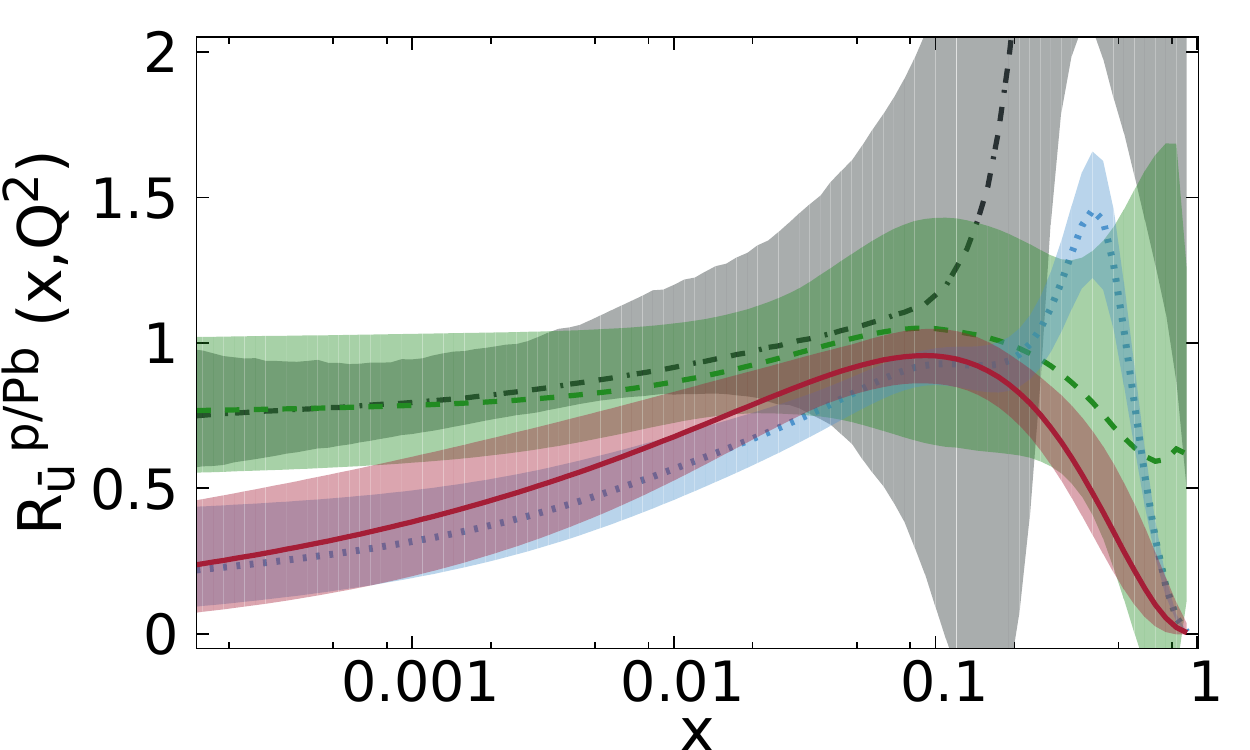}}
          \subfigure{\includegraphics[width=0.32\textwidth]{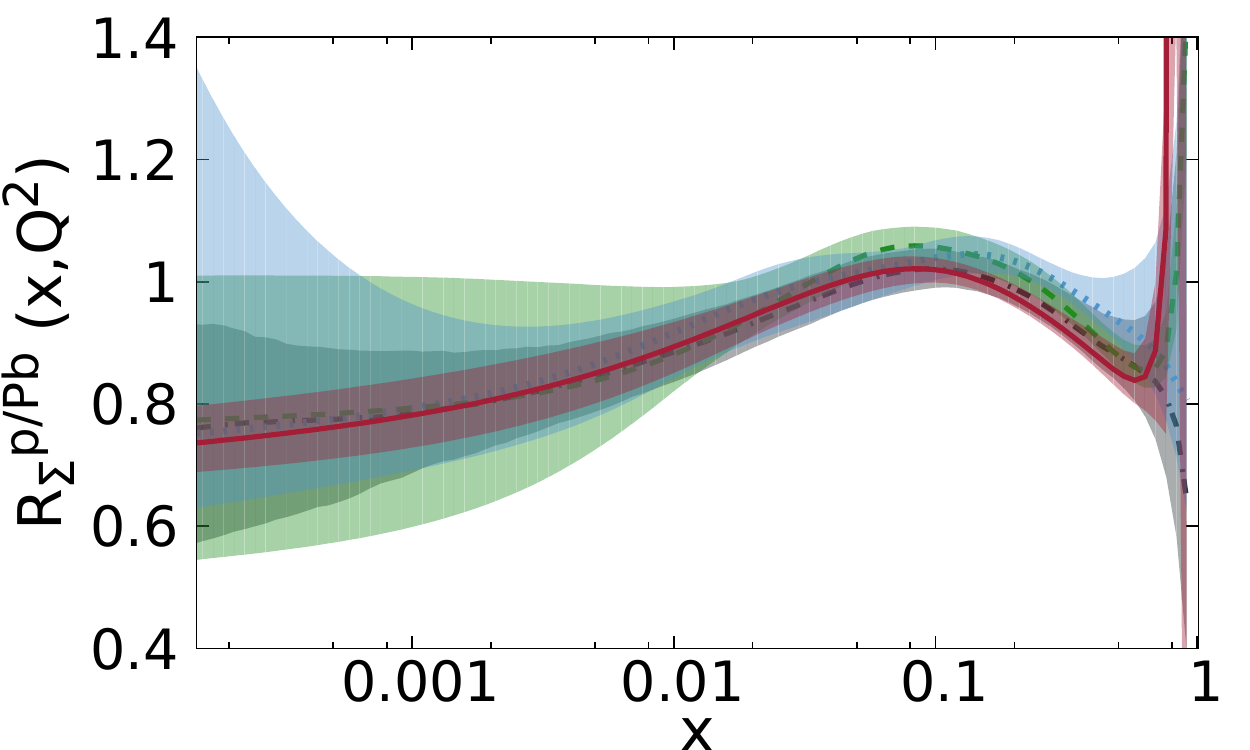}} 
          \subfigure{\includegraphics[width=0.32\textwidth]{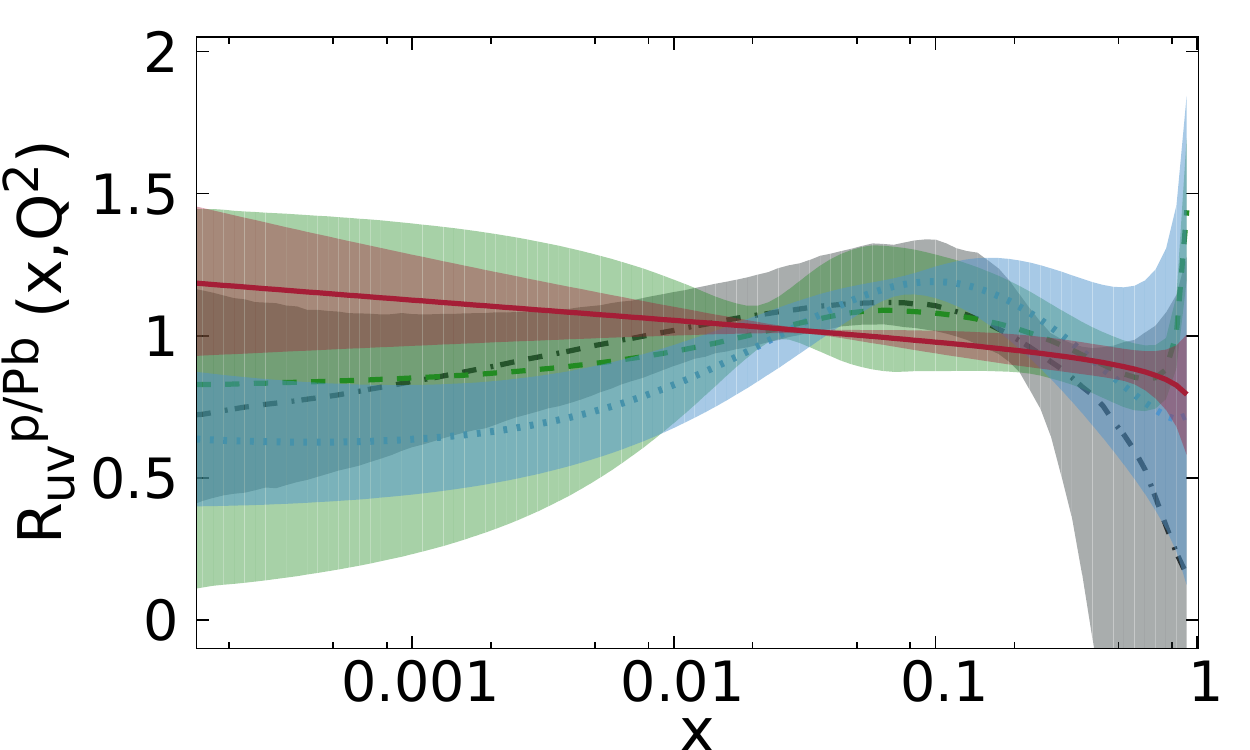}}
          \subfigure{\includegraphics[width=0.32\textwidth]{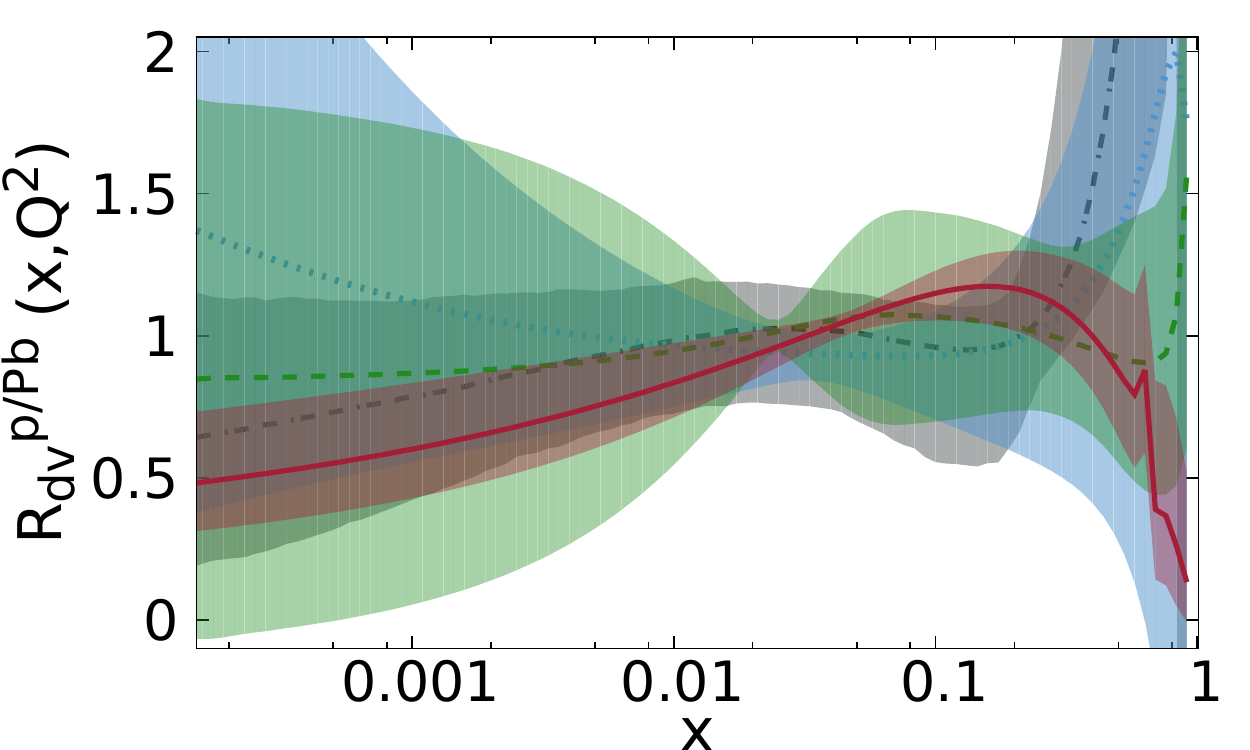}}
          \subfigure{\includegraphics[width=0.32\textwidth]{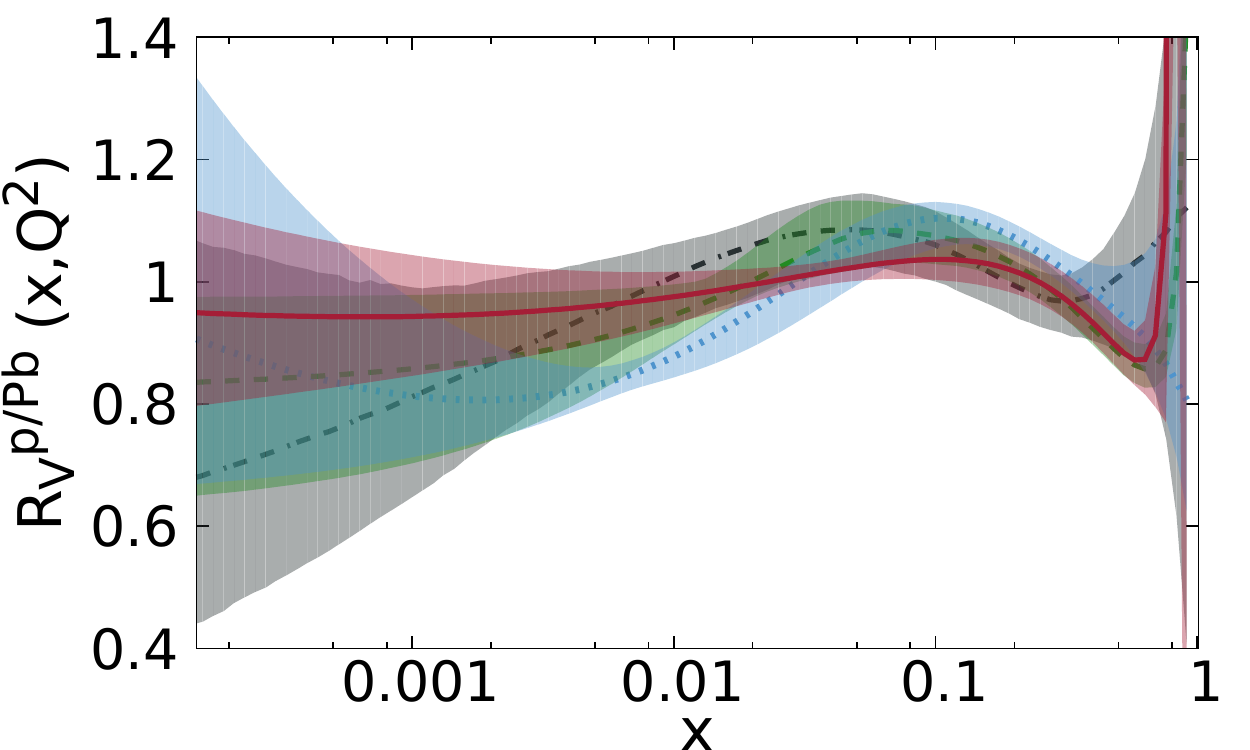}} 
          \end{center} 
\caption{Ratios $R_i^{\mathrm{p/Pb}}$ of PDFs in a proton bound in a lead nucleus compared to the PDFs in a free proton for TUJU21, nCTEQ15wz \cite{Kusina:2020lyz}, EPPS16 \cite{Eskola:2016oht} and nNNPDF2.0 \cite{AbdulKhalek:2020yuc}, shown at $Q^2=100\,\mathrm{GeV}^2$.}
\label{fig-nPDF-comparisonQ2-NLO-ALL}
\end{figure*}

We next compare the nuclear PDFs obtained in this work to those of other nPDF analyses. Figure \ref{fig-nPDF-comparisonQ2-NLO-ALL} presents the nuclear modification ratios for $g,\bar{u},u_{\mathrm{v}},d_{\mathrm{v}},\mathrm{{\mathrm{v}}}$ (total valence) and $\Sigma\equiv u+\bar{u}+d+\bar{d}+s+\bar{s}$ at $Q^2 = 100~\text{GeV}^2$ for bound protons in a lead nucleus from EPPS16~\cite{Eskola:2016oht}, nCTEQ15wz~\cite{Kusina:2020lyz}, nNNPDF2.0~\cite{AbdulKhalek:2020yuc} and from this work, TUJU21. For our comparisons we stick to the more recent sets of nPDFs; similar comparisons could be carried out to earlier sets such as EPS09~\cite{Eskola:2009uj} or DSSZ~\cite{deFlorian:2011fp}. Notice that the uncertainty bands we provide in the figure do not include the uncertainty of the respective proton baselines. In the case of nNNPDF2.0 the proton PDF uncertainty would even affect the central result of the fit as the latter is obtained as the median of the ratios of replica sets. All analyses shown include some data from p+Pb collisions at the LHC, including EW boson production and also dijet production in case of EPPS16. Overall, some level of qualitative agreement can be observed within the given uncertainties, but notable differences persist even now that data from the LHC have been included. For example, a tendency for opposite valence quark modifications for $u$ and $d$ are found for nCTEQ15wz and TUJU21, whereas the wide uncertainty bands in nNNPDF2.0 and EPPS16 cover all possibilities, reflecting the fact that individual valence flavors are not well constrained by the available data as only certain combinations of them can be accessed in nuclear collisions. By contrast, the total valence nuclear modification is rather well under control for $0.01 < x < 0.5$. All the considered analyses support small-$x$ gluon shadowing, though with varying uncertainty. Also gluon anti-shadowing, enhancement somewhere around $0.05 < x < 0.5$, is present in all analyses, but the precise location in $x$ varies among the sets, with TUJU21 and nNNPDF2.0 being more in line. The largest differences arise for the $\bar{u}$ distribution. TUJU21 and nCTEQ15wz show a strong small-$x$ shadowing, whereas EPPS16 and nNNPDF2.0 have only modest suppression at $x<0.01$. The reason for these differences is likely associated with the different treatment of flavor dependence of sea quarks in the various analyses. We note that when considering the nuclear effects for the singlet distribution, $\Sigma$, all analyses are very well in line with each other, and also the uncertainties are well matched for $0.01 < x < 0.5$. 

\begin{figure*}[tb!]
     \begin{center}
          \subfigure{\includegraphics[width=0.237\textwidth]{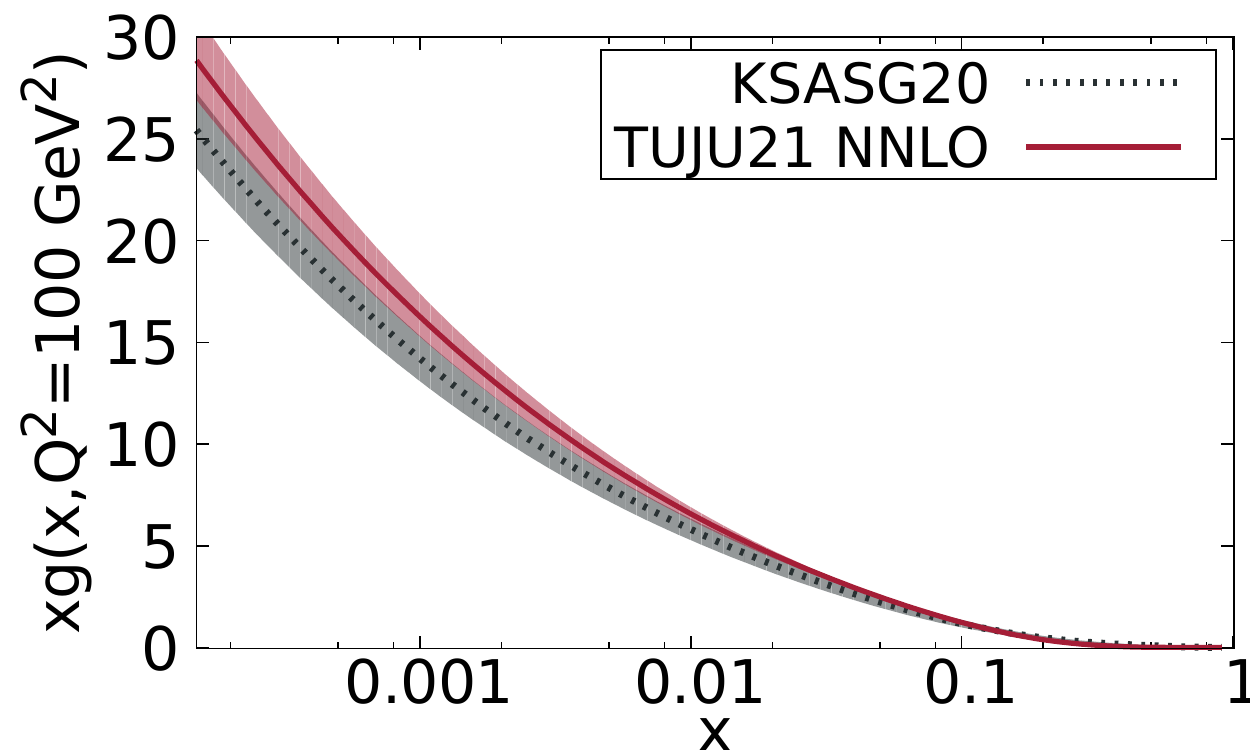}}
          \subfigure{\includegraphics[width=0.237\textwidth]{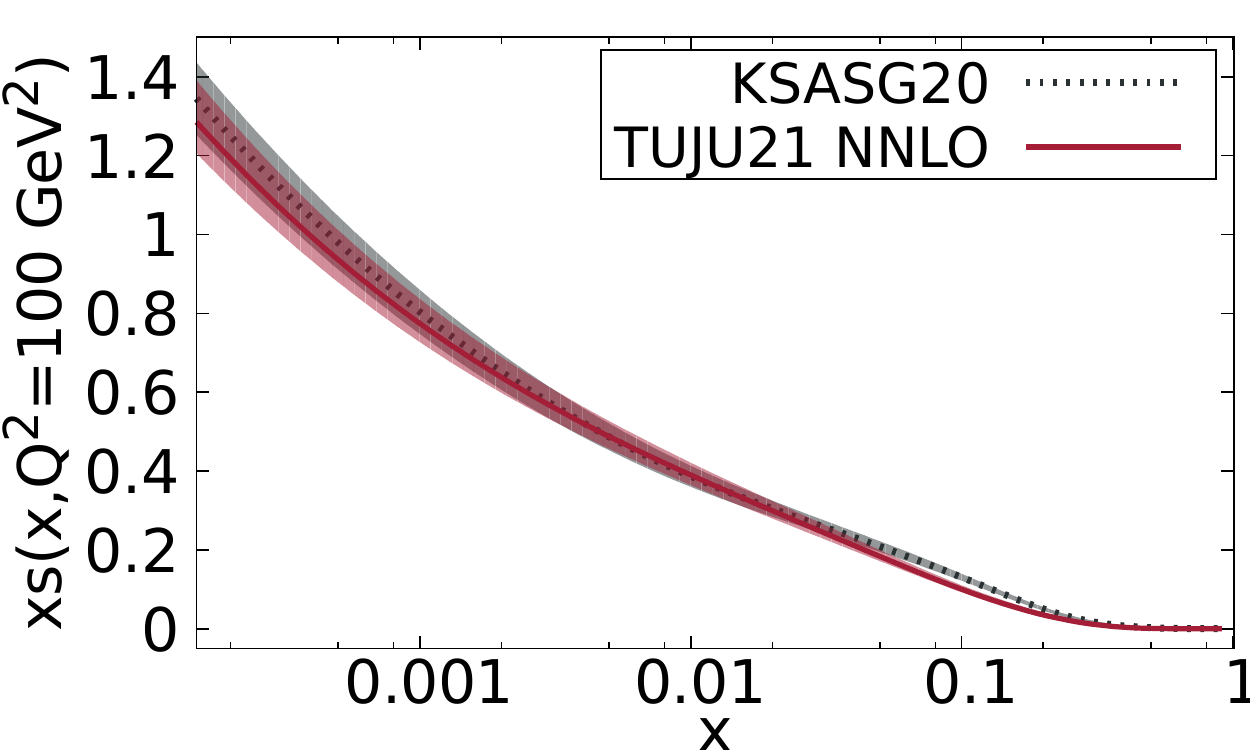}}
          \subfigure{\includegraphics[width=0.237\textwidth]{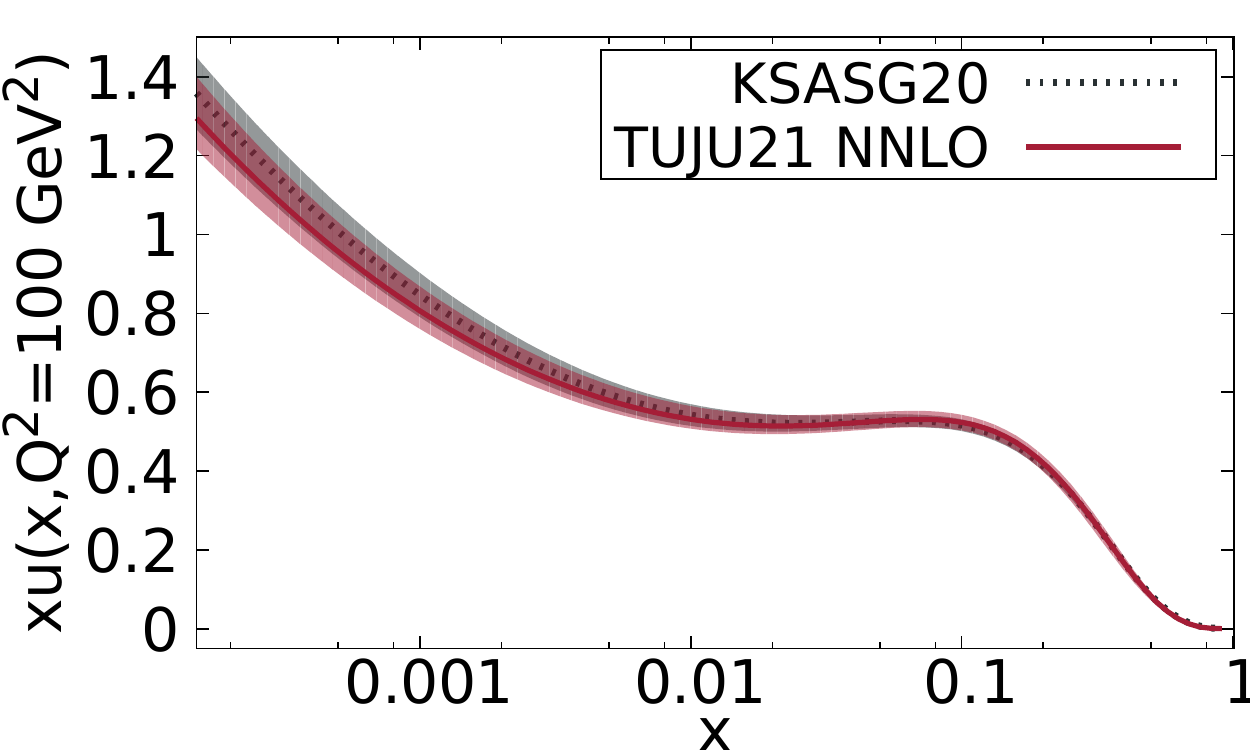}}
          \subfigure{\includegraphics[width=0.237\textwidth]{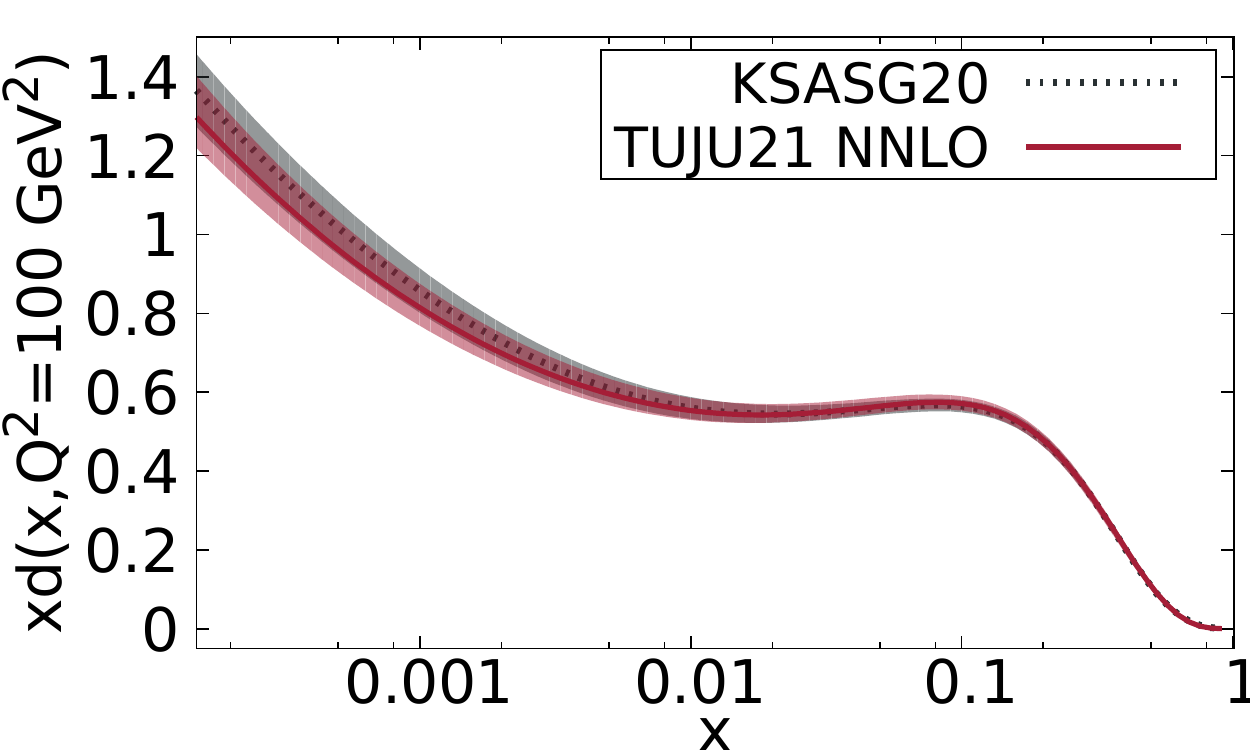}}
          \end{center} 
    \caption{Nuclear parton distribution functions TUJU21 in lead at NNLO compared to the KSASG20 results \cite{Khanpour:2020zyu}, shown at $Q^2=100\,\mathrm{GeV}^2$. The comparison is presented for the distribution functions $xf_i(x,Q^2)$ with $i=g,\,s=\bar{s}=\bar{u}=\bar{d},\,u,\,d$ for a proton bound in a lead nucleus.}
\label{fig-nPDF-comparison-NNLO}    
    \end{figure*}

The only other recent nPDF fit performed at NNLO (apart from nNNPDF1.0 which is now superseded by nNNPDF2.0), is the KSASG20 analysis \cite{Khanpour:2020zyu}. However, no LHC data have been included for this set. As the provided LHAPDF grids include only limited information about the uncertainties of the individual distributions, we show only the comparisons to the full-nucleus PDFs. These are provided in Fig. \ref{fig-nPDF-comparison-NNLO} for $g,s,u,d$, where $s$ represents a generic sea quark distribution since a full flavor separation for the sea quarks is not available in either of these analyses. Overall, we find reasonable agreement, although generally the gluons tend to be higher and the different quark PDFs lower in TUJU21 than in KSASG20. Notably there is some discrepancy for the sea quarks at $x\approx 0.1$, where the KSASG20 fit shows a particularly large sea-quark antishadowing.

\section{Applications to EW-boson production in nuclear collisions}

\label{sec:applications}

\subsection{EW-boson production in Pb+Pb}

Since the $W^{\pm}$ and $Z$ bosons do not experience the strong interactions, it is expected that once they are produced in a high-energy heavy-ion collision they will be able to exit from the collision without much attenuation, even in the presence of a quark-gluon plasma that has been formed. Therefore such processes can be applied to study initial-state effects and to test the Glauber model required to normalize the measured centrality-dependent yields and to convert minimum-bias results into cross sections. No deviations from the theoretical predictions with nuclear PDFs were observed for the Run-I LHC Pb+Pb data at $\sqrt{s_{\mathrm{NN}}} = 2.76~\text{TeV}$ \cite{CMS:2011zfr, CMS:2012fgk, ATLAS:2012qdj, ATLAS:2014sic} but uncertainties in these data were fairly sizable. However, the more recent high-precision Run II data at $\sqrt{s_{\mathrm{NN}}} = 5.02~\text{TeV}$ by ATLAS \cite{ATLAS:2019maq, ATLAS:2019ibd} do show some difference in normalization when compared to NNLO calculations with NNPDF3.1 NNLO proton PDFs and EPPS16 NLO nuclear modifications \cite{Eskola:2020lee}. Similarly the recent Run-II CMS data for $Z$ boson production in Pb+Pb collisions seem to sit at the upper edge of nPDF-based predictions based on an NLO computation matched to a parton shower \cite{CMS:2021kvd}, the latter providing an approximation to leading-logarithmic resummation. Unlike ATLAS, CMS has not relied on the Glauber model to convert the measured minimum bias yield to a cross section, but utilized the measured luminosity instead. A surprising feature is that the centrality dependence of these two measurements is opposite: CMS finds a decreasing trend for the normalized yield towards more peripheral collisions, whereas in the ATLAS data it increases with centrality. The former has been explained with different possible biases in centrality classification in high-scale processes \cite{Loizides:2017sqq}, and the latter could be due to nuclear shadowing for $\sigma^{\mathrm{inel}}_{\mathrm{NN}}$ \cite{Eskola:2020lee} or anchor-point bias \cite{Jonas:2021xju}. Here we, however, confront the minimum bias data with our nPDFs constrained by the $W^{\pm}$ and $Z$ boson production in p+Pb at NLO and NNLO, in order to investigate whether these data are compatible with our nPDFs, and whether they are mutually consistent.

\begin{figure*}[tb!]
     \begin{center}
          \subfigure{\includegraphics[width=0.325\textwidth]{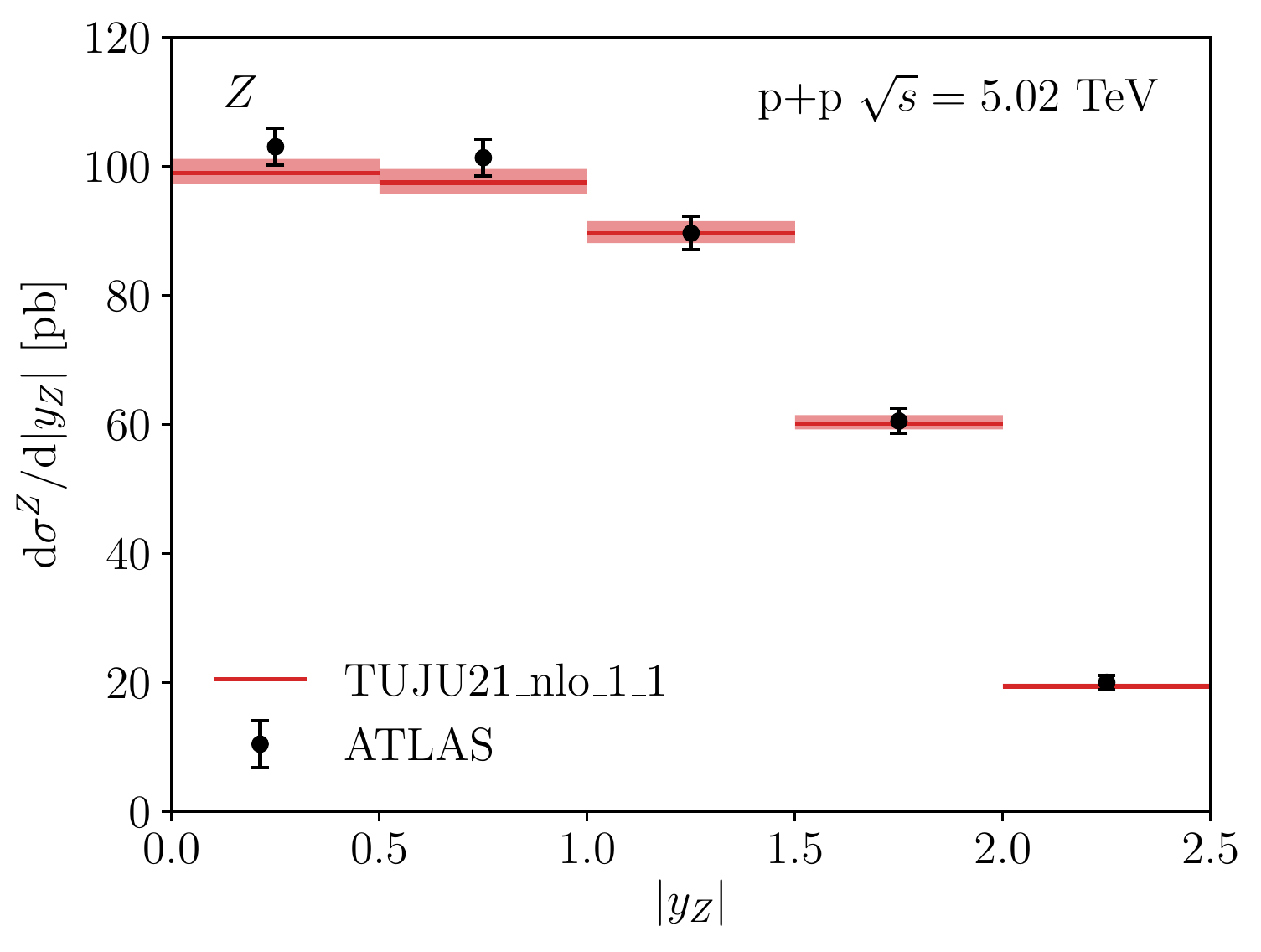}}
           \subfigure{\includegraphics[width=0.325\textwidth]{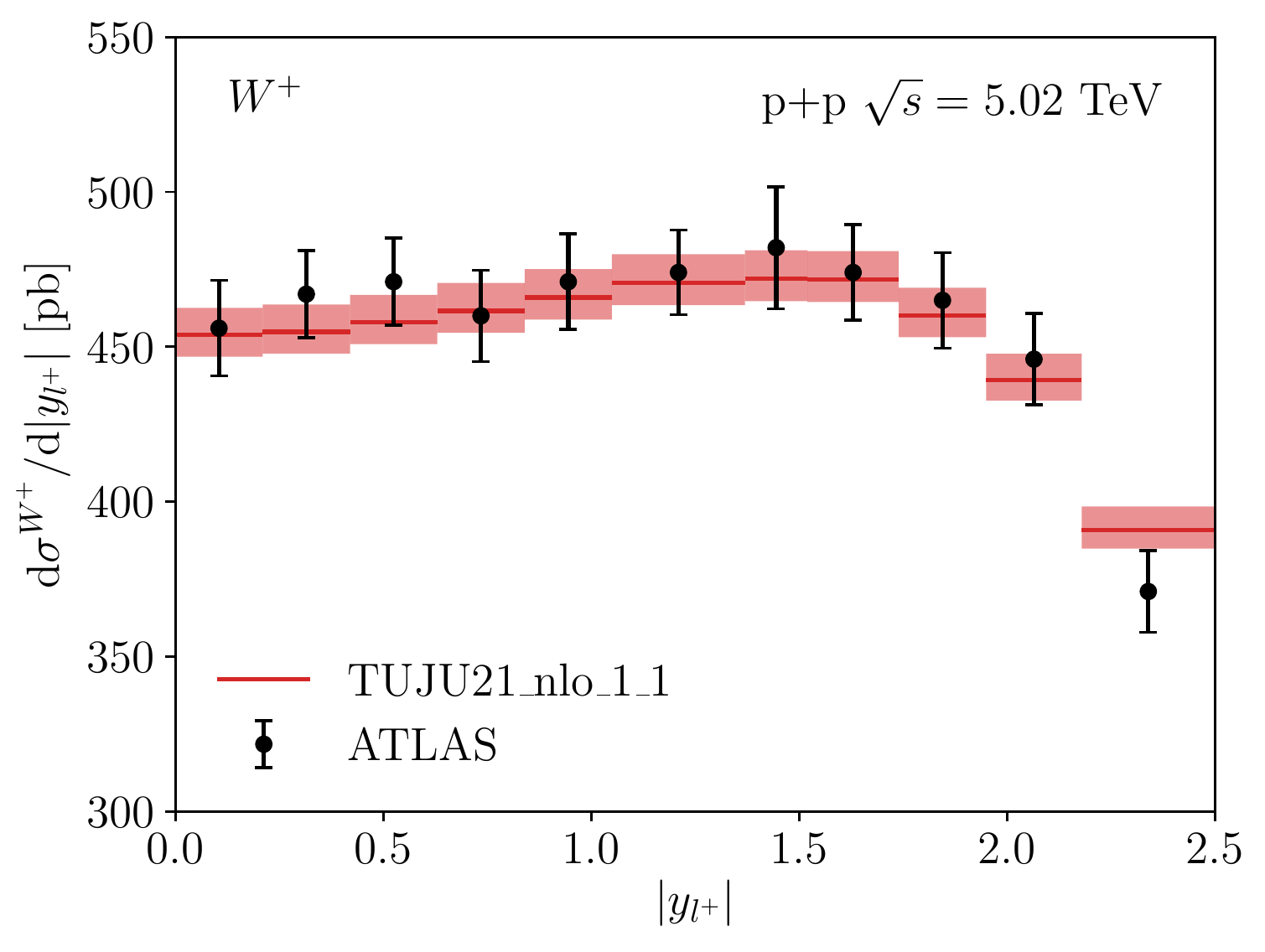}}
           \subfigure{\includegraphics[width=0.325\textwidth]{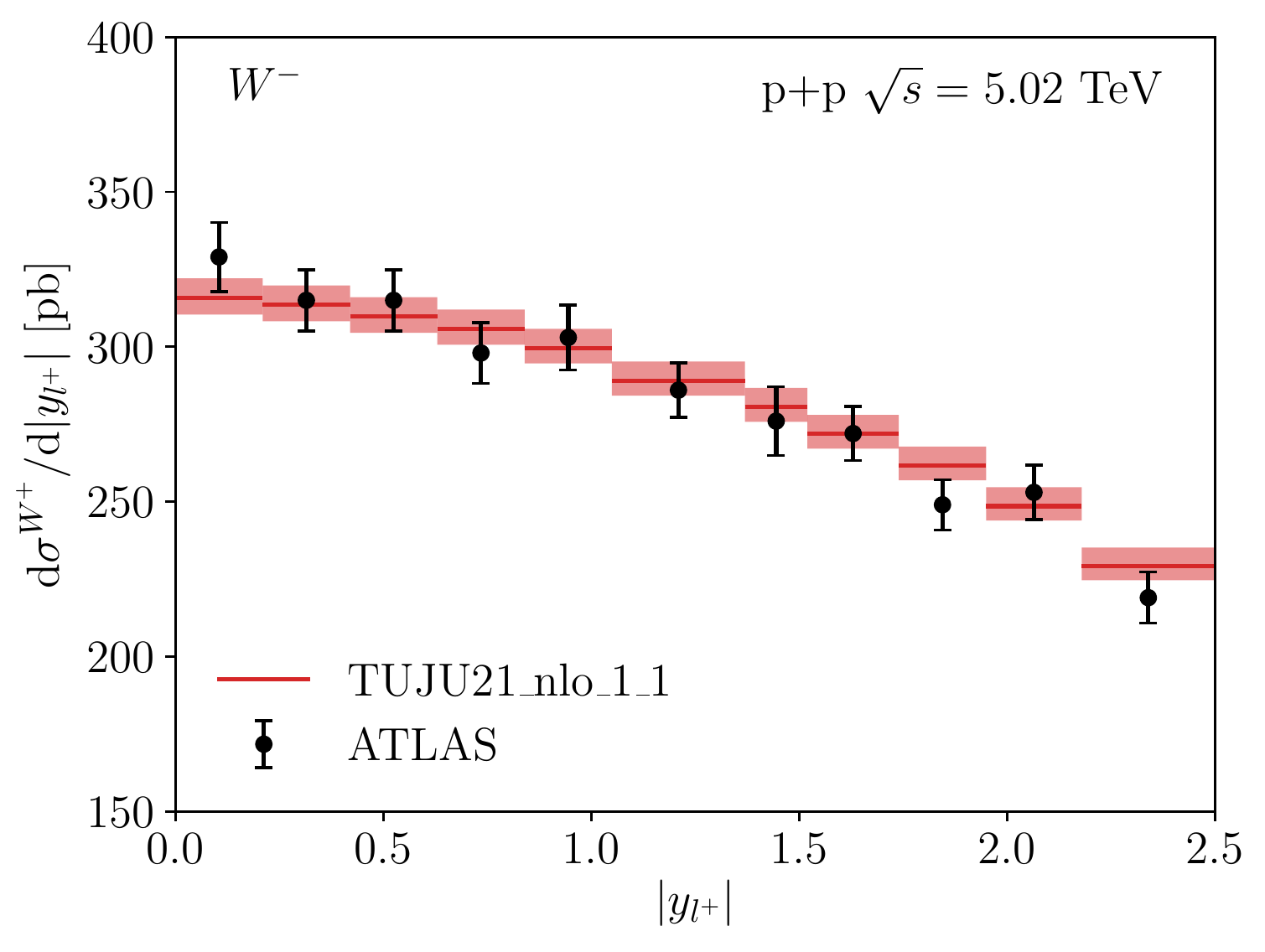}}          
          \subfigure{\includegraphics[width=0.325\textwidth]{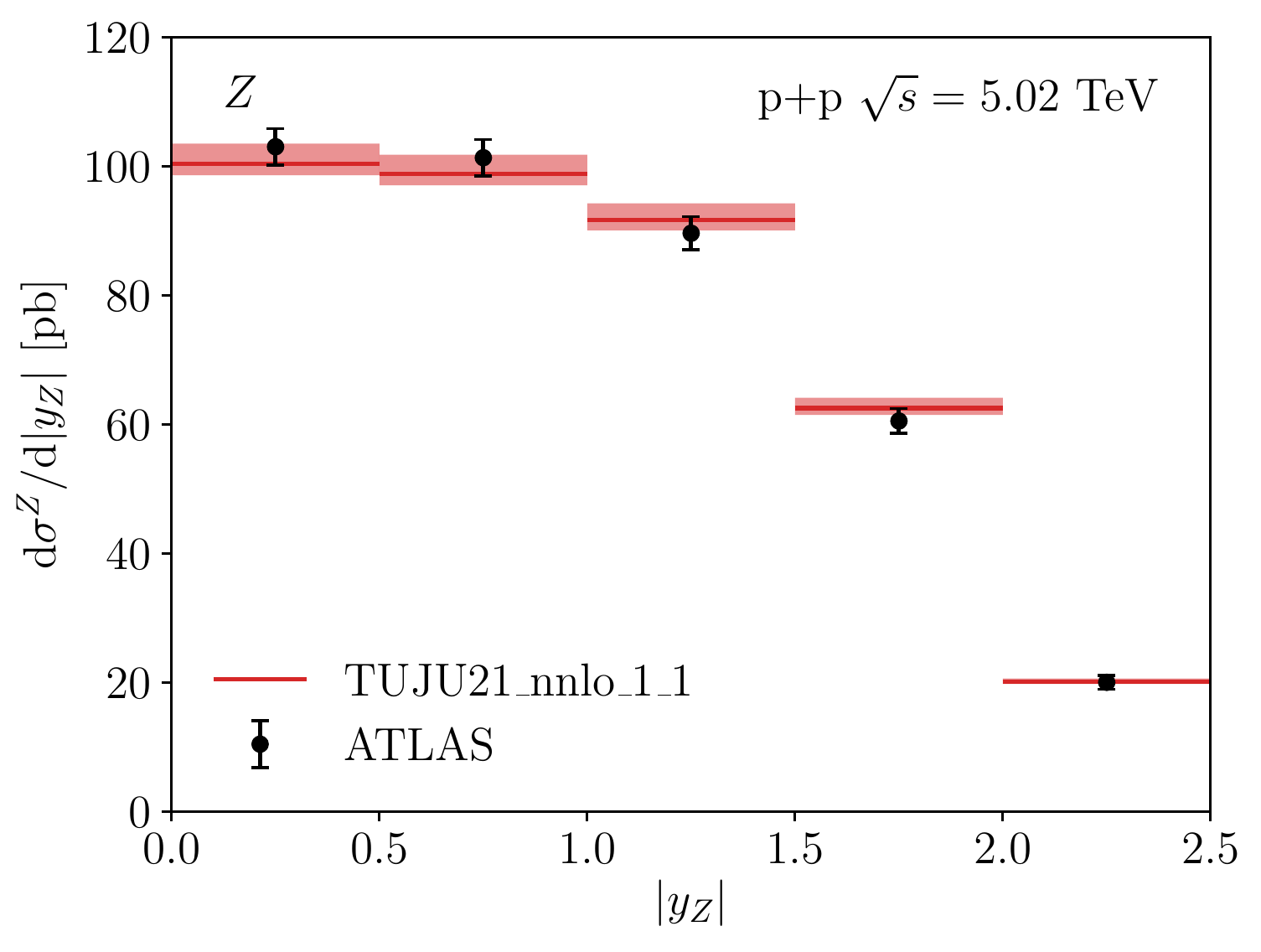}} 
           \subfigure{\includegraphics[width=0.325\textwidth]{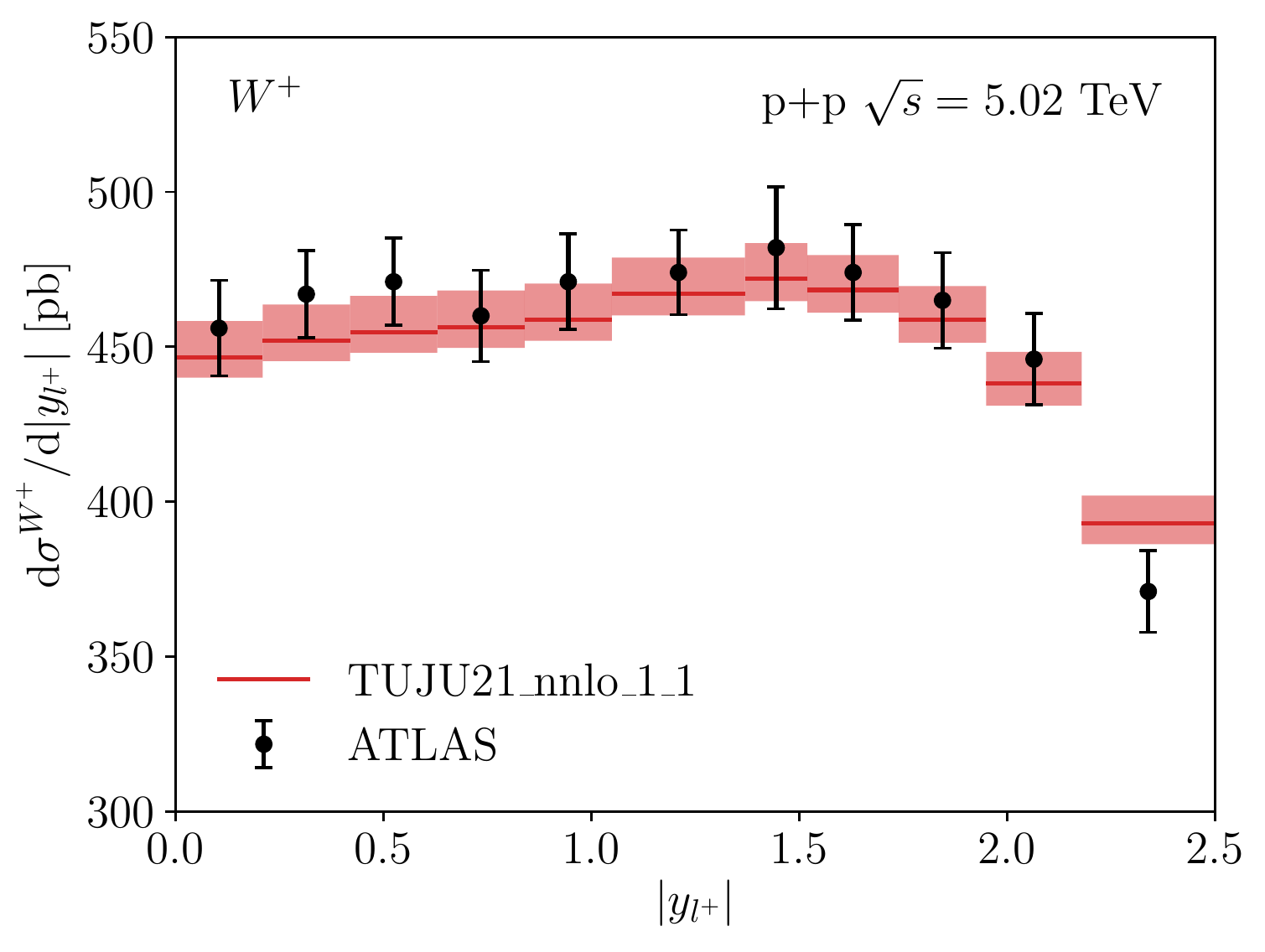}}
           \subfigure{\includegraphics[width=0.325\textwidth]{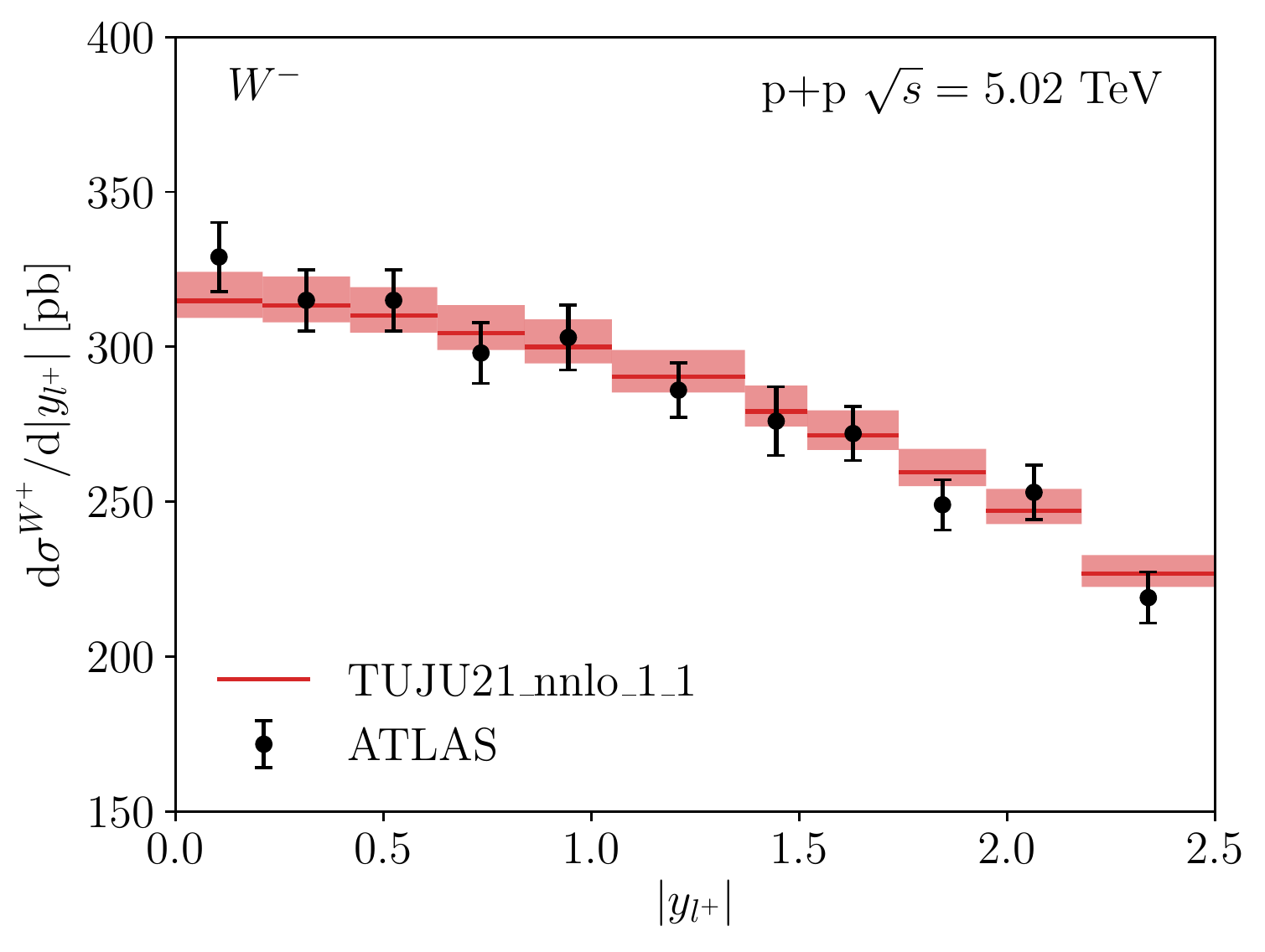}}  
          \end{center} 
\caption{EW boson production cross sections in p+p collisions at $\sqrt{s} = 5.02~\text{TeV}$ for $Z$ (left), $W^+$ (center) and $W^-$ (right), calculated at NLO (upper panels) and NNLO (lower panels) and compared to data from ATLAS \cite{ATLAS:2018pyl}.}
\label{fig:applicationspp}    
\end{figure*}

To calculate the EW boson production cross section at NLO and NNLO in QCD we use the \textsc{MCFM} code version 10.1 which includes an improved calculation of PDF-uncertainties from Ref.~\cite{Campbell:2019dru}. Before applying the setup to heavy-ion collisions, we validate our computation against the p+p data taken by ATLAS \cite{ATLAS:2018pyl} at the same energy, $\sqrt{s} = 5.02~\text{TeV}$, using our proton baseline PDFs that actually did not include these data sets. The comparisons for $Z$, $W^+$ and $W^-$ bosons are separately shown in Fig. \ref{fig:applicationspp} at NLO and NNLO. We find a very good agreement in all cases, which can be expected since similar data at other energies were used in our proton baseline fit.

\begin{figure*}[tb!]
     \begin{center}
           \subfigure{\includegraphics[width=0.325\textwidth]{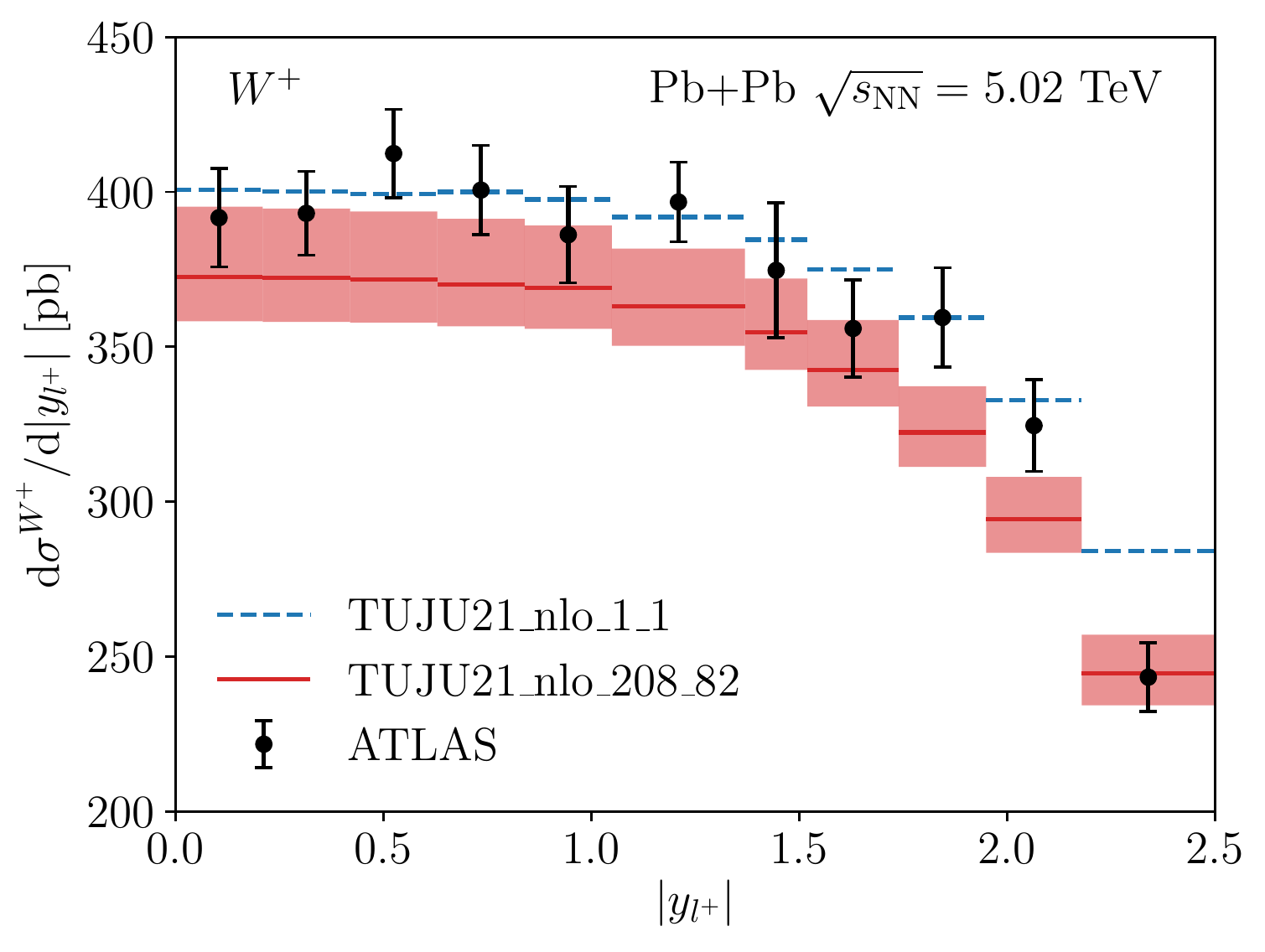}}
          \subfigure{\includegraphics[width=0.325\textwidth]{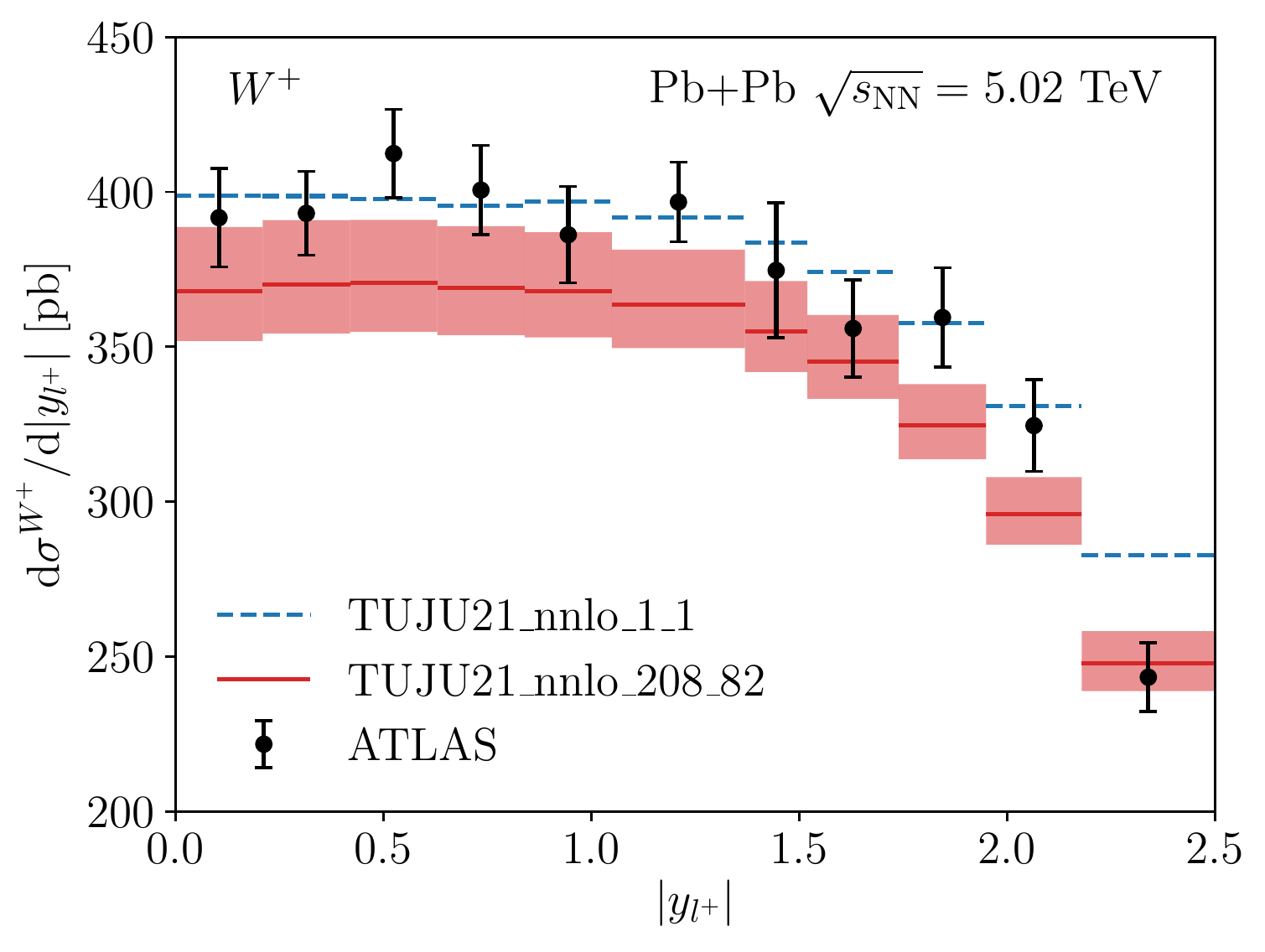}}
          \subfigure{\includegraphics[width=0.325\textwidth]{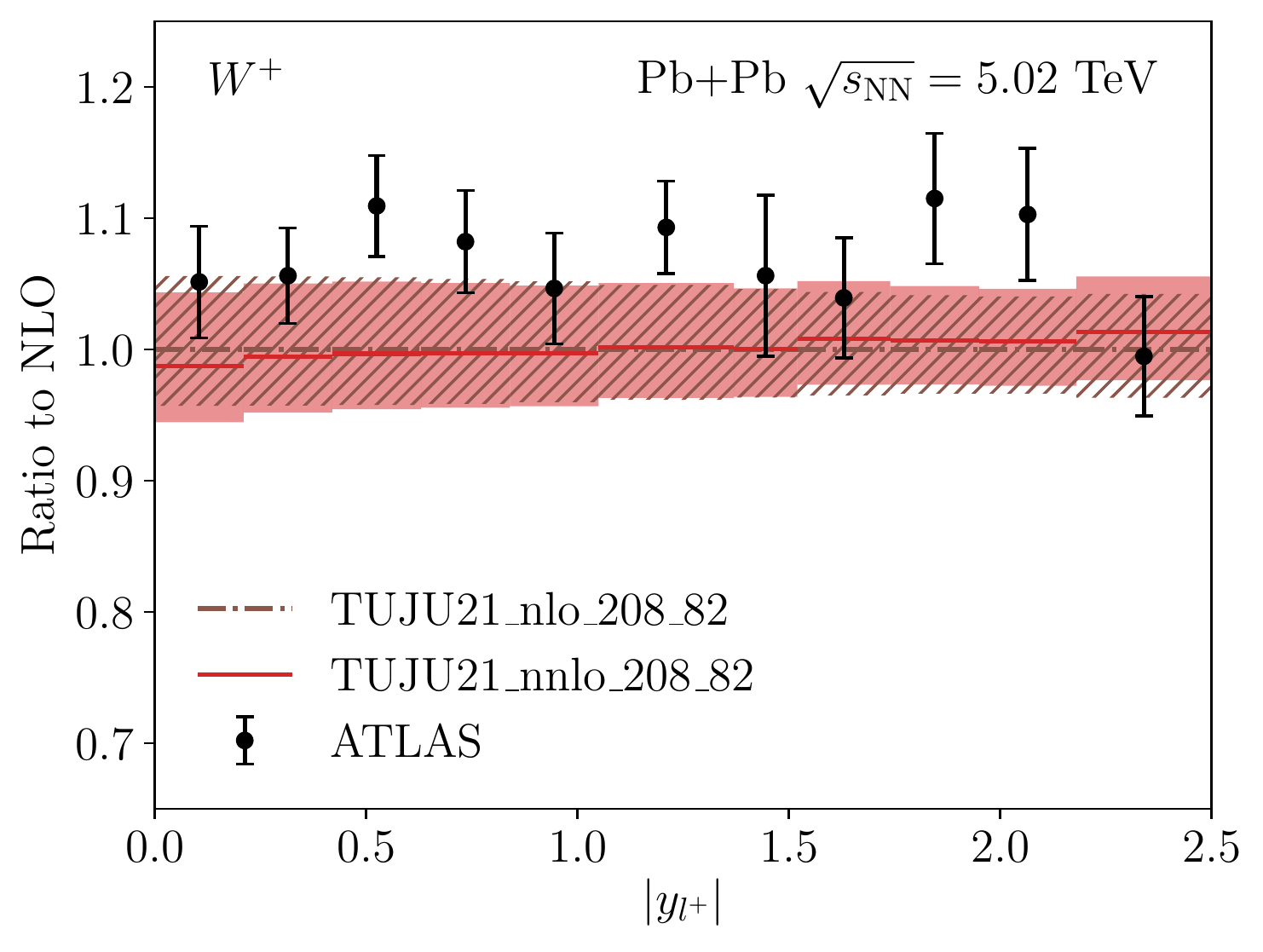}}
           \subfigure{\includegraphics[width=0.325\textwidth]{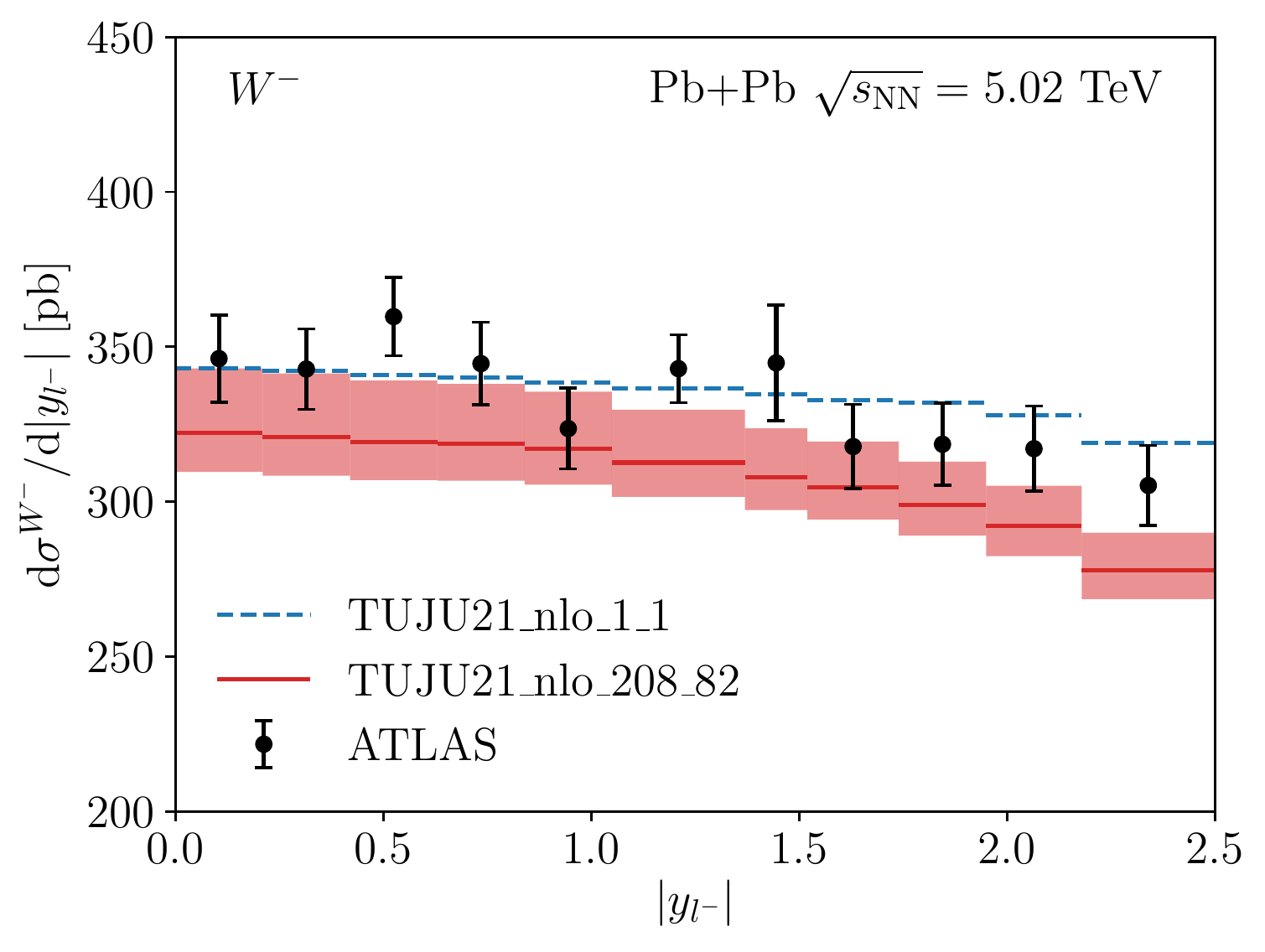}}          
           \subfigure{\includegraphics[width=0.325\textwidth]{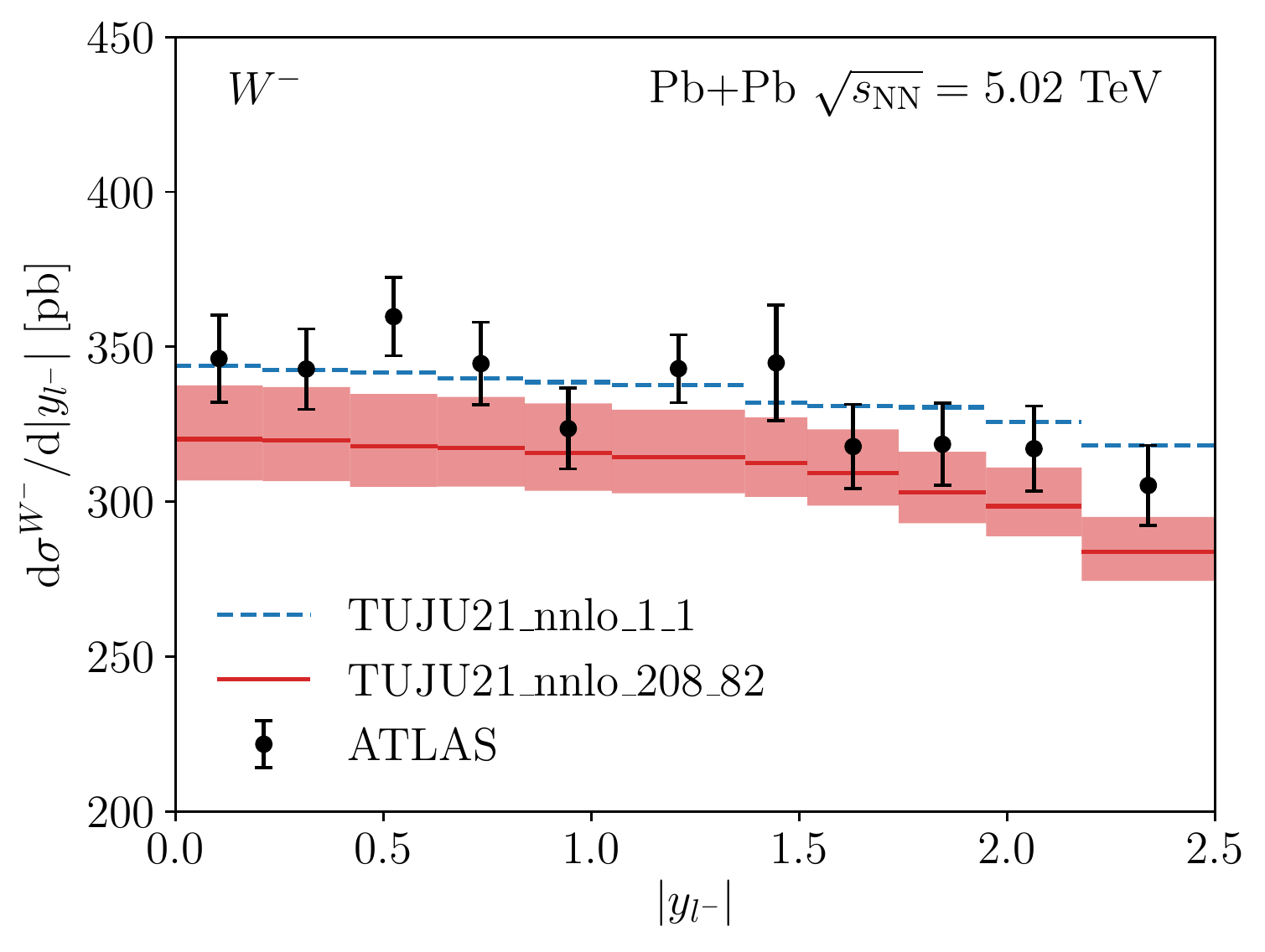}}
          \subfigure{\includegraphics[width=0.325\textwidth]{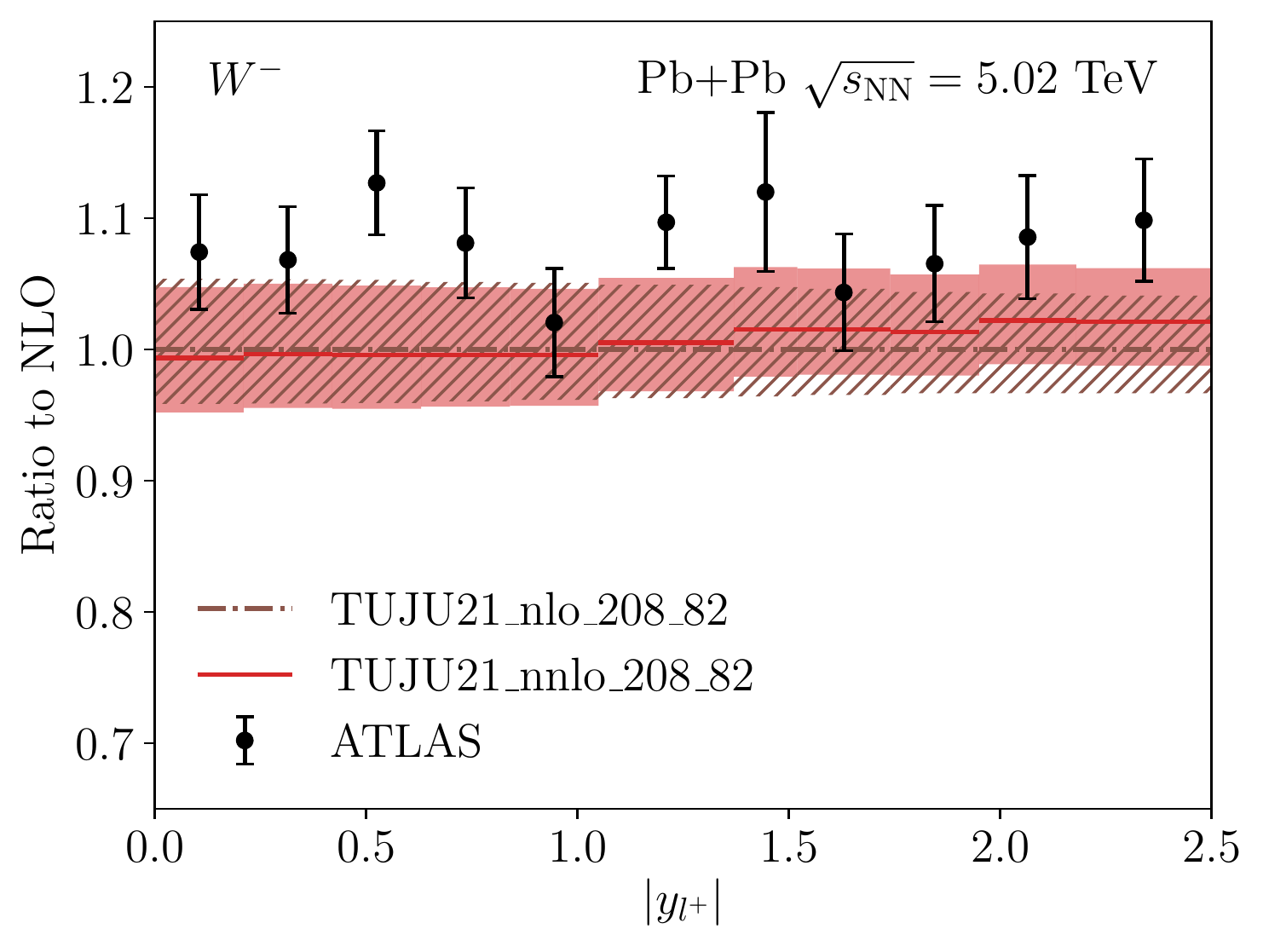}}
          \end{center} 
\caption{Comparison of $W^+$ (top) and $W^-$ (bottom) boson production in Pb+Pb collisions at $\sqrt{s_{\mathrm{NN}}} = 5.02~\text{TeV}$ at NLO (left) and NNLO (center) with (solid with uncertainty band) and without (dashed) nuclear PDF modifications to the ATLAS data \cite{ATLAS:2019ibd}. In the right part we plot the ratios of the NNLO (red with uncertainty) and NLO (dot-dashed brown with hatched uncertainty) together with the data.}
\label{fig:applicationsPbPb-W}    
\end{figure*}

In Fig.~\ref{fig:applicationsPbPb-W} we present a comparison of our NLO and NNLO calculations of $W^{\pm}$ production in Pb+Pb collisions at $\sqrt{s_{\mathrm{NN}}} = 5.02~\text{TeV}$ to the measurement by ATLAS \cite{ATLAS:2019ibd}. Here we find that for both $W^+$ and $W^-$ our calculations with nuclear PDFs tend to undershoot the data, both at NLO and NNLO. Accounting for all the uncertainties, including also the normalization uncertainty, there is rough qualitative agreement, but the overall trend is that the data is consistently above the calculation. This is in line with the observation made in \cite{Eskola:2020lee}. Interestingly the calculation with only proton PDFs (but accounting for isospin effects) shows a better agreement with the data. This is unexpected when keeping in mind the strong preference for nPDF effects that we found in our analysis of p+Pb collisions. To quantify the impact of the NNLO corrections we also plot ratios between the NNLO and NLO results and compare them with the data in Fig.~\ref{fig:applicationsPbPb-W}. For this observable the NNLO corrections are only of the order few percent and do not significantly improve the agreement with the data.

\begin{figure*}[tb!]
     \begin{center}
          \subfigure{\includegraphics[width=0.325\textwidth]{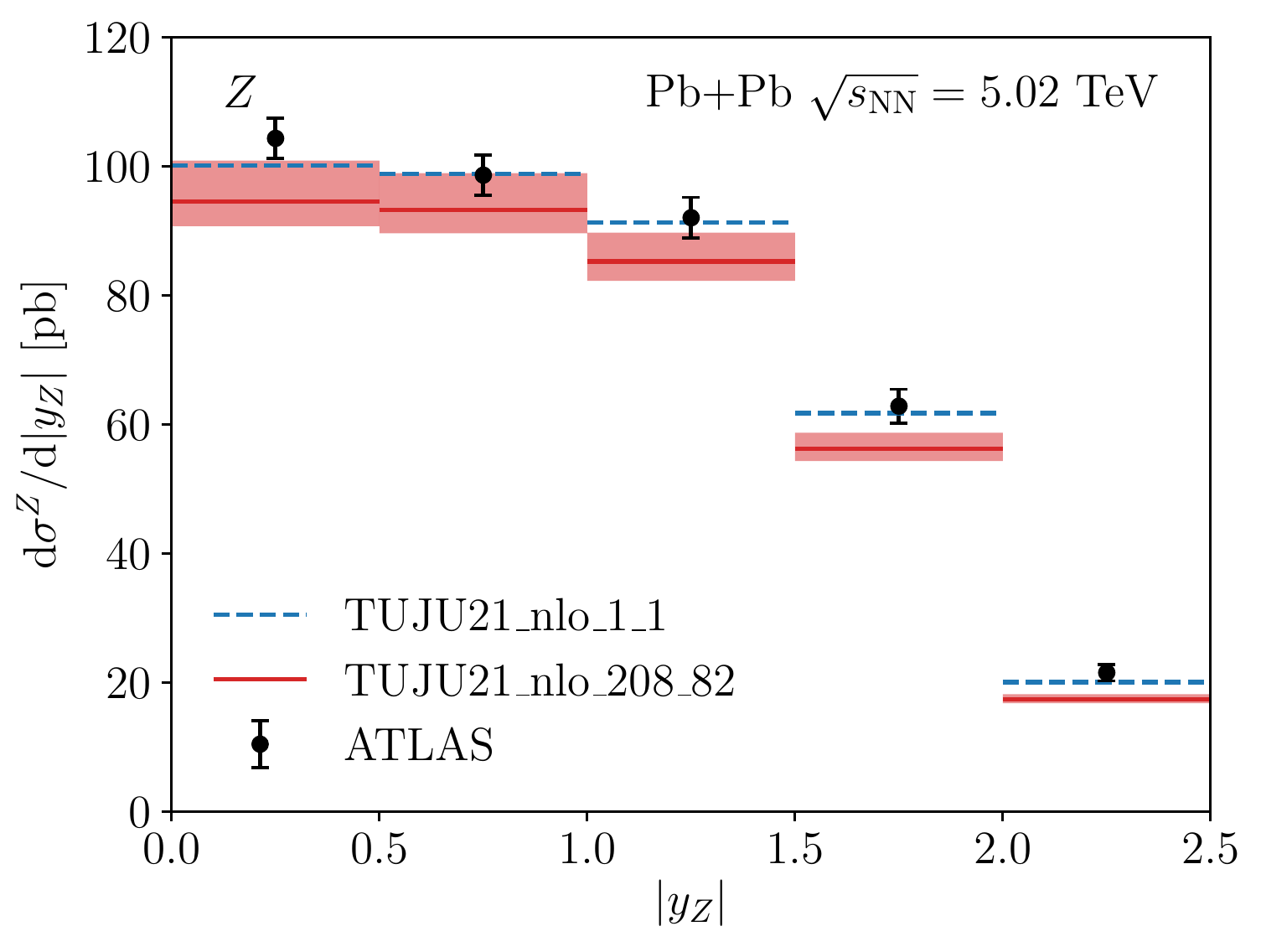}}
          \subfigure{\includegraphics[width=0.325\textwidth]{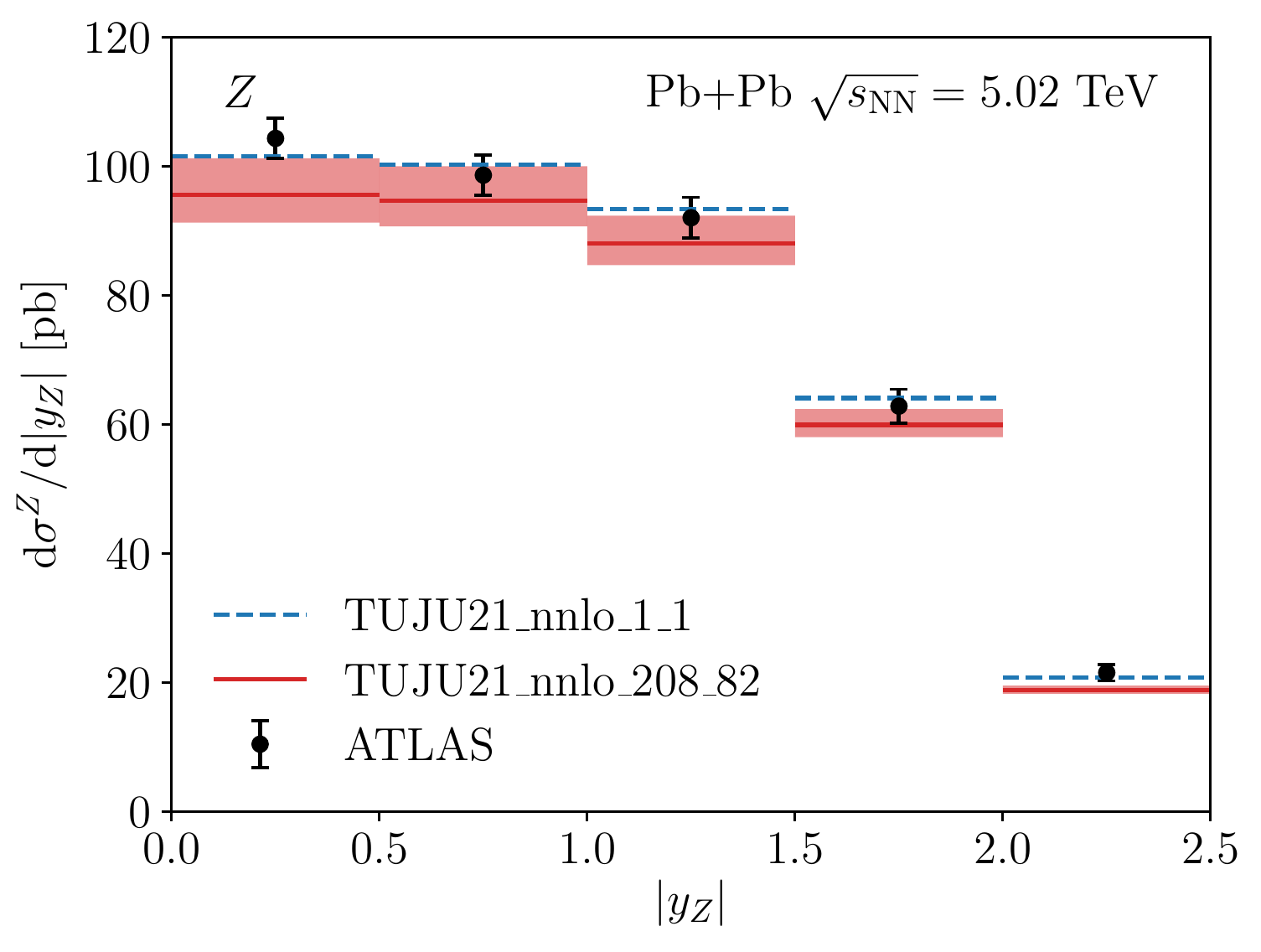}}
          \subfigure{\includegraphics[width=0.325\textwidth]{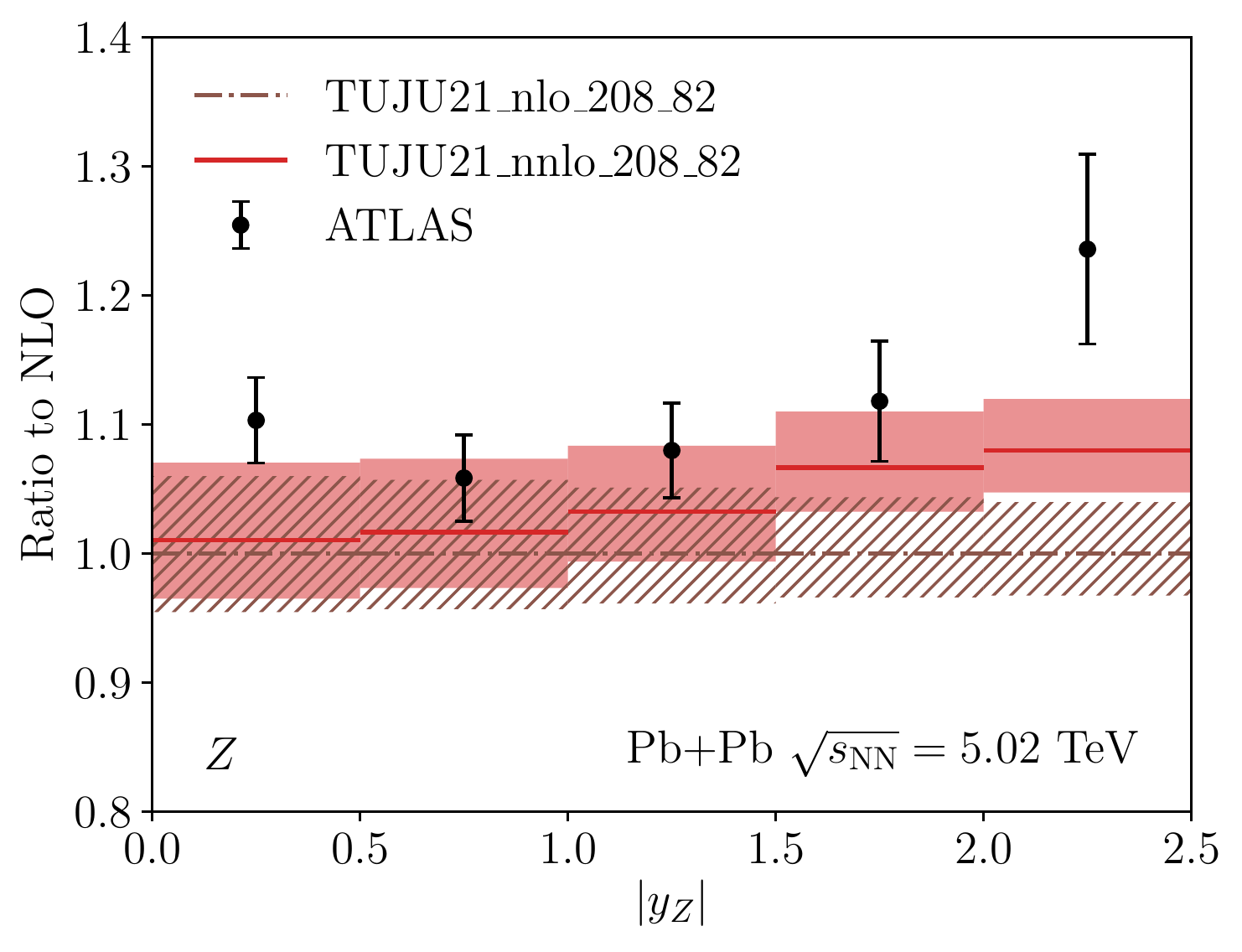}}
          \subfigure{\includegraphics[width=0.325\textwidth]{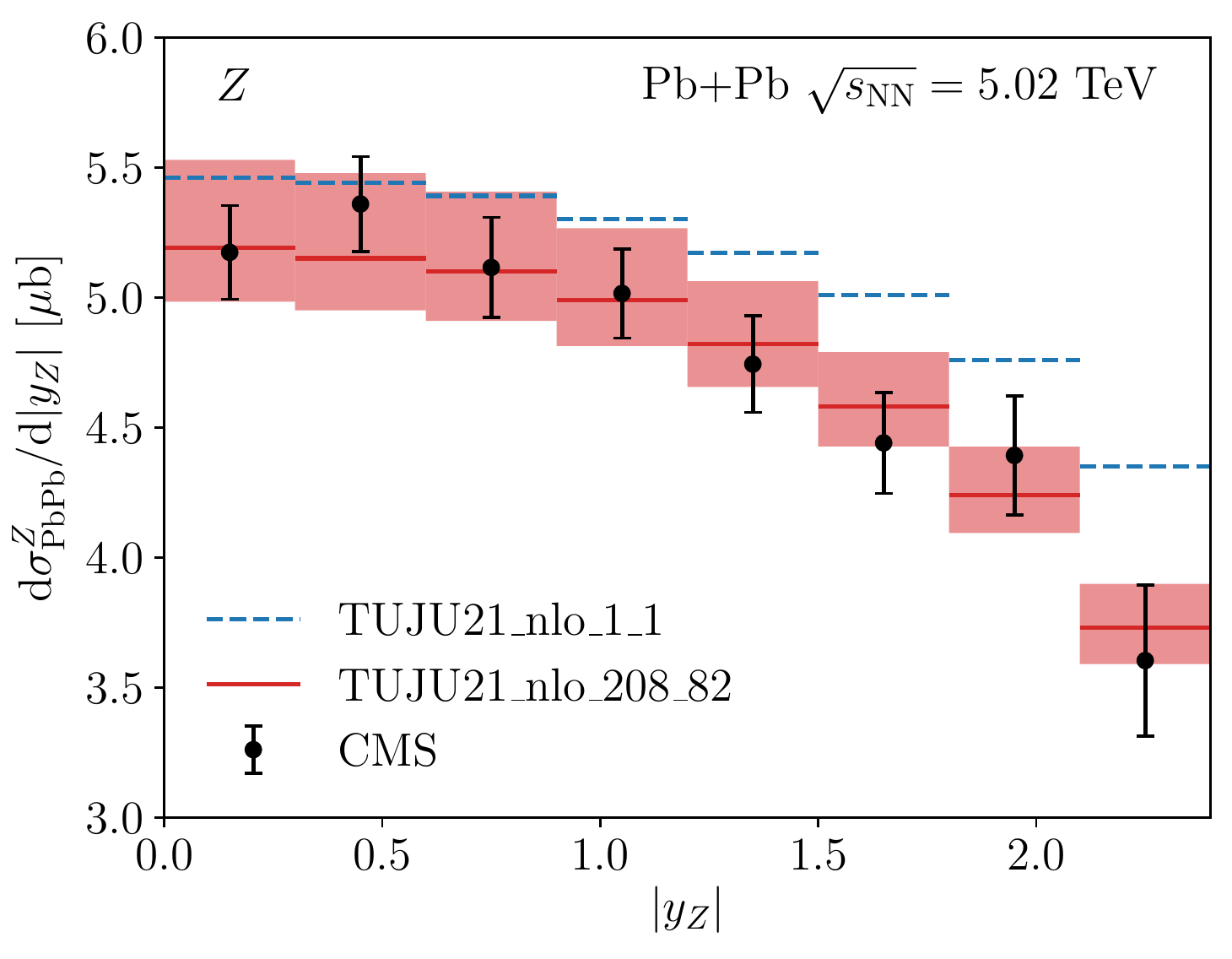}}
          \subfigure{\includegraphics[width=0.325\textwidth]{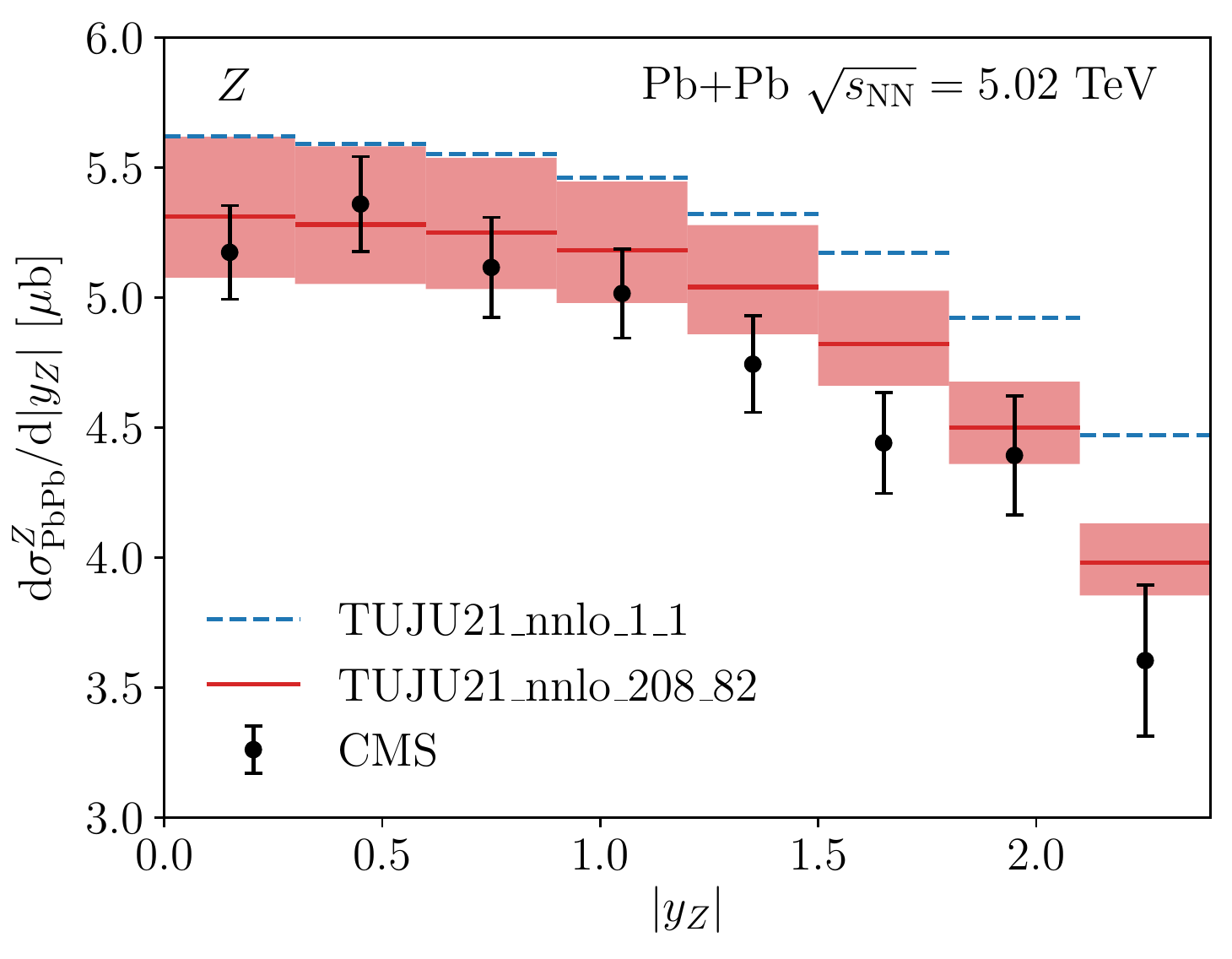}}
          \subfigure{\includegraphics[width=0.325\textwidth]{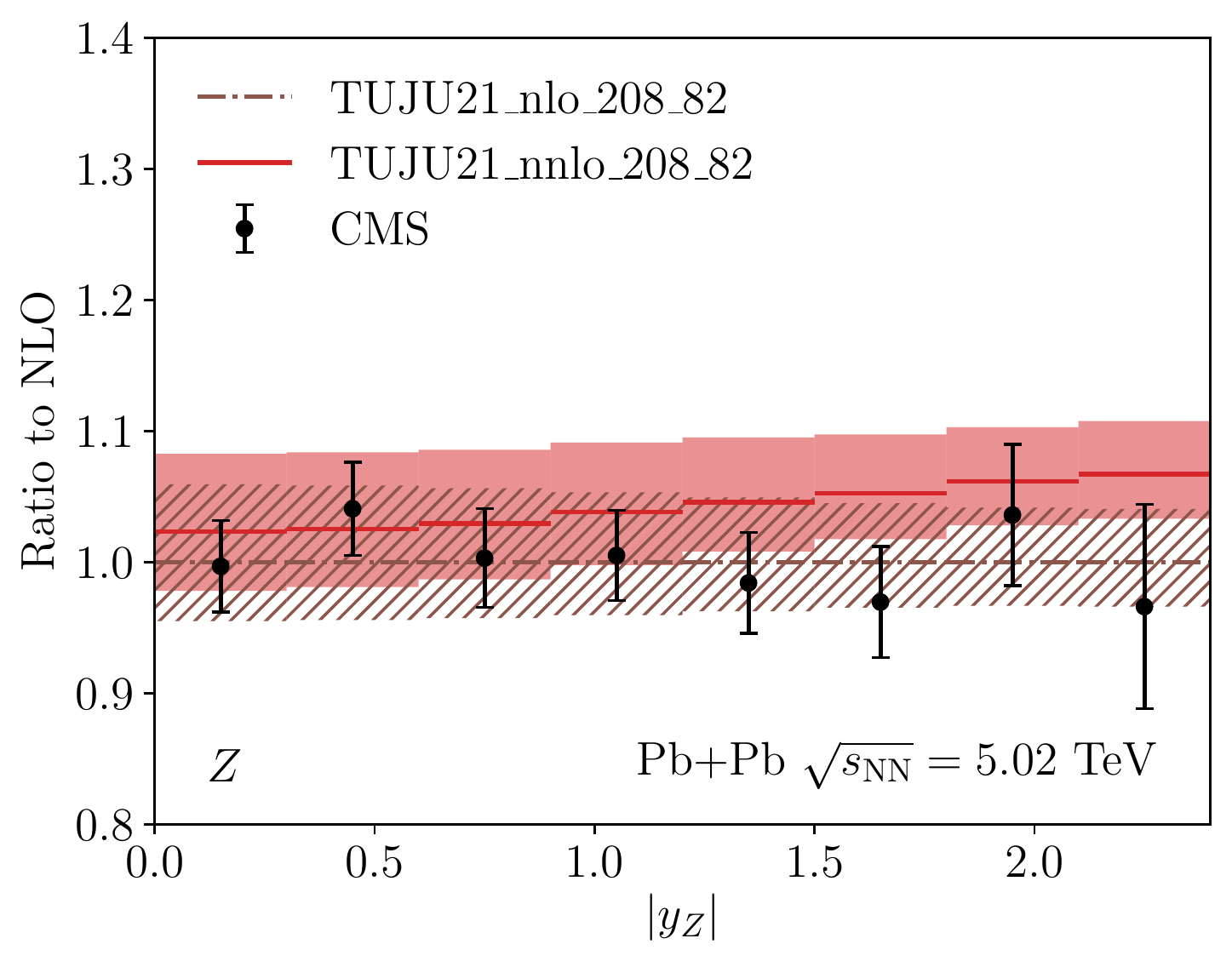}}
          \end{center} 
\caption{Comparison of $Z$ boson production in Pb+Pb collisions at $\sqrt{s_{\mathrm{NN}}} = 5.02~\text{TeV}$ at NLO (left) and NNLO (center) with (solid with uncertainty band) and without (dashed) nuclear PDF modifications to ATLAS \cite{ATLAS:2019maq} (upper panels) and CMS \cite{CMS:2021kvd} (lower panels) data. In the right part we plot the ratios of the NNLO (red with uncertainty) and NLO (dot-dashed brown with hatched uncertainty) together with the data.}
\label{fig:applicationsPbPb-Z}
\end{figure*}

The same trend is visible in Fig.~\ref{fig:applicationsPbPb-Z} for the $Z$ boson production data in Pb+Pb collisions when comparing the NLO and NNLO calculations to the ATLAS data \cite{ATLAS:2019maq}. However, the recent CMS data for the same observable and the same collision energy \cite{CMS:2021kvd}, also shown in the figure, is well in line with our calculations with nuclear PDFs. At NNLO, it even appears that the calculated cross section is somewhat above the data, contrary to the ATLAS comparison. The differences are well visible also in the plots showing the ratio between the NNLO and NLO results together with the data in Fig.~\ref{fig:applicationsPbPb-Z}. There, the NNLO corrections grow with $y_Z$ and are of the order 10\% at the largest rapidities. The ATLAS data seem to agree with the NNLO result, whereas the CMS results seem to fall a bit below the NNLO calculation at larger rapidities and are better in line with the NLO result. Therefore our results point to possible tensions between the two data sets. That said, one should keep in mind that there are some differences in the experimental analyses: In case of ATLAS, the Glauber model was used to calculate the normalization, whereas CMS applied the measured luminosity. Also, ATLAS provides the result only in the fiducial phase-space region, while the CMS data has been corrected to include also the phase space removed by cuts on the final-state leptons. These features make direct comparisons of the two data sets difficult.

\subsection{DY production in p+Pb}
\label{sec:DYpPb}

A recent dataset that has proved difficult to include in an nPDF analysis at the NLO is the CMS DY production in p+Pb collisions \cite{CMS:2021ynu}. It has been anticipated that for the lower-mass bin ($15 < M < 60$ GeV) the NNLO corrections could be significant \cite{Khalek:2022zqe} and for the higher-mass bin ($60 < M < 120$ GeV) it has been noted that due to large fluctuations at the mid-rapidity it is difficult to have acceptable $\chi^2$ values with any PDF-based calculation \cite{Eskola:2021nhw}. Here we quantify the impact of the NNLO corrections on these data to study whether these could explain the observed differences in the low-mass bin.

\begin{figure*}[tb!]
     \begin{center}
          \subfigure{\includegraphics[width=0.325\textwidth]{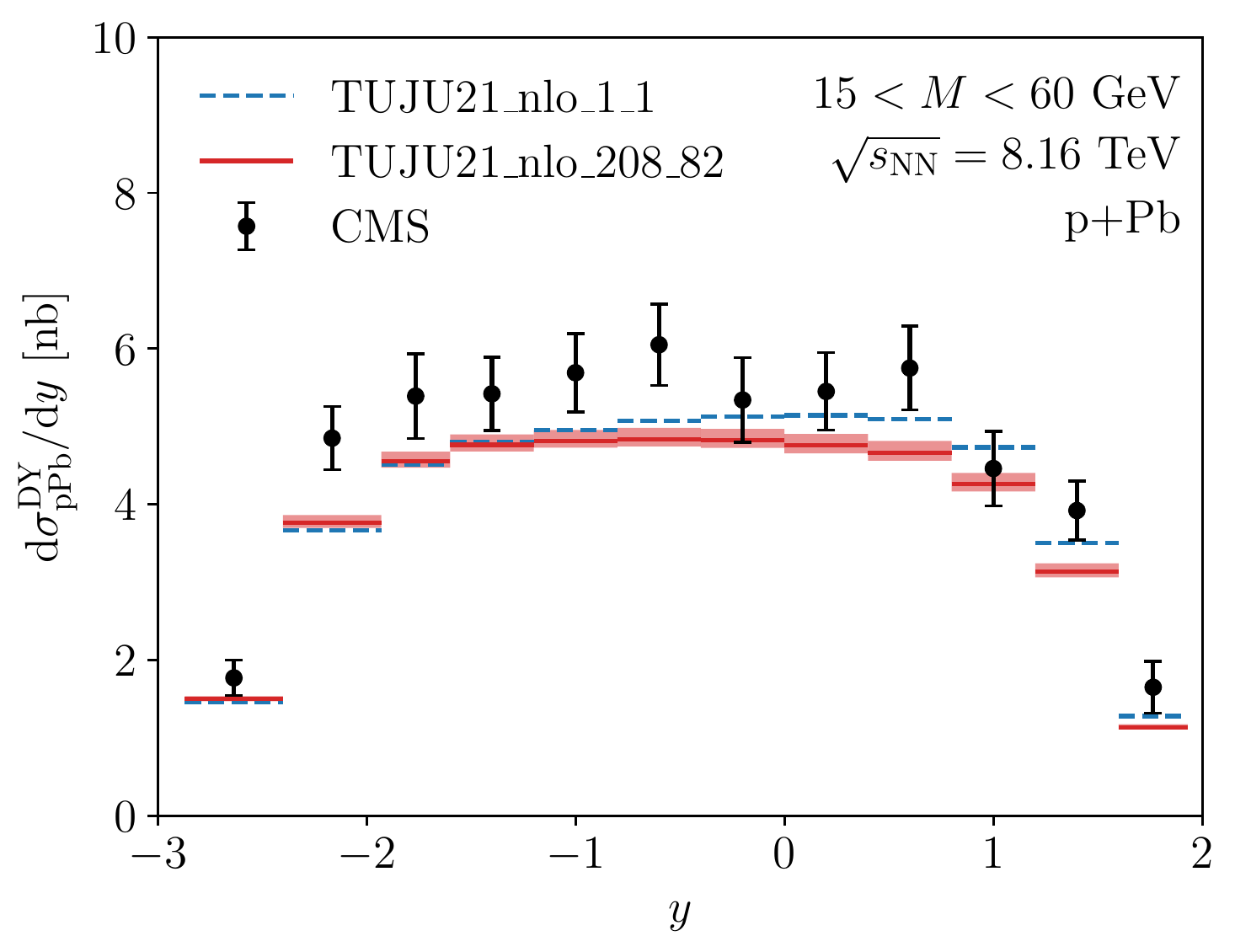}}
          \subfigure{\includegraphics[width=0.325\textwidth]{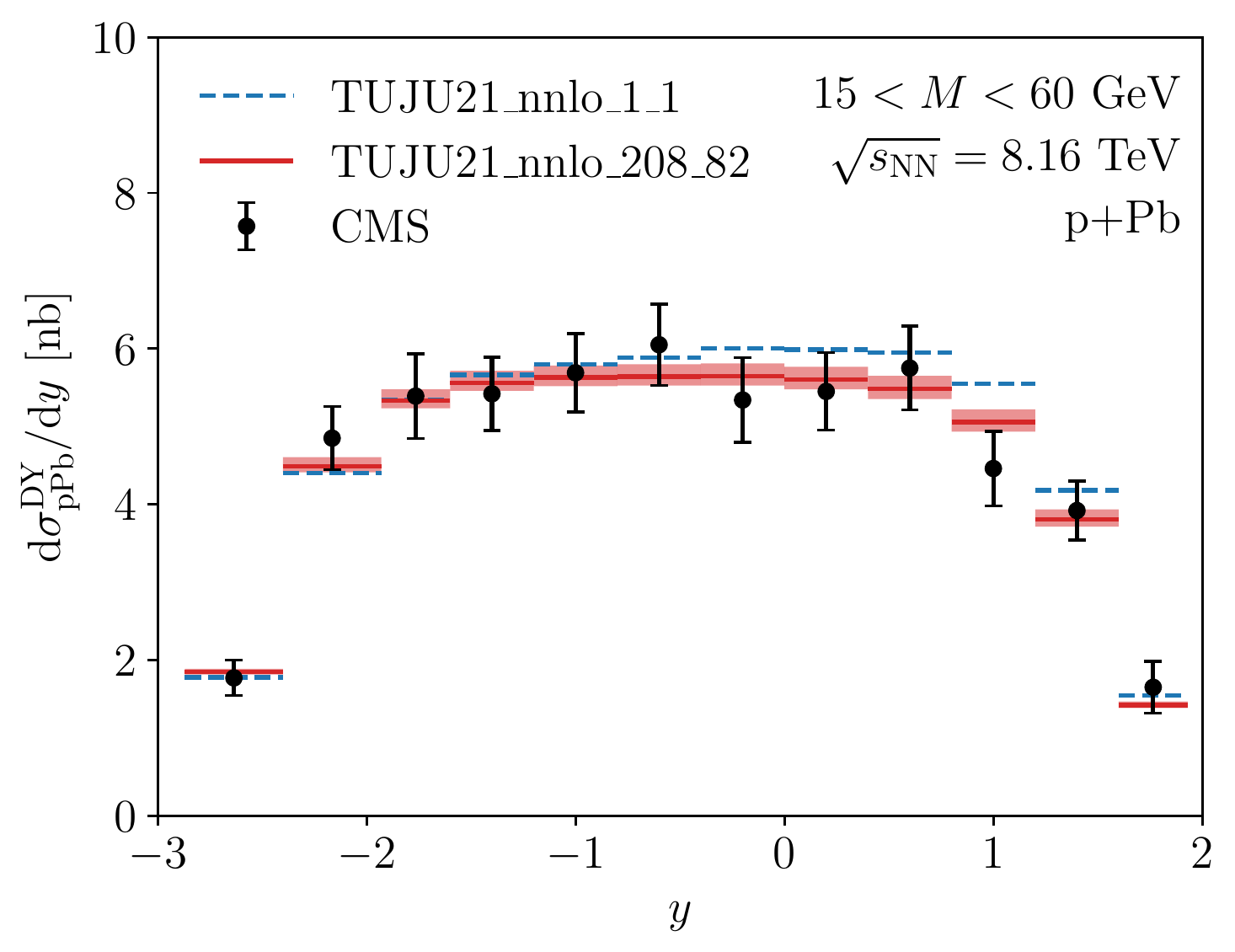}}
          \subfigure{\includegraphics[width=0.325\textwidth]{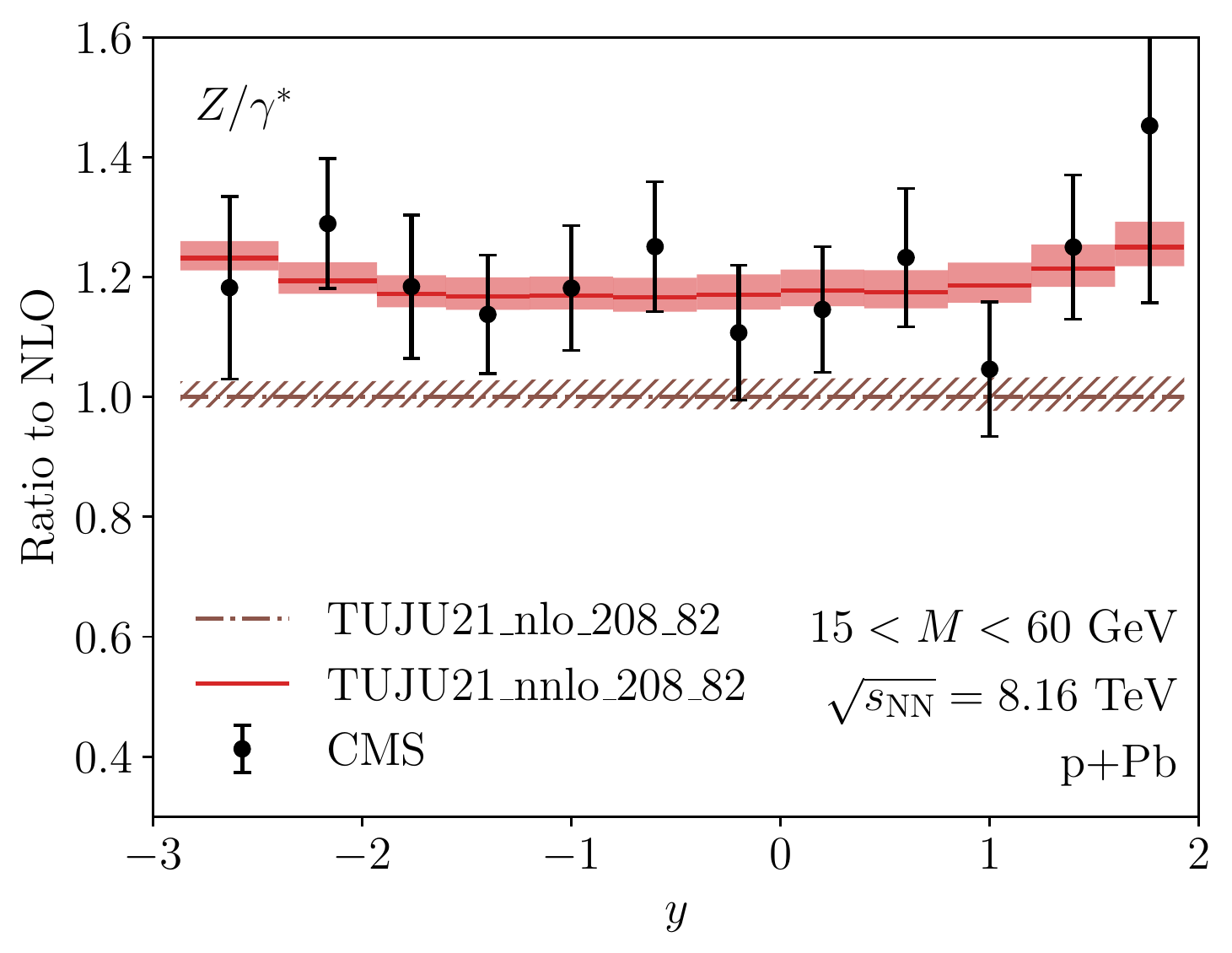}}
          \subfigure{\includegraphics[width=0.325\textwidth]{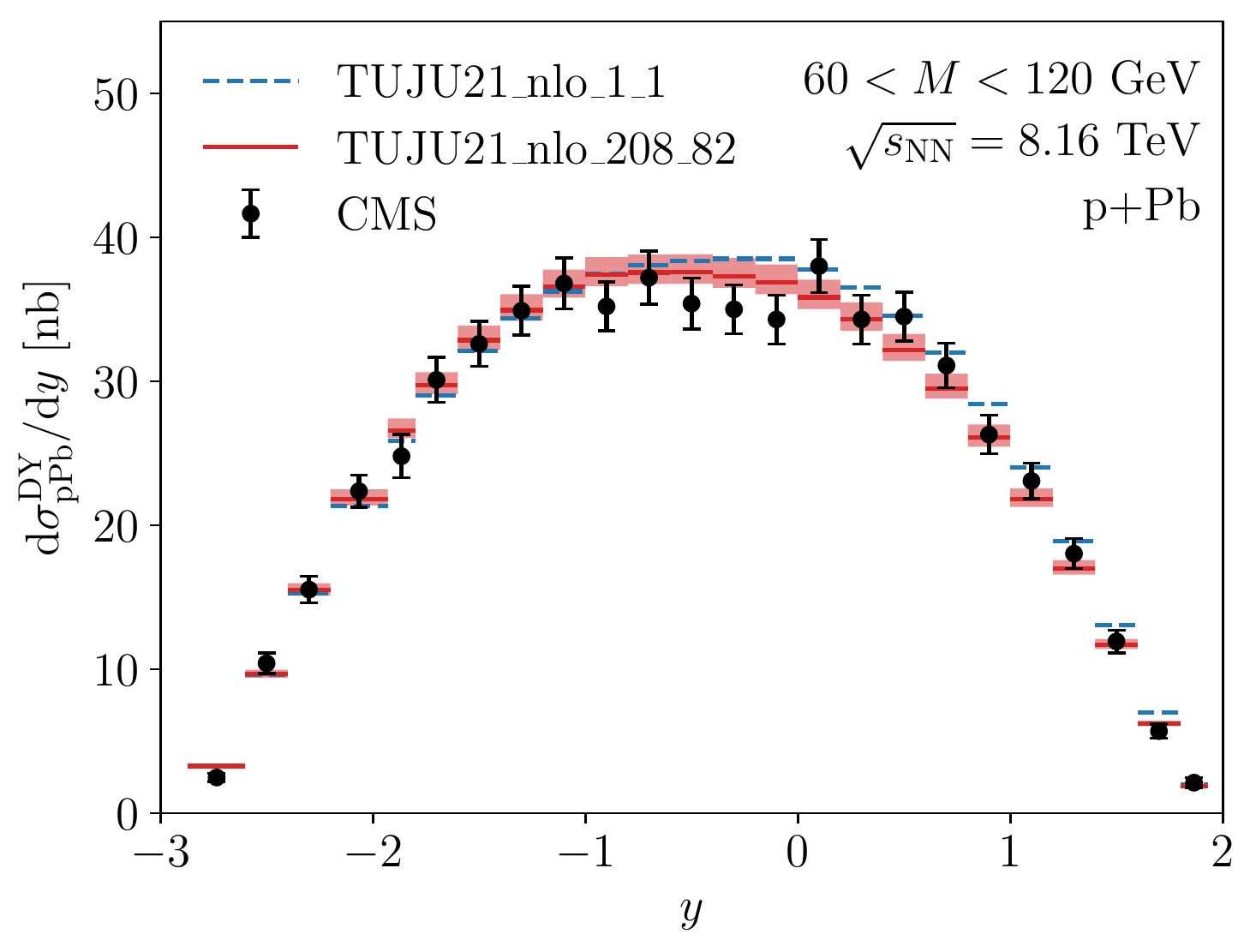}}
          \subfigure{\includegraphics[width=0.325\textwidth]{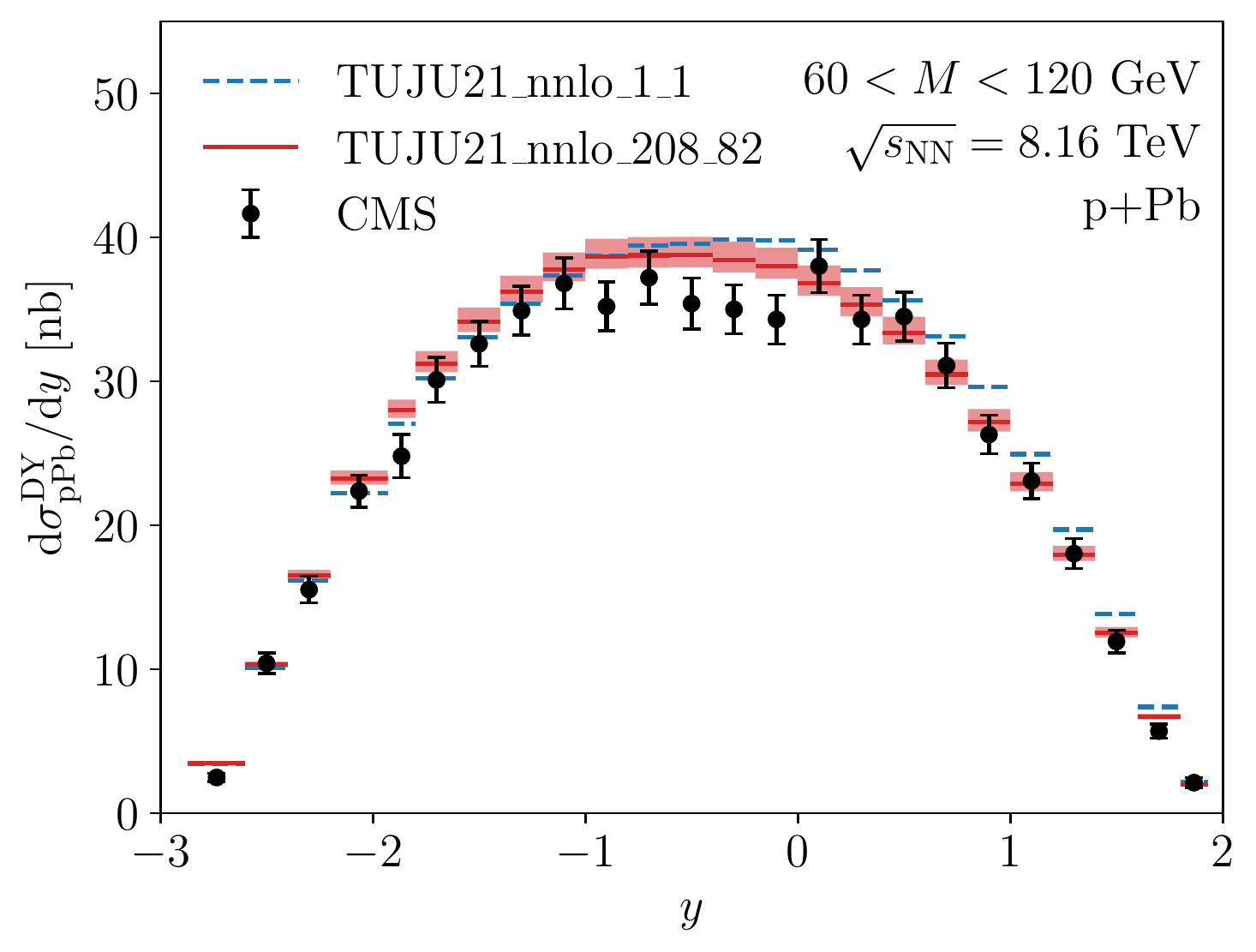}}
          \subfigure{\includegraphics[width=0.325\textwidth]{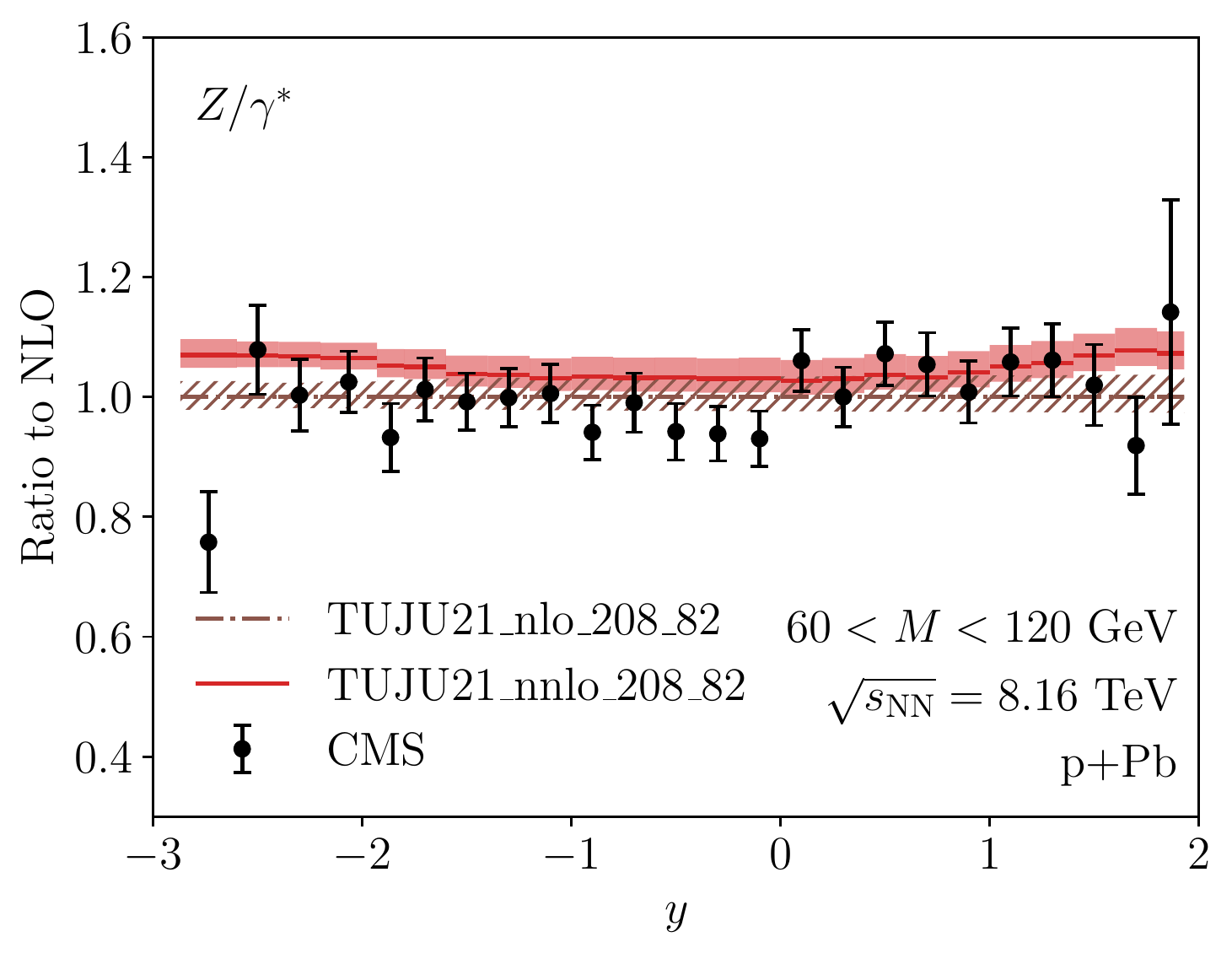}}
          \end{center} 
\caption{Comparison of DY production in p+Pb collisions at $\sqrt{s_{\mathrm{NN}}} = 8.16~\text{TeV}$ at NLO (left) and NNLO (center) results with (solid with uncertainty band) and without (dashed) nuclear PDF modifications in two invariant mass bins, $15 < M < 60$ GeV (upper panels) and $60 < M < 120$ GeV (lower panels) to CMS data \cite{CMS:2021ynu}. In the right part we plot the ratios of the NNLO (red with uncertainty) and NLO (dot-dashed brown with hatched uncertainty) together with the data.}
\label{fig:applicationspPb-DY}
\end{figure*}

Comparison with this CMS data is presented in Fig.~\ref{fig:applicationspPb-DY} for both mass windows as a function of the rapidity $y$ for the dilepton pair including also the ratios between the NNLO and NLO results. The comparisons are made for the fiducial cross section that has not been corrected for the limited acceptance. Here we notice that the NNLO corrections are rather mild, around 5\% for the high-mass bin but become significant for the low-mass bin, reaching 20\% at the largest (absolute) rapidities. To further quantify this effect we have calculated the $\chi^2/N_{\mathrm{dp}}$ values for these data at NLO and NNLO, shown in Table~\ref{tab:chi2DY} separately for the low- and high-mass bins and for the combination of these two. The common luminosity uncertainty is not included in the data uncertainties for $\chi^2$ calculation but has been accounted for by finding a common normalization factor that minimizes the combined $\chi^2/N_{\mathrm{dp}}$. In both cases the scaling factor is consistent with the quoted luminosity uncertainty of $3.5\%$. For the high-mass bin the data actually seem to be better described by the NLO calculation at negative rapidites whereas for positive rapidities it is in good agreement with the NNLO result. For the lower mass bin it seems clear that the NNLO corrections are needed to have a good agreement with this data and also the combined $\chi^2/N_{\mathrm{dp}}$ is significantly smaller at NNLO (1.554) than at NLO (2.261).
\begin{center}
\renewcommand{\arraystretch}{1.25}
\begin{table}[tb!]
\caption{The values of $\chi^2/N_{\mathrm{dp}}$ for the CMS DY data in Fig.~\ref{fig:applicationspPb-DY}. The data points have been scaled by a factor that minimizes $\chi^2$ to account for the correlated luminosity uncertainty ($3.5\%$).}
\label{tab:chi2DY}
\begin{tabular}{l|c|c}
 & $\chi^2/N_{\mathrm{dp}}$(NLO) & $\chi^2/N_{\mathrm{dp}}$(NNLO) \\ 
\hline
$15 < M < 60 $ GeV & 3.002 & 0.735 \\ 
$60 < M < 120 $ GeV & 1.894 & 2.009 \\ 
\hline
combined & 2.261 & 1.554 \\
\hline
\hline
Scaling factor & 0.989 & 1.037 \\ 
\end{tabular}

\end{table}
\end{center}

\section{Summary and outlook}
\label{sec:summary}

We have presented new analyses of nuclear PDFs at NLO and NNLO, TUJU21. We have adopted the same framework as in our previous TUJU19 analysis, but in addition to neutral-current DIS and charged-current neutrino DIS data we have now also included new electroweak-boson production data from the LHC, both for our proton baseline fit and for the nuclear modifications. The resulting nPDFs provide a fully consistent setup for cross section calculations at NNLO in nuclear collisions, for the first time incorporating the LHC data in a full NNLO analysis for nuclear PDFs. The comparisons to the existing nPDF sets demonstrate a reasonable agreement within the error bands, although some discrepancies in flavor dependence were observed. We do point out, however, that the adopted parametrization is rather restrictive, likely resulting in uncertainties that are underestimated in the region $x < 0.001$ where no data have been included. The resulting cross sections show very good agreement with the included experimental data, as confirmed by the total $\Chi^2/N_{\mathrm{dp}}<1.0$ for the nuclear part of the analysis. In the presented framework, the fit performed at NNLO was found to have a significantly lower $\Chi^2/N_{\mathrm{dp}}$ value than the NLO one, 0.84 instead of 0.94. The resulting PDFs will become available in LHAPDF6 format from the LHAPDF home page\footnote{https://lhapdf.hepforge.org/} or by request from the authors.

As an application, we have studied EW-boson production in Pb+Pb collisions at the LHC and have compared the results to recent ATLAS and CMS data. We found that for ATLAS both the NLO and the NNLO computation with the fitted nuclear PDFs tends to be below the data, even though the cross sections were well reproduced for p+Pb collisions. We find better agreement when comparing to the very recent CMS data for $Z$ boson production \cite{CMS:2021kvd}, which hints at a possible tension between the two experimental data sets. We compare our results also to the recent CMS data for DY dilepton production in p+Pb collisions \cite{CMS:2021ynu} that were not included in the presented analysis. Here we find that the NNLO corrections are significant, especially for the lower mass bin, and necessary to have a good description of the data. This demonstrates that for some observables the NNLO corrections can be larger than the uncertainties in nuclear PDF analyses and that these corrections should be taken into account when considering such data.

A possible future improvement would be to analyze the $W^{\pm}$-boson production data removing the constraint $\bar{u}=\bar{d}=\bar{s}$. However, a release of that constraint will increase the number of free parameters and have an impact on the convergence of the fit, likely requiring an extension of the analysis. Another avenue for a step forward will be a combined analysis of proton and nuclear PDFs. The ensuing doubling of the number of fit parameters will demand a full re-thinking of the minimization and PDF determination procedure.

Clearly, our results -- especially the comparisons to other sets of nPDFs -- show that we still have a long way to go until we can be confident to have a good understanding of nuclear modifications of parton distributions. Despite the still large uncertainties, we are encouraged by the improvement that the inclusion of NNLO corrections appears to provide. Ultimately, we hope that the future Electron Ion Collider (EIC) will provide precision constraints on the nuclear PDFs. On the time scale of the EIC, we also expect new analysis technologies to become available that offer new methods for extending the possibilities of theoretical investigations. Among them could be physics simulations on a quantum computer. As an example, a recent study \cite{Lamm:2019uyc} presents an algorithm for computing predictions of parton distribution functions and the hadronic tensor, where it is also speculated that ``quantum supremacy'' could possibly be demonstrated in this area of research as such a task has proven very difficult with classical computers. As another example, a recent study has presented a proof-of-principle demonstration of the determination of proton PDFs with a quantum computer \cite{Perez-Salinas:2020nem}, paving the way for further applications to, for example, protons bound in a nucleus. These new methods and new technologies will be most relevant when aiming at the highest possible perturbative precision, and we are confident that our NNLO analysis will provide a solid reference for such future studies.

\section*{Acknowledgments}

The authors acknowledge support by the state of Baden-W\"urttemberg through bwHPC providing the possibility to run the computational calculations on the high-performance cluster and also wish to acknowledge CSC – IT Center for Science, Finland, for additional computational resources through project jyy2580. Also the support from the Academy of Finland, project numbers 308301 and 331545, and through the CoE in Quark Matter, is acknowledged (I.H.). Furthermore, the authors thank HQS Quantum Simulations GmbH for the supportive working environment allowing the finalization of the publication of the presented results. This work was also supported in part by the Bundesministerium f\"{u}r Bildung und Forschung (BMBF), grant 05P18VTCA1. We thank the \textsc{xFitter} team for the support and I. Novikov for the help with the \textsc{APPLgrid} interfacing. Thanks also to H. Paukkunen for many useful discussions.

\appendix

\section{PDF parameters}
\label{app-pdf-params}

Here we collect the input parameters obtained for the proton and nuclear parton distribution functions presented in section \ref{sec:results}. The naming convention corresponds to the PDF parameterization given in Eqs. (\ref{pdf-parameterization}) and (\ref{coeff-A}). Table \ref{tab-results-NLO} provides the NLO parameters, while Table \ref{tab-results-NNLO} presents the NNLO ones. Some of the parameters were manually set to zero if the data that were used did not provide enough sensitivity to constrain them without large uncertainty. The $A$-dependence was implemented for a subset of parameters, again selected such that the data provided enough sensitivity to result in a converged fit.
\begin{center}
\renewcommand{\arraystretch}{1.25}
\begin{table}[h]
\caption{{Values of the NLO fit parameters at the initial scale, $Q_0^2=1.69\,\mathrm{GeV^2}$. (SR) means that the normalization for that particular parton is fixed by the momentum and valence number sum rules. A dash indicates that this parameter was excluded from the fit. Parameter values for the sea quarks, apart from $\bar{u}$, were derived from the applied constraints $\bar{s}=s=\bar{d}=\bar{u}$.}}
\label{tab-results-NLO} 
\scriptsize
\begin{tabular}{lc|lc|lc|lc}
\hline
$g$ & value & $u_{\mathrm{v}}$ & value & $d_{\mathrm{v}}$ & value & $\bar{u}$ & value \\ \hline
\hline
$c^g_{0,0}$& 8.9596 & $c^{u_v}_{0,0}$ & (SR) & $c^{dv}_{0,0}$ & (SR) & $c^{\bar{u}}_{0,0}$& (SR)  \\ 
$c^g_{1,0}$& 0.3270 & $c^{u_v}_{1,0}$ & 0.7121 & $c^{dv}_{1,0}$ & 0.7629 & $c^{\bar{u}}_{1,0}$& -0.1815 \\ 
$c^g_{2,0}$& 13.438 & $c^{u_v}_{2,0}$ & 3.4290 & $c^{dv}_{2,0}$ & 2.0996 & $c^{\bar{u}}_{2,0}$& 5.2593 \\ 
$c^g_{3,0}$& 6.4371 & $c^{u_v}_{3,0}$ & 1.4506 & $c^{dv}_{3,0}$ & -1.4391 & $c^{\bar{u}}_{3,0}$& 2.4151 \\ 
$c^g_{4,0}$& - & $c^{u_v}_{4,0}$ & - & $c^{dv}_{4,0}$ & - & $c^{\bar{u}}_{4,0}$& - \\  \hline
$c^g_{1,1}$& -5.4728 & $c^{u_v}_{1,1}$ & -0.0462 & $c^{dv}_{1,1}$ & -19.16 & $c^{\bar{u}}_{1,1}$& 251.91 \\ 
$c^g_{1,2}$& -0.0013 & $c^{u_v}_{1,2}$ & 0.3411 & $c^{dv}_{1,2}$ & -0.0026 & $c^{\bar{u}}_{1,2}$& 0.0002 \\ 
$c^g_{2,1}$& -2.000 & $c^{u_v}_{2,1}$ & 4.2325 & $c^{dv}_{2,1}$ & 1.2264 & $c^{\bar{u}}_{2,1}$& -276.53 \\ 
$c^g_{2,2}$& 0.3695 & $c^{u_v}_{2,2}$ & 0.0025 & $c^{dv}_{2,2}$ & 0.4273 & $c^{\bar{u}}_{2,2}$& -0.0017 \\ \hline \hline
\end{tabular} 
\end{table}
\end{center}
\begin{center}
\renewcommand{\arraystretch}{1.25}
\begin{table}[h]
\caption{{Same as Table \ref{tab-results-NLO}, but at NNLO.}}
\label{tab-results-NNLO} 
\scriptsize
\begin{tabular}{lc|lc|lc|lc}
\hline
$g$ & value & $u_{\mathrm{v}}$ & value & $d_{\mathrm{v}}$ & value & $\bar{u}$ & value \\ \hline
$c^g_{0,0}$& 6.4747 & $c^{u_v}_{0,0}$ & (SR) & $c^{dv}_{0,0}$ & (SR) & $c^{\bar{u}}_{0,0}$& (SR) \\ 
$c^g_{1,0}$& 0.2858 & $c^{u_v}_{1,0}$ & 0.7157 & $c^{dv}_{1,0}$ & 0.9101 & $c^{\bar{u}}_{1,0}$& -0.1197
\\ 
$c^g_{2,0}$& 7.6890 & $c^{u_v}_{2,0}$ & 3.6964 & $c^{dv}_{2,0}$ & 3.8936 & $c^{\bar{u}}_{2,0}$& 8.0188 \\ 
$c^g_{3,0}$& -0.0413 & $c^{u_v}_{3,0}$ & 2.5811 & $c^{dv}_{3,0}$ & -0.5844 & $c^{\bar{u}}_{3,0}$& - \\ 
$c^g_{4,0}$& - & $c^{u_v}_{4,0}$ & - & $c^{dv}_{4,0}$ & - & $c^{\bar{u}}_{4,0}$&  11.960 \\  \hline
$c^g_{1,1}$& 2.9882 & $c^{u_v}_{1,1}$ & -0.0235 & $c^{dv}_{1,1}$ & -0.6681 & $c^{\bar{u}}_{1,1}$& -85.228 \\ 
$c^g_{1,2}$& 0.0003 & $c^{u_v}_{1,2}$ & 0.6564 & $c^{dv}_{1,2}$ & -0.0376 & $c^{\bar{u}}_{1,2}$& -0.0005 \\ 
$c^g_{2,1}$& -0.6166 & $c^{u_v}_{2,1}$ & 15.614 & $c^{dv}_{2,1}$ &  1.2905 & $c^{\bar{u}}_{2,1}$& -0.1323 \\ 
$c^g_{2,2}$& 0.4518 & $c^{u_v}_{2,2}$ & -0.0011 & $c^{dv}_{2,2}$ & 0.3396 & $c^{\bar{u}}_{2,2}$& -0.4051 \\ 
\hline \hline
\end{tabular}
\end{table}
\end{center}

\newpage

\bibliography{Paper-nPDFs}

\begin{thebibliography}{100}%
\makeatletter
\providecommand \@ifxundefined [1]{%
 \@ifx{#1\undefined}
}%
\providecommand \@ifnum [1]{%
 \ifnum #1\expandafter \@firstoftwo
 \else \expandafter \@secondoftwo
 \fi
}%
\providecommand \@ifx [1]{%
 \ifx #1\expandafter \@firstoftwo
 \else \expandafter \@secondoftwo
 \fi
}%
\providecommand \natexlab [1]{#1}%
\providecommand \enquote  [1]{``#1''}%
\providecommand \bibnamefont  [1]{#1}%
\providecommand \bibfnamefont [1]{#1}%
\providecommand \citenamefont [1]{#1}%
\providecommand \href@noop [0]{\@secondoftwo}%
\providecommand \href [0]{\begingroup \@sanitize@url \@href}%
\providecommand \@href[1]{\@@startlink{#1}\@@href}%
\providecommand \@@href[1]{\endgroup#1\@@endlink}%
\providecommand \@sanitize@url [0]{\catcode `\\12\catcode `\$12\catcode
  `\&12\catcode `\#12\catcode `\^12\catcode `\_12\catcode `\%12\relax}%
\providecommand \@@startlink[1]{}%
\providecommand \@@endlink[0]{}%
\providecommand \url  [0]{\begingroup\@sanitize@url \@url }%
\providecommand \@url [1]{\endgroup\@href {#1}{\urlprefix }}%
\providecommand \urlprefix  [0]{URL }%
\providecommand \Eprint [0]{\href }%
\providecommand \doibase [0]{http://dx.doi.org/}%
\providecommand \selectlanguage [0]{\@gobble}%
\providecommand \bibinfo  [0]{\@secondoftwo}%
\providecommand \bibfield  [0]{\@secondoftwo}%
\providecommand \translation [1]{[#1]}%
\providecommand \BibitemOpen [0]{}%
\providecommand \bibitemStop [0]{}%
\providecommand \bibitemNoStop [0]{.\EOS\space}%
\providecommand \EOS [0]{\spacefactor3000\relax}%
\providecommand \BibitemShut  [1]{\csname bibitem#1\endcsname}%
\let\auto@bib@innerbib\@empty
\bibitem [{\citenamefont {Collins}\ \emph {et~al.}(1989)\citenamefont
  {Collins}, \citenamefont {Soper},\ and\ \citenamefont
  {Sterman}}]{Collins:1989gx}%
  \BibitemOpen
  \bibfield  {author} {\bibinfo {author} {\bibfnamefont {J.~C.}\ \bibnamefont
  {Collins}}, \bibinfo {author} {\bibfnamefont {D.~E.}\ \bibnamefont {Soper}},
  \ and\ \bibinfo {author} {\bibfnamefont {G.~F.}\ \bibnamefont {Sterman}},\
  }\href {\doibase 10.1142/9789814503266_0001} {\bibfield  {journal} {\bibinfo
  {journal} {Adv. Ser. Direct. High Energy Phys.}\ }\textbf {\bibinfo {volume}
  {5}},\ \bibinfo {pages} {1} (\bibinfo {year} {1989})},\ \Eprint
  {http://arxiv.org/abs/hep-ph/0409313} {arXiv:hep-ph/0409313 [hep-ph]}
  \BibitemShut {NoStop}%
\bibitem [{\citenamefont {Gao}\ \emph {et~al.}(2018)\citenamefont {Gao},
  \citenamefont {Harland-Lang},\ and\ \citenamefont {Rojo}}]{Gao:2017yyd}%
  \BibitemOpen
  \bibfield  {author} {\bibinfo {author} {\bibfnamefont {J.}~\bibnamefont
  {Gao}}, \bibinfo {author} {\bibfnamefont {L.}~\bibnamefont {Harland-Lang}}, \
  and\ \bibinfo {author} {\bibfnamefont {J.}~\bibnamefont {Rojo}},\ }\href
  {\doibase 10.1016/j.physrep.2018.03.002} {\bibfield  {journal} {\bibinfo
  {journal} {Phys. Rept.}\ }\textbf {\bibinfo {volume} {742}},\ \bibinfo
  {pages} {1} (\bibinfo {year} {2018})},\ \Eprint
  {http://arxiv.org/abs/1709.04922} {arXiv:1709.04922 [hep-ph]} \BibitemShut
  {NoStop}%
\bibitem [{\citenamefont {Aaij}\ \emph {et~al.}(2017)\citenamefont {Aaij} \emph
  {et~al.}}]{LHCb:2017yua}%
  \BibitemOpen
  \bibfield  {author} {\bibinfo {author} {\bibfnamefont {R.}~\bibnamefont
  {Aaij}} \emph {et~al.} (\bibinfo {collaboration} {LHCb}),\ }\href {\doibase
  10.1007/JHEP10(2017)090} {\bibfield  {journal} {\bibinfo  {journal} {JHEP}\
  }\textbf {\bibinfo {volume} {10}},\ \bibinfo {pages} {090} (\bibinfo {year}
  {2017})},\ \Eprint {http://arxiv.org/abs/1707.02750} {arXiv:1707.02750
  [hep-ex]} \BibitemShut {NoStop}%
\bibitem [{\citenamefont {Eskola}\ \emph
  {et~al.}(2020{\natexlab{a}})\citenamefont {Eskola}, \citenamefont {Helenius},
  \citenamefont {Paakkinen},\ and\ \citenamefont {Paukkunen}}]{Eskola:2019bgf}%
  \BibitemOpen
  \bibfield  {author} {\bibinfo {author} {\bibfnamefont {K.~J.}\ \bibnamefont
  {Eskola}}, \bibinfo {author} {\bibfnamefont {I.}~\bibnamefont {Helenius}},
  \bibinfo {author} {\bibfnamefont {P.}~\bibnamefont {Paakkinen}}, \ and\
  \bibinfo {author} {\bibfnamefont {H.}~\bibnamefont {Paukkunen}},\ }\href
  {\doibase 10.1007/JHEP05(2020)037} {\bibfield  {journal} {\bibinfo  {journal}
  {JHEP}\ }\textbf {\bibinfo {volume} {05}},\ \bibinfo {pages} {037} (\bibinfo
  {year} {2020}{\natexlab{a}})},\ \Eprint {http://arxiv.org/abs/1906.02512}
  {arXiv:1906.02512 [hep-ph]} \BibitemShut {NoStop}%
\bibitem [{\citenamefont {Aad}\ \emph {et~al.}(2015{\natexlab{a}})\citenamefont
  {Aad} \emph {et~al.}}]{Aad:2015gta}%
  \BibitemOpen
  \bibfield  {author} {\bibinfo {author} {\bibfnamefont {G.}~\bibnamefont
  {Aad}} \emph {et~al.} (\bibinfo {collaboration} {ATLAS}),\ }\href {\doibase
  10.1103/PhysRevC.92.044915} {\bibfield  {journal} {\bibinfo  {journal} {Phys.
  Rev.}\ }\textbf {\bibinfo {volume} {C92}},\ \bibinfo {pages} {044915}
  (\bibinfo {year} {2015}{\natexlab{a}})},\ \Eprint
  {http://arxiv.org/abs/1507.06232} {arXiv:1507.06232 [hep-ex]} \BibitemShut
  {NoStop}%
\bibitem [{\citenamefont {Khachatryan}\ \emph
  {et~al.}(2015{\natexlab{a}})\citenamefont {Khachatryan} \emph
  {et~al.}}]{Khachatryan:2015hha}%
  \BibitemOpen
  \bibfield  {author} {\bibinfo {author} {\bibfnamefont {V.}~\bibnamefont
  {Khachatryan}} \emph {et~al.} (\bibinfo {collaboration} {CMS}),\ }\href
  {\doibase 10.1016/j.physletb.2015.09.057} {\bibfield  {journal} {\bibinfo
  {journal} {Phys. Lett.}\ }\textbf {\bibinfo {volume} {B750}},\ \bibinfo
  {pages} {565} (\bibinfo {year} {2015}{\natexlab{a}})},\ \Eprint
  {http://arxiv.org/abs/1503.05825} {arXiv:1503.05825 [nucl-ex]} \BibitemShut
  {NoStop}%
\bibitem [{\citenamefont {Khachatryan}\ \emph
  {et~al.}(2016{\natexlab{a}})\citenamefont {Khachatryan} \emph
  {et~al.}}]{Khachatryan:2015pzs}%
  \BibitemOpen
  \bibfield  {author} {\bibinfo {author} {\bibfnamefont {V.}~\bibnamefont
  {Khachatryan}} \emph {et~al.} (\bibinfo {collaboration} {CMS}),\ }\href
  {\doibase 10.1016/j.physletb.2016.05.044} {\bibfield  {journal} {\bibinfo
  {journal} {Phys. Lett.}\ }\textbf {\bibinfo {volume} {B759}},\ \bibinfo
  {pages} {36} (\bibinfo {year} {2016}{\natexlab{a}})},\ \Eprint
  {http://arxiv.org/abs/1512.06461} {arXiv:1512.06461 [hep-ex]} \BibitemShut
  {NoStop}%
\bibitem [{\citenamefont {Paukkunen}\ and\ \citenamefont
  {Salgado}(2011)}]{Paukkunen:2010qg}%
  \BibitemOpen
  \bibfield  {author} {\bibinfo {author} {\bibfnamefont {H.}~\bibnamefont
  {Paukkunen}}\ and\ \bibinfo {author} {\bibfnamefont {C.~A.}\ \bibnamefont
  {Salgado}},\ }\href {\doibase 10.1007/JHEP03(2011)071} {\bibfield  {journal}
  {\bibinfo  {journal} {JHEP}\ }\textbf {\bibinfo {volume} {03}},\ \bibinfo
  {pages} {071} (\bibinfo {year} {2011})},\ \Eprint
  {http://arxiv.org/abs/1010.5392} {arXiv:1010.5392 [hep-ph]} \BibitemShut
  {NoStop}%
\bibitem [{\citenamefont {Kusina}\ \emph {et~al.}(2020)\citenamefont {Kusina}
  \emph {et~al.}}]{Kusina:2020lyz}%
  \BibitemOpen
  \bibfield  {author} {\bibinfo {author} {\bibfnamefont {A.}~\bibnamefont
  {Kusina}} \emph {et~al.},\ }\href {\doibase 10.1140/epjc/s10052-020-08532-4}
  {\bibfield  {journal} {\bibinfo  {journal} {Eur. Phys. J. C}\ }\textbf
  {\bibinfo {volume} {80}},\ \bibinfo {pages} {968} (\bibinfo {year} {2020})},\
  \Eprint {http://arxiv.org/abs/2007.09100} {arXiv:2007.09100 [hep-ph]}
  \BibitemShut {NoStop}%
\bibitem [{\citenamefont {Sirunyan}\ \emph {et~al.}(2018)\citenamefont
  {Sirunyan} \emph {et~al.}}]{CMS:2018jpl}%
  \BibitemOpen
  \bibfield  {author} {\bibinfo {author} {\bibfnamefont {A.~M.}\ \bibnamefont
  {Sirunyan}} \emph {et~al.} (\bibinfo {collaboration} {CMS}),\ }\href
  {\doibase 10.1103/PhysRevLett.121.062002} {\bibfield  {journal} {\bibinfo
  {journal} {Phys. Rev. Lett.}\ }\textbf {\bibinfo {volume} {121}},\ \bibinfo
  {pages} {062002} (\bibinfo {year} {2018})},\ \Eprint
  {http://arxiv.org/abs/1805.04736} {arXiv:1805.04736 [hep-ex]} \BibitemShut
  {NoStop}%
\bibitem [{\citenamefont {Eskola}\ \emph {et~al.}(2019)\citenamefont {Eskola},
  \citenamefont {Paakkinen},\ and\ \citenamefont {Paukkunen}}]{Eskola:2019dui}%
  \BibitemOpen
  \bibfield  {author} {\bibinfo {author} {\bibfnamefont {K.~J.}\ \bibnamefont
  {Eskola}}, \bibinfo {author} {\bibfnamefont {P.}~\bibnamefont {Paakkinen}}, \
  and\ \bibinfo {author} {\bibfnamefont {H.}~\bibnamefont {Paukkunen}},\
  }\href@noop {} {\bibfield  {journal} {\bibinfo  {journal} {Eur. Phys. J.}\
  }\textbf {\bibinfo {volume} {C79}},\ \bibinfo {pages} {511} (\bibinfo {year}
  {2019})},\ \Eprint {http://arxiv.org/abs/1903.09832} {arXiv:1903.09832
  [hep-ph]} \BibitemShut {NoStop}%
\bibitem [{\citenamefont {Aaboud}\ \emph
  {et~al.}(2019{\natexlab{a}})\citenamefont {Aaboud} \emph
  {et~al.}}]{ATLAS:2019ery}%
  \BibitemOpen
  \bibfield  {author} {\bibinfo {author} {\bibfnamefont {M.}~\bibnamefont
  {Aaboud}} \emph {et~al.} (\bibinfo {collaboration} {ATLAS}),\ }\href
  {\doibase 10.1016/j.physletb.2019.07.031} {\bibfield  {journal} {\bibinfo
  {journal} {Phys. Lett. B}\ }\textbf {\bibinfo {volume} {796}},\ \bibinfo
  {pages} {230} (\bibinfo {year} {2019}{\natexlab{a}})},\ \Eprint
  {http://arxiv.org/abs/1903.02209} {arXiv:1903.02209 [nucl-ex]} \BibitemShut
  {NoStop}%
\bibitem [{\citenamefont {Acharya}\ \emph {et~al.}(2018)\citenamefont {Acharya}
  \emph {et~al.}}]{ALICE:2018vhm}%
  \BibitemOpen
  \bibfield  {author} {\bibinfo {author} {\bibfnamefont {S.}~\bibnamefont
  {Acharya}} \emph {et~al.} (\bibinfo {collaboration} {ALICE}),\ }\href
  {\doibase 10.1140/epjc/s10052-018-6013-8} {\bibfield  {journal} {\bibinfo
  {journal} {Eur. Phys. J. C}\ }\textbf {\bibinfo {volume} {78}},\ \bibinfo
  {pages} {624} (\bibinfo {year} {2018})},\ \Eprint
  {http://arxiv.org/abs/1801.07051} {arXiv:1801.07051 [nucl-ex]} \BibitemShut
  {NoStop}%
\bibitem [{\citenamefont {Adam}\ \emph {et~al.}(2016)\citenamefont {Adam} \emph
  {et~al.}}]{ALICE:2016dei}%
  \BibitemOpen
  \bibfield  {author} {\bibinfo {author} {\bibfnamefont {J.}~\bibnamefont
  {Adam}} \emph {et~al.} (\bibinfo {collaboration} {ALICE}),\ }\href {\doibase
  10.1016/j.physletb.2016.07.050} {\bibfield  {journal} {\bibinfo  {journal}
  {Phys. Lett. B}\ }\textbf {\bibinfo {volume} {760}},\ \bibinfo {pages} {720}
  (\bibinfo {year} {2016})},\ \Eprint {http://arxiv.org/abs/1601.03658}
  {arXiv:1601.03658 [nucl-ex]} \BibitemShut {NoStop}%
\bibitem [{\citenamefont {Acharya}\ \emph {et~al.}(2022)\citenamefont {Acharya}
  \emph {et~al.}}]{ALICE:2021est}%
  \BibitemOpen
  \bibfield  {author} {\bibinfo {author} {\bibfnamefont {S.}~\bibnamefont
  {Acharya}} \emph {et~al.} (\bibinfo {collaboration} {ALICE}),\ }\href
  {\doibase 10.1016/j.physletb.2022.136943} {\bibfield  {journal} {\bibinfo
  {journal} {Phys. Lett. B}\ }\textbf {\bibinfo {volume} {827}},\ \bibinfo
  {pages} {136943} (\bibinfo {year} {2022})},\ \Eprint
  {http://arxiv.org/abs/2104.03116} {arXiv:2104.03116 [nucl-ex]} \BibitemShut
  {NoStop}%
\bibitem [{\citenamefont {Eskola}\ \emph {et~al.}(2009)\citenamefont {Eskola},
  \citenamefont {Paukkunen},\ and\ \citenamefont {Salgado}}]{Eskola:2009uj}%
  \BibitemOpen
  \bibfield  {author} {\bibinfo {author} {\bibfnamefont {K.~J.}\ \bibnamefont
  {Eskola}}, \bibinfo {author} {\bibfnamefont {H.}~\bibnamefont {Paukkunen}}, \
  and\ \bibinfo {author} {\bibfnamefont {C.~A.}\ \bibnamefont {Salgado}},\
  }\href {\doibase 10.1088/1126-6708/2009/04/065} {\bibfield  {journal}
  {\bibinfo  {journal} {JHEP}\ }\textbf {\bibinfo {volume} {04}},\ \bibinfo
  {pages} {065} (\bibinfo {year} {2009})},\ \Eprint
  {http://arxiv.org/abs/0902.4154} {arXiv:0902.4154 [hep-ph]} \BibitemShut
  {NoStop}%
\bibitem [{\citenamefont {de~Florian}\ \emph {et~al.}(2012)\citenamefont
  {de~Florian}, \citenamefont {Sassot}, \citenamefont {Zurita},\ and\
  \citenamefont {Stratmann}}]{deFlorian:2011fp}%
  \BibitemOpen
  \bibfield  {author} {\bibinfo {author} {\bibfnamefont {D.}~\bibnamefont
  {de~Florian}}, \bibinfo {author} {\bibfnamefont {R.}~\bibnamefont {Sassot}},
  \bibinfo {author} {\bibfnamefont {P.}~\bibnamefont {Zurita}}, \ and\ \bibinfo
  {author} {\bibfnamefont {M.}~\bibnamefont {Stratmann}},\ }\href {\doibase
  10.1103/PhysRevD.85.074028} {\bibfield  {journal} {\bibinfo  {journal} {Phys.
  Rev.}\ }\textbf {\bibinfo {volume} {D85}},\ \bibinfo {pages} {074028}
  (\bibinfo {year} {2012})},\ \Eprint {http://arxiv.org/abs/1112.6324}
  {arXiv:1112.6324 [hep-ph]} \BibitemShut {NoStop}%
\bibitem [{\citenamefont {Eskola}\ \emph {et~al.}(2017)\citenamefont {Eskola},
  \citenamefont {Paakkinen}, \citenamefont {Paukkunen},\ and\ \citenamefont
  {Salgado}}]{Eskola:2016oht}%
  \BibitemOpen
  \bibfield  {author} {\bibinfo {author} {\bibfnamefont {K.~J.}\ \bibnamefont
  {Eskola}}, \bibinfo {author} {\bibfnamefont {P.}~\bibnamefont {Paakkinen}},
  \bibinfo {author} {\bibfnamefont {H.}~\bibnamefont {Paukkunen}}, \ and\
  \bibinfo {author} {\bibfnamefont {C.~A.}\ \bibnamefont {Salgado}},\ }\href
  {\doibase 10.1140/epjc/s10052-017-4725-9} {\bibfield  {journal} {\bibinfo
  {journal} {Eur. Phys. J.}\ }\textbf {\bibinfo {volume} {C77}},\ \bibinfo
  {pages} {163} (\bibinfo {year} {2017})},\ \Eprint
  {http://arxiv.org/abs/1612.05741} {arXiv:1612.05741 [hep-ph]} \BibitemShut
  {NoStop}%
\bibitem [{\citenamefont {Kovarik}\ \emph {et~al.}(2016)\citenamefont {Kovarik}
  \emph {et~al.}}]{Kovarik:2015cma}%
  \BibitemOpen
  \bibfield  {author} {\bibinfo {author} {\bibfnamefont {K.}~\bibnamefont
  {Kovarik}} \emph {et~al.},\ }\href {\doibase 10.1103/PhysRevD.93.085037}
  {\bibfield  {journal} {\bibinfo  {journal} {Phys. Rev.}\ }\textbf {\bibinfo
  {volume} {D93}},\ \bibinfo {pages} {085037} (\bibinfo {year} {2016})},\
  \Eprint {http://arxiv.org/abs/1509.00792} {arXiv:1509.00792 [hep-ph]}
  \BibitemShut {NoStop}%
\bibitem [{\citenamefont {Abdul~Khalek}\ \emph {et~al.}(2020)\citenamefont
  {Abdul~Khalek}, \citenamefont {Ethier}, \citenamefont {Rojo},\ and\
  \citenamefont {van Weelden}}]{AbdulKhalek:2020yuc}%
  \BibitemOpen
  \bibfield  {author} {\bibinfo {author} {\bibfnamefont {R.}~\bibnamefont
  {Abdul~Khalek}}, \bibinfo {author} {\bibfnamefont {J.~J.}\ \bibnamefont
  {Ethier}}, \bibinfo {author} {\bibfnamefont {J.}~\bibnamefont {Rojo}}, \ and\
  \bibinfo {author} {\bibfnamefont {G.}~\bibnamefont {van Weelden}},\ }\href
  {\doibase 10.1007/JHEP09(2020)183} {\bibfield  {journal} {\bibinfo  {journal}
  {JHEP}\ }\textbf {\bibinfo {volume} {09}},\ \bibinfo {pages} {183} (\bibinfo
  {year} {2020})},\ \Eprint {http://arxiv.org/abs/2006.14629} {arXiv:2006.14629
  [hep-ph]} \BibitemShut {NoStop}%
\bibitem [{\citenamefont {Duwent\"aster}\ \emph {et~al.}(2021)\citenamefont
  {Duwent\"aster}, \citenamefont {Husov\'a}, \citenamefont {Je\v{z}o},
  \citenamefont {Klasen}, \citenamefont {Kova\v{r}\'\i{}k}, \citenamefont
  {Kusina}, \citenamefont {Muzakka}, \citenamefont {Olness}, \citenamefont
  {Schienbein},\ and\ \citenamefont {Yu}}]{Duwentaster:2021ioo}%
  \BibitemOpen
  \bibfield  {author} {\bibinfo {author} {\bibfnamefont {P.}~\bibnamefont
  {Duwent\"aster}}, \bibinfo {author} {\bibfnamefont {L.~A.}\ \bibnamefont
  {Husov\'a}}, \bibinfo {author} {\bibfnamefont {T.}~\bibnamefont {Je\v{z}o}},
  \bibinfo {author} {\bibfnamefont {M.}~\bibnamefont {Klasen}}, \bibinfo
  {author} {\bibfnamefont {K.}~\bibnamefont {Kova\v{r}\'\i{}k}}, \bibinfo
  {author} {\bibfnamefont {A.}~\bibnamefont {Kusina}}, \bibinfo {author}
  {\bibfnamefont {K.~F.}\ \bibnamefont {Muzakka}}, \bibinfo {author}
  {\bibfnamefont {F.~I.}\ \bibnamefont {Olness}}, \bibinfo {author}
  {\bibfnamefont {I.}~\bibnamefont {Schienbein}}, \ and\ \bibinfo {author}
  {\bibfnamefont {J.~Y.}\ \bibnamefont {Yu}},\ }\href {\doibase
  10.1103/PhysRevD.104.094005} {\bibfield  {journal} {\bibinfo  {journal}
  {Phys. Rev. D}\ }\textbf {\bibinfo {volume} {104}},\ \bibinfo {pages}
  {094005} (\bibinfo {year} {2021})},\ \Eprint
  {http://arxiv.org/abs/2105.09873} {arXiv:2105.09873 [hep-ph]} \BibitemShut
  {NoStop}%
\bibitem [{\citenamefont {Eskola}\ \emph {et~al.}(2021)\citenamefont {Eskola},
  \citenamefont {Paakkinen}, \citenamefont {Paukkunen},\ and\ \citenamefont
  {Salgado}}]{Eskola:2021nhw}%
  \BibitemOpen
  \bibfield  {author} {\bibinfo {author} {\bibfnamefont {K.~J.}\ \bibnamefont
  {Eskola}}, \bibinfo {author} {\bibfnamefont {P.}~\bibnamefont {Paakkinen}},
  \bibinfo {author} {\bibfnamefont {H.}~\bibnamefont {Paukkunen}}, \ and\
  \bibinfo {author} {\bibfnamefont {C.~A.}\ \bibnamefont {Salgado}},\
  }\href@noop {} {\  (\bibinfo {year} {2021})},\ \Eprint
  {http://arxiv.org/abs/2112.12462} {arXiv:2112.12462 [hep-ph]} \BibitemShut
  {NoStop}%
\bibitem [{\citenamefont {Khalek}\ \emph {et~al.}(2022)\citenamefont {Khalek},
  \citenamefont {Gauld}, \citenamefont {Giani}, \citenamefont {Nocera},
  \citenamefont {Rabemananjara},\ and\ \citenamefont {Rojo}}]{Khalek:2022zqe}%
  \BibitemOpen
  \bibfield  {author} {\bibinfo {author} {\bibfnamefont {R.~A.}\ \bibnamefont
  {Khalek}}, \bibinfo {author} {\bibfnamefont {R.}~\bibnamefont {Gauld}},
  \bibinfo {author} {\bibfnamefont {T.}~\bibnamefont {Giani}}, \bibinfo
  {author} {\bibfnamefont {E.~R.}\ \bibnamefont {Nocera}}, \bibinfo {author}
  {\bibfnamefont {T.~R.}\ \bibnamefont {Rabemananjara}}, \ and\ \bibinfo
  {author} {\bibfnamefont {J.}~\bibnamefont {Rojo}},\ }\href@noop {} {\
  (\bibinfo {year} {2022})},\ \Eprint {http://arxiv.org/abs/2201.12363}
  {arXiv:2201.12363 [hep-ph]} \BibitemShut {NoStop}%
\bibitem [{\citenamefont {Abdul~Khalek}\ \emph {et~al.}(2019)\citenamefont
  {Abdul~Khalek}, \citenamefont {Ethier},\ and\ \citenamefont
  {Rojo}}]{AbdulKhalek:2019mzd}%
  \BibitemOpen
  \bibfield  {author} {\bibinfo {author} {\bibfnamefont {R.}~\bibnamefont
  {Abdul~Khalek}}, \bibinfo {author} {\bibfnamefont {J.~J.}\ \bibnamefont
  {Ethier}}, \ and\ \bibinfo {author} {\bibfnamefont {J.}~\bibnamefont {Rojo}}
  (\bibinfo {collaboration} {NNPDF}),\ }\href {\doibase
  10.1140/epjc/s10052-019-6983-1} {\bibfield  {journal} {\bibinfo  {journal}
  {Eur. Phys. J.}\ }\textbf {\bibinfo {volume} {C79}},\ \bibinfo {pages} {471}
  (\bibinfo {year} {2019})},\ \Eprint {http://arxiv.org/abs/1904.00018}
  {arXiv:1904.00018 [hep-ph]} \BibitemShut {NoStop}%
\bibitem [{\citenamefont {Walt}\ \emph {et~al.}(2019)\citenamefont {Walt},
  \citenamefont {Helenius},\ and\ \citenamefont {Vogelsang}}]{Walt:2019slu}%
  \BibitemOpen
  \bibfield  {author} {\bibinfo {author} {\bibfnamefont {M.}~\bibnamefont
  {Walt}}, \bibinfo {author} {\bibfnamefont {I.}~\bibnamefont {Helenius}}, \
  and\ \bibinfo {author} {\bibfnamefont {W.}~\bibnamefont {Vogelsang}},\ }\href
  {\doibase 10.1103/PhysRevD.100.096015} {\bibfield  {journal} {\bibinfo
  {journal} {Phys. Rev.}\ }\textbf {\bibinfo {volume} {D100}},\ \bibinfo
  {pages} {096015} (\bibinfo {year} {2019})},\ \Eprint
  {http://arxiv.org/abs/1908.03355} {arXiv:1908.03355 [hep-ph]} \BibitemShut
  {NoStop}%
\bibitem [{\citenamefont {Khanpour}\ \emph {et~al.}(2021)\citenamefont
  {Khanpour}, \citenamefont {Soleymaninia}, \citenamefont {Atashbar~Tehrani},
  \citenamefont {Spiesberger},\ and\ \citenamefont {Guzey}}]{Khanpour:2020zyu}%
  \BibitemOpen
  \bibfield  {author} {\bibinfo {author} {\bibfnamefont {H.}~\bibnamefont
  {Khanpour}}, \bibinfo {author} {\bibfnamefont {M.}~\bibnamefont
  {Soleymaninia}}, \bibinfo {author} {\bibfnamefont {S.}~\bibnamefont
  {Atashbar~Tehrani}}, \bibinfo {author} {\bibfnamefont {H.}~\bibnamefont
  {Spiesberger}}, \ and\ \bibinfo {author} {\bibfnamefont {V.}~\bibnamefont
  {Guzey}},\ }\href {\doibase 10.1103/PhysRevD.104.034010} {\bibfield
  {journal} {\bibinfo  {journal} {Phys. Rev. D}\ }\textbf {\bibinfo {volume}
  {104}},\ \bibinfo {pages} {034010} (\bibinfo {year} {2021})},\ \Eprint
  {http://arxiv.org/abs/2010.00555} {arXiv:2010.00555 [hep-ph]} \BibitemShut
  {NoStop}%
\bibitem [{\citenamefont {Aad}\ \emph {et~al.}(2019)\citenamefont {Aad} \emph
  {et~al.}}]{ATLAS:2019ibd}%
  \BibitemOpen
  \bibfield  {author} {\bibinfo {author} {\bibfnamefont {G.}~\bibnamefont
  {Aad}} \emph {et~al.} (\bibinfo {collaboration} {ATLAS}),\ }\href {\doibase
  10.1140/epjc/s10052-019-7439-3} {\bibfield  {journal} {\bibinfo  {journal}
  {Eur. Phys. J. C}\ }\textbf {\bibinfo {volume} {79}},\ \bibinfo {pages} {935}
  (\bibinfo {year} {2019})},\ \Eprint {http://arxiv.org/abs/1907.10414}
  {arXiv:1907.10414 [nucl-ex]} \BibitemShut {NoStop}%
\bibitem [{\citenamefont {Aad}\ \emph {et~al.}(2020)\citenamefont {Aad} \emph
  {et~al.}}]{ATLAS:2019maq}%
  \BibitemOpen
  \bibfield  {author} {\bibinfo {author} {\bibfnamefont {G.}~\bibnamefont
  {Aad}} \emph {et~al.} (\bibinfo {collaboration} {ATLAS}),\ }\href {\doibase
  10.1016/j.physletb.2020.135262} {\bibfield  {journal} {\bibinfo  {journal}
  {Phys. Lett. B}\ }\textbf {\bibinfo {volume} {802}},\ \bibinfo {pages}
  {135262} (\bibinfo {year} {2020})},\ \Eprint
  {http://arxiv.org/abs/1910.13396} {arXiv:1910.13396 [nucl-ex]} \BibitemShut
  {NoStop}%
\bibitem [{\citenamefont {Sirunyan}\ \emph
  {et~al.}(2021{\natexlab{a}})\citenamefont {Sirunyan} \emph
  {et~al.}}]{CMS:2021kvd}%
  \BibitemOpen
  \bibfield  {author} {\bibinfo {author} {\bibfnamefont {A.~M.}\ \bibnamefont
  {Sirunyan}} \emph {et~al.} (\bibinfo {collaboration} {CMS}),\ }\href
  {\doibase 10.1103/PhysRevLett.127.102002} {\bibfield  {journal} {\bibinfo
  {journal} {Phys. Rev. Lett.}\ }\textbf {\bibinfo {volume} {127}},\ \bibinfo
  {pages} {102002} (\bibinfo {year} {2021}{\natexlab{a}})},\ \Eprint
  {http://arxiv.org/abs/2103.14089} {arXiv:2103.14089 [hep-ex]} \BibitemShut
  {NoStop}%
\bibitem [{\citenamefont {Eskola}\ \emph
  {et~al.}(2020{\natexlab{b}})\citenamefont {Eskola}, \citenamefont {Helenius},
  \citenamefont {Kuha},\ and\ \citenamefont {Paukkunen}}]{Eskola:2020lee}%
  \BibitemOpen
  \bibfield  {author} {\bibinfo {author} {\bibfnamefont {K.~J.}\ \bibnamefont
  {Eskola}}, \bibinfo {author} {\bibfnamefont {I.}~\bibnamefont {Helenius}},
  \bibinfo {author} {\bibfnamefont {M.}~\bibnamefont {Kuha}}, \ and\ \bibinfo
  {author} {\bibfnamefont {H.}~\bibnamefont {Paukkunen}},\ }\href {\doibase
  10.1103/PhysRevLett.125.212301} {\bibfield  {journal} {\bibinfo  {journal}
  {Phys. Rev. Lett.}\ }\textbf {\bibinfo {volume} {125}},\ \bibinfo {pages}
  {212301} (\bibinfo {year} {2020}{\natexlab{b}})},\ \Eprint
  {http://arxiv.org/abs/2003.11856} {arXiv:2003.11856 [hep-ph]} \BibitemShut
  {NoStop}%
\bibitem [{\citenamefont {Drell}\ and\ \citenamefont
  {Yan}(1970)}]{Drell:1970wh}%
  \BibitemOpen
  \bibfield  {author} {\bibinfo {author} {\bibfnamefont {S.~D.}\ \bibnamefont
  {Drell}}\ and\ \bibinfo {author} {\bibfnamefont {T.-M.}\ \bibnamefont
  {Yan}},\ }\href {\doibase 10.1103/PhysRevLett.25.316,
  10.1103/PhysRevLett.25.902.2} {\bibfield  {journal} {\bibinfo  {journal}
  {Phys. Rev. Lett.}\ }\textbf {\bibinfo {volume} {25}},\ \bibinfo {pages}
  {316} (\bibinfo {year} {1970})},\ \bibinfo {note} {[Erratum: Phys. Rev.
  Lett.25,902(1970)]}\BibitemShut {NoStop}%
\bibitem [{\citenamefont {Bodwin}(1985)}]{Bodwin:1984hc}%
  \BibitemOpen
  \bibfield  {author} {\bibinfo {author} {\bibfnamefont {G.~T.}\ \bibnamefont
  {Bodwin}},\ }\href {\doibase 10.1103/PhysRevD.34.3932,
  10.1103/PhysRevD.31.2616} {\bibfield  {journal} {\bibinfo  {journal} {Phys.
  Rev.}\ }\textbf {\bibinfo {volume} {D31}},\ \bibinfo {pages} {2616} (\bibinfo
  {year} {1985})},\ \bibinfo {note} {[Erratum: Phys.
  Rev.D34,3932(1986)]}\BibitemShut {NoStop}%
\bibitem [{\citenamefont {Collins}\ \emph {et~al.}(1985)\citenamefont
  {Collins}, \citenamefont {Soper},\ and\ \citenamefont
  {Sterman}}]{Collins:1985ue}%
  \BibitemOpen
  \bibfield  {author} {\bibinfo {author} {\bibfnamefont {J.~C.}\ \bibnamefont
  {Collins}}, \bibinfo {author} {\bibfnamefont {D.~E.}\ \bibnamefont {Soper}},
  \ and\ \bibinfo {author} {\bibfnamefont {G.~F.}\ \bibnamefont {Sterman}},\
  }\href {\doibase 10.1016/0550-3213(85)90565-6} {\bibfield  {journal}
  {\bibinfo  {journal} {Nucl. Phys. B}\ }\textbf {\bibinfo {volume} {261}},\
  \bibinfo {pages} {104} (\bibinfo {year} {1985})}\BibitemShut {NoStop}%
\bibitem [{\citenamefont {Collins}\ \emph {et~al.}(1988)\citenamefont
  {Collins}, \citenamefont {Soper},\ and\ \citenamefont
  {Sterman}}]{Collins:1988ig}%
  \BibitemOpen
  \bibfield  {author} {\bibinfo {author} {\bibfnamefont {J.~C.}\ \bibnamefont
  {Collins}}, \bibinfo {author} {\bibfnamefont {D.~E.}\ \bibnamefont {Soper}},
  \ and\ \bibinfo {author} {\bibfnamefont {G.~F.}\ \bibnamefont {Sterman}},\
  }\href {\doibase 10.1016/0550-3213(88)90130-7} {\bibfield  {journal}
  {\bibinfo  {journal} {Nucl. Phys. B}\ }\textbf {\bibinfo {volume} {308}},\
  \bibinfo {pages} {833} (\bibinfo {year} {1988})}\BibitemShut {NoStop}%
\bibitem [{\citenamefont {Ellis}\ \emph {et~al.}(1996)\citenamefont {Ellis},
  \citenamefont {Stirling},\ and\ \citenamefont {Webber}}]{ellis_qcd_1996}%
  \BibitemOpen
  \bibfield  {author} {\bibinfo {author} {\bibfnamefont {R.~K.}\ \bibnamefont
  {Ellis}}, \bibinfo {author} {\bibfnamefont {W.~J.}\ \bibnamefont {Stirling}},
  \ and\ \bibinfo {author} {\bibfnamefont {B.~R.}\ \bibnamefont {Webber}},\
  }\href@noop {} {\emph {\bibinfo {title} {{QCD} and {Collider} {Physics}}}}\
  (\bibinfo  {publisher} {Cambridge University Press},\ \bibinfo {year}
  {1996})\BibitemShut {NoStop}%
\bibitem [{\citenamefont {Sterman~at al.}(2001)}]{sterman_handbook_2001}%
  \BibitemOpen
  \bibfield  {author} {\bibinfo {author} {\bibfnamefont {G.}~\bibnamefont
  {Sterman~at al.}},\ }\href
  {http://www.e-booksdirectory.com/details.php?ebook=5644} {\emph {\bibinfo
  {title} {Handbook of {Perturbative} {QCD}}}}\ (\bibinfo  {publisher} {College
  Park, Maryland: CTEQ},\ \bibinfo {year} {2001})\BibitemShut {NoStop}%
\bibitem [{\citenamefont {Altarelli}\ \emph {et~al.}(1979)\citenamefont
  {Altarelli}, \citenamefont {Ellis},\ and\ \citenamefont
  {Martinelli}}]{Altarelli:1979ub}%
  \BibitemOpen
  \bibfield  {author} {\bibinfo {author} {\bibfnamefont {G.}~\bibnamefont
  {Altarelli}}, \bibinfo {author} {\bibfnamefont {R.~K.}\ \bibnamefont
  {Ellis}}, \ and\ \bibinfo {author} {\bibfnamefont {G.}~\bibnamefont
  {Martinelli}},\ }\href {\doibase 10.1016/0550-3213(79)90116-0} {\bibfield
  {journal} {\bibinfo  {journal} {Nucl. Phys. B}\ }\textbf {\bibinfo {volume}
  {157}},\ \bibinfo {pages} {461} (\bibinfo {year} {1979})}\BibitemShut
  {NoStop}%
\bibitem [{\citenamefont {Hamberg}\ \emph {et~al.}(1991)\citenamefont
  {Hamberg}, \citenamefont {van Neerven},\ and\ \citenamefont
  {Matsuura}}]{Hamberg:1990np}%
  \BibitemOpen
  \bibfield  {author} {\bibinfo {author} {\bibfnamefont {R.}~\bibnamefont
  {Hamberg}}, \bibinfo {author} {\bibfnamefont {W.~L.}\ \bibnamefont {van
  Neerven}}, \ and\ \bibinfo {author} {\bibfnamefont {T.}~\bibnamefont
  {Matsuura}},\ }\href {\doibase 10.1016/0550-3213(91)90064-5} {\bibfield
  {journal} {\bibinfo  {journal} {Nucl. Phys. B}\ }\textbf {\bibinfo {volume}
  {359}},\ \bibinfo {pages} {343} (\bibinfo {year} {1991})},\ \bibinfo {note}
  {[Erratum: Nucl.Phys.B 644, 403--404 (2002)]}\BibitemShut {NoStop}%
\bibitem [{\citenamefont {Harlander}\ and\ \citenamefont
  {Kilgore}(2002)}]{Harlander:2002wh}%
  \BibitemOpen
  \bibfield  {author} {\bibinfo {author} {\bibfnamefont {R.~V.}\ \bibnamefont
  {Harlander}}\ and\ \bibinfo {author} {\bibfnamefont {W.~B.}\ \bibnamefont
  {Kilgore}},\ }\href {\doibase 10.1103/PhysRevLett.88.201801} {\bibfield
  {journal} {\bibinfo  {journal} {Phys. Rev. Lett.}\ }\textbf {\bibinfo
  {volume} {88}},\ \bibinfo {pages} {201801} (\bibinfo {year} {2002})},\
  \Eprint {http://arxiv.org/abs/hep-ph/0201206} {arXiv:hep-ph/0201206}
  \BibitemShut {NoStop}%
\bibitem [{\citenamefont {Anastasiou}\ \emph {et~al.}(2003)\citenamefont
  {Anastasiou}, \citenamefont {Dixon}, \citenamefont {Melnikov},\ and\
  \citenamefont {Petriello}}]{Anastasiou:2003yy}%
  \BibitemOpen
  \bibfield  {author} {\bibinfo {author} {\bibfnamefont {C.}~\bibnamefont
  {Anastasiou}}, \bibinfo {author} {\bibfnamefont {L.~J.}\ \bibnamefont
  {Dixon}}, \bibinfo {author} {\bibfnamefont {K.}~\bibnamefont {Melnikov}}, \
  and\ \bibinfo {author} {\bibfnamefont {F.}~\bibnamefont {Petriello}},\ }\href
  {\doibase 10.1103/PhysRevLett.91.182002} {\bibfield  {journal} {\bibinfo
  {journal} {Phys. Rev. Lett.}\ }\textbf {\bibinfo {volume} {91}},\ \bibinfo
  {pages} {182002} (\bibinfo {year} {2003})},\ \Eprint
  {http://arxiv.org/abs/hep-ph/0306192} {arXiv:hep-ph/0306192} \BibitemShut
  {NoStop}%
\bibitem [{\citenamefont {Anastasiou}\ \emph {et~al.}(2004)\citenamefont
  {Anastasiou}, \citenamefont {Dixon}, \citenamefont {Melnikov},\ and\
  \citenamefont {Petriello}}]{Anastasiou:2003ds}%
  \BibitemOpen
  \bibfield  {author} {\bibinfo {author} {\bibfnamefont {C.}~\bibnamefont
  {Anastasiou}}, \bibinfo {author} {\bibfnamefont {L.~J.}\ \bibnamefont
  {Dixon}}, \bibinfo {author} {\bibfnamefont {K.}~\bibnamefont {Melnikov}}, \
  and\ \bibinfo {author} {\bibfnamefont {F.}~\bibnamefont {Petriello}},\ }\href
  {\doibase 10.1103/PhysRevD.69.094008} {\bibfield  {journal} {\bibinfo
  {journal} {Phys. Rev. D}\ }\textbf {\bibinfo {volume} {69}},\ \bibinfo
  {pages} {094008} (\bibinfo {year} {2004})},\ \Eprint
  {http://arxiv.org/abs/hep-ph/0312266} {arXiv:hep-ph/0312266} \BibitemShut
  {NoStop}%
\bibitem [{\citenamefont {Catani}\ \emph {et~al.}(2009)\citenamefont {Catani},
  \citenamefont {Cieri}, \citenamefont {Ferrera}, \citenamefont {de~Florian},\
  and\ \citenamefont {Grazzini}}]{Catani:2009sm}%
  \BibitemOpen
  \bibfield  {author} {\bibinfo {author} {\bibfnamefont {S.}~\bibnamefont
  {Catani}}, \bibinfo {author} {\bibfnamefont {L.}~\bibnamefont {Cieri}},
  \bibinfo {author} {\bibfnamefont {G.}~\bibnamefont {Ferrera}}, \bibinfo
  {author} {\bibfnamefont {D.}~\bibnamefont {de~Florian}}, \ and\ \bibinfo
  {author} {\bibfnamefont {M.}~\bibnamefont {Grazzini}},\ }\href {\doibase
  10.1103/PhysRevLett.103.082001} {\bibfield  {journal} {\bibinfo  {journal}
  {Phys. Rev. Lett.}\ }\textbf {\bibinfo {volume} {103}},\ \bibinfo {pages}
  {082001} (\bibinfo {year} {2009})},\ \Eprint {http://arxiv.org/abs/0903.2120}
  {arXiv:0903.2120 [hep-ph]} \BibitemShut {NoStop}%
\bibitem [{\citenamefont {Gavin}\ \emph {et~al.}(2011)\citenamefont {Gavin},
  \citenamefont {Li}, \citenamefont {Petriello},\ and\ \citenamefont
  {Quackenbush}}]{Gavin:2010az}%
  \BibitemOpen
  \bibfield  {author} {\bibinfo {author} {\bibfnamefont {R.}~\bibnamefont
  {Gavin}}, \bibinfo {author} {\bibfnamefont {Y.}~\bibnamefont {Li}}, \bibinfo
  {author} {\bibfnamefont {F.}~\bibnamefont {Petriello}}, \ and\ \bibinfo
  {author} {\bibfnamefont {S.}~\bibnamefont {Quackenbush}},\ }\href {\doibase
  10.1016/j.cpc.2011.06.008} {\bibfield  {journal} {\bibinfo  {journal}
  {Comput. Phys. Commun.}\ }\textbf {\bibinfo {volume} {182}},\ \bibinfo
  {pages} {2388} (\bibinfo {year} {2011})},\ \Eprint
  {http://arxiv.org/abs/1011.3540} {arXiv:1011.3540 [hep-ph]} \BibitemShut
  {NoStop}%
\bibitem [{\citenamefont {Gribov}\ and\ \citenamefont
  {Lipatov}(1972)}]{Gribov:1972ri}%
  \BibitemOpen
  \bibfield  {author} {\bibinfo {author} {\bibfnamefont {V.~N.}\ \bibnamefont
  {Gribov}}\ and\ \bibinfo {author} {\bibfnamefont {L.~N.}\ \bibnamefont
  {Lipatov}},\ }\href@noop {} {\bibfield  {journal} {\bibinfo  {journal} {Sov.
  J. Nucl. Phys.}\ }\textbf {\bibinfo {volume} {15}},\ \bibinfo {pages} {438}
  (\bibinfo {year} {1972})},\ \bibinfo {note} {[Yad.
  Fiz.15,781(1972)]}\BibitemShut {NoStop}%
\bibitem [{\citenamefont {Lipatov}(1975)}]{Lipatov:1974qm}%
  \BibitemOpen
  \bibfield  {author} {\bibinfo {author} {\bibfnamefont {L.~N.}\ \bibnamefont
  {Lipatov}},\ }\href@noop {} {\bibfield  {journal} {\bibinfo  {journal} {Sov.
  J. Nucl. Phys.}\ }\textbf {\bibinfo {volume} {20}},\ \bibinfo {pages} {94}
  (\bibinfo {year} {1975})},\ \bibinfo {note} {[Yad.
  Fiz.20,181(1974)]}\BibitemShut {NoStop}%
\bibitem [{\citenamefont {Altarelli}\ and\ \citenamefont
  {Parisi}(1977)}]{Altarelli:1977zs}%
  \BibitemOpen
  \bibfield  {author} {\bibinfo {author} {\bibfnamefont {G.}~\bibnamefont
  {Altarelli}}\ and\ \bibinfo {author} {\bibfnamefont {G.}~\bibnamefont
  {Parisi}},\ }\href {\doibase 10.1016/0550-3213(77)90384-4} {\bibfield
  {journal} {\bibinfo  {journal} {Nucl. Phys.}\ }\textbf {\bibinfo {volume}
  {B126}},\ \bibinfo {pages} {298} (\bibinfo {year} {1977})}\BibitemShut
  {NoStop}%
\bibitem [{\citenamefont {Dokshitzer}(1977)}]{Dokshitzer:1977sg}%
  \BibitemOpen
  \bibfield  {author} {\bibinfo {author} {\bibfnamefont {Y.~L.}\ \bibnamefont
  {Dokshitzer}},\ }\href@noop {} {\bibfield  {journal} {\bibinfo  {journal}
  {Sov. Phys. JETP}\ }\textbf {\bibinfo {volume} {46}},\ \bibinfo {pages} {641}
  (\bibinfo {year} {1977})},\ \bibinfo {note} {[Zh. Eksp. Teor.
  Fiz.73,1216(1977)]}\BibitemShut {NoStop}%
\bibitem [{\citenamefont {Cacciari}\ \emph {et~al.}(1998)\citenamefont
  {Cacciari}, \citenamefont {Greco},\ and\ \citenamefont
  {Nason}}]{Cacciari:1998it}%
  \BibitemOpen
  \bibfield  {author} {\bibinfo {author} {\bibfnamefont {M.}~\bibnamefont
  {Cacciari}}, \bibinfo {author} {\bibfnamefont {M.}~\bibnamefont {Greco}}, \
  and\ \bibinfo {author} {\bibfnamefont {P.}~\bibnamefont {Nason}},\ }\href
  {\doibase 10.1088/1126-6708/1998/05/007} {\bibfield  {journal} {\bibinfo
  {journal} {JHEP}\ }\textbf {\bibinfo {volume} {05}},\ \bibinfo {pages} {007}
  (\bibinfo {year} {1998})},\ \Eprint {http://arxiv.org/abs/hep-ph/9803400}
  {arXiv:hep-ph/9803400 [hep-ph]} \BibitemShut {NoStop}%
\bibitem [{\citenamefont {Forte}\ \emph {et~al.}(2010)\citenamefont {Forte},
  \citenamefont {Laenen}, \citenamefont {Nason},\ and\ \citenamefont
  {Rojo}}]{Forte:2010ta}%
  \BibitemOpen
  \bibfield  {author} {\bibinfo {author} {\bibfnamefont {S.}~\bibnamefont
  {Forte}}, \bibinfo {author} {\bibfnamefont {E.}~\bibnamefont {Laenen}},
  \bibinfo {author} {\bibfnamefont {P.}~\bibnamefont {Nason}}, \ and\ \bibinfo
  {author} {\bibfnamefont {J.}~\bibnamefont {Rojo}},\ }\href {\doibase
  10.1016/j.nuclphysb.2010.03.014} {\bibfield  {journal} {\bibinfo  {journal}
  {Nucl. Phys.}\ }\textbf {\bibinfo {volume} {B834}},\ \bibinfo {pages} {116}
  (\bibinfo {year} {2010})},\ \Eprint {http://arxiv.org/abs/1001.2312}
  {arXiv:1001.2312 [hep-ph]} \BibitemShut {NoStop}%
\bibitem [{\citenamefont {Pumplin}\ \emph
  {et~al.}(2001{\natexlab{a}})\citenamefont {Pumplin}, \citenamefont {Stump},
  \citenamefont {Brock}, \citenamefont {Casey}, \citenamefont {Huston},
  \citenamefont {Kalk}, \citenamefont {Lai},\ and\ \citenamefont
  {Tung}}]{Pumplin:2001ct}%
  \BibitemOpen
  \bibfield  {author} {\bibinfo {author} {\bibfnamefont {J.}~\bibnamefont
  {Pumplin}}, \bibinfo {author} {\bibfnamefont {D.}~\bibnamefont {Stump}},
  \bibinfo {author} {\bibfnamefont {R.}~\bibnamefont {Brock}}, \bibinfo
  {author} {\bibfnamefont {D.}~\bibnamefont {Casey}}, \bibinfo {author}
  {\bibfnamefont {J.}~\bibnamefont {Huston}}, \bibinfo {author} {\bibfnamefont
  {J.}~\bibnamefont {Kalk}}, \bibinfo {author} {\bibfnamefont {H.~L.}\
  \bibnamefont {Lai}}, \ and\ \bibinfo {author} {\bibfnamefont {W.~K.}\
  \bibnamefont {Tung}},\ }\href {\doibase 10.1103/PhysRevD.65.014013}
  {\bibfield  {journal} {\bibinfo  {journal} {Phys. Rev.}\ }\textbf {\bibinfo
  {volume} {D65}},\ \bibinfo {pages} {014013} (\bibinfo {year}
  {2001}{\natexlab{a}})},\ \Eprint {http://arxiv.org/abs/hep-ph/0101032}
  {arXiv:hep-ph/0101032 [hep-ph]} \BibitemShut {NoStop}%
\bibitem [{\citenamefont {Pumplin}\ \emph
  {et~al.}(2001{\natexlab{b}})\citenamefont {Pumplin}, \citenamefont {Stump},\
  and\ \citenamefont {Tung}}]{Pumplin:2000vx}%
  \BibitemOpen
  \bibfield  {author} {\bibinfo {author} {\bibfnamefont {J.}~\bibnamefont
  {Pumplin}}, \bibinfo {author} {\bibfnamefont {D.~R.}\ \bibnamefont {Stump}},
  \ and\ \bibinfo {author} {\bibfnamefont {W.~K.}\ \bibnamefont {Tung}},\
  }\href {\doibase 10.1103/PhysRevD.65.014011} {\bibfield  {journal} {\bibinfo
  {journal} {Phys. Rev.}\ }\textbf {\bibinfo {volume} {D65}},\ \bibinfo {pages}
  {014011} (\bibinfo {year} {2001}{\natexlab{b}})},\ \Eprint
  {http://arxiv.org/abs/hep-ph/0008191} {arXiv:hep-ph/0008191 [hep-ph]}
  \BibitemShut {NoStop}%
\bibitem [{\citenamefont {Zenaiev}(2016)}]{Zenaiev:2016jnq}%
  \BibitemOpen
  \bibfield  {author} {\bibinfo {author} {\bibfnamefont {O.}~\bibnamefont
  {Zenaiev}} (\bibinfo {collaboration} {xFitter team}),\ }\href {\doibase
  10.22323/1.265.0033} {\bibfield  {journal} {\bibinfo  {journal} {PoS}\
  }\textbf {\bibinfo {volume} {DIS2016}},\ \bibinfo {pages} {033} (\bibinfo
  {year} {2016})}\BibitemShut {NoStop}%
\bibitem [{\citenamefont {Bertone}\ \emph {et~al.}(2018)\citenamefont {Bertone}
  \emph {et~al.}}]{Bertone:2017tig}%
  \BibitemOpen
  \bibfield  {author} {\bibinfo {author} {\bibfnamefont {V.}~\bibnamefont
  {Bertone}} \emph {et~al.} (\bibinfo {collaboration} {xFitter Developers'
  Team}),\ }\href {\doibase 10.22323/1.297.0203} {\bibfield  {journal}
  {\bibinfo  {journal} {PoS}\ }\textbf {\bibinfo {volume} {DIS2017}},\ \bibinfo
  {pages} {203} (\bibinfo {year} {2018})},\ \Eprint
  {http://arxiv.org/abs/1709.01151} {arXiv:1709.01151 [hep-ph]} \BibitemShut
  {NoStop}%
\bibitem [{\citenamefont {Carli}\ \emph {et~al.}(2010)\citenamefont {Carli},
  \citenamefont {Clements}, \citenamefont {Cooper-Sarkar}, \citenamefont
  {Gwenlan}, \citenamefont {Salam}, \citenamefont {Siegert}, \citenamefont
  {Starovoitov},\ and\ \citenamefont {Sutton}}]{Carli:2010rw}%
  \BibitemOpen
  \bibfield  {author} {\bibinfo {author} {\bibfnamefont {T.}~\bibnamefont
  {Carli}}, \bibinfo {author} {\bibfnamefont {D.}~\bibnamefont {Clements}},
  \bibinfo {author} {\bibfnamefont {A.}~\bibnamefont {Cooper-Sarkar}}, \bibinfo
  {author} {\bibfnamefont {C.}~\bibnamefont {Gwenlan}}, \bibinfo {author}
  {\bibfnamefont {G.~P.}\ \bibnamefont {Salam}}, \bibinfo {author}
  {\bibfnamefont {F.}~\bibnamefont {Siegert}}, \bibinfo {author} {\bibfnamefont
  {P.}~\bibnamefont {Starovoitov}}, \ and\ \bibinfo {author} {\bibfnamefont
  {M.}~\bibnamefont {Sutton}},\ }\href {\doibase
  10.1140/epjc/s10052-010-1255-0} {\bibfield  {journal} {\bibinfo  {journal}
  {Eur. Phys. J.}\ }\textbf {\bibinfo {volume} {C66}},\ \bibinfo {pages} {503}
  (\bibinfo {year} {2010})},\ \Eprint {http://arxiv.org/abs/0911.2985}
  {arXiv:0911.2985 [hep-ph]} \BibitemShut {NoStop}%
\bibitem [{\citenamefont {Brun}\ and\ \citenamefont
  {Rademakers}(1997)}]{Brun:1997pa}%
  \BibitemOpen
  \bibfield  {author} {\bibinfo {author} {\bibfnamefont {R.}~\bibnamefont
  {Brun}}\ and\ \bibinfo {author} {\bibfnamefont {F.}~\bibnamefont
  {Rademakers}},\ }\href {\doibase 10.1016/S0168-9002(97)00048-X} {\bibfield
  {journal} {\bibinfo  {journal} {Nucl. Instrum. Meth. A}\ }\textbf {\bibinfo
  {volume} {389}},\ \bibinfo {pages} {81} (\bibinfo {year} {1997})}\BibitemShut
  {NoStop}%
\bibitem [{\citenamefont {Buckley}\ \emph {et~al.}(2015)\citenamefont
  {Buckley}, \citenamefont {Ferrando}, \citenamefont {Lloyd}, \citenamefont
  {Nordström}, \citenamefont {Page}, \citenamefont {Rüfenacht}, \citenamefont
  {Schönherr},\ and\ \citenamefont {Watt}}]{Buckley:2014ana}%
  \BibitemOpen
  \bibfield  {author} {\bibinfo {author} {\bibfnamefont {A.}~\bibnamefont
  {Buckley}}, \bibinfo {author} {\bibfnamefont {J.}~\bibnamefont {Ferrando}},
  \bibinfo {author} {\bibfnamefont {S.}~\bibnamefont {Lloyd}}, \bibinfo
  {author} {\bibfnamefont {K.}~\bibnamefont {Nordström}}, \bibinfo {author}
  {\bibfnamefont {B.}~\bibnamefont {Page}}, \bibinfo {author} {\bibfnamefont
  {M.}~\bibnamefont {Rüfenacht}}, \bibinfo {author} {\bibfnamefont
  {M.}~\bibnamefont {Schönherr}}, \ and\ \bibinfo {author} {\bibfnamefont
  {G.}~\bibnamefont {Watt}},\ }\href {\doibase 10.1140/epjc/s10052-015-3318-8}
  {\bibfield  {journal} {\bibinfo  {journal} {Eur. Phys. J.}\ }\textbf
  {\bibinfo {volume} {C75}},\ \bibinfo {pages} {132} (\bibinfo {year}
  {2015})},\ \Eprint {http://arxiv.org/abs/1412.7420} {arXiv:1412.7420
  [hep-ph]} \BibitemShut {NoStop}%
\bibitem [{\citenamefont {Falkowski}\ \emph {et~al.}(2013)\citenamefont
  {Falkowski}, \citenamefont {Mangano}, \citenamefont {Martin}, \citenamefont
  {Perez},\ and\ \citenamefont {Winter}}]{Falkowski:2012cu}%
  \BibitemOpen
  \bibfield  {author} {\bibinfo {author} {\bibfnamefont {A.}~\bibnamefont
  {Falkowski}}, \bibinfo {author} {\bibfnamefont {M.~L.}\ \bibnamefont
  {Mangano}}, \bibinfo {author} {\bibfnamefont {A.}~\bibnamefont {Martin}},
  \bibinfo {author} {\bibfnamefont {G.}~\bibnamefont {Perez}}, \ and\ \bibinfo
  {author} {\bibfnamefont {J.}~\bibnamefont {Winter}},\ }\href {\doibase
  10.1103/PhysRevD.87.034039} {\bibfield  {journal} {\bibinfo  {journal} {Phys.
  Rev.}\ }\textbf {\bibinfo {volume} {D87}},\ \bibinfo {pages} {034039}
  (\bibinfo {year} {2013})},\ \Eprint {http://arxiv.org/abs/1212.4003}
  {arXiv:1212.4003 [hep-ph]} \BibitemShut {NoStop}%
\bibitem [{\citenamefont {Campbell}\ \emph {et~al.}(2015)\citenamefont
  {Campbell}, \citenamefont {Ellis},\ and\ \citenamefont
  {Giele}}]{Campbell:2015qma}%
  \BibitemOpen
  \bibfield  {author} {\bibinfo {author} {\bibfnamefont {J.~M.}\ \bibnamefont
  {Campbell}}, \bibinfo {author} {\bibfnamefont {R.~K.}\ \bibnamefont {Ellis}},
  \ and\ \bibinfo {author} {\bibfnamefont {W.~T.}\ \bibnamefont {Giele}},\
  }\href {\doibase 10.1140/epjc/s10052-015-3461-2} {\bibfield  {journal}
  {\bibinfo  {journal} {Eur. Phys. J.}\ }\textbf {\bibinfo {volume} {C75}},\
  \bibinfo {pages} {246} (\bibinfo {year} {2015})},\ \Eprint
  {http://arxiv.org/abs/1503.06182} {arXiv:1503.06182} \BibitemShut {NoStop}%
\bibitem [{\citenamefont {Boughezal}\ \emph {et~al.}(2017)\citenamefont
  {Boughezal}, \citenamefont {Campbell}, \citenamefont {Ellis}, \citenamefont
  {Focke}, \citenamefont {Giele}, \citenamefont {Liu}, \citenamefont
  {Petriello},\ and\ \citenamefont {Williams}}]{Boughezal:2016wmq}%
  \BibitemOpen
  \bibfield  {author} {\bibinfo {author} {\bibfnamefont {R.}~\bibnamefont
  {Boughezal}}, \bibinfo {author} {\bibfnamefont {J.~M.}\ \bibnamefont
  {Campbell}}, \bibinfo {author} {\bibfnamefont {R.~K.}\ \bibnamefont {Ellis}},
  \bibinfo {author} {\bibfnamefont {C.}~\bibnamefont {Focke}}, \bibinfo
  {author} {\bibfnamefont {W.}~\bibnamefont {Giele}}, \bibinfo {author}
  {\bibfnamefont {X.}~\bibnamefont {Liu}}, \bibinfo {author} {\bibfnamefont
  {F.}~\bibnamefont {Petriello}}, \ and\ \bibinfo {author} {\bibfnamefont
  {C.}~\bibnamefont {Williams}},\ }\href {\doibase
  10.1140/epjc/s10052-016-4558-y} {\bibfield  {journal} {\bibinfo  {journal}
  {Eur. Phys. J.}\ }\textbf {\bibinfo {volume} {C77}},\ \bibinfo {pages} {7}
  (\bibinfo {year} {2017})},\ \Eprint {http://arxiv.org/abs/1605.08011}
  {arXiv:1605.08011 [hep-ph]} \BibitemShut {NoStop}%
\bibitem [{xfi()}]{xfitter_link}%
  \BibitemOpen
  \href {https://www.xfitter.org/xFitter} {\enquote {\bibinfo {title} {xfitter:
  https://www.xfitter.org/xfitter},}\ }\BibitemShut {NoStop}%
\bibitem [{\citenamefont {Aad}\ \emph {et~al.}(2013{\natexlab{a}})\citenamefont
  {Aad} \emph {et~al.}}]{Aad:2013iua}%
  \BibitemOpen
  \bibfield  {author} {\bibinfo {author} {\bibfnamefont {G.}~\bibnamefont
  {Aad}} \emph {et~al.} (\bibinfo {collaboration} {ATLAS}),\ }\href {\doibase
  10.1016/j.physletb.2013.07.049} {\bibfield  {journal} {\bibinfo  {journal}
  {Phys. Lett.}\ }\textbf {\bibinfo {volume} {B725}},\ \bibinfo {pages} {223}
  (\bibinfo {year} {2013}{\natexlab{a}})},\ \Eprint
  {http://arxiv.org/abs/1305.4192} {arXiv:1305.4192 [hep-ex]} \BibitemShut
  {NoStop}%
\bibitem [{\citenamefont {Aad}\ \emph {et~al.}(2014)\citenamefont {Aad} \emph
  {et~al.}}]{Aad:2014qja}%
  \BibitemOpen
  \bibfield  {author} {\bibinfo {author} {\bibfnamefont {G.}~\bibnamefont
  {Aad}} \emph {et~al.} (\bibinfo {collaboration} {ATLAS}),\ }\href {\doibase
  10.1007/JHEP06(2014)112} {\bibfield  {journal} {\bibinfo  {journal} {JHEP}\
  }\textbf {\bibinfo {volume} {06}},\ \bibinfo {pages} {112} (\bibinfo {year}
  {2014})},\ \Eprint {http://arxiv.org/abs/1404.1212} {arXiv:1404.1212
  [hep-ex]} \BibitemShut {NoStop}%
\bibitem [{\citenamefont {Aad}\ \emph {et~al.}(2012)\citenamefont {Aad} \emph
  {et~al.}}]{Aad:2012sb}%
  \BibitemOpen
  \bibfield  {author} {\bibinfo {author} {\bibfnamefont {G.}~\bibnamefont
  {Aad}} \emph {et~al.} (\bibinfo {collaboration} {ATLAS}),\ }\href {\doibase
  10.1103/PhysRevLett.109.012001} {\bibfield  {journal} {\bibinfo  {journal}
  {Phys. Rev. Lett.}\ }\textbf {\bibinfo {volume} {109}},\ \bibinfo {pages}
  {012001} (\bibinfo {year} {2012})},\ \Eprint {http://arxiv.org/abs/1203.4051}
  {arXiv:1203.4051 [hep-ex]} \BibitemShut {NoStop}%
\bibitem [{\citenamefont {Aaboud}\ \emph {et~al.}(2017)\citenamefont {Aaboud}
  \emph {et~al.}}]{Aaboud:2016btc}%
  \BibitemOpen
  \bibfield  {author} {\bibinfo {author} {\bibfnamefont {M.}~\bibnamefont
  {Aaboud}} \emph {et~al.} (\bibinfo {collaboration} {ATLAS}),\ }\href
  {\doibase 10.1140/epjc/s10052-017-4911-9} {\bibfield  {journal} {\bibinfo
  {journal} {Eur. Phys. J.}\ }\textbf {\bibinfo {volume} {C77}},\ \bibinfo
  {pages} {367} (\bibinfo {year} {2017})},\ \Eprint
  {http://arxiv.org/abs/1612.03016} {arXiv:1612.03016 [hep-ex]} \BibitemShut
  {NoStop}%
\bibitem [{\citenamefont {Benvenuti}\ \emph {et~al.}(1989)\citenamefont
  {Benvenuti} \emph {et~al.}}]{Benvenuti:1989rh}%
  \BibitemOpen
  \bibfield  {author} {\bibinfo {author} {\bibfnamefont {A.~C.}\ \bibnamefont
  {Benvenuti}} \emph {et~al.} (\bibinfo {collaboration} {BCDMS}),\ }\href
  {\doibase 10.1016/0370-2693(89)91637-7} {\bibfield  {journal} {\bibinfo
  {journal} {Phys. Lett.}\ }\textbf {\bibinfo {volume} {B223}},\ \bibinfo
  {pages} {485} (\bibinfo {year} {1989})}\BibitemShut {NoStop}%
\bibitem [{\citenamefont {Abramowicz}\ \emph {et~al.}(2015)\citenamefont
  {Abramowicz} \emph {et~al.}}]{Abramowicz:2015mha}%
  \BibitemOpen
  \bibfield  {author} {\bibinfo {author} {\bibfnamefont {H.}~\bibnamefont
  {Abramowicz}} \emph {et~al.} (\bibinfo {collaboration} {H1, ZEUS}),\ }\href
  {\doibase 10.1140/epjc/s10052-015-3710-4} {\bibfield  {journal} {\bibinfo
  {journal} {Eur. Phys. J.}\ }\textbf {\bibinfo {volume} {C75}},\ \bibinfo
  {pages} {580} (\bibinfo {year} {2015})},\ \Eprint
  {http://arxiv.org/abs/1506.06042} {arXiv:1506.06042 [hep-ex]} \BibitemShut
  {NoStop}%
\bibitem [{\citenamefont {Arneodo}\ \emph {et~al.}(1997)\citenamefont {Arneodo}
  \emph {et~al.}}]{Arneodo:1996qe}%
  \BibitemOpen
  \bibfield  {author} {\bibinfo {author} {\bibfnamefont {M.}~\bibnamefont
  {Arneodo}} \emph {et~al.} (\bibinfo {collaboration} {New Muon}),\ }\href
  {\doibase 10.1016/S0550-3213(96)00538-X} {\bibfield  {journal} {\bibinfo
  {journal} {Nucl. Phys.}\ }\textbf {\bibinfo {volume} {B483}},\ \bibinfo
  {pages} {3} (\bibinfo {year} {1997})},\ \Eprint
  {http://arxiv.org/abs/hep-ph/9610231} {arXiv:hep-ph/9610231 [hep-ph]}
  \BibitemShut {NoStop}%
\bibitem [{\citenamefont {Khachatryan}\ \emph
  {et~al.}(2016{\natexlab{b}})\citenamefont {Khachatryan} \emph
  {et~al.}}]{Khachatryan:2016pev}%
  \BibitemOpen
  \bibfield  {author} {\bibinfo {author} {\bibfnamefont {V.}~\bibnamefont
  {Khachatryan}} \emph {et~al.} (\bibinfo {collaboration} {CMS}),\ }\href
  {\doibase 10.1140/epjc/s10052-016-4293-4} {\bibfield  {journal} {\bibinfo
  {journal} {Eur. Phys. J.}\ }\textbf {\bibinfo {volume} {C76}},\ \bibinfo
  {pages} {469} (\bibinfo {year} {2016}{\natexlab{b}})},\ \Eprint
  {http://arxiv.org/abs/1603.01803} {arXiv:1603.01803 [hep-ex]} \BibitemShut
  {NoStop}%
\bibitem [{\citenamefont {Sirunyan}\ \emph {et~al.}(2020)\citenamefont
  {Sirunyan} \emph {et~al.}}]{CMS:2019leu}%
  \BibitemOpen
  \bibfield  {author} {\bibinfo {author} {\bibfnamefont {A.~M.}\ \bibnamefont
  {Sirunyan}} \emph {et~al.} (\bibinfo {collaboration} {CMS}),\ }\href
  {\doibase 10.1016/j.physletb.2019.135048} {\bibfield  {journal} {\bibinfo
  {journal} {Phys. Lett. B}\ }\textbf {\bibinfo {volume} {800}},\ \bibinfo
  {pages} {135048} (\bibinfo {year} {2020})},\ \Eprint
  {http://arxiv.org/abs/1905.01486} {arXiv:1905.01486 [hep-ex]} \BibitemShut
  {NoStop}%
\bibitem [{\citenamefont {Adam}\ \emph {et~al.}(2017)\citenamefont {Adam} \emph
  {et~al.}}]{ALICE:2016rzo}%
  \BibitemOpen
  \bibfield  {author} {\bibinfo {author} {\bibfnamefont {J.}~\bibnamefont
  {Adam}} \emph {et~al.} (\bibinfo {collaboration} {ALICE}),\ }\href {\doibase
  10.1007/JHEP02(2017)077} {\bibfield  {journal} {\bibinfo  {journal} {JHEP}\
  }\textbf {\bibinfo {volume} {02}},\ \bibinfo {pages} {077} (\bibinfo {year}
  {2017})},\ \Eprint {http://arxiv.org/abs/1611.03002} {arXiv:1611.03002
  [nucl-ex]} \BibitemShut {NoStop}%
\bibitem [{\citenamefont {Aaij}\ \emph {et~al.}(2014)\citenamefont {Aaij} \emph
  {et~al.}}]{LHCb:2014jgh}%
  \BibitemOpen
  \bibfield  {author} {\bibinfo {author} {\bibfnamefont {R.}~\bibnamefont
  {Aaij}} \emph {et~al.} (\bibinfo {collaboration} {LHCb}),\ }\href {\doibase
  10.1007/JHEP09(2014)030} {\bibfield  {journal} {\bibinfo  {journal} {JHEP}\
  }\textbf {\bibinfo {volume} {09}},\ \bibinfo {pages} {030} (\bibinfo {year}
  {2014})},\ \Eprint {http://arxiv.org/abs/1406.2885} {arXiv:1406.2885
  [hep-ex]} \BibitemShut {NoStop}%
\bibitem [{\citenamefont {Khachatryan}\ \emph
  {et~al.}(2015{\natexlab{b}})\citenamefont {Khachatryan} \emph
  {et~al.}}]{CMS:2015ehw}%
  \BibitemOpen
  \bibfield  {author} {\bibinfo {author} {\bibfnamefont {V.}~\bibnamefont
  {Khachatryan}} \emph {et~al.} (\bibinfo {collaboration} {CMS}),\ }\href
  {\doibase 10.1016/j.physletb.2015.09.057} {\bibfield  {journal} {\bibinfo
  {journal} {Phys. Lett. B}\ }\textbf {\bibinfo {volume} {750}},\ \bibinfo
  {pages} {565} (\bibinfo {year} {2015}{\natexlab{b}})},\ \Eprint
  {http://arxiv.org/abs/1503.05825} {arXiv:1503.05825 [nucl-ex]} \BibitemShut
  {NoStop}%
\bibitem [{\citenamefont {Alde}\ \emph {et~al.}(1990)\citenamefont {Alde} \emph
  {et~al.}}]{Alde:1990im}%
  \BibitemOpen
  \bibfield  {author} {\bibinfo {author} {\bibfnamefont {D.~M.}\ \bibnamefont
  {Alde}} \emph {et~al.},\ }\href {\doibase 10.1103/PhysRevLett.64.2479}
  {\bibfield  {journal} {\bibinfo  {journal} {Phys. Rev. Lett.}\ }\textbf
  {\bibinfo {volume} {64}},\ \bibinfo {pages} {2479} (\bibinfo {year}
  {1990})}\BibitemShut {NoStop}%
\bibitem [{\citenamefont {Vasilev}\ \emph {et~al.}(1999)\citenamefont {Vasilev}
  \emph {et~al.}}]{NuSea:1999egr}%
  \BibitemOpen
  \bibfield  {author} {\bibinfo {author} {\bibfnamefont {M.~A.}\ \bibnamefont
  {Vasilev}} \emph {et~al.} (\bibinfo {collaboration} {NuSea}),\ }\href
  {\doibase 10.1103/PhysRevLett.83.2304} {\bibfield  {journal} {\bibinfo
  {journal} {Phys. Rev. Lett.}\ }\textbf {\bibinfo {volume} {83}},\ \bibinfo
  {pages} {2304} (\bibinfo {year} {1999})},\ \Eprint
  {http://arxiv.org/abs/hep-ex/9906010} {arXiv:hep-ex/9906010} \BibitemShut
  {NoStop}%
\bibitem [{\citenamefont {Schmookler}\ \emph {et~al.}(2019)\citenamefont
  {Schmookler} \emph {et~al.}}]{CLAS:2019vsb}%
  \BibitemOpen
  \bibfield  {author} {\bibinfo {author} {\bibfnamefont {B.}~\bibnamefont
  {Schmookler}} \emph {et~al.} (\bibinfo {collaboration} {CLAS}),\ }\href
  {\doibase 10.1038/s41586-019-0925-9} {\bibfield  {journal} {\bibinfo
  {journal} {Nature}\ }\textbf {\bibinfo {volume} {566}},\ \bibinfo {pages}
  {354} (\bibinfo {year} {2019})},\ \Eprint {http://arxiv.org/abs/2004.12065}
  {arXiv:2004.12065 [nucl-ex]} \BibitemShut {NoStop}%
\bibitem [{\citenamefont {Arneodo}\ \emph {et~al.}(1990)\citenamefont {Arneodo}
  \emph {et~al.}}]{Arneodo:1989sy}%
  \BibitemOpen
  \bibfield  {author} {\bibinfo {author} {\bibfnamefont {M.}~\bibnamefont
  {Arneodo}} \emph {et~al.} (\bibinfo {collaboration} {European Muon}),\ }\href
  {\doibase 10.1016/0550-3213(90)90221-X} {\bibfield  {journal} {\bibinfo
  {journal} {Nucl. Phys.}\ }\textbf {\bibinfo {volume} {B333}},\ \bibinfo
  {pages} {1} (\bibinfo {year} {1990})}\BibitemShut {NoStop}%
\bibitem [{\citenamefont {Airapetian}\ \emph {et~al.}(2002)\citenamefont
  {Airapetian} \emph {et~al.}}]{Airapetian:2002fx}%
  \BibitemOpen
  \bibfield  {author} {\bibinfo {author} {\bibfnamefont {A.}~\bibnamefont
  {Airapetian}} \emph {et~al.} (\bibinfo {collaboration} {HERMES}),\
  }\href@noop {} {\  (\bibinfo {year} {2002})},\ \Eprint
  {http://arxiv.org/abs/hep-ex/0210068} {arXiv:hep-ex/0210068 [hep-ex]}
  \BibitemShut {NoStop}%
\bibitem [{\citenamefont {Amaudruz}\ \emph {et~al.}(1995)\citenamefont
  {Amaudruz} \emph {et~al.}}]{Amaudruz:1995tq}%
  \BibitemOpen
  \bibfield  {author} {\bibinfo {author} {\bibfnamefont {P.}~\bibnamefont
  {Amaudruz}} \emph {et~al.} (\bibinfo {collaboration} {New Muon}),\ }\href
  {\doibase 10.1016/0550-3213(94)00023-9} {\bibfield  {journal} {\bibinfo
  {journal} {Nucl. Phys.}\ }\textbf {\bibinfo {volume} {B441}},\ \bibinfo
  {pages} {3} (\bibinfo {year} {1995})},\ \Eprint
  {http://arxiv.org/abs/hep-ph/9503291} {arXiv:hep-ph/9503291 [hep-ph]}
  \BibitemShut {NoStop}%
\bibitem [{\citenamefont {Gomez}\ \emph {et~al.}(1994)\citenamefont {Gomez}
  \emph {et~al.}}]{Gomez:1993ri}%
  \BibitemOpen
  \bibfield  {author} {\bibinfo {author} {\bibfnamefont {J.}~\bibnamefont
  {Gomez}} \emph {et~al.},\ }\href {\doibase 10.1103/PhysRevD.49.4348}
  {\bibfield  {journal} {\bibinfo  {journal} {Phys. Rev.}\ }\textbf {\bibinfo
  {volume} {D49}},\ \bibinfo {pages} {4348} (\bibinfo {year}
  {1994})}\BibitemShut {NoStop}%
\bibitem [{\citenamefont {Arneodo}\ \emph {et~al.}(1995)\citenamefont {Arneodo}
  \emph {et~al.}}]{Arneodo:1995cs}%
  \BibitemOpen
  \bibfield  {author} {\bibinfo {author} {\bibfnamefont {M.}~\bibnamefont
  {Arneodo}} \emph {et~al.} (\bibinfo {collaboration} {New Muon}),\ }\href
  {\doibase 10.1016/0550-3213(95)00023-2} {\bibfield  {journal} {\bibinfo
  {journal} {Nucl. Phys.}\ }\textbf {\bibinfo {volume} {B441}},\ \bibinfo
  {pages} {12} (\bibinfo {year} {1995})},\ \Eprint
  {http://arxiv.org/abs/hep-ex/9504002} {arXiv:hep-ex/9504002 [hep-ex]}
  \BibitemShut {NoStop}%
\bibitem [{\citenamefont {Arneodo}\ \emph
  {et~al.}(1996{\natexlab{a}})\citenamefont {Arneodo} \emph
  {et~al.}}]{Arneodo:1996rv}%
  \BibitemOpen
  \bibfield  {author} {\bibinfo {author} {\bibfnamefont {M.}~\bibnamefont
  {Arneodo}} \emph {et~al.} (\bibinfo {collaboration} {New Muon}),\ }\href
  {\doibase 10.1016/S0550-3213(96)90117-0} {\bibfield  {journal} {\bibinfo
  {journal} {Nucl. Phys.}\ }\textbf {\bibinfo {volume} {B481}},\ \bibinfo
  {pages} {3} (\bibinfo {year} {1996}{\natexlab{a}})}\BibitemShut {NoStop}%
\bibitem [{\citenamefont {Adams}\ \emph {et~al.}(1995)\citenamefont {Adams}
  \emph {et~al.}}]{Adams:1995is}%
  \BibitemOpen
  \bibfield  {author} {\bibinfo {author} {\bibfnamefont {M.~R.}\ \bibnamefont
  {Adams}} \emph {et~al.} (\bibinfo {collaboration} {E665}),\ }\href {\doibase
  10.1007/BF01624583} {\bibfield  {journal} {\bibinfo  {journal} {Z. Phys.}\
  }\textbf {\bibinfo {volume} {C67}},\ \bibinfo {pages} {403} (\bibinfo {year}
  {1995})},\ \Eprint {http://arxiv.org/abs/hep-ex/9505006}
  {arXiv:hep-ex/9505006 [hep-ex]} \BibitemShut {NoStop}%
\bibitem [{\citenamefont {Ashman}\ \emph {et~al.}(1988)\citenamefont {Ashman}
  \emph {et~al.}}]{Ashman:1988bf}%
  \BibitemOpen
  \bibfield  {author} {\bibinfo {author} {\bibfnamefont {J.}~\bibnamefont
  {Ashman}} \emph {et~al.} (\bibinfo {collaboration} {European Muon}),\ }\href
  {\doibase 10.1016/0370-2693(88)91872-2} {\bibfield  {journal} {\bibinfo
  {journal} {Phys. Lett.}\ }\textbf {\bibinfo {volume} {B202}},\ \bibinfo
  {pages} {603} (\bibinfo {year} {1988})}\BibitemShut {NoStop}%
\bibitem [{\citenamefont {Dasu}\ \emph {et~al.}(1994)\citenamefont {Dasu} \emph
  {et~al.}}]{Dasu:1993vk}%
  \BibitemOpen
  \bibfield  {author} {\bibinfo {author} {\bibfnamefont {S.}~\bibnamefont
  {Dasu}} \emph {et~al.},\ }\href {\doibase 10.1103/PhysRevD.49.5641}
  {\bibfield  {journal} {\bibinfo  {journal} {Phys. Rev.}\ }\textbf {\bibinfo
  {volume} {D49}},\ \bibinfo {pages} {5641} (\bibinfo {year}
  {1994})}\BibitemShut {NoStop}%
\bibitem [{\citenamefont {Berge}\ \emph {et~al.}(1991)\citenamefont {Berge}
  \emph {et~al.}}]{Berge:1989hr}%
  \BibitemOpen
  \bibfield  {author} {\bibinfo {author} {\bibfnamefont {J.~P.}\ \bibnamefont
  {Berge}} \emph {et~al.},\ }\href {\doibase 10.1007/BF01555493} {\bibfield
  {journal} {\bibinfo  {journal} {Z. Phys.}\ }\textbf {\bibinfo {volume}
  {C49}},\ \bibinfo {pages} {187} (\bibinfo {year} {1991})}\BibitemShut
  {NoStop}%
\bibitem [{\citenamefont {Ashman}\ \emph {et~al.}(1993)\citenamefont {Ashman}
  \emph {et~al.}}]{Ashman:1992kv}%
  \BibitemOpen
  \bibfield  {author} {\bibinfo {author} {\bibfnamefont {J.}~\bibnamefont
  {Ashman}} \emph {et~al.} (\bibinfo {collaboration} {European Muon}),\ }\href
  {\doibase 10.1007/BF01565050} {\bibfield  {journal} {\bibinfo  {journal} {Z.
  Phys.}\ }\textbf {\bibinfo {volume} {C57}},\ \bibinfo {pages} {211} (\bibinfo
  {year} {1993})}\BibitemShut {NoStop}%
\bibitem [{\citenamefont {Arneodo}\ \emph
  {et~al.}(1996{\natexlab{b}})\citenamefont {Arneodo} \emph
  {et~al.}}]{Arneodo:1996ru}%
  \BibitemOpen
  \bibfield  {author} {\bibinfo {author} {\bibfnamefont {M.}~\bibnamefont
  {Arneodo}} \emph {et~al.} (\bibinfo {collaboration} {New Muon}),\ }\href
  {\doibase 10.1016/S0550-3213(96)90119-4} {\bibfield  {journal} {\bibinfo
  {journal} {Nucl. Phys.}\ }\textbf {\bibinfo {volume} {B481}},\ \bibinfo
  {pages} {23} (\bibinfo {year} {1996}{\natexlab{b}})}\BibitemShut {NoStop}%
\bibitem [{\citenamefont {Adams}\ \emph {et~al.}(1992)\citenamefont {Adams}
  \emph {et~al.}}]{Adams:1992nf}%
  \BibitemOpen
  \bibfield  {author} {\bibinfo {author} {\bibfnamefont {M.~R.}\ \bibnamefont
  {Adams}} \emph {et~al.} (\bibinfo {collaboration} {E665}),\ }\href {\doibase
  10.1103/PhysRevLett.68.3266} {\bibfield  {journal} {\bibinfo  {journal}
  {Phys. Rev. Lett.}\ }\textbf {\bibinfo {volume} {68}},\ \bibinfo {pages}
  {3266} (\bibinfo {year} {1992})}\BibitemShut {NoStop}%
\bibitem [{\citenamefont {Onengut}\ \emph {et~al.}(2006)\citenamefont {Onengut}
  \emph {et~al.}}]{Onengut:2005kv}%
  \BibitemOpen
  \bibfield  {author} {\bibinfo {author} {\bibfnamefont {G.}~\bibnamefont
  {Onengut}} \emph {et~al.} (\bibinfo {collaboration} {CHORUS}),\ }\href
  {\doibase 10.1016/j.physletb.2005.10.062} {\bibfield  {journal} {\bibinfo
  {journal} {Phys. Lett.}\ }\textbf {\bibinfo {volume} {B632}},\ \bibinfo
  {pages} {65} (\bibinfo {year} {2006})}\BibitemShut {NoStop}%
\bibitem [{\citenamefont {Chatrchyan}\ \emph {et~al.}(2011)\citenamefont
  {Chatrchyan} \emph {et~al.}}]{CMS:2011zfr}%
  \BibitemOpen
  \bibfield  {author} {\bibinfo {author} {\bibfnamefont {S.}~\bibnamefont
  {Chatrchyan}} \emph {et~al.} (\bibinfo {collaboration} {CMS}),\ }\href
  {\doibase 10.1103/PhysRevLett.106.212301} {\bibfield  {journal} {\bibinfo
  {journal} {Phys. Rev. Lett.}\ }\textbf {\bibinfo {volume} {106}},\ \bibinfo
  {pages} {212301} (\bibinfo {year} {2011})},\ \Eprint
  {http://arxiv.org/abs/1102.5435} {arXiv:1102.5435 [nucl-ex]} \BibitemShut
  {NoStop}%
\bibitem [{\citenamefont {Chatrchyan}\ \emph {et~al.}(2012)\citenamefont
  {Chatrchyan} \emph {et~al.}}]{CMS:2012fgk}%
  \BibitemOpen
  \bibfield  {author} {\bibinfo {author} {\bibfnamefont {S.}~\bibnamefont
  {Chatrchyan}} \emph {et~al.} (\bibinfo {collaboration} {CMS}),\ }\href
  {\doibase 10.1016/j.physletb.2012.07.025} {\bibfield  {journal} {\bibinfo
  {journal} {Phys. Lett. B}\ }\textbf {\bibinfo {volume} {715}},\ \bibinfo
  {pages} {66} (\bibinfo {year} {2012})},\ \Eprint
  {http://arxiv.org/abs/1205.6334} {arXiv:1205.6334 [nucl-ex]} \BibitemShut
  {NoStop}%
\bibitem [{\citenamefont {Aad}\ \emph {et~al.}(2013{\natexlab{b}})\citenamefont
  {Aad} \emph {et~al.}}]{ATLAS:2012qdj}%
  \BibitemOpen
  \bibfield  {author} {\bibinfo {author} {\bibfnamefont {G.}~\bibnamefont
  {Aad}} \emph {et~al.} (\bibinfo {collaboration} {ATLAS}),\ }\href {\doibase
  10.1103/PhysRevLett.110.022301} {\bibfield  {journal} {\bibinfo  {journal}
  {Phys. Rev. Lett.}\ }\textbf {\bibinfo {volume} {110}},\ \bibinfo {pages}
  {022301} (\bibinfo {year} {2013}{\natexlab{b}})},\ \Eprint
  {http://arxiv.org/abs/1210.6486} {arXiv:1210.6486 [hep-ex]} \BibitemShut
  {NoStop}%
\bibitem [{\citenamefont {Aad}\ \emph {et~al.}(2015{\natexlab{b}})\citenamefont
  {Aad} \emph {et~al.}}]{ATLAS:2014sic}%
  \BibitemOpen
  \bibfield  {author} {\bibinfo {author} {\bibfnamefont {G.}~\bibnamefont
  {Aad}} \emph {et~al.} (\bibinfo {collaboration} {ATLAS}),\ }\href {\doibase
  10.1140/epjc/s10052-014-3231-6} {\bibfield  {journal} {\bibinfo  {journal}
  {Eur. Phys. J. C}\ }\textbf {\bibinfo {volume} {75}},\ \bibinfo {pages} {23}
  (\bibinfo {year} {2015}{\natexlab{b}})},\ \Eprint
  {http://arxiv.org/abs/1408.4674} {arXiv:1408.4674 [hep-ex]} \BibitemShut
  {NoStop}%
\bibitem [{\citenamefont {Loizides}\ and\ \citenamefont
  {Morsch}(2017)}]{Loizides:2017sqq}%
  \BibitemOpen
  \bibfield  {author} {\bibinfo {author} {\bibfnamefont {C.}~\bibnamefont
  {Loizides}}\ and\ \bibinfo {author} {\bibfnamefont {A.}~\bibnamefont
  {Morsch}},\ }\href {\doibase 10.1016/j.physletb.2017.09.002} {\bibfield
  {journal} {\bibinfo  {journal} {Phys. Lett. B}\ }\textbf {\bibinfo {volume}
  {773}},\ \bibinfo {pages} {408} (\bibinfo {year} {2017})},\ \Eprint
  {http://arxiv.org/abs/1705.08856} {arXiv:1705.08856 [nucl-ex]} \BibitemShut
  {NoStop}%
\bibitem [{\citenamefont {Jonas}\ and\ \citenamefont
  {Loizides}(2021)}]{Jonas:2021xju}%
  \BibitemOpen
  \bibfield  {author} {\bibinfo {author} {\bibfnamefont {F.}~\bibnamefont
  {Jonas}}\ and\ \bibinfo {author} {\bibfnamefont {C.}~\bibnamefont
  {Loizides}},\ }\href {\doibase 10.1103/PhysRevC.104.044905} {\bibfield
  {journal} {\bibinfo  {journal} {Phys. Rev. C}\ }\textbf {\bibinfo {volume}
  {104}},\ \bibinfo {pages} {044905} (\bibinfo {year} {2021})},\ \Eprint
  {http://arxiv.org/abs/2104.14903} {arXiv:2104.14903 [nucl-ex]} \BibitemShut
  {NoStop}%
\bibitem [{\citenamefont {Campbell}\ and\ \citenamefont
  {Neumann}(2019)}]{Campbell:2019dru}%
  \BibitemOpen
  \bibfield  {author} {\bibinfo {author} {\bibfnamefont {J.}~\bibnamefont
  {Campbell}}\ and\ \bibinfo {author} {\bibfnamefont {T.}~\bibnamefont
  {Neumann}},\ }\href {\doibase 10.1007/JHEP12(2019)034} {\bibfield  {journal}
  {\bibinfo  {journal} {JHEP}\ }\textbf {\bibinfo {volume} {12}},\ \bibinfo
  {pages} {034} (\bibinfo {year} {2019})},\ \Eprint
  {http://arxiv.org/abs/1909.09117} {arXiv:1909.09117 [hep-ph]} \BibitemShut
  {NoStop}%
\bibitem [{\citenamefont {Aaboud}\ \emph
  {et~al.}(2019{\natexlab{b}})\citenamefont {Aaboud} \emph
  {et~al.}}]{ATLAS:2018pyl}%
  \BibitemOpen
  \bibfield  {author} {\bibinfo {author} {\bibfnamefont {M.}~\bibnamefont
  {Aaboud}} \emph {et~al.} (\bibinfo {collaboration} {ATLAS}),\ }\href
  {\doibase 10.1140/epjc/s10052-019-6622-x} {\bibfield  {journal} {\bibinfo
  {journal} {Eur. Phys. J. C}\ }\textbf {\bibinfo {volume} {79}},\ \bibinfo
  {pages} {128} (\bibinfo {year} {2019}{\natexlab{b}})},\ \bibinfo {note}
  {[Erratum: Eur.Phys.J.C 79, 374 (2019)]},\ \Eprint
  {http://arxiv.org/abs/1810.08424} {arXiv:1810.08424 [hep-ex]} \BibitemShut
  {NoStop}%
\bibitem [{\citenamefont {Sirunyan}\ \emph
  {et~al.}(2021{\natexlab{b}})\citenamefont {Sirunyan} \emph
  {et~al.}}]{CMS:2021ynu}%
  \BibitemOpen
  \bibfield  {author} {\bibinfo {author} {\bibfnamefont {A.~M.}\ \bibnamefont
  {Sirunyan}} \emph {et~al.} (\bibinfo {collaboration} {CMS}),\ }\href
  {\doibase 10.1007/JHEP05(2021)182} {\bibfield  {journal} {\bibinfo  {journal}
  {JHEP}\ }\textbf {\bibinfo {volume} {05}},\ \bibinfo {pages} {182} (\bibinfo
  {year} {2021}{\natexlab{b}})},\ \Eprint {http://arxiv.org/abs/2102.13648}
  {arXiv:2102.13648 [hep-ex]} \BibitemShut {NoStop}%
\bibitem [{\citenamefont {Lamm}\ \emph {et~al.}(2020)\citenamefont {Lamm},
  \citenamefont {Lawrence},\ and\ \citenamefont {Yamauchi}}]{Lamm:2019uyc}%
  \BibitemOpen
  \bibfield  {author} {\bibinfo {author} {\bibfnamefont {H.}~\bibnamefont
  {Lamm}}, \bibinfo {author} {\bibfnamefont {S.}~\bibnamefont {Lawrence}}, \
  and\ \bibinfo {author} {\bibfnamefont {Y.}~\bibnamefont {Yamauchi}} (\bibinfo
  {collaboration} {NuQS}),\ }\href {\doibase 10.1103/PhysRevResearch.2.013272}
  {\bibfield  {journal} {\bibinfo  {journal} {Phys. Rev. Res.}\ }\textbf
  {\bibinfo {volume} {2}},\ \bibinfo {pages} {013272} (\bibinfo {year}
  {2020})},\ \Eprint {http://arxiv.org/abs/1908.10439} {arXiv:1908.10439
  [hep-lat]} \BibitemShut {NoStop}%
\bibitem [{\citenamefont {P\'erez-Salinas}\ \emph {et~al.}(2021)\citenamefont
  {P\'erez-Salinas}, \citenamefont {Cruz-Martinez}, \citenamefont {Alhajri},\
  and\ \citenamefont {Carrazza}}]{Perez-Salinas:2020nem}%
  \BibitemOpen
  \bibfield  {author} {\bibinfo {author} {\bibfnamefont {A.}~\bibnamefont
  {P\'erez-Salinas}}, \bibinfo {author} {\bibfnamefont {J.}~\bibnamefont
  {Cruz-Martinez}}, \bibinfo {author} {\bibfnamefont {A.~A.}\ \bibnamefont
  {Alhajri}}, \ and\ \bibinfo {author} {\bibfnamefont {S.}~\bibnamefont
  {Carrazza}},\ }\href {\doibase 10.1103/PhysRevD.103.034027} {\bibfield
  {journal} {\bibinfo  {journal} {Phys. Rev. D}\ }\textbf {\bibinfo {volume}
  {103}},\ \bibinfo {pages} {034027} (\bibinfo {year} {2021})},\ \Eprint
  {http://arxiv.org/abs/2011.13934} {arXiv:2011.13934 [hep-ph]} \BibitemShut
  {NoStop}%
\end{thebibliography}%

\end{document}